%% file: main.tex
\documentclass{book}

\title{{\sffamily\bfseries\fontsize{29}{60}\selectfont EQUILIBRIA IN MULTIPLAYER GRAPH GAMES} \\ {\sffamily\Large\bfseries AN ALGORITHMIC STUDY}}
\author{Léonard Brice}
\date{Academic year 2024-2025}

\input{0amacros}

\begin{document}

\input{Couverture_these_sciences_FR}

\maketitle

\chapter*{Abstract}
    To verify the robustness of a program or protocol, it is common in the computer science community to rely on the theoretical framework of game theory.
    In particular, if one seeks to enforce a desired property, or \emph{specification}, despite an unpredictable environment, a useful abstraction is to model the situation as a two-player zero-sum game.
    The goal is then to find a strategy for the system that guarantees the specification against any strategy of the environment.

However, to model more complex situations, such as multiple systems with different objectives or an environment composed of various agents, the richer framework of multiplayer games must be considered.
In this setting, a natural question is to identify \emph{equilibria}, i.e., strategy profiles that are robust in the sense that no player has an incentive to deviate.
The most well-known equilibrium concept is the \emph{Nash equilibrium}, but several alternatives exist.
We study five such notions and, for each of them, we provide complexity results for the \emph{constrained existence problem}, which consists of deciding whether a given game contains an equilibrium that ensures each player a payoff within a specified interval.

Regarding Nash equilibria in particular, we prove the following results: the constrained existence problem is undecidable (but recursively enumerable) in energy games.
In discounted-sum games, that same problem is co-recursively enumerable, but at least as hard as the \emph{target discounted-sum problem}, whose decidability remains a long-standing open question.

The main part of our contribution focuses on \emph{subgame-perfect equilibria} (SPEs), a refinement of Nash equilibria in sequential games that eliminates \emph{non-credible threats} by requiring that the planned strategies of all players still form a Nash equilibrium after any possible history.
We establish connections between SPEs and a function on vertex labelings with useful properties, the \emph{negotiation function}, to prove that the constrained existence problem for SPEs is $\NP$-complete in co-Büchi, parity, and mean-payoff games.
Furthermore, we demonstrate that the same results as those established for Nash equilibria in energy games and discounted-sum games hold for SPEs, except for recursive enumerability in energy games, which remains an open question.
We then explore the relationship between the constrained existence problem and a related question, \emph{rational verification}, which involves verifying whether a given strategy enforces a specified property against all \emph{rational} responses---defined in terms of Nash equilibria or SPEs.
In cases where rational responses are not guaranteed to exist, such as SPEs in mean-payoff games, we propose an alternative definition of rational verification.
Here, the specification must hold against every response that is \emph{as rational as possible}, using the notion of $\epsilon$-SPE.
We show that this problem is $\P^\NP$-complete in mean-payoff games.

A third notion that we investigate is the \emph{strong secure equilibrium}, where no coalition of players can harm another player without also harming at least one of its own members.
We argue that strong secure equilibria provide a suitable model for safe protocols among untrusted agents.
We further show that the constrained existence problem for these equilibria is $\PSpace$-complete in parity games and $\ExpTime$-complete in Boolean games where winning conditions are defined by parity automata.

Finally, we study the case of stochastic games, where players are allowed to randomize their strategies.
In this context, the constrained existence problem for Nash or subgame-perfect equilibria is known to be undecidable, even in games with rewards on terminal vertices, if each player seeks to maximize their expected payoff.
We therefore investigate \emph{risk-sensitive equilibria}, where players instead maximize a \emph{risk measure} that accounts for their inclination or aversion to risk.
In particular, we prove that if all players exhibit extreme risk aversion or extreme risk inclination, the constrained existence problem becomes $\NP$-complete, and even $\PTime$-complete if they all have extreme risk inclination.

\chapter*{Acknowledgements}
\input{0aaAckowledgments}

\tableofcontents

\part*{\MakeUppercase{Introduction}}\label{part:intro} 
\chapter{Introduction}
\input{0bIntro}

\part{\MakeUppercase{Background}}\label{part:background}
\chapter{Background} 
\label{chap:background}
\input{0cBackground}

\part{\MakeUppercase{Nash equilibria}}\label{part:nash}
\chapter{Nash equilibria} 
\label{chap:nash}
\input{1NE}

\part{\MakeUppercase{Subgame-perfect equilibria}}\label{part:SPE}

\chapter{Subgame-perfect equilibria and negotiation} 
\label{chap:nego}
\input{2aNego}

\chapter{Parity games} 
\label{chap:parity}
\input{2bParity}

\chapter{Mean-payoff games} 
\label{chap:MP}
\input{2cMP}

\chapter{Energy and discounted-sum games} 
\label{chap:energy&DS}
\input{2dEnergy_DS}

\chapter{About rational verification and its limits} 
\label{chap:verif}
\input{2eVerif}

\part{\MakeUppercase{Other equilibria}}\label{part:other_equilibria}

\chapter{Strong secure equilibria and their applications} 
\label{chap:SSE}
\input{3aSSE}

\chapter{Risk-sensitive equilibria} 
\label{chap:RSE}
\input{3bRSE}

\chapter{Extreme risk-sensitive equilibria} 
\label{chap:XRSE}
\input{3cXRSE}

\part*{\MakeUppercase{Discussion}} 
\label{part:discussion}
\chapter*{Discussion}
\input{4Discussion}

\bibliography{bibli}

\end{document}

%% file: 0aMacros.tex
\usepackage{amssymb,amsthm}
\usepackage{amsmath}
\usepackage{stmaryrd,wasysym}
\usepackage{graphicx}
\usepackage{subcaption}
\usepackage{tikz}
\usetikzlibrary{positioning, arrows, calc}
\usetikzlibrary{automata}
\usepackage{algorithm}
\usepackage{algpseudocode}
\usepackage{array}
\usepackage{hyperref}
\usepackage{cleveref}
\usepackage{amsthm}
\usepackage{setspace}
\usepackage{multirow}
\usepackage{booktabs}
\usepackage{xargs}
\usepackage{fontawesome}
\usepackage{tcolorbox} 
\usepackage{wrapfig}
\usepackage{times}

\usepackage{pgfplots}
\pgfplotsset{compat=1.17}

\usetikzlibrary{automata, arrows, positioning, decorations.markings, decorations.pathreplacing}
\usetikzlibrary {decorations.pathmorphing, decorations.shapes}
\usetikzlibrary{backgrounds,fit,shapes.geometric}
\usetikzlibrary{graphs, quotes}
\usetikzlibrary{calc}
\usetikzlibrary{shapes,shapes.geometric,fit}
\usepackage[new]{old-arrows}
\usepackage{forest}
\usetikzlibrary{graphs,shapes.geometric,quotes}
\usetikzlibrary{decorations.pathreplacing}
\usetikzlibrary{arrows,automata,decorations.pathmorphing}
\usetikzlibrary{shapes,shapes.geometric,fit,calc,automata}

\usetikzlibrary{shapes}

\RequirePackage[T1]{fontenc}
\RequirePackage[tt=false, type1=true]{libertine} 
\RequirePackage[varqu]{zi4}
\RequirePackage[libertine]{newtxmath}

\usepackage[a4paper,margin=1.2in]{geometry}
\usepackage{microtype}

\usepackage{titlesec}
\titleformat{\part}[block]{\sffamily\Huge\bfseries}{\MakeUppercase{Part~\thepart:}\\}{2pc}{}
[\vspace{1ex}]

\titleformat{\chapter}[hang]{\sffamily\huge\bfseries\MakeUppercase}{Chapter~\thechapter:}{1.5pc}{}

\titleformat{\section}[hang]{\sffamily\Large\bfseries\MakeUppercase}{\thesection}{1.5pc}{}

\titleformat{\subsection}[hang]{\sffamily\large\bfseries}{\thesubsection}{1pc}{}

\titleformat{\paragraph}[hang]{\sffamily\normalsize\bfseries}
  {\hspace{15pt}$\blacktriangleright$}{1pc}{}  

\titleformat{\subparagraph}[runin]{\sffamily\normalsize\bfseries}{}{0pt}{}
\newcommand{\subp}[1]{\subparagraph{#1.}}

\theoremstyle{plain}

\newtheorem{thm}{\textsf{Theorem}}
\newtheorem{cor}{\textsf{Corollary}}
\newtheorem{lem}{\textsf{Lemma}}
\newtheorem{slem}{\textsf{Sublemma}}
\newtheorem{prop}{\textsf{Proposition}}
\newtheorem{inv}{\textsf{Invariant}}

\theoremstyle{definition}

\newtheorem{rem}{\textsf{Remark}}

\newtheorem{exa}{\textsf{Example}}

\newtheorem{defi}{\textsf{Definition}}

\newtheorem{conj}{\textsf{Conjecture}}
\newtheorem{prob}{\textsf{Problem}}
\newtheorem{oprob}{\textsf{Open Problem}}

\counterwithout{table}{chapter}
\counterwithout{figure}{chapter}

\crefname{algorithm}{Algorithm}{Algorithms}
\crefname{figure}{Figure}{Figures}
\crefname{lem}{Lemma}{Lemmas}
\crefname{slem}{Sublemma}{Sublemmas}
\crefname{inv}{Invariant}{Invariants}
\crefname{thm}{Theorem}{Theorems}
\crefname{prop}{Proposition}{Propositions}
\crefname{cor}{Corollary}{Corollaries}
\crefname{defi}{Definition}{Definitions}
\crefname{rem}{Remark}{Remarks}
\crefname{clm}{Claim}{Claims}
\crefname{exa}{Example}{Examples}
\crefname{part}{Part}{Parts}
\crefname{chapter}{Chapter}{Chapters}
\crefname{section}{Section}{Sections}
\crefname{subsection}{Subsection}{Subsections}
\crefname{oprob}{Open Problem}{Open Problems}
\crefname{conj}{Conjecture}{Conjectures}

\makeatletter
\def\thm@space@setup{\thm@preskip=8pt \thm@postskip=8pt}

\makeatother

\makeatletter
\newlength{\defaultparindent} 
\AtBeginDocument{\setlength{\defaultparindent}{\parindent}} 

\renewenvironment{proof}[1][\proofname]{%
  \par\pushQED{\qed}
  \normalfont
  \begin{list}{}{%
    \setlength{\leftmargin}{0pt} 
    \setlength{\labelwidth}{0pt} 
    \setlength{\labelsep}{15pt}   
    \addtolength{\leftmargin}{\defaultparindent} 
  }
  \setlength{\parindent}{\defaultparindent} 
  \item[\itshape #1\@addpunct{.}]\ignorespaces
}{%
  \popQED
  \end{list}
}
\makeatother

\renewcommand{\Circle}{\mathbin{\raisebox{-0.34ex}{\scalebox{1.5}{$\circ$}}}}

\renewcommand{\l}{\ell}
\renewcommand{\phi}{\varphi}
\renewcommand{\epsilon}{\varepsilon}
\renewcommand{\|}{}
\newcommand{\triangleone}{\triangledown\!_1}

\newcommand{\btd}{\triangledown}

\renewcommand{\Game}{\mathcal{G}}
\newcommand{\Gamebis}{\mathcal{H}}
\newcommand{\Mem}{\mathcal{M}}
\newcommand{\Kount}{\mathcal{K}}
\newcommand{\auto}{\mathcal{A}}
\newcommand{\TA}{\mathcal{T}}

\newcommand{\Eq}{\mathcal{E}}
\newcommand{\Class}{\mathcal{C}}

\newcommand{\NN}{\mathbb{N}}
\newcommand{\ZZ}{\mathbb{Z}}
\newcommand{\RR}{\mathbb{R}}
\newcommand{\QQ}{\mathbb{Q}}
\newcommand{\EE}{\mathbb{E}}

\newcommand{\PP}{\mathbb{P}}

\newcommand{\M}{\mathbb{M}}
\newcommand{\X}{\mathbb{X}}
\newcommand{\OM}{\mathbb{OM}}
\newcommand{\PM}{\mathbb{PM}}

\newcommand{\dseal}{\!\,^\llcorner}
\newcommand{\<}{\langle}
\renewcommand{\>}{\rangle}
\newcommand{\lv}{\lVert}
\newcommand{\rv}{\rVert}
\renewcommand{\|}{\upharpoonright}
\newcommand{\tpl}[1]{\left( #1 \right)}

\newcommand{\anch}{\scalebox{0.5}{\text{\faAnchor}}}
\newcommand{\bad}{{\frownie}}

\newcommand{\hC}{\hat{C}}

\newcommand{\bsigma}{{\bar{\sigma}}}
\newcommand{\btau}{{\bar{\tau}}}

\newcommand{\ba}{\bar{a}}
\newcommand{\bx}{\bar{x}}
\newcommand{\by}{\bar{y}}
\newcommand{\bz}{\bar{z}}
\newcommand{\bbx}{\bar{\bar{x}}}
\newcommand{\bzero}{\bar{0}}
\newcommand{\bq}{\bar{q}}
\newcommand{\bM}{\bar{M}}
\newcommand{\brho}{\bar{\rho}}

\newcommand{\bbalpha}{\bar{\bar{\alpha}}}

\newcommand{\dpi}{\dot{\pi}}
\newcommand{\dchi}{\dot{\chi}}
\newcommand{\dxi}{\dot{\xi}}

\newcommand{\dg}{\dot{g}}
\newcommand{\dc}{\dot{c}}
\newcommand{\dd}{\dot{d}}

\newcommand{\tpi}{\tilde{\pi}}
\renewcommand{\th}{\tilde{h}}

\newcommand{\hkappa}{\hat{\kappa}}

\newcommand{\noDev}{\Delta_\emptyset}

\usepackage{amsmath}

\DeclareMathOperator{\essinf}{\mathsf{ess\,inf}}
\DeclareMathOperator{\esssup}{\mathsf{ess\,sup}}
\DeclareMathOperator*{\minop}{\mathsf{min}}
\DeclareMathOperator*{\maxop}{\mathsf{max}}
\DeclareMathOperator*{\infop}{\mathsf{inf}}
\DeclareMathOperator*{\supop}{\mathsf{sup}}
\DeclareMathOperator*{\liminfop}{\mathsf{lim\,inf}}
\DeclareMathOperator*{\limsupop}{\mathsf{lim\,sup}}
\DeclareMathOperator*{\limop}{\mathsf{lim}}

\renewcommand{\min}{\minop}

\renewcommand{\max}{\maxop}

\renewcommand{\inf}{\infop}

\renewcommand{\sup}{\supop}

\renewcommand{\liminf}{\liminfop}

\renewcommand{\limsup}{\limsupop}

\renewcommand{\lim}{\limop}

\renewcommand{\log}{\mathsf{log}}

\newcommand{\SPE}{\mathsf{SPE}}
\newcommand{\SPR}{\mathsf{SPR}}
\newcommand{\card}{\mathsf{card}}
\newcommand{\SC}{\mathsf{SCyc}}
\newcommand{\Plays}{\mathsf{Plays}}
\newcommand{\Hist}{\mathsf{Hist}}

\newcommand{\Conv}{\mathsf{Conv}}

\newcommand{\nego}{\mathsf{nego}}
\newcommand{\lCons}{\lambda\mathsf{Cons}}
\newcommand{\lRat}{\lambda\mathsf{Rat}}

\newcommand{\Prop}{\mathsf{Prop}}
\newcommand{\Acc}{\mathsf{Acc}}
\newcommand{\Dev}{\mathsf{Dev}}
\newcommand{\dev}{\mathsf{dev}}
\newcommand{\EDev}{\mathsf{edev}}
\newcommand{\Stat}{\mathsf{Sta}}

\newcommand{\Rat}{\mathsf{Rat}}
\newcommand{\abs}{\mathsf{abs}}
\newcommand{\conc}{\mathsf{conc}}

\newcommand{\val}{\mathsf{val}}
\newcommand{\Sol}{\mathsf{Sol}}
\newcommand{\red}{\mathsf{red}}
\newcommand{\last}{\mathsf{last}}
\newcommand{\first}{\mathsf{first}}
\newcommand{\Occ}{\mathsf{Occ}}
\newcommand{\Inf}{\mathsf{Inf}}
\newcommand{\MP}{\mathsf{mp}}
\newcommand{\MPi}{\text{\underline{\smash{$\MP$}}}}

\newcommand{\ds}{\mathsf{ds}}

\newcommand{\Parity}{\mathsf{Parity}}
\newcommand{\EL}{\mathsf{el}}
\newcommand{\f}{\mathsf{f}}
\renewcommand{\a}{\mathsf{a}}
\newcommand{\con}{\mathsf{c}}
\renewcommand{\r}{\mathsf{r}}
\newcommand{\s}{\mathsf{s}}
\newcommand{\cd}{\mathsf{d}}
\newcommand{\ecd}{\mathsf{e}}
\newcommand{\Strat}{\mathsf{Strat}}
\newcommand{\B}{\mathsf{B}}
\newcommand{\SConn}{\mathsf{SConn}}
\newcommand{\Ind}{\mathsf{Ind}}
\newcommand{\subgamep}{\mathsf{SubgamePerfect}}
\newcommand{\nash}{\mathsf{Nash}}
\newcommand{\itm}{\mathsf{item}}
\newcommand{\D}{\mathsf{D}}
\newcommand{\Lang}{\mathsf{L}}
\newcommand{\Supp}{\mathsf{Supp}}
\newcommand{\p}{\mathsf{p}}
\newcommand{\punish}{\mathsf{punish}}
\newcommand{\anchor}{\mathsf{anchor}}
\renewcommand{\d}{\mathsf{d}}
\newcommand{\Attr}{\mathsf{Attr}}
\newcommand{\Yes}{\mathsf{Yes}}
\newcommand{\No}{\mathsf{No}}

\newcommand{\modifiedreward}[2]{f_{#1 #2}}

\newcommand{\NP}{\mathsf{NP}}
\newcommand{\ExpTime}{\mathsf{EXPTIME}}
\renewcommand{\P}{\mathsf{P}}
\newcommand{\PTime}{\P}
\newcommand{\coNP}{\mathsf{coNP}}
\newcommand{\PSpace}{\mathsf{PSPACE}}
\newcommand{\NPSpace}{\mathsf{NPSPACE}}
\newcommand{\ExpSpace}{\mathsf{EXPSPACE}}

\newcommand{\RE}{\mathsf{RE}}
\newcommand{\coRE}{\mathsf{coRE}}

\newcommand{\SAT}{\mathrm{\textsc{Sat}}}
\newcommand{\QSAT}{\mathrm{\textsc{QSat}}}
\newcommand{\THREESAT}{3\SAT}
\newcommand{\TDS}{\mathrm{\textsc{TDS}}}
\newcommand{\coTDS}{\mathrm{\textsc{coTDS}}}

\newcommand{\Solver}{\mathfrak{S}}
\newcommand{\Prover}{\mathfrak{P}}
\newcommand{\Challenger}{\mathfrak{C}}
\newcommand{\Witness}{{\mathfrak{W}}}
\newcommand{\Leader}{{\mathfrak{L}}}
\newcommand{\Demon}{{\mathfrak{D}}}
\newcommand{\adel}{{\mathfrak{A}}}
\newcommand{\bart}{{\mathfrak{B}}}
\newcommand{\capu}{{\mathfrak{C}}}
\newcommand{\ttp}{{\mathfrak{T}}}

\newcommand{\opponent}{{\mathfrak{O}}}

\newcommand{\obj}[2]{\mbox{\tikz{\node[font=\tiny,inner sep=1.5pt,draw,shape=rectangle,solid,minimum width=0pt,fill=white,rounded corners=0pt] {$#1\rightarrow#2$};}}}
\newcommand{\iab}{\obj{\adel}{\bart}}
\newcommand{\iba}{\obj{\bart}{\adel}}
\newcommand{\iac}{\obj{\adel}{\capu}}
\newcommand{\ica}{\obj{\capu}{\adel}}
\newcommand{\ibc}{\obj{\bart}{\capu}}
\newcommand{\icb}{\obj{\capu}{\bart}}

\tikzset{every node/.append style={initial text={}}}
\tikzset{every edge/.append style={->, >=latex}}

\tikzstyle{abstractstate}=[vert,shape=rectangle,rounded corners=7pt]
\tikzstyle{aliceMove}=[solid,red,thick]
\tikzstyle{bobMove}=[dashed,blue,thick]
\tikzstyle{ttpMove}=[dotted,green!60!black,thick]
\tikzstyle{infraMove}=[dashdotted,orange,thick]

\tikzstyle{player1}=[regular polygon,regular polygon sides=4,draw,inner sep=0.5pt]
\tikzstyle{player2}=[circle,draw,inner sep=2pt]
\tikzstyle{player3}=[regular polygon,regular polygon sides=6,,draw,inner sep=0.5pt]
\tikzstyle{initial}+=[initial by arrow,initial text={}]

\newcommand{\playerOneSymb}{\bgroup\mathchoice{%
        \tikz{\node[player1,minimum size=9pt]{~};}
    }{%
        \tikz{\node[player1,minimum size=9pt]{~};}
    }{%
        \tikz{\useasboundingbox (-3pt,-3pt) rectangle (3pt,4pt); \node[player1,minimum size=6pt]{~};}
    }{%
        \tikz{\node[player1,minimum size=3pt]{~};}
    }\egroup}

\newcommand{\playerTwoSymb}{\bgroup\mathchoice{%
        \tikz{\useasboundingbox (-3.5pt,-3pt) rectangle (3.5pt,4pt);\node[player2,minimum size=7pt]{~};}
    }{%
        \tikz{\useasboundingbox (-3.5pt,-3pt) rectangle (3.5pt,4pt);\node[player2,minimum size=7pt]{~};}
    }{%
        \tikz{\useasboundingbox (-2pt,-3pt) rectangle (2pt,4pt);\node[player2,minimum size=4pt,inner sep=0pt]{};}
    }{%
        \tikz{\useasboundingbox (-1.5pt,-2pt) rectangle (1.5pt,2.5pt);\node[player2,inner sep=0pt,minimum size=3pt]{};}
    }\egroup}
\newcommand{\playerThreeSymb}{\bgroup\mathchoice{%
        \tikz{\node[player3,minimum size=8pt]{~};}
    }{%
        \tikz{\node[player3,minimum size=8pt]{~};}
    }{%
        \tikz{\useasboundingbox (-2pt,-2pt) rectangle (2.5pt,4pt);\node[player3,minimum size=5pt]{~};}
    }{%
        \tikz{\useasboundingbox (-1.25pt,-2pt) rectangle (1.5pt,2.5pt);\node[player3,minimum size=3.5pt]{~};}
    }\egroup}

\tikzset{stoch/.style={state, fill=black, inner sep=0.5pt, text=white, minimum size=0pt}}

\tikzstyle{initial}+=[initial by arrow,initial text={}]

\tikzset{vert/.style={state,minimum size=20pt,fill, top color=white, bottom color=black!20}}

\tikzset{proverVert/.style={state, rectangle, rounded corners, minimum size=20pt,fill, top color=white, bottom color=blue!20}}

\tikzset{challengerVert/.style={state, rectangle, rounded corners, minimum size=20pt,fill, top color=white, bottom color=orange!20}}

\tikzset{demonVert/.style={state, rectangle, rounded corners, minimum size=20pt, text=white, fill, top color=red!70!black, bottom color=black}}

\tikzset{leaderVert/.style={state, rectangle, rounded corners, minimum size=20pt, fill, top color=violet!20, bottom color=violet!40}}

\tikzstyle{mealyTrans}=[->,font=\scriptsize]

\tikzstyle{ownedVert}[$i$]=[vert,%
    prefix after command={\pgfextra{\tikzset{every label/.style={
                    fill=black,text=white,rounded corners=3pt,outer sep=0pt,font=\scriptsize,label distance=-3pt,inner sep=2pt
    }}}},
    label=-60:{#1}]

    \tikzstyle{nobodyState}=[vert,rectangle,inner sep=3pt,minimum size=5pt]
    \tikzstyle{literalState}[$i$]=[nobodyState,%
    prefix after command={\pgfextra{\tikzset{every label/.style={
                    fill=black,text=white,rounded corners=3pt,outer sep=0pt,font=\scriptsize,label distance=-3pt,inner sep=2pt
    }}}},
    label=-60:{#1}]
    \tikzstyle{opponentState}=[vert,regular polygon,regular polygon sides=6,inner sep=0pt,minimum size=5pt]
    \tikzstyle{solverState}=[vert,circle,inner sep=1pt,minimum size=5pt]
    \tikzstyle{loseState}=[shape=cloud,inner sep=1pt,fill=red!80!black,text=white,font=\scriptsize,cloud ignores aspect]
    \newcommand{\cobuchiLabel}[3][]{\node[anchor=west,loseState,xshift=-2pt,#1] at (#2.east) {#3};}

    \newcommand{\cobuchiLabelLeft}[3][]{\node[anchor=east,loseState,xshift=2pt,#1] at (#2.west) {#3};}

    \tikzstyle{automatonState}=[vert,circle,inner sep=1pt,minimum size=5pt,fill, bottom color=white, top color=black!20]

    \tikzstyle{vanishingPattern}=[dash pattern=on 5pt off 2pt on 2pt off 2pt on 2pt off 2pt]

    \pgfdeclarelayer{arrows}
    \pgfsetlayers{arrows,main}

\makeatletter
\def\myvdots{\vbox{\baselineskip4\p@ \lineskiplimit\z@
    \hbox{.}\hbox{.}\hbox{.}}}
\def\myddots{\mathinner{\mkern1mu\raise7\p@\vbox{\hbox{.}}\mkern2mu
    \raise4\p@\hbox{.}\mkern2mu\raise\p@\hbox{.}\mkern1mu}}
\def\myiddots{\mathinner{\mkern1mu\raise-6\p@\vbox{\hbox{.}}\mkern2mu
    \raise-3\p@\hbox{.}\mkern2mu\raise\p@\hbox{.}\mkern1mu}}
\makeatother


%% file: Couverture_these_sciences_FR.tex
\definecolor{darkblue}{HTML}{004C93} 
\tcbset{
    center title,
		top=1cm,
		right=0cm,
		left=0cm,
    colback=darkblue,
    colframe=white,
    width=15.85cm,
		height=5.62cm,
		halign=right,
		valign=top,
		flush right,
		sharp corners=all
    }
\thispagestyle{empty}

\includegraphics[height=3.07cm]{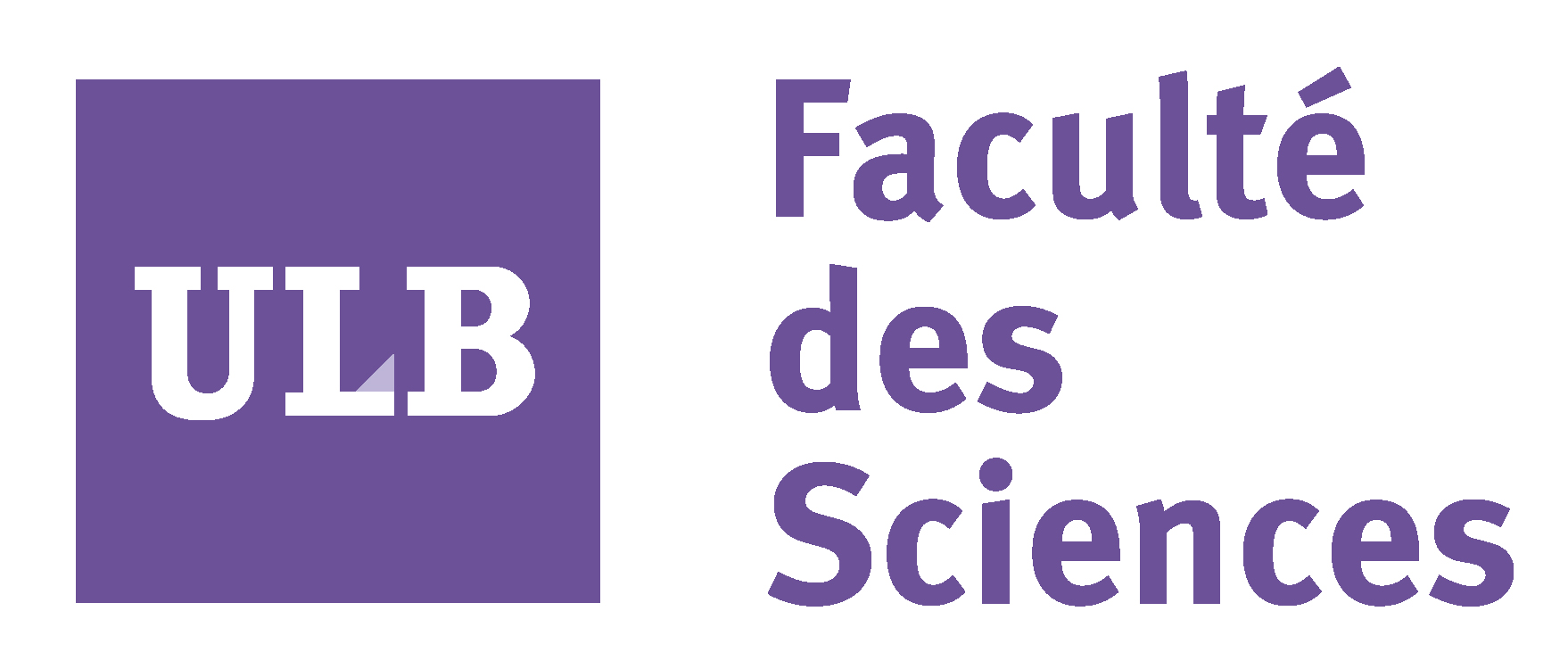} \\ \\

\begin{tcolorbox}
\color[rgb]{1,1,1}
\Large{\textbf{Equilibria in Multiplayer Graph Games}} \\
\large{\textit{An Algorithmic Study}}
\end{tcolorbox} 

\begin{tcolorbox}[colback=white, halign=left]
\color{darkblue}
\large{\textbf{thèse présentée par Léonard \textsc{Brice}}} \\
\color{black}
\large{en vue de l'obtention du grade académique de Doctorat en Sciences \\
Année académique 2024-2025}
 
\vspace{2cm}

\begin{flushright}
\color{darkblue}
Sous la direction du Professeur Jean-François \textsc{Raskin}, \ promoteur  \\
et de la Professeure Marie \textsc{van den Bogaard},  \ co-promotrice
\end{flushright}
\end{tcolorbox}
\vspace{3cm}

\noindent
\textbf{Jury de thèse : } \\ 
Jean \textsc{Cardinal} (Université libre de Bruxelles, Président) \\
Jean-François \textsc{Raskin} (Université libre de Bruxelles)\\
Marie \textsc{van den Bogaard} (Université Gustave Eiffel) \\
Emmanuel \textsc{Filiot} (Université libre de Bruxelles) \\
Patricia \textsc{Bouyer-Decître} (Université Paris-Saclay)\\
Orna \textsc{Kupferman} (Université hébraïque de Jérusalem)\\
Thomas \textsc{Henzinger} (Institute of Science and Technology Austria)

%% file: 0aaAckowledgments.tex
To express my feelings as sincerely as possible, I trust the reader will allow me to use my mother tongue.

La rédaction d'une thèse est un grand mensonge, qui consiste à faire passer pour une matière froide et lisse le résultat chaotique et bouillonnant d'une aventure humaine. Ce document n'aurait jamais vu le jour, ou du moins n'aurait pas été le même, sans le concours d'un nombre considérable de personnes qui m'ont accompagné dans ce petit bout de chemin.
En premier lieu, je veux bien sûr remercier ma promotrice et mon promoteur, Marie van den Bogaard et Jean-François Raskin. Votre présence rassurante, votre disponibilité, vos conseils avisés et votre empathie pendant ces quatre années ont constitué le socle le plus solide que je pouvais espérer pour commencer cette carrière de chercheur.

Je remercie ensuite les personnes qui ont accepté de participer à mon jury de thèse, tâche dont je mesure l'ampleur. J'ai fait mon possible pour rendre la lecture de cette thèse la moins pénible possible en dépit de sa longueur, aggravée par mon incapacité maladive à faire le tri entre résultats importants et curiosités dispensables. J'espère que vous y trouverez une matière intéressante.

Je remercie également tous mes autres coauteurs et coautrice, aux côtés desquels nous avons produit ce travail collectif qu'ils et elle m'ont permis de signer de mon nom. Merci à Anirban et Thomas Bruss d'avoir attiré mon attention sur le problème de Robbin, collaboration qui a abouti à un article dont les résultats ne figurent pas ici. Merci à Guillaume de m'avoir permis de mettre un petit pied dans la communauté des protocoles et de la sécurité. Merci à Mathieu pour ces longues heures de réflexion dans notre bureau commun (et pour tes efforts quotidiens pour ne pas faire de bruit à l'heure de la sieste). Merci à Thomas Henzinger de m'avoir accueilli à l'IST Austria, et d'avoir accepté de prolonger cette collaboration après le dépôt de cette thèse. Merci à Thejaswini de m'avoir fait découvrir l'Autriche---j'ai hâte que nous puissions de nouveau \emph{brainstormer} dans ces fauteuils horriblement inconfortables où nous avons établi notre camp de base.

Je veux aussi remercier Véronique et Emmanuel, qui ont complété cet accompagnement en participant à mon comité de thèse. Et, plus largement, toutes les personnes avec qui j'ai partagé quotidiennement les pauses-café, les déjeuners et bien plus pendant ces quatre ans~: Allen, Ayrat, Damien, Debraj, Gilles, Mrudula, Sarah, Sayan et Thierry à Bruxelles, Ahad, Ali, Ana, Ehsan, Emily, Konstantin, Krishnendu, Mahyar, Marek, Maximilian, Mehrdad, Nicolas, Pawel, Raimundo, Stefanie, Valentin et Yakub à Vienne, ainsi que Aaron, Claire, Léo, Nadime, Victor ou encore Zéphyr à Marne-la-Vallée. J'ajoute un remerciement à tous les étudiants et étudiantes qui ont eu la patience de subir mes explications maladroites dans le cadre de mes tâches d'enseignement~: transmettre des connaissances reste le meilleur moyen de prendre du recul sur elles, et à ce titre, j'ai appris de vous au moins autant que vous de moi. Enfin, je n'oublie pas Véronique, Maryka et Marie, nos extraordinaires secrétaires de département, sans qui rien de tout cela ne serait possible.

Comme tout travail, la recherche ne se dissocie jamais parfaitement des autres aspects de nos vies. Il y a par conséquent nombre d'autres personnes, hors des murs de l'Académie, qui m'ont accompagné pendant l'écriture de cette thèse, généralement sans comprendre un traître mot de ce que j'en racontais quand je m'essayais à la tâche. Parmi celles qui m'ont permis de me sentir chez moi en Belgique, je pense à ma marraine, Pajka, à sa fille, Alicia, ou encore à mon cousin Raoul et ma belle-cousine Ilektra, qui ont rendu ces années plus belles en amenant deux nouveaux petits membres à notre grande famille.
Je pense à mes colocataires, avec lesquels j'aurais aimé pouvoir passer plus de temps---promis, dès que cette thèse est soumise, je sors la poubelle de verre.
Je pense également à mes amis et amies de longue date, en particulier les deux Marie, Mathilde, Solène, Pierre ou encore Zoé qui, malgré la distance, ont invariablement été là pour moi dès que le besoin s'en faisait ressentir.
Et je pense, bien sûr, à toutes ces personnes formidables avec lesquelles j'ai passé une large partie de mon temps extra-professionnel, parce que nous partagions les mêmes combats.
Je n'en énumérerai pas la liste, elle serait fort longue, et je crois de toute manière qu'elles se reconnaîtront. Je me crois cependant fondé à faire une exception pour Raphaël, dont la perte brutale, quelques jours après son vingtième anniversaire, m'a profondément meurtri. Tu nous as beaucoup apporté, et nous n'avons pas eu le temps de te remercier.

Je pense, enfin, à ma famille, le roc insubmersible sur lequel je n'ai fait que me hisser. Je remercie tous mes oncles, tantes, cousins et grands-parents que je ne vois pas suffisamment souvent, mais qui m'apportent à chaque fois beaucoup de bonheur. Je veux ici avoir une pensée particulière pour ma grand-mère, qui vient de nous quitter~: je mesure aujourd'hui la chance incroyable que nous avons eue de t'avoir dans nos vies. Je remercie mon frère, Philémon, et ma s\oe ur, Louve, pour toute cette complicité et cette solidarité que nous avons su construire ensemble~: je n'ai pas suffisamment eu l'occasion de le dire, mais je suis immensément fier des adultes que vous êtes devenus. Et bien sûr, je remercie mes parents, qui ont toujours su, avec courage, patience et bienveillance, entendre et soutenir mes choix de vie dans tout ce qu'ils avaient d'improbable et de douteux.

Je pose le point final à cette thèse dans une époque pleine d'incertitudes. Les récentes avancées de l'informatique, fruit du travail de cette communauté scientifique dans laquelle j'essaie modestement de me faire une place, sont aujourd'hui perçues comme une menace plus que comme une promesse de progrès, parce qu'elles sont mises au service d'intérêts économiques et d'idéologies réactionnaires---les mêmes qui font la sourde oreille lorsque la science sonne l'alerte sur la destruction de nos environnements. Dans mon pays d'origine, dans celui où je vis et dans celui où je m'apprête à partir, les sphères du pouvoir sont petit à petit gangrenées par une extrême droite qui répand ses idées rétrogrades et obscurantistes. Mais je ne crois pas que l'humanité soit condamnée à la barbarie. Mes derniers remerciements iront donc à toutes celles et tous ceux qui résistent et qui, à contre-courant, continuent à tisser d'autres futurs.

%% file: 0bIntro.tex
The world is an arena, and life is a game.

Every day, we interact with an environment composed of agents.
They may be similar to us or vastly different, whether in terms of the range of actions they can take to influence the physical world, or the motivations that drive them---be they other human beings, animals, institutions, or machines.

Given the omnipresence of interactions between such agents, each defined by their \emph{objectives} and their ability to choose among multiple \emph{strategies}, scientists have developed \emph{game theory}---a broad conceptual framework designed to model such interactions using mathematical structures.

Historically, the first advances in game theory were focused on economics, a field that remains one of its most significant, if not primary, areas of application. Naturally, its influence extended to social and political sciences, and it is frequently used to analyze international relations.
However, its applications have also reached fields that might initially seem far removed from its origins. In biology, for instance, game theory is now commonly used to model and study how species adapt to their evolving environments.

Finally, in recent decades, there has been a growing interest in game theory within computer science---a field to which this document belongs.

\section{Two-player zero-sum games} \label{sec:intro_2players}

When a computer-controlled system interacts with an unpredictable environment, one of the most prominent approaches today is \emph{machine learning}, often referred to---somewhat imprecisely---as \emph{artificial intelligence}.   
For an introduction to machine learning, see~\cite{Bishop2006}.
This approach enables the system to \emph{learn} how to interact with the environment in order to maximize the satisfaction of human designers.  
Such a learning process builds on a substantial amount of data, which may be collected from previous similar interactions, or generated by the system itself through self-testing.  
For instance, while writing this document, the author occasionally used ChatGPT, a tool developed by the company OpenAI, to rephrase certain sentences in more correct English.
This tool does not rely on a fixed definition of correct English, but instead interpolates it from a large set of texts generally regarded as correct.
However, it is crucial to remember that every dataset represents only a partial observation of the physical world.  
Consequently, machine learning can at best lead to \emph{statistical} satisfaction. 

Therefore, this approach is not suitable for all computer-controlled systems.  
For instance, in critical infrastructures, where errors could have severe consequences, such as in the case of a hydraulic dam, relying solely on machine learning may not be viable: such a system must absolutely guarantee certain safety properties---called its \emph{specification}---regardless of the environment's behavior.
It is then convenient to model the environment as an adversarial \emph{player}, whose goal is to breach the specification.

Thus, this situation can be abstracted as a two-player zero-sum game: the goal is to design a strategy for the system that ensures compliance with the specification, considered as its objective, against any strategy the opponent might adopt---like a chess player that, at least in theory, aims to play in a way that guarantees victory no matter how their opponent responds.
This explains why the most classical applications of game theory in computer science consider two-player zero-sum games.

\section{Games played on graphs}

Game theory provides a wide range of mathematical models for games, each with its own advantages and limitations.
A simple yet classical example is the model of \emph{matrix games}, also known as \emph{strategic games}.
In this framework, each player selects a \emph{strategy} from a (often finite) set of strategies, treated here as abstract choices.
Each possible combination of strategies yields a given payoff to each player.
A canonical example of a game that can be represented in such a form is the \emph{prisoner's dilemma}: two prisoners are arrested after committing a crime.
They must choose simultaneously between confessing and denying.
If one denies and the other confesses, the prisoner who denies is sentenced to 10 years in jail, while the one who confesses is immediately released as a reward.
If both confess, the penalty is shared: each is sentenced to 5 years in jail.
If both deny, they each spend only 1 year in jail, as a precautionary measure\footnote{Although this is the usual way of presenting the prisoner's dilemma, it is worth noting that such preventive penalties contradict the principle of the presumption of innocence, as guaranteed by the Universal Declaration of Human Rights and the International Covenant on Civil and Political Rights.}.
The matrix describing this game is given in Table~\ref{tab:prisoners}.
A pair $(x, y)$ indicates that prisoner~1 is sentenced to $x$ years, and prisoner~2 to $y$ years.

\begin{table}
    \centering
\begin{tabular}{c|c|c|}
    & Prisoner 1 confesses & Prisoner 1 denies \\
    \hline
    Prisoner 2 confesses & $(5, 5)$ & $(10, 0)$ \\
    \hline
    Prisoner 2 denies & $(0, 10)$ & $(1, 1)$ \\
    \hline
\end{tabular}
    \caption{The prisoner's dilemma}
    \label{tab:prisoners}
\end{table}

However, this model fails to capture sequential dynamics, where one player's action creates new possibilities for another, and so on.
Such sequential interactions are better represented by \emph{games played on graphs}, or \emph{graph games}.

In this model, at each time step, the game transitions to a new state, represented by a vertex in a graph, with transitions depicted as edges.
A graph game is therefore similar to a snakes-and-ladders game (or a \emph{Jeu de l'oie} in the French-speaking world), where all players collectively move a single token.
The games considered in this document are \emph{turn-based}, meaning that players never act simultaneously.
Each vertex in the game is \emph{controlled} by exactly one player, who decides which edge to follow when the token is located on their vertex.

\begin{figure}
    \centering
    \begin{tikzpicture}[node distance = 3cm]
        \node[vert, rectangle, initial left] (low) {\textsf{Low}};
        \node[vert, rectangle, right of=low] (medium) {\textsf{Medium}};
        \node[vert, right of=medium] (high) {$\begin{matrix}
            \textsf{High}\\
            \textsf{Release?}
        \end{matrix}$};
        \node[vert, rectangle, right of=high] (veryhigh) {\textsf{High}};
        \node[vert, rectangle, right of=veryhigh] (flood) {\textsf{Flood}};
        \node[vert, above of=medium] (mediumReq) {$\begin{matrix}
            \textsf{Medium}\\
            \textsf{Supply?}
        \end{matrix}$};
        \node[vert, below of=veryhigh] (veryhighReq) {$\begin{matrix}
            \textsf{High}\\
            \textsf{Supply?}
        \end{matrix}$};

        \path (low) edge node[above] {\textsf{rain}} (medium);
        \path (medium) edge[bend left] node[above] {\textsf{rain}} (high);
        \path (high) edge node[above] {\textsf{no release}} (veryhigh);
        \path (veryhigh) edge node[above] {\textsf{rain}} (flood);
        \path (medium) edge[bend left] node[left] {\textsf{request}} (mediumReq);
        \path (mediumReq) edge[bend right] node[above left] {\textsf{supply}} (low);
        \path (veryhigh) edge[bend left] node[right] {\textsf{request}} (veryhighReq);
        \path (veryhighReq) edge[bend left] node[below left] {\textsf{supply}} (medium);
        \path (high) edge[bend left] node[below] {\textsf{release}} (medium);
        \path (mediumReq) edge[bend left] node[right] {\textsf{ignore}} (medium);
        \path (veryhighReq) edge[bend left] node[left] {\textsf{ignore}} (veryhigh);
        \path (flood) edge[loop right] (flood);
        \path (low) edge[loop below] node[below] {\textsf{drought}} (low);
        \path (medium) edge[loop below] node[below] {\textsf{drought}} (medium);
        \path (veryhigh) edge[loop above] node[above] {\textsf{drought}} (veryhigh);
    \end{tikzpicture}
    \caption{Game associated with a hydroelectric dam}
    \label{fig:dam}
\end{figure}
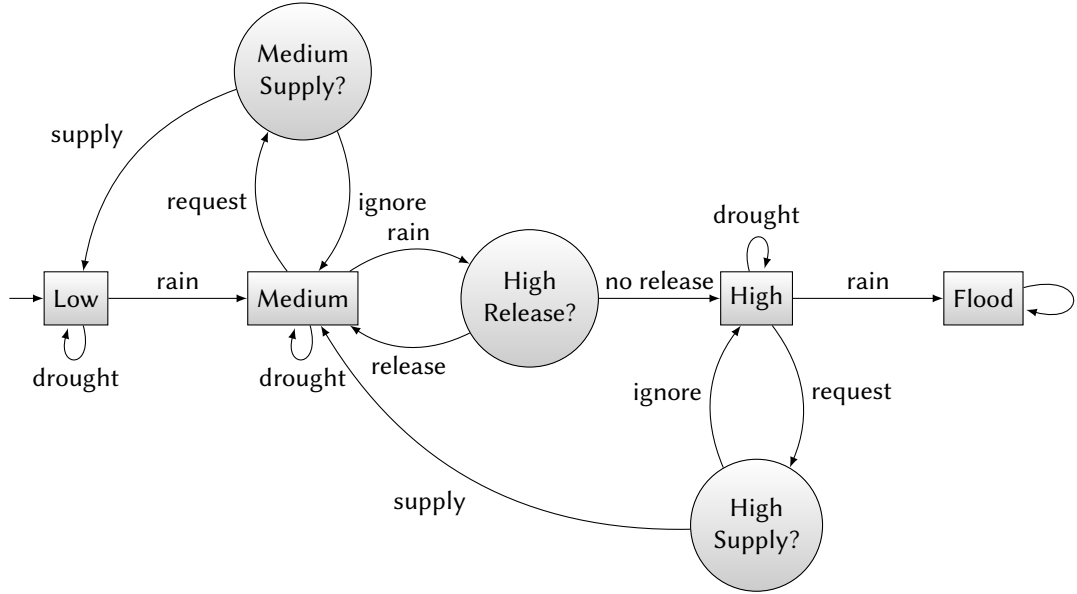

As an example, \cref{fig:dam} illustrates a (highly simplified) abstraction of a game modeling the decisions an automated dam must make to manage the water level.
Circled vertices represent configurations where the system can act and choose among multiple (or sometimes only one) possible actions.
Rectangular vertices, on the other hand, represent states where the dam is waiting for some event: a power request from the grid operator, or weather events. 
Thus, circled vertices are controlled by the player representing the dam, and rectangular ones are controlled by the antagonist.
For example, from the vertex \textsf{High Release?}, the dam player chooses between moving the token to the vertex \textsf{Medium} or to the vertex \textsf{High}.
If she chooses the latter, then the antagonist chooses between moving it to the vertex \textsf{Flood}, to the vertex \textsf{High Supply?}, or keeping it in the vertex \textsf{High}.

A first natural objective for the dam player is to avoid the vertex \textsf{Flood}.
To do so, she must also avoid the vertex \textsf{High}.
Indeed, from such a vertex, she might be temporarily protected by a drought, or saved by a request that will enable her to get back to a medium level of water by supplying electricity, but in the event of a rain, she will not be able to prevent a flood: hence the necessity of considering that this vertex is controlled by an antagonistic player, who will then move to the vertex \textsf{Flood}.
Thus, a winning strategy for the dam player consists in, whenever the vertex \textsf{High Release?} is reached, releasing water, and going back to the vertex \textsf{Medium}.
And this is indeed what we want dams to do.

Note that, even though the game may be understood as over when that vertex is reached, there is still a loop from the vertex to itself.
This is because the main model we will consider assumes that games have an \emph{infinite horizon}.
Although this might seem like a bold abstraction---since no system operates indefinitely---it is a common one in computer science, especially when specifications are properties that must be enforced over the long run.

For instance, another objective for the dam player in this scenario would be to always meet every power request.
From the vertex \textsf{Medium Supply?} and the vertex \textsf{High Supply?}, a winning strategy would then always supply electricity, and go back to the vertex \textsf{Low} or \textsf{Medium}.
This condition applies to every request, and it is difficult to place an upper bound on the number of such requests a dam will receive over its operational lifetime---hence the infinite horizon abstraction.

\section{Multiplayer games}

Even though many situations can be captured by two-player zero-sum games, recent developments in computer science have led to an increasing need for more sophisticated models.
The rise of the Internet and the proliferation of automated systems in our daily lives call for models capable of capturing interactions between an arbitrary number of agents, each driven by objectives that may not align yet are not necessarily in direct opposition.

To illustrate this, let us move from the perspective of the hydraulic dam to an analysis of a complete energy system.
A grid operator interacts with a large number of electricity consumers as well as multiple producers.
Maintaining a constant balance between production and consumption is a challenge that must be met continuously, requiring a level of responsiveness that necessarily relies on automated processes.
A sudden surge in consumption or the unexpected shutdown of a power plant must immediately trigger the activation of a standby producer capable of bridging the gap, such as our dam---or, when necessary, the temporary shutdown of non-essential consumption.

However, the agents involved each have their own interests in this process: consumers generally dislike power outages, and companies operating hydroelectric dams---when they are privately owned rather than public utilities---lose money if they are not given sufficient opportunities to sell their electricity.
Such a situation, therefore, requires a well-defined protocol with strong guarantees ensuring that all agents adhere to it.
For instance, it should not be profitable for a company operating a dam to activate it without being requested to, merely to sell electricity at the expense of a competitor.
This example highlights the need for a rigorous study of \emph{equilibria}, that is, agreements among all agents that remain stable against the temptation of individual agents (or sometimes coalitions of agents) to deviate for personal gain.

\section{Equilibria} \label{ssec:intro_equilibria}

The literature on multiplayer games offers a wide range of equilibrium concepts.
Here, we mention the ones that are studied in this work.

\subp{Nash equilibria}
The most fundamental notion of equilibrium (though not the only one, as we will see) is the Nash equilibrium, which is among the foundational concepts of game theory as a scientific field. 
It requires each player's strategy to be such that no alternative strategy would provide a better payoff for that player, assuming all other players maintain their chosen strategies.

\subp{Subgame-perfect equilibria}
However, in sequential games, Nash equilibria suffer from a significant limitation known as \emph{non-credible threats}: players may threaten each other with irrational behavior in the event of a deviation, even when such behavior contradicts their own objectives.
Consider a simple game in which two players, \emph{Adélaïde} and \emph{Barthélémy}, are sequentially given the opportunity to sign a document.
Both earn money if and only if they have both signed; otherwise, they earn nothing.
A clear Nash equilibrium in this game is the one where both players sign.
However, if both adopt the strategy of never signing, they receive no money---but improving their situation requires them to deviate together.
This makes the strategy profile a Nash equilibrium as well.
Yet, this outcome is highly counterintuitive, as it assumes that even if Adélaïde were to sign, Barthélémy would still refuse---which is precisely what we call a non-credible threat.

To eliminate such inconsistencies, the stronger notion of \emph{subgame-perfect equilibrium} (SPE) is often more relevant.
A strategy profile is an SPE if, in every subgame---that is, after every possible sequence of actions---the strategies planned by the players still form a Nash equilibrium.
Thus, players can still threaten each other, but only with strategies that they would have no incentive to deviate from if they actually had to follow through on them.
For instance, in the game described above, only the strategy profile where both players sign is an SPE.

\subp{Strong secure equilibria}
In some contexts, the mere absence of profitable deviations might not be a sufficiently strong guarantee.
If the primary concern is the \emph{safety} of agents, then a more appropriate concept may be that of \emph{secure equilibria}, where no player can deviate to harm another without also harming themselves.
In the game described above, both Nash equilibria are also secure equilibria.
On the other hand, considering only individual deviations may not always be sufficient, as players with overlapping interests can form coalitions.
This leads to the notion of \emph{strong equilibria}, which are Nash equilibria where no coalition of players has a collectively profitable deviation\footnote{The definition of \emph{collectively profitable} is not consistent in the literature: some definitions consider that it needs only to be profitable for one player, whereas others also require that it does not reduce any coalition member's payoff. This ambiguity does not impact our point.}.
In the game described above, the strategy profile that consists in both signing is a strong equilibrium, but the one that consists in never signing is not: the coalition made of Adélaïde and Barthélémy has a profitable deviation.

In this work, we do not consider secure equilibria or strong equilibria independently.
Instead, we introduce \emph{strong secure equilibria}, combining these two notions, and argue that this concept is useful for modeling safe protocols, particularly in the field of digital security.
The strategy profile where both players sign is also a strong secure equilibrium.

\subp{Risk-sensitive equilibria}
Finally, when randomness is involved, Nash equilibria are typically defined using the expected payoff.
This approach is justified, for instance, when the game is likely to be repeated many times, as the law of large numbers ensures that a player's average payoff will be close to their expected payoff.
However, outside of such cases, this definition has a major limitation: it does not account for the risk aversion that some players may exhibit.
Assume, for instance, that a player is proposed either (1) to win \texteuro 1, or (2) to play a lottery in which she loses \texteuro 100 with probability $\frac{99}{100}$, and wins \texteuro 10000 with probability $\frac{1}{100}$.
Then, the two options yield the same expected payoff, but the author would definitely prefer option~(1).
This motivates the introduction of \emph{risk-sensitive equilibria}, where the expectation operator is replaced by a well-chosen \emph{risk measure}, which generalizes it.
Many risk measures have been defined and studied in economics and finance.

One particularly interesting example is the \emph{entropic risk measure}, which offers useful properties, especially its flexibility due to the presence of a \emph{risk parameter} $\rho$ that can take any real value: the case $\rho = 0$ corresponds to the classical expected payoff,
the limit case $\rho = +\infty$ represents extreme risk aversion (or extreme \emph{pessimism}), and the dual case $\rho = -\infty$ represents extreme risk-seeking behavior (or extreme \emph{optimism}).
In those two last cases, playing the lottery described above would result in a perceived payoff of \texteuro$-100$ for an extreme pessimist and of \texteuro$10000$ for an extreme optimist.

We will argue in \cref{chap:XRSE} that these extreme cases are particularly interesting to study, as extreme pessimists can model systems of which we know that they are designed to be completely secure (say, hydroelectric dams), while extreme optimists model agents that can restart the interaction as often as they desire (a pirate trying to hack a system, for instance).
Another argument in this direction is that these extreme cases define equilibrium notions where the associated decision problems become decidable---whereas they are usually undecidable with the classical expected payoff, as well as with risk entropy, as soon as randomness is involved.

\section{The complexity of constrained existence}

Throughout this document, we will often focus on computing the complexity of the following problem, referred to as the \emph{constrained existence problem}:

\begin{quote}
Given a game $\Game$ and two payoff vectors $\bx$ and $\by$ (representing one payoff for each player), does there exist an equilibrium in $\Game$ that guarantees, for each player $i$, a payoff within the interval $[x_i, y_i]$?
\end{quote}
To derive variations of this problem, we will explore different aspects: varying the class of games in which $\Game$ evolves, particularly by modifying the functions that define the players' objectives; or considering different equilibrium notions among those introduced in \cref{ssec:intro_equilibria}.

By \emph{computing the complexity}, we mean determining the problem's placement within standard complexity classes, such as $\PTime$ (problems solvable in polynomial time by a deterministic algorithm), $\NP$ (problems solvable in polynomial time by a non-deterministic algorithm, i.e., an algorithm that can make guesses about the direction in which it should search), $\PSpace$ (problems solvable using a memory of polynomial size), etc.
Additionally, when possible, we aim to identify a class for which the problem is \emph{complete}, meaning that the problem belongs to the class \emph{and} is at least as hard as every other problem in the class---thus establishing that there is no hope that the algorithm belongs to a lower complexity class, unless the classes are actually equal.

This approach comes with certain limitations.
First, the constrained existence problem is only one of many decision problems that could be studied in relation to equilibria.
Second, determining the complexity class for which a problem is complete may provide only a partial view of its practical difficulty.
In real-world scenarios, many exponential-time algorithms are frequently used---even in cases where polynomial-time algorithms exist but are more cumbersome to implement, or turn out to be very slow on small instances.

Nonetheless, we adopt this perspective based on the conviction that characterizing the complexity of the constrained existence problem (and sometimes some of its variants or subcases) for a given equilibrium notion in a given game class, necessitates the development of techniques that provide deeper insights into these concepts.
Beyond direct applications of these results, we believe this contributes to a better understanding of equilibria.
We hope that the techniques presented in this work will convince the reader of this claim.

\section{Related Works}

Non-zero-sum infinite-duration games have attracted significant attention in recent years, particularly due to their applications in reactive synthesis problems.
For a comprehensive overview, we refer the reader to the survey papers~\cite{DBLP:conf/lata/BrenguierCHPRRS16,DBLP:conf/dlt/Bruyere17,bouyer:LIPIcs.TIME.2019.3} and the references therein.

From a historical perspective, the main seminal paper of this field is probably that of Nash~\cite{nash1951noncooperative} in which he formalizes Nash equilibria.
As for SPEs, the conceptual roots of subgame-perfection can be traced back to Zermelo's analysis of finite, perfect-information games~\cite{zermelo1913}, where a backward reasoning procedure is used to determine winning strategies.
However, it was Kuhn~\cite{kuhn1953extensive} who formally framed this reasoning within extensive-form games, and Selten~\cite{selten} who ultimately defined subgame-perfect equilibrium as a refinement of Nash equilibrium applicable to more general dynamic strategic settings.
Below, we discuss the contributions most relevant to our work.

\subp{The complexity of Nash and subgame-perfect equilibria}
Bouyer-Decître et al.~\cite{DBLP:journals/corr/BouyerBMU15} study (pure, i.e. non-randomized) Nash equilibria in \emph{concurrent} games with $\omega$-regular objectives, where players make choice simultaneously.
In this work, we only consider turn-based game, which can be seen as a particular case of concurrent games.

Brihaye et al.~\cite{DBLP:conf/lfcs/BrihayePS13} characterize Nash equilibria in quantitative games with cost-prefix-linear payoff functions based on the adversarial value.
This framework encompasses parity, mean-payoff, discounted-sum objectives, and simple quantitative functions.

Meunier~\cite{thesis_noemie} develops a method based on Prover-Challenger games to decide the existence of SPEs in games with a finite number of possible payoffs. While this method results in high complexity for parity games and is inapplicable to mean-payoff and discounted-sum games (where payoffs form an infinite set), it inspired the \emph{concrete negotiation game}, which we introduce in \cref{chap:nego}.

Le Roux and Pauly~\cite{10.1145/2603088.2603120} study conditions for the existence of Nash equilibria, $\epsilon$-Nash equilibria, SPEs, and $\epsilon$-SPEs.
Flesch et al.~\cite{DBLP:journals/mor/FleschKMSSV10} prove that $\epsilon$-SPEs always exist when the payoff functions are \emph{lower-semicontinuous}. Flesch and Predtetchinski~\cite{DBLP:journals/mor/FleschP17} propose an alternative characterization of SPEs in games with finitely many payoffs, based on a game structure that we refer to as the \emph{abstract negotiation game}, without formalizing the notion of negotiation function.

Brihaye et al.~\cite{DBLP:conf/concur/BrihayeBGRB19} analyze the constrained existence problem in quantitative reachability games and show that it is $\PSpace$-complete. Their proof introduces a preliminary version of what we call the \emph{negotiation function}.

Ummels and Wojtczak~\cite{UW11} prove that the constrained existence problem for Nash equilibria in mean-payoff games is undecidable when players can randomize strategies and $\NP$-complete when they cannot. Furthermore, they show that the same problem remains undecidable in games with rewards on terminal vertices if stochastic vertices are allowed~\cite{Ummels2011}.

For SPEs, Ummels and Grädel~\cite{GU08} establish their existence in games with $\omega$-regular objectives and provides an $\ExpTime$ algorithm based on tree automata to solve the constrained SPE existence problem in parity games. However, this algorithm only places the problem in the class $\ExpTime$ while proving it is $\NP$-hard. We prove $\NP$-completeness in \cref{chap:parity}.

Brihaye et al.~\cite{DBLP:conf/csl/BrihayeBMR15} introduce \emph{weak subgame-perfect equilibria}, a relaxation of classical SPEs. This weakening is equivalent to standard SPEs when the payoff function is \emph{continuous}, which holds for quantitative reachability and discounted-sum objectives but not for parity and mean-payoff objectives. The techniques introduced in~\cite{DBLP:conf/csl/BrihayeBMR15} and extended in~\cite{DBLP:conf/fossacs/Bruyere0PR17} cannot be used to characterize SPEs in non-continuous settings: this sequence of papers leaves therefore open the complexity of SPEs in games such as mean-payoff games, which will be one of our main contributions.

Strategy logics, as studied for instance in~\cite{DBLP:journals/iandc/ChatterjeeHP10}, can encode SPEs for LTL objectives, a strict subset of $\omega$-regular objectives.

\subp{Techniques for some game classes}
Chatterjee et al.~\cite{DBLP:conf/concur/ChatterjeeDEHR10} study mean-payoff \emph{automata} and provide results that can be translated into an expression of all possible payoff vectors in a mean-payoff game. Brenguier and Raskin~\cite{DBLP:conf/cav/BrenguierR15} propose an algorithm to compute the Pareto curve of multi-dimensional two-player zero-sum mean-payoff games. Techniques from these papers are used in several technical steps of our algorithm presented in \cref{chap:MP}.

Energy objectives, widely studied in the context of vector addition systems with states and Petri nets, are predominantly considered in two-player zero-sum settings~\cite{DBLP:conf/formats/BouyerFLMS08, DBLP:journals/iandc/VelnerC0HRR15, DBLP:journals/amai/KupfermanPV16,RASKIN200569}. Discounted-sum objectives, introduced by Zwick and Paterson~\cite{ZwickPaterson}, have also been primarily analyzed in the two-player zero-sum framework. Their connection to the target discounted-sum problem, a long-standing open problem, is discussed by Boker, Henzinger, and Otop~\cite{DBLP:conf/lics/BokerHO15}. To the best of our knowledge, no algorithmic results exist for those objectives in multiplayer non-zero-sum settings prior to this work.

\subp{Rational verification, rational synthesis}
\emph{Rational synthesis} was first introduced in a cooperative setting by Fishman, Kupferman, and Lustig~\cite{10.1007/978-3-642-12002-2_16}, and later extended to a non-cooperative setting---the one we refer to in \cref{chap:verif}---by Perelli, Kupferman, and Vardi~\cite{DBLP:journals/amai/KupfermanPV16}.
Non-cooperative rational synthesis can be seen as an adaptation of the classical notion of \emph{Stackelberg games}~\cite{stackelberg1934} to game-theoretic problems arising in computer science.
Filiot, Gentilini, and Raskin~\cite{filiot_et_al:LIPIcs:2020:12534} also study rational synthesis (under the name \emph{Stackelberg value}) in two-player mean-payoff and discounted-sum games, where a leader plays optimally and the follower responds with a best response (or best response up to $\epsilon \geq 0$).
They consider a single follower, making their setting a special case of rational synthesis as defined in~\cite{DBLP:journals/amai/KupfermanPV16}.

The notion of \emph{rational verification} is explored in~\cite{DBLP:journals/amai/GutierrezNPW23}, where Gutierrez, Najib, Perelli, and Wooldridge analyze the complexity of related problems. Their setting differs from what we call rational verification in \cref{chap:verif}: they study whether all Nash equilibria (or at least one) satisfy a given specification, without any player representing the system (Leader in our setting)---such a problem is closer to what we call the constrained existence problem.
Nevertheless, as we show in Corollary~\ref{cor:reductions}, those two problems are closely related.

\subp{Secure protocols and strong secure equilibria}
Several works rely on notions of equilibria to argue for the resilience of protocols, such as the RatFish peer-to-peer protocol~\cite{BackesCiobotaruKrohmer10}, although they rely on 2-player Nash equilibria.
Two-player games are also considered in~\cite{KremerRaskin2003} to evaluate the fairness of an exchange protocol, through the logic ATL.
The authors hint at using coalitions, as they solve several two-player games to take into account the fact that the infrastructure may side with one agent or the other to facilitate their cheating.
In contrast, strong secure equilibria, which we study in \cref{chap:SSE}, syntactically consider all possible coalitions.

In~\cite{AbrahamDolevGonenHalpern06}, the authors use resilient strong Nash equilibria to show that their protocol does not require a TTP as long as the agents are all rational, but they do not provide a general framework.
A more general formalism to handle rational fairness is laid out in~\cite{ButtyanHubauxCapkun04}, where the authors point out the limit of fairness as a strong trace property.
They also choose the game modeling approach and propose a variant of Nash equilibria as the appropriate solution concept to capture rationality in protocols.
However, their notion of rationality of the agents is \emph{optimistic}: they are not assumed to be willing to deviate from the protocol only to affect negatively the others, it must at the very least impact them strictly positively. 
Furthermore, since the model they choose is strictly two-player, there is no notion of robustness against malignant coalitions. Thus, they favor a sort of \emph{best} Nash equilibria as a rational solution concept.
Finally, while they propose a relevant general framework to model protocols in terms of multiplayer games, they provide no implementations or decision procedure.

At the other end of the spectrum, a very general approach uses \emph{hyperproperties}~\cite{ClarksonSchneider10,RakotonirinaBartheSchneidewind24}.
These allow to express properties of sets of traces, and are therefore well suited to the verification of protocols.
The drawback of such a general approach is that decision problems are often not solvable algorithmically, or non-elementary for relevant fragments; in comparison, the complexity of the most difficult problems in our formalism for safe protocols, studied in \cref{chap:SSE}, is "only" $\ExpTime$.

Although a wide array of notions of rationality in games have been defined, few works focus on the notion of secure equilibria.
Two notable exceptions are~\cite{ChatterjeeHenzingerJurdzinski05,10.1007/978-3-642-27940-9_11} which actually also motivates the use of secure equilibria with the need to model malicious agents, and~\cite{DBLP:conf/csl/BruyereMR14}, for quantitative objectives.
Both however only consider the case of two-player games.

Secure equilibria are also investigated in~\cite{DBLP:conf/csl/BruyereMR14} by Bruyère, Meunier, and Raskin, and in~\cite{CHATTERJEE200667} by Chatterjee, Henzinger, and Jurdzi\'{n}ski.
In~\cite{DBLP:conf/fossacs/Brenguier16}, Romain Brenguier studies \emph{$(k, t)$-robust equilibria}, a generalization of strong equilibria.

\subp{Risk measures and stochastic systems}
Risk measures generalize expected payoffs by capturing players' perceptions of uncertainty. Widely used in economics and finance, these include expected shortfall, value at risk~\cite{Aue18}, variance~\cite{Bra99}, and entropic risk measures~\cite{FS02}. Their application to Markov decision processes has been extensively studied, particularly regarding variance-penalized risk measures~\cite{FK89, PSB22,MT11}, expected shortfall~\cite{RRS15,KM18,Meg22}, and entropic risk~\cite{How72,BR14,BCMP24}. Among these, entropic risk stands out due to its tractability: unlike expected shortfall and variance-penalized measures, which require exponential memory~\cite{HK15} and computational resources~\cite{PSB22}, entropic risk allows for optimal positional strategies in Markov decision processes~\cite{How72}, making it a prime candidate for multi-agent settings.

\section{Structure and contributions}

In \cref{chap:background}, we introduce the necessary background, particularly the game model that we use throughout this document.

In \cref{chap:nash}, we study Nash equilibria.
Although extensively covered in the literature, we introduce several new results for discounted-sum games and energy games.
For the former, we remark that the constrained existence problem of Nash equilibria is at least as hard as the \emph{target discounted-sum problem}, a decision problem whose decidability status has remained open to date.
However, we prove that this problem is co-recursively enumerable.
Similarly, we establish that the constrained existence problem of SPEs in energy games is undecidable, but recursively enumerable.
This chapter also serves as an introduction to the classes of games studied in the next part of this work.

All of \cref{part:SPE} is dedicated to subgame-perfect equilibria, the notion of equilibrium to which most of our work has been devoted.

In \cref{chap:nego}, we introduce SPEs, and present the \emph{negotiation function}, a key tool used in the following two chapters.
The negotiation function maps every vertex labeling $\lambda: V \to \RR \cup \{\pm \infty\}$ to a new labeling $\nego(\lambda)$, where for each vertex $v$, the value $\nego(\lambda)(v)$ represents the best payoff that the player controlling $v$ can enforce from $v$, assuming that the other players act rationally to minimize that player's payoff, while still satisfying their own payoff requirements, as defined by $\lambda$.
We show that the negotiation function can be computed using auxiliary two-player zero-sum game structures, and that its fixed points characterize the SPEs of a game.

In \cref{chap:parity}, we apply these tools to prove that the constrained existence problem of SPEs in parity games is $\NP$-complete, thereby resolving the complexity gap for this problem, which was previously known to be $\NP$-hard and $\ExpTime$-easy.

In \cref{chap:MP}, we adapt these techniques to show that the same problem is $\NP$-complete in mean-payoff games.
This result requires several intermediary results, such as a polynomial bound on the size of the least fixed point of the negotiation function, making this result the most involved of \cref{part:SPE}, if not of this whole document.

In \cref{chap:energy&DS}, we extend our results on Nash equilibria in discounted-sum and energy games to SPEs.
However, unlike the case of Nash equilibria, we are unable to establish whether the constrained existence problem in energy games is recursively enumerable.
We leave this as an open question.

In \cref{chap:verif}, we switch to another decision problem: \emph{rational verification}, which consists in verifying that a given strategy for one player (called \emph{Leader}) guarantees a payoff above some threshold against every response to that strategy that is \emph{rational}---rationality being defined in terms of NEs or SPEs.
We show that this problem is strongly linked to the constrained existence problem, and deduce its complexity in all the game classes we mentioned above.
Then, we highlight a limitation to the classical definition of rational verification, which results in counter-intuitive outcomes when no rational response exists: we then propose an alternative definition, called \emph{achaotic rational verification}, which avoids that issue.
We show that this new decision problem is $\P^\NP$-complete in mean-payoff games with SPEs as a rationality concept, and coincides with classical rational verification in all the other cases.

\cref{part:other_equilibria} explores more specialized notions of equilibria.
In \cref{chap:SSE}, we study strong secure equilibria in parity games, and in games where each player's objective is specified by a parity automaton.
We show that the constrained existence problem is $\PSpace$-complete in the former and $\ExpTime$-complete in the latter.

In \cref{chap:RSE}, we consider settings involving randomness, either because the game itself contains stochastic vertices or because players are allowed to randomize their strategies.
We recall several known undecidability results in such settings and introduce \emph{entropic risk-sensitive equilibria}, which account for players' risk aversion or attraction.
We show that the constrained existence problem of entropic risk-sensitive equilibria is undecidable even in \emph{simple} stochastic games, where payoffs are determined by the \emph{terminal vertex} reached by a play.
However, we establish some decidability results when players are restricted to pure, stationary, or positional strategies.

Finally, in \cref{chap:XRSE}, we introduce \emph{extreme risk-sensitive equilibria}, where players interpret randomness with extreme pessimism or optimism.
We prove that the constrained existence problem for such equilibria is $\NP$-complete, and even $\PTime$-complete when all players are optimists.
This result establishes, to our knowledge, the first decidable fragment of the constrained existence problem for equilibria in stochastic games, without restrictions on the number of players or on their strategies.

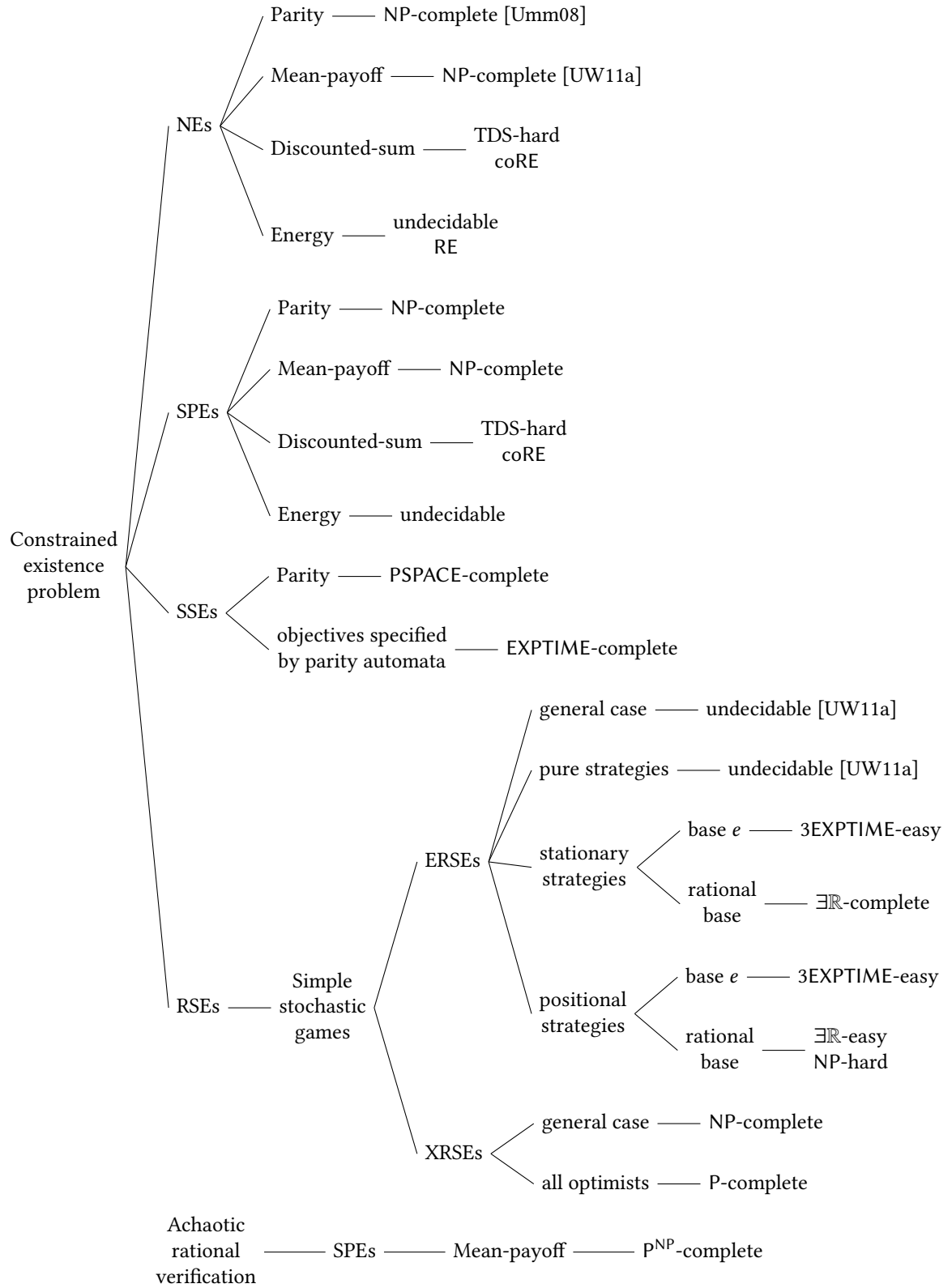
\begin{figure}
\centering
\begin{forest}
for tree={
    grow=east,
    parent anchor=east,
    child anchor=west,
    thick,
    l sep=7mm, 
    s sep=3mm, 
    anchor=west,
    align=center,
}
[Constrained\\
existence\\
problem
    [RSEs
        [Simple\\stochastic\\games
            [XRSEs
                [all optimists
                    [$\P$-complete]
                ]
                [general case
                    [$\NP$-complete]
                ]
            ]
            [ERSEs
                [positional\\strategies
                    [rational\\base
                        [$\exists\RR$-easy\\$\NP$-hard]
                    ]
                    [base $e$
                        [$3\ExpTime$-easy]
                    ]
                ]
                [stationary\\strategies
                    [rational\\base
                        [$\exists\RR$-complete]
                    ]
                    [base $e$
                        [$3\ExpTime$-easy]
                    ]
                ]
                [pure strategies
                    [undecidable~\cite{Ummels2011}]
                ]
                [general case
                    [undecidable~\cite{Ummels2011}]
                ]
            ]
        ]
    ]
    [SSEs
        [objectives specified\\by parity automata
            [$\ExpTime$-complete]
        ]
        [Parity
            [$\PSpace$-complete]
        ]
    ]
    [SPEs
        [Energy
            [undecidable]
        ]
        [Discounted-sum
            [$\TDS$-hard\\$\coRE$]
        ]
        [Mean-payoff
            [$\NP$-complete]
        ]
        [Parity
            [$\NP$-complete]
        ]
    ]
    [NEs
        [Energy
            [undecidable\\$\RE$]
        ]
        [Discounted-sum
            [$\TDS$-hard\\$\coRE$]
        ]
        [Mean-payoff
            [$\NP$-complete~\cite{Ummels2011}]
        ]
        [Parity
            [$\NP$-complete~\cite{DBLP:conf/fossacs/Ummels08}]
        ]
    ]
]
\end{forest}
\begin{forest}
for tree={
    grow=east,
    parent anchor=east,
    child anchor=west,
    thick,
    l sep=10mm, 
    s sep=3mm, 
    anchor=west,
    align=center,
}
[Achaotic\\rational\\verification
    [SPEs
        [Mean-payoff
            [$\P^\NP$-complete]
        ]
    ]
]
\end{forest}
\caption{Our main complexity results}
\label{fig:results}
\end{figure}

\Cref{fig:results} summarizes results about the complexity of the constrained existence problem.

\section{Previous publications and co-authorship}

The results presented in \cref{chap:nego} and part of those in \cref{chap:MP} were previously published in~\cite{DBLP:conf/concur/BriceRB21} and its journal version~\cite{DBLP:journals/lmcs/BriceBR23}.  
The results in \cref{chap:parity} appeared in~\cite{DBLP:conf/csl/BriceRB22}, while the main findings from \cref{chap:MP} were published in~\cite{DBLP:conf/icalp/BriceRB22}.  
The results from \cref{chap:nash,chap:energy&DS,chap:verif} were presented in~\cite{DBLP:conf/mfcs/BriceRB23}.  
All the aforementioned articles are joint works with Jean-François Raskin and Marie van den Bogaard.  

The results in \cref{chap:SSE} stem from a collaboration with Jean-François Raskin, Mathieu Sassolas, Guillaume Scerri, and Marie van den Bogaard.  
They were published in~\cite{DBLP:journals/corr/abs-2405-18958}.  
Finally, \cref{chap:RSE,chap:XRSE} compile results from a joint work with Thomas Henzinger and K.S. Thejaswini, published in~\cite{brice2025findingequilibriasimplerpessimists}.

\section{How to read this thesis}

This document has been written with full awareness that few people have the time or inclination to read an entire PhD thesis.
For those who do, we hope it serves as a valuable, albeit necessarily incomplete, resource for developing a strong understanding of the algorithmic aspects of equilibria in multiplayer games.

More realistically, many readers will approach this document in search of specific results or proofs.
While all the results presented here have previously appeared in conference papers, this version may be more convenient: some proofs have been revisited, and the material is structured in a more pedagogical manner than the rapid, often fragmented presentation imposed by conference proceeding formats.
For such readers, it may be useful to know that some chapters are independent of others.

\subp{Dependences between chapters}
This introduction provides an overview of where this work fits within theoretical computer science and game theory, but it is not required reading before diving into a specific chapter.
\cref{chap:background} collects definitions used throughout the document.
Readers already well-versed in the field may choose to skim it or skip it entirely, but they may still find it useful as a reference whenever notation or concepts become unclear.

Similarly, \cref{chap:nash} presents results on Nash equilibria but also introduces the game classes studied in later chapters.
Those already familiar with parity, mean-payoff, discounted-sum, or energy games may skip it, but for others, reading it before \cref{part:SPE} is strongly recommended.

Within \cref{part:SPE}, \cref{chap:nego} should be read before \cref{chap:parity}, which in turn should be read before \cref{chap:MP}.
\cref{chap:energy&DS} is more self-contained, at least for a reader who is familiar with SPEs---otherwise, the reader should read \cref{chap:nego} beforehand.
\cref{chap:verif} builds on results from previous chapters, especially \cref{chap:MP}, but can be read independently from \cref{chap:energy&DS}, provided the reader accepts the stated results without proof.

\cref{chap:SSE} is relatively independent, though reading \cref{chap:nash} beforehand is advisable, especially for those unfamiliar with parity objectives.
Finally, \cref{chap:RSE,chap:XRSE} are independent of the rest of the document, but \cref{chap:RSE} should be read before \cref{chap:XRSE}.

\begin{figure}
    \centering
    \begin{tikzpicture}[node distance = 3cm]
        \node[state, rectangle, rounded corners, fill, top color=white, bottom color=blue!30] (nash) {\cref{chap:nash}};
        \node[state, rectangle, rounded corners, below of=nash, fill, top color=white, bottom color=red!30] (parity) {\cref{chap:parity}};
        \node[state, rectangle, rounded corners, right of=nash, fill, top color=white, bottom color=red!30] (nego) {\cref{chap:nego}};
        \node[state, rectangle, rounded corners, right of=parity, fill, top color=white, bottom color=red!30] (MP) {\cref{chap:MP}};
        \node[state, rectangle, rounded corners, right of=MP, fill, top color=white, bottom color=red!30] (energyDS) {\cref{chap:energy&DS}};
        \node[state, rectangle, rounded corners, below of=parity, fill, top color=white, bottom color=red!30] (verif) {\cref{chap:verif}};
        \node[state, rectangle, rounded corners, right of=verif, fill, top color=white, bottom color=green!30] (SSE) {\cref{chap:SSE}};
        \node[state, rectangle, rounded corners, right of=SSE, fill, top color=white, bottom color=green!30] (RSE) {\cref{chap:RSE}};
        \node[state, rectangle, rounded corners, right of=RSE, fill, top color=white, bottom color=green!30] (XRSE) {\cref{chap:XRSE}};

        \path (nash) edge (parity);
        \path (nego) edge (parity);
        \path (parity) edge (MP);
        \path (nash) edge (energyDS);
        \path (MP) edge (verif);
        \path[dotted] (energyDS) edge (verif);
        \path[dotted] (nash) edge (SSE);
        \path (RSE) edge (XRSE);
        \path[dotted] (nego) edge (energyDS);
    \end{tikzpicture}
    \caption{Dependences between chapters}
    \label{fig:dependence_chapters}
\end{figure}
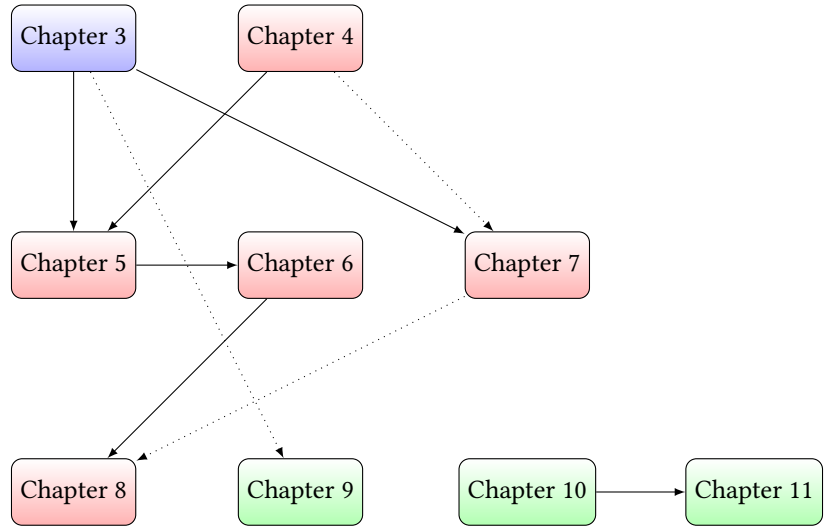

Those dependences are depicted by \Cref{fig:dependence_chapters}: a chapter is blue if it belongs to \cref{part:nash}, red if it belongs to \cref{part:SPE}, and green if it belongs to \cref{part:other_equilibria}, and an arrow between Chapter $i$ and Chapter $j$ indicates that Chapter $i$ should be read before reading Chapter $j$---a dotted arrow indicates that reading is advised but not essential.
The transitivity of the relation \emph{"should be read before"} has been used to omit some arrows.

\subp{Proofs}
Driven by the conviction that the proof is always at least as interesting as the result itself, we have chosen to present complete proofs directly alongside each stated result. Of course, skipping a proof is always an option. Though we have done our best to avoid it, there are instances where a proof references another result or construction from within a different proof. In such cases, we explicitly indicate which proof should be read for a full understanding.

The proofs in this document fall into four broad categories:
\begin{enumerate}
    \item\label{itm:pf_short_useless} short proofs that merely verify a result without introducing particularly interesting arguments, such as that of \cref{lm_lasso};
    
    \item\label{itm:pf_short_useful} short proofs that present key arguments, as in \cref{thm:ds_ne_hardness}, or that conclude previous results, as in \cref{lm_parity_check_winning};
    
    \item\label{itm:pf_long_useless} long (and sometimes intricate) proofs that establish an unsurprising result using unsurprising arguments, but require significant technical development, such as \cref{thm:abstract}, or, to an extreme degree, \cref{lm:size_lambda};
    
    \item\label{itm:pf_long_useful} long proofs that are genuinely worth reading, as they introduce original ideas and non-trivial arguments.
\end{enumerate}

Readers may prefer to skip proofs in categories~\ref{itm:pf_short_useless} and especially~\ref{itm:pf_long_useless}, while focusing on those in categories~\ref{itm:pf_short_useful} and~\ref{itm:pf_long_useful}.
Naturally, some proofs fall between these categories or contain sections of varying interest.
Typically, hardness proofs based on reductions are most valuable for their construction steps, which we always aim to present in an intuitive manner, allowing the reader to grasp the essence of the reduction without needing to follow the full proof.
The latter part of such proofs, establishing the correctness of the reduction, almost always falls into category~\ref{itm:pf_long_useless}.
Subjectively, the author would consider the nucleus of category~\ref{itm:pf_long_useful} to include the proofs of \cref{thm:spe_parity_np,thm:spe_mp_np_complete,thm:energy_spe_undec,thm:mp_ach_verif,thm:sseExistencePspaceEasy,thm:XRSEexists,thm:XRSE_NPc,thm:XRSE_Pcomplete}.

A practical heuristic for distinguishing between proofs in categories~\ref{itm:pf_long_useless} and~\ref{itm:pf_long_useful} without reading them in full is as follows: the more a proof is filled with mathematical symbols and equations, the more technical it tends to be---and, in most cases, the less it introduces new conceptual insights.
However, exceptions exist, and this heuristic may not align perfectly with every reader's perspective.

Now that all these introductory remarks have been made, we wish the reader a pleasant journey.

%% file: 0cBackground.tex
\section{Writing conventions}\label{sec:conventions}

From this point, we assume that the reader is familiar with the basics of algebra, language theory, graph theory, algorithmics, and complexity theory.
However, we recall some fundamental notions to establish consistency in notation.

\subparagraph*{Sets of numbers.}
We write $\NN, \ZZ, \QQ$, and $\RR$ for the sets of, respectively, natural numbers, integers, rational numbers, and real numbers, all defined with Zermelo-Fraenkel's theory of sets.
The definition of the set $\NN$ is classically contentious: although the English-speaking world, and specifically secondary education, often considers that natural numbers start with $1$, the convention used in the French-speaking world, according to which $0$ is also a natural number, tends to take over in the field of computer science and combinatorics.
We will therefore use the latter, and write $\NN = \{0, 1, 2, \dots\}$.

\subparagraph*{Tuples.}
For a set $X$, and an indexing (usually finite, but not always) set $I$, when we have defined an element $x_i \in X$ for every $i \in I$, we usually use the bar notation $\bx_I$, or simply $\bx$, for the tuple $(x_i)_{i \in I} \in X^I$.
Conversely, when we introduce a tuple $\bx_I$, or $\bx$, the notation $x_i$ will refer to the element of index $i \in I$.
In some cases, we will need the double-bar $\bbx$ to denote tuples of tuples.
We will implicitly use pointwise comparisons: when we write $\bx \leq \by$, for two tuples $\bx, \by \in X^I$, the reader should read $x_i \leq y_i$ for each $i \in I$.
Tuples are often called \emph{vectors}, typically when they have real values, or \emph{profiles} when the set $I$ is a set of players.
When the set $I$ is clear from the context, we write $\bzero$ for the vector $(0)_{i \in I}$.

\subparagraph*{Words.}
A tuple $\bx \in X^n$ (resp. $\bx \in X^\NN$), for a given natural integer $n$, is also called \emph{finite word} (resp. \emph{infinite word}) over $X$.
Then, we use the formalisms of language theory: the set $X$ is called \emph{alphabet}, and the word $\bx$ can be written $x = x_0 \dots x_{n-1}$ (resp. $x = x_0 x_1 \dots$).
The integer $n$ is called \emph{length} of $x$, and written $|x|$; the length of an infinite word is $+\infty$.
To avoid confusion between length and cardinality, the cardinality of a set $X$ will always be written $\card X$.

When a tuple is introduced as a (finite or infinite) word $x$ (or a specific type of word, such as paths, histories, or plays, which will be defined later), we will still write $x_k$ for the $(k+1)$th element of $x$.
Moreover, we will write $x_{\leq k}$, or $x_{<k+1}$, for the (finite) prefix $x_0 \dots x_k$, and $x_{\geq k}$, or $x_{>k-1}$, for the (finite or infinite) suffix $x_k x_{k+1} \dots$, for every integer $k$.

We write $X^*$ for the set of finite words, and $X^\omega$ for the set of infinite words, over the alphabet $X$.

\subparagraph*{Complexity classes.}
We will make an extensive use of the classical complexity classes $\P$, $\NP$, $\coNP$, $\PSpace$, $\ExpTime$, $\P^\NP$, assuming that the reader is familiar with their definitions---the reader who is not will find a very complete overview in~\cite{DBLP:books/daglib/0086373}.
All complexity classes are written in sans-serif capital letters, while problems are denoted using serif small capitals---for instance, the satisfiability problem is written as $\SAT$.
For a given complexity class $\Class$, we call \emph{$\Class$-easy} a decision problem that belongs to $\Class$ and \emph{$\Class$-hard} a decision problem to which every $\Class$-easy problem can be reduced in polynomial time.
For a given decision problem $P$, we will write \emph{$P$-hard} instead of $\{P\}$-hard.

\subparagraph{Size.}
The complexity of an algorithm is defined as a function of the size $\lv x \rv$ of its input $x$, itself defined as the number of bits that are necessary to describe $x$ by a word on the alphabet $\{\mathtt{0, 1}\}$.
Most of the time, we will not explain how the objects we consider are encoded by such a word, considering that there are canonical encodings with which the reader is familiar: for example, an integer $n$ is represented by its binary encoding, which uses $\lceil \log_2(n+1) \rceil$ bits.
We give here some first definitions of sizes that may not be straightforward.
The size of a rational number $r = \frac{p}{q}$, where $p, q \in \ZZ$ are co-prime, is the quantity $\Vert r\Vert = 1 + \lceil \log_2(|p|+1) \rceil + \lceil \log_2(|q|+1) \rceil$.
The size of an irrational number is $+\infty$.
The size of infinite numbers is $\Vert +\infty \Vert = \Vert -\infty \Vert = 1$.
The size of a tuple $\bx \in O^I$, where $I$ is an index set and $O$ is a set of objects for which the notion of size has been defined, is the quantity $\card I + \sum_{i \in I} \Vert x_i \Vert$.
Similarly, the size of a function $f: I \to X$ is the quantity $\card I + \sum_{i \in I} \Vert f(i) \Vert$, and the size of a set $X \subseteq O$ is the quantity $\card X + \sum_{x \in X} \Vert x \Vert$.

\subp{Players}
We will give below a definition of \emph{games}, in which \emph{players} are abstract objects from an abstract set.
However, in game theory, it is common to refer to players as persons rather than objects.
We will therefore often abuse language by saying, for example, that a player \emph{wishes} or \emph{intends} to maximize some quantity, to say that the mentioned quantity is their payoff function.
One advantage of this convention, aside from aiding intuition, is that it allows the writer to take full advantage of the flexibility of the English language, where people have genders, whereas objects do not.
Thus, in examples and proofs, a gender will be implicitly assigned to all well-defined players.
For instance, when an example involves players denoted as $\Circle, \Square$, or $\Diamond$, players $\Circle$ and $\Diamond$ will be grammatically female, while player $\Box$ will be grammatically male.
Finally, when we refer to some undefined player of which we do not know the gender, for example when we write \emph{"let $i \in \{\Circle, \Square, \Diamond\}$"}, we will use the neutral pronoun \emph{they} to emphasize the fact that we are referring to some player, but not to an abstract object, for which we would have used the pronoun \emph{it}.

\subp{Middle dot}
We sometimes use the middle dot $\cdot$ to denote an unspecified element.
For example, if the pair $\bz$ is known to belong to the cartesian product $X \times Y$, then we write $\bz = (x, \cdot)$ to mean that there exists $y \in Y$ such that $\bz = (x, y)$.

\subparagraph*{Usual notations.}
We will endeavor to make sure that every mathematical notation is well defined or clear from the context.
However, if we fail in this objective, the reader may refer to Table~\ref{tab:notations}: for every line, when a mathematical object is written with the symbol presented on the left side, it is probably of the type indicated on the right side.
Note that gothic capital letters are only used to denote players, while cursive capital letters are used for computing structures---games, memory structures, counter machines, automata.

\begin{table}
    \centering
\begin{tabular}{c|c}
    $a, b, c, d, \dots$ & vertices in a graph (constants) or real numbers \\
    $c, d$ & cycles \\
    $e$ & Euler's number $e = 2.71828\dots$ \\
    $f$ & function \\
    $g, h$ & histories \\
    $i, j$ & players (variables) \\
    $k, \l, m, n$ & integers \\
    $p, q, r$ & states of a machine, integers, or probabilities \\
    $r$ & reward function \\
    $t$ & real number (target or threshold) or terminal vertex \\
    $u, v, w$ & vertices in a graph (variables) \\
    $x, y, z$ & real numbers, variables of a formula, or abstract objects \\
    $E$ & set of edges \\
    $K$ & strongly connected component in a graph \\
    $M$ & risk measure \\
    $Q$ & set of states of a machine \\
    $T$ & tree, set of terminal vertices \\
    $U, V, W$ & sets of vertices \\
    $X, Y, Z$ & abstract sets, sets of reals, or random variables \\
    $\alpha$ & path \\
    $\alpha, \beta, \gamma$ & real numbers \\
    $\beta$ & discount factor, or base of risk entropy \\
    $\delta, \epsilon$ & (small) real numbers \\
    $\delta$ & transition function of an automaton \\
    $\kappa$ & color function \\
    $\lambda$ & requirement \\
    $\lambda_0$ & the vacuous requirement $\lambda_0: v \mapsto -\infty$ \\
    $\mu, \nu$ & payoff functions \\
    $\nu$ & valuation \\
    $\pi, \chi, \xi$ & plays \\
    $\rho, \eta$ & runs of a machine \\
    $\rho$ & risk parameter or rationality concept \\
    $\sigma_i, \tau_i$ & strategies \\
    $\phi, \psi$ & formulae or functions \\
    $\omega$ & the first transfinite ordinal \\
    $\Delta$ & set of transitions of a machine \\
    $\p$ & probability function of a stochastic game \\
    $\auto$ & automaton \\
    $\Game, \Gamebis$ & games \\
    $\Kount$ & counter machine \\
    $\Mem$ & memory structure \\
    $\Prover, \Challenger, \Solver, \Witness, \Leader, \Demon, \adel, \bart, \ttp, \opponent$ & players (constants) \\
    $\Circle, \Square, \Diamond$ & players (constants) \\
\end{tabular}
    \caption{Notations used in this thesis}
    \label{tab:notations}
\end{table}

\subp{Operators}
Operators are typically written in sans-serif typeface; for instance, the set of plays in a game $\Game_{\|v_0}$ is denoted by $\Plays\Game_{\|v_0}$.
They start with an uppercase letter when their output is considered as a set, and with a lowercase letter otherwise.
A notable exception is probabilistic operators, like the expectation $\EE$, for which we use the blackboard uppercase notation.
A list of operators used in this thesis, along with the locations where they are introduced, can be found in Table~\ref{tab:operators}.

\begin{table}
    \centering
\begin{tabular}{c|c|c}
    $\abs_{\lambda i}(\Game)_{\|v_0}$ & abstract negotiation game & \cref{def:abstract} \\
    $\card X$ & cardinal & \cref{sec:conventions} \\
    $\conc_{\lambda i}(\Game)_{\|v_0^\con}$ & concrete negotiation game & \cref{def:concrete} \\
    $\lCons\Game_{\|v}$ (or $\lCons(v)$) & $\lambda$-consistent plays & \cref{def:lambda_cons} \\
    $\Conv X$ & convex hull & \cref{ssec:achievable_payoffs} \\
    $\dev_{\bx\phi}(\Game)_{\|v_0^\cd}$ & deviator game & \cref{def:deviator_game} \\
    $\EL_{r_i}(h)$ (or $\EL_i(h)$) & energy level & \cref{def:energy} \\
    $\ds^\beta_{r_i}(\pi)$ (or $\ds_i(\pi))$ & discounted sum & \cref{def:ds} \\
    $\first(\alpha)$ & first vertex & \cref{sec:graph_games} \\
    $\Hist\Game$ & histories & \cref{sec:graph_games} \\
    $\Hist_i\Game$ & histories ending in $V_i$ & \cref{sec:graph_games} \\
    $\Ind_{\|v_0}(\Mem)$ & induced strategies (or strategy profiles) & \cref{def:memory_structure} \\
    $\Inf(\alpha)$ & vertices occurring infinitely often & \cref{sec:graph_games} \\
    $\last(h)$ & last vertex & \cref{sec:graph_games} \\
    $\MP_{r_i}(h)$ (or $\MP_i(h)$) & mean-payoff & \cref{def:mp} \\
    $\MPi_{r_i}(\pi)$ (or $\MPi_i(\pi)$) & mean-payoff & \cref{def:mp} \\
    $\nego$ & negotiation function & \cref{def_nego} \\
    $\Occ(\alpha)$ & vertices occurring & \cref{sec:graph_games} \\
    $\Plays\Game$ & plays & \cref{sec:graph_games} \\
    $\lRat_{-i}\Game_{\|v}$ (or $\lRat(v)$) & $\lambda$-rational strategy profiles & \cref{def:lambda_rat} \\
    $\red^\beta_{\lambda i}(\Game)_{\|v_0^\r}$ & reduced negotiation game & \cref{def_reduced_game} (or \cref{ssec:parity_representatives}) \\
    $\SConn(V, E)$ & strongly connected components & \cref{sec:graph_games} \\
    $\SC(V, E)$ & simple cycles & \cref{sec:graph_games} \\
    $\Stat_i\Game_{\|v_0}$ & stationary strategies & \cref{sec:stationary_finite_memory} \\
    $\Stat_P\Game_{\|v_0}$ & stationary strategy profiles  & \cref{sec:stationary_finite_memory} \\
    $\Strat_i\Game_{\|v_0}$ & strategies & \cref{def:strategies} \\
    $\Strat_P\Game_{\|v_0}$ & strategy profiles & \cref{def:strategies} \\
    $\val_i\Game_{\|v_0}$ (or $\val(v_0)$) & value & \cref{sec:two_players} \\
    $\dseal X$ & downward sealing & \cref{def_dseal} \\
    $\EE^\PP[X]$ & expectation & \cref{ssec:proba} \\
    $\OM^\PP[X]$ & optimistic risk measure  & \cref{def:opt_pess} \\
    $\M^\PP_{\beta\rho}[X]$ & entropic risk measure & \cref{def:er} \\
    $\PP(E)$ & probability & \cref{ssec:proba} \\
    $\PM^\PP[X]$ & pessimistic risk measure & \cref{def:opt_pess} \\
    $\X_i[X]$ & extreme risk measure & \cref{ssec:def_xr}
\end{tabular}
    \caption{Operators}
    \label{tab:operators}
\end{table}

\section{Graphs games} \label{sec:graph_games}

Throughout this thesis, we will use the word \emph{game} for infinite-duration turn-based quantitative games with complete information played on \emph{graphs}.

\subparagraph*{Graphs and paths.}
A \emph{graph} is an ordered pair $(V, E)$, where $V$ is a (often finite, but not always) set of \emph{vertices} and $E \subseteq V \times V$ is a set of \emph{edges}.
For the simplicity of writing, an edge $(v, w) \in E$ will often be written $vw$.
A \emph{path} in $(V, E)$ is a word $\alpha = \alpha_0 \alpha_1 \dots$, finite or infinite, over the alphabet $V$, such that for every $k$ such that $\alpha_k$ and $\alpha_{k+1}$ exist, we have $\alpha_k \alpha_{k+1} \in E$.
Given a path $\alpha$, we write $\Occ(\alpha)$ for the set of vertices that appear in $\alpha$, and $\Inf(\alpha)$ for the set of vertices that appear infinitely often in $\alpha$ (which is empty if $\alpha$ is finite).
We write $\first(\alpha)$ for the first vertex of $\alpha$, and $\last(\alpha)$ for the last vertex of $\alpha$ (if $\alpha$ is finite).

We say that the path $\alpha$ is \emph{simple} if no vertex occur more than once.
It is a \emph{cycle} if it is finite, and if we have $\last(\alpha)\first(\alpha) \in E$.
A \emph{lasso} is a path of the form $\alpha c^\omega$, where $\alpha$ is finite and $c$ is a cycle.
A \emph{simple lasso} is a lasso $\alpha c^\omega$ such that the path $\alpha c$ is simple.

Given a graph $(V, E)$, we denote by $\SC(V, E)$ the set of simple cycles in $(V, E)$.
Similarly, we denote by $\SConn(V, E)$ the set of \emph{strongly connected components} in $(V, E)$, i.e., the set of graphs $(V', E')$ with $V' \subseteq V$ and $E' = E \cap (V' \times V')$ such that for every two vertices $u, v \in V'$, there is a finite path $\alpha$ with $|\alpha| \geq 2$ from $u$ to $v$.

\subparagraph*{Games.} A game is a graph equipped with \emph{players}, each of them \emph{controlling} some of the vertices, and expressing preferences using a \emph{payoff function}.
A \emph{play} is then an infinite path in the graph, which can be seen as an infinite sequence of moves of a token on its vertices: when the token is on some vertex, the player controlling that vertex chooses to which vertex it moves, following one of the outgoing edges of that vertex, and so on, each play being associated with a payoff for each player.

\begin{defi}[Non-initialized game]\label{defi_game}
	A \emph{non-initialized game} is a tuple
	$\Game = \left(\Pi, V, (V_i)_{i \in \Pi}, E, \mu\right)$, with:
	
	\begin{itemize}
		\item a finite set $\Pi$ of \emph{players};
		
		\item a graph $(V, E)$, called the \emph{underlying graph} of $\Game$, in which every vertex has at least one outgoing edge;

		\item a partition $(V_i)_{i \in \Pi}$ of $V$, in which $V_i$ is the set of vertices \emph{controlled} by player $i$;
		
		\item a mapping $\mu$ called \emph{payoff function}, that maps each infinite path $\pi$ to the tuple $\mu(\pi) = (\mu_i(\pi))_{i \in \Pi} \in \RR^\Pi$ of the players' \emph{payoffs}.
	\end{itemize}
\end{defi}

A game is called \emph{prefix-independent} if for every finite path $h$ and every infinite path $\pi$ satisfying $\last(h)\first(\pi) \in E$, we have $\mu(h\pi) = \mu(\pi)$.
It is called \emph{Boolean} if for every infinite path $\pi$ and each player $i$, we have $\mu(\pi) \in \{0, 1\}$.
In such a context, we say that player $i$ \emph{wins} the play $\pi$ if $\mu_i(\pi) = 1$, and \emph{loses} otherwise.
When a game $\Game$ is given, we will often use the notations $\Pi$, $V$, $E$, $(V_i)_i$, and $\mu$, without necessarily recalling them.
We sometimes abuse notations and assimilate a game to its underlying graph, for example writing $\SC(\Game)$ instead of $\SC(V, E)$.

\begin{defi}[Initialized game]
	An \emph{initialized game} is a tuple $(\Game, v_0)$, often written $\Game_{\|v_0}$, where $\Game$ is a non-initialized game and $v_0 \in V$ is a vertex called \emph{initial vertex}.
\end{defi}

When the context is clear, we often use the word \emph{game} for both initialized and non-initialized games.

\begin{exa}\label{ex:first_game}
    We give in \Cref{fig:first_example_game} a first example of game with two players, player $\Circle$ and player $\Box$.
    Circled vertices belong to the former, squared ones to the latter.
    In this example, both players get the payoff $0$ in the path $a^\omega$ and in every path of the form $a^k b^\omega$, and the payoff $1$ in paths of the form $a^k b^\l c^\omega$.

    \begin{figure}
    \centering
    \begin{tikzpicture}[node distance=3cm]
        \node[vert,rectangle,rounded corners=7pt,dotted,minimum size=2.5cm] (aa) {$\begin{matrix}~\\~\\~\\~\\\mu_\circ = \mu_\Box = 0\end{matrix}$};
        \node[vert, initial] (a) at (0, 0) {$a$};
        \node[vert,rectangle,rounded corners=7pt,dotted,minimum size=2.5cm, right of=a] (bb) {$\begin{matrix}~\\~\\~\\~\\\mu_\circ = \mu_\Box = 0\end{matrix}$};
        \node[vert, rectangle, right of=a] (b) {$b$};
        \node[vert,rectangle,rounded corners=7pt,dotted,minimum size=2.5cm, right of=b] (cc) {$\begin{matrix}~\\~\\~\\~\\\mu_\circ = \mu_\Box = 1\end{matrix}$};
        \node[vert, right of=b] (c) {$c$};
        \path (a) edge[loop above] (a);
        \path (b) edge[loop above] (b);
        \path (c) edge[loop above] (c);
        \path (a) edge (b);
        \path (b) edge (c);
    \end{tikzpicture}
    \caption{A game}
    \label{fig:first_example_game}
\end{figure}
\end{exa}

\begin{defi}[Play, history]
	A \emph{play} (resp. \emph{history}) in the non-initialized game $\Game$ is an infinite (resp. finite) path in the graph $(V, E)$.
	A play (resp. history) in the initialized game $\Game_{\|v_0}$ is a play (resp. history) path in $\Game$ whose first vertex is $v_0$.
\end{defi}

	The set of plays (resp. histories) in the game $\Game$ (resp. the initialized game $\Game_{\|v_0}$) is denoted by $\Plays \Game$ (resp. $\Plays \Game_{\|v_0}, \Hist \Game, \Hist \Game_{\|v_0}$).
	We write $\Hist_i \Game$ (resp. $\Hist_i \Game_{\|v_0}$) for the set of histories in $\Game$ (resp. $\Game_{\|v_0}$) of the form $hv$, where $v$ is a vertex controlled by player $i$.
    We also write $\Hist_P \Game = \bigcup_{i \in P} \Hist_i \Game$ (and $\Hist_P \Game_{\|v_0} = \bigcup_{i \in P} \Hist_i \Game_{\|v_0}$) for every $P \subseteq \Pi$.

In order to obtain the plays they desire, the players choose their actions following a given \emph{strategy}.

\section{Strategies, strategy profiles}

\begin{defi}[Strategy, strategy profile]\label{def:strategies}
	A \emph{strategy} for player $i$ in the initialized game $\Game_{\|v_0}$ is a mapping $\sigma_i: \Hist_i \Game_{\|v_0} \to V$, such that $v\sigma_i(hv)$ is an edge of $(V, E)$ for every history $hv$.
	A history $h$ is \emph{compatible} with a strategy $\sigma_i$ if and only if $h_{k+1} = \sigma_i(h_0 \dots h_k)$ for all $k$ such that $h_k \in V_i$.
    A play $\pi$ is compatible with $\sigma_i$ if all its finite prefixes are.

	A \emph{strategy profile} for $P \subseteq \Pi$ is a tuple $\bsigma_P = (\sigma_i)_{i \in P}$, where each $\sigma_i$ is a strategy for player $i$ in $\Game_{\|v_0}$.
    A play or a history is \emph{compatible} with $\bsigma_P$ if it is compatible with each $\sigma_i$ for $i \in P$.
    A \emph{complete} strategy profile, typically written $\bsigma$, is a strategy profile for $\Pi$.
    Exactly one play is compatible with the strategy profile $\bsigma$: we call it its \emph{outcome} and write it $\< \bsigma \>$.
\end{defi}

When $i$ is a player and when the context is clear, we will often write $-i$ for the set $\Pi \setminus \{i\}$.
When $\btau_P$ and $\btau'_{P'}$ are two strategy profiles with $P \cap P' = \emptyset$, we write $(\btau_P, \btau'_{P'})$ for the strategy profile $\bsigma_{P \cup P'}$ such that $\sigma_i = \tau_i$ for $i \in P$, and $\sigma_i = \tau'_i$ for $i \in P'$.
In a strategy profile $\bsigma_P$, the $\sigma_i$'s domains are pairwise disjoint.
Therefore, we can consider $\bsigma_P$ as one function: for $hv \in \Hist \Game_{\|v_0}$ such that $v \in \bigcup_{i \in P} V_i$, we liberally write $\bsigma_P(hv)$ for $\sigma_i(hv)$ with $i$ such that $v \in V_i$.
The set of strategies for player $i$ (resp. of strategy profiles for the set $P$) in $\Game_{\|v_0}$ is written $\Strat_i\Game_{\|v_0}$ (resp. $\Strat_P\Game_{\|v_0}$).

For a given strategy profile $\bsigma$ for $\Pi$, for a given player $i$, a strategy $\sigma'_i$ will often be called \emph{deviation} for player $i$ from $\bsigma$, or sometimes simply \emph{deviation from $\sigma_i$}, when it is seen as an alternative to $\sigma_i$.

Note that our formalism does not allow the players to use randomization to decide which edge they take.
Randomized strategies will only be considered in \cref{chap:RSE,chap:XRSE}, in which we will give a more general definition.

\section{Stationary and finite-memory strategies} \label{sec:stationary_finite_memory}

As we have defined them, strategies can use all the information of a history to decide which edge must be taken: they are not limited in the \emph{memory} they use.
Such strategies cannot, in general, be simulated perfectly by a physical computer; and they are often hard to handle in algorithms, since there is no straightforward way to describe them with a finite number of bits.
In contrast, \emph{stationary strategies} are much simpler objects.

\begin{defi}
    A strategy $\sigma_i$ is \emph{stationary} if for every two histories $gu, hu \in \Hist_i\Game_{\|v_0}$, we have $\sigma_i(gu) = \sigma_i(hu)$.
\end{defi}

When the strategy $\sigma_i$ is stationary, we will often consider that it is defined in every game $\Game_{\|u}$, and simply write $\sigma_i(u)$ for $\sigma_i(hu)$.
We will also write $\Game[\sigma_i]$ for the non-initialized game obtained by removing all edges $uv$ with $u \in V_i$ and $v \neq \sigma_i(v_i)$, and $\Game_{\|v_0}[\sigma_i]$ for the same game, but initialized in $v_0$, and where all vertices that are no longer accessible from $v_0$ have been omitted.
Finally, for every game $\Game_{\|v_0}$ and each player $i$, we write $\Stat_i\Game_{\|v_0}$, or $\Stat\Game_{\|v_0}$ when the context is clear, for the set of stationary strategies for player $i$ in $\Game_{\|v_0}$.

As an intermediary object, we sometimes consider \emph{finite-memory strategies}, i.e., strategies that are induced by a \emph{memory structure}.

\begin{defi}[Memory structure]\label{def:memory_structure}
		A \emph{memory structure for player $i$} on a game $\Game$ is a tuple $\Mem = (Q, q_0, \Delta)$, where $Q$ is a finite set of \emph{states}, where $q_0 \in Q$ is the \emph{initial state}, and where $\Delta \subseteq (Q \times V_{-i} \times Q) \cup (Q \times V_i \times Q \times V)$ is a finite set of \emph{transitions}, such that:
        \begin{itemize}
            \item for every $(p, u, q, v) \in \Delta$, we have $uv \in E$;

            \item and for every $p \in Q$ and $u \in V$, there exists a transition $(p, u, q)$ or $(p, u, q, v) \in \Delta$.
            The memory structure $\Mem$ is \emph{deterministic} if for each $p$ and $u$, there exists exactly one such transition.
        \end{itemize}
		
		A strategy $\sigma_i$ in $\Game_{\|v_0}$ is \emph{induced} by $\Mem$ if there exists a mapping $h \mapsto q_h$ that maps every history $h$ in $\Game_{\|v_0}$ to a state $q_h \in Q$, such that for every $hv \in \Hist_{-i} \Game_{\|v_0}$, we have $(q_h, v, q_{hv}) \in \Delta$, and for every $hv \in \Hist_i \Game_{\|v_0}$, we have $(q_h, v, q_{hv}, \sigma_i(hv)) \in \Delta$.
		The set of strategies in $\Game_{\|v_0}$ induced by $\Mem$ is written $\Ind_{\|v_0}(\Mem)$.
		If $\Mem$ is deterministic, then there is exactly one strategy induced by $\Mem$; we call it a \emph{finite-memory} strategy.
\end{defi}

\begin{rem}
Stationary strategies are exactly the strategies that are induced by a memory structure with one state.
\end{rem}

Specialist readers may have noted that the definition of deterministic memory structures, outside of the context of game theory, corresponds to the classical notion of \emph{Mealy machine}.
Results about deterministic memory structures can be applied to \emph{programs}, which are supposed to run deterministically; we chose to take a more general definition to capture also \emph{protocols}, which may be given to an agent who would still have some room for manoeuvre in how they apply it.

We define analogously memory structures that induce a set of strategy profiles for several players, including for the whole set $\Pi$.
Note that, from such a memory structure inducing a set of strategy profiles $S$, one can extract a memory structure inducing the projection $\{\sigma_i \mid \bsigma \in S\}$ for some given player $i$, by replacing all transitions of the form $(p, u, q, v)$ with $u \not\in V_i$ by the transition $(p, u, q)$.
Note also that every memory structure $\Mem$ can be encoded with a finite number of bits: we write $\lv \Mem \rv$ for that number.

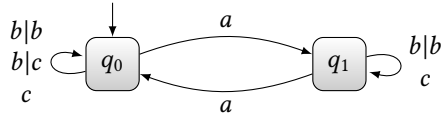
\begin{figure}
			\centering
			\begin{tikzpicture}
				\node[vert, rectangle, rounded corners, initial above] (q0) at (0, 0) {$q_0$};
				\node[vert, rectangle, rounded corners] (q1) at (3, 0) {$q_1$};
				
				\path (q0) edge[bend left=20] node[above] {$a$} (q1);
                \path (q1) edge[bend left=20] node[below] {$a$} (q0);
				\path (q0) edge[loop left] node[left] {$\begin{matrix} b|b\\ b|c\\ c \end{matrix}$} (q0);
                \path (q1) edge[loop right] node[right] {$\begin{matrix} b|b\\ c \end{matrix}$} (q1);
			\end{tikzpicture}
			\caption{A non-deterministic one-player memory structure} \label{fig:ex_1player_machine}
		\end{figure}

\begin{figure}
			\centering
			\begin{tikzpicture}
				\node[vert, rectangle, rounded corners, initial above] (q0) at (0, 0) {$q_0$};
				\node[vert, rectangle, rounded corners] (q1) at (3, 0) {$q_1$};
				
				\path (q0) edge node[above] {$b|b$} (q1);
				\path (q0) edge[loop left] node[left] {$\begin{matrix} a|a\\ c|c \end{matrix}$} (q0);
                \path (q1) edge[loop right] node[right] {$\begin{matrix} a|a\\ b|c\\ c|c \end{matrix}$} (q1);
			\end{tikzpicture}
			\caption{A deterministic multiplayer memory structure} \label{fig:ex_multiplayer_machine}
		\end{figure}
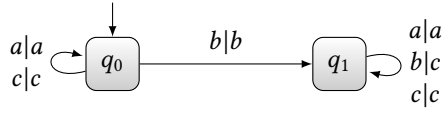

\begin{exa}
Figure~\ref{fig:ex_1player_machine} depicts a one-player memory structure on the game of Figure~\ref{fig:first_example_game}.
        Each arrow from a state $p$ to a state $q$ labeled $u|v$ denotes the existence of a transition $(p, u, q, v)$ (from the state $p$, the machine reads the vertex $u$, switches to the state $q$ and outputs the vertex $v$).
        Each arrow from a state $p$ to a state $q$ labeled $u$ denotes the existence of a transition $(p, u, q)$ (from $p$, the machine reads $u$, switches to $q$ and outputs nothing).
        It is a machine for player $\Box$, that is not deterministic: from the state $q_0$, reading the vertex $b$, the machine stays in $q_0$ but it can output either $b$ or $c$.
        The strategies that are compatible with it can be described as follows: when player $\Box$ has to play, if the vertex $a$ was seen an odd number of times, then he stays in $b$; in the opposite case, he can either stay in $b$ or eventually go to $c$.

Figure~\ref{fig:ex_multiplayer_machine} depicts a deterministic multiplayer memory structure on the same game.
The strategy profile that is compatible with it loops on the vertex $a$, and after a possible deviation of player $\Circle$ that would lead to the vertex $b$, loops once on $b$, before going to $c$.
\end{exa}
	

        \section{Two-player zero-sum games}\label{sec:two_players}

Although this document studies multiplayer games, we will sometimes use methods that bring us back to the more classical framework of two-player zero-sum games.
We will therefore need the following notions and results.

\begin{defi}[Zero-sum game]
	A game $\Game$, with $\Pi = \{i, j\}$, is \emph{zero-sum} if $\mu_j = -\mu_i$.
\end{defi}

Most of the games we will consider, be they zero-sum or not, are \emph{Borel games}.
	
\begin{defi}[Borel game]
	A game $\Game$ is \emph{Borel} if the function $\mu$, from the set $V^\omega$ equipped with the product topology to the Euclidian space $\RR^\Pi$, is Borel, i.e., if for every Borel set $B \subseteq \RR^\Pi$, the set $\mu^{-1}(B)$ is Borel.
\end{defi}

Two-players zero-sum Borel games have the following important property, called \emph{determinacy}.
	
\begin{lem}[\cite{BorelDeterminacy}] \label{lm_borel_determinacy}
	Let $\Game_{\|v_0}$ be a zero-sum Borel game, with $\Pi = \{i, j\}$. Then, we have the following equality:
	$$\sup_{\sigma_i} ~ \inf_{\sigma_j} ~ \mu_i\< \bsigma \> = \inf_{\sigma_j} ~ \sup_{\sigma_i} ~ \mu_i\< \bsigma \>.$$
\end{lem}

That quantity is called \emph{adversarial value}, or simply \emph{value}, of the game $\Game_{\|v_0}$, denoted by $\val_i\Game_{\|v_0}$.
\emph{Solving} the game $\Game_{\|v_0}$ means computing its value.
The concept of value can also be extended to multiplayer games: when $\Game_{\|v_0}$ is a multiplayer game and $i$ is a player, the quantity $\val_i\Game_{\|v_0}$ is the value of the game obtained by merging all the other players into one fictional \emph{adversarial} player, i.e. a player whose payoff function would be the opposite of player $i$'s one.
In other words, the quantity $\val_i\Game_{\|v_0}$ is the best payoff that player $i$ can ensure in the game $\Game_{\|v_0}$, whatever the other players do.
When $\Game$ is clear from the context and when $v_0 \in V_i$, we write $\val(v_0)$ for $\val_i\Game_{\|v_0}$, and call it simply \emph{value of the vertex} $v_0$.
When it exists, a strategy that ensures the payoff $\val_i\Game_{\|v_0}$ for player $i$ is called \emph{optimal}.
If $\Game_{\|v_0}$ is a Boolean game and if $\val_i\Game_{\|v_0} = 1$, it will also be called a \emph{winning} strategy.

Solving two-player zero-sum games will often be much simpler in games where stationary optimal strategies exist.
Let us give a sufficient condition to identify such games.

\begin{defi}[Shuffling]
    Let $\pi, \chi$ and $\xi$ be three plays in a game $\Game$.
    The play $\xi$ is a \emph{shuffling} of $\pi$ and $\chi$ if there exist two infinite sequences of indices $k_0 < k_1 < \dots$ and $\l_0 < \l_1 < \dots$ such that $\chi_0 = \pi_{k_0} = \chi_{\l_0} = \pi_{k_1} = \dots$, and:
    $$\xi = \pi_0 \dots \pi_{k_0-1} \chi_0 \dots \chi_{\l_0-1} \pi_{k_0} \dots \pi_{k_1-1} \chi_{\l_0} \dots \chi_{\l_1-1} \dots.$$
\end{defi}

\begin{defi}[Convexity, concavity]
    A payoff function $\mu_i: \Plays \Game \to \RR$ is \emph{convex} if for every shuffling $\xi$ of two plays $\pi$ and $\chi$, we have $\mu_i(\xi) \geq \min\{\mu_i(\pi), \mu_i(\chi)\}$.
    It is \emph{concave} if the function $-\mu_i$ is convex.
\end{defi}

\begin{lem} \label{lm:concave}
    In a two-player zero-sum game played on a finite graph, every player whose payoff function is concave has an optimal strategy that is stationary.
\end{lem}

\begin{proof}
    According to \cite{DBLP:conf/icalp/Kopczynski06}, this result is true for Boolean objectives.
    It follows that for every $x \in \RR$, if a player $i$, whose payoff function is concave, has a strategy that ensures $\mu_i(\pi) \geq x$ (understood as a Boolean objective), then they have a stationary one.
    Hence the equality:
    $$\val_i\Game_{\|v_0} = \sup_{\sigma_i \in \Stat\Game_{\|v_0}} ~ \inf_{\sigma_j} ~ \mu_i\< \bsigma \>.$$
    Since the underlying graph $(V, E)$ is assumed to be finite, there exists a finite number of stationary strategies, hence there exists a stationary strategy $\sigma_i$ that realizes the infimum above.
\end{proof}

\section{Equilibria and constrained existence problem}

In a game, among the strategy profiles that are available for all players, we will be interested in specific classes of strategy profiles that can be considered as more stable than others, because of guarantees they maintain when players deviate from their strategies.
Such classes of strategy profiles are usually called \emph{equilibria}.
Our contribution focuses mostly on \emph{Nash equilibria} and \emph{subgame-perfect equilibria}, and later on \emph{strong secure equilibria} and \emph{risk-sensitive equilibria}.
We will also study those notions in different classes of games.

However, the question that we ask will almost always be the same: does there exist an equilibrium, in a given game, that generates a payoff in a given interval for each player?
We give here a formal definition of that problem for some class $\Class$ of games, and for some general class $\Eq$ of strategy profiles, that must be understood as some notion of equilibrium.

\begin{prob}[Constrained existence problem]
    Given a game $\Game_{\|v_0} \in \Class$ and two vectors $\bx, \by \in (\QQ \cup \{\pm \infty\})^\Pi$, does there exist a strategy profile $\bsigma \in \Eq$ such that the inequality $\bx \leq \mu\< \bsigma \> \leq \by$ holds?
\end{prob}

The quantities $x_i$ are called \emph{lower thresholds}.
They are said to be \emph{effective} when they are different from $-\infty$.  
Similarly, the quantities $y_i$ are called \emph{upper thresholds}, and are considered \emph{effective} when they are different from $+\infty$.  
Note that thresholds are assumed to be rational, even in games where payoffs may not be, so that they can be represented using a finite number of bits.  
Most of this document is dedicated to studying the complexity of this problem, for various equilibrium notions $\Eq$ and game classes $\Class$.

%% file: 1NE.tex
The concept of \emph{Nash equilibria}, though under a different name, appears to have first been introduced by the French mathematician Antoine-Augustin Cournot~\cite{cournot1838recherches}.
Later, John Nash provided a more general definition~\cite{nash1951noncooperative} in a setting quite different from ours, as he was not considering graph games but matrix games, in the sense of \cref{ssec:intro_equilibria}.
In such games, he proved that what is now known as a \emph{Nash equilibrium} always exists, provided that players are allowed to \emph{randomize} over multiple actions---a behavior that is implicitly prohibited in our formalism.
This result is now widely recognized as \emph{Nash's theorem}.

Both Cournot's and Nash's work were motivated by economics, where their insights remain extensively used today.
For instance, if a strategy profile represents the behavior of competing firms, a Nash equilibrium corresponds to a stable situation: if those companies want to coordinate but do not trust each other, following an agreement that constitutes a Nash equilibrium ensures that no company will have an incentive to deviate from that agreement.

As multiplayer games became a topic of interest in computer science, researchers naturally focused on this notion.
The complexity of the constrained existence problem of Nash equilibria was therefore already known for several of the most classical classes of games that we study here.
We use however this chapter to present some of those results (\cref{thm:NE_parity,thm:NE_MP}), prove new ones (\cref{thm:ds_ne_hardness,thm:ds_ne_easiness,thm:NE_energy}), and introduce the classes of games that we will also study in \cref{part:SPE}.

\section{Definition}

Nash equilibria are strategy profiles from which no player can increase their payoff by deviating unilaterally.

\begin{defi}[Nash equilibrium]
	Let $\Game_{\|v_0}$ be a game.
    A strategy profile $\bsigma$ is a \emph{Nash equilibrium}, or \emph{NE} for short, in the game $\Game_{\|v_0}$ if and only if for each player $i$ and for every deviation $\sigma'_i$ for player $i$, we have the inequality $\mu_i\< \bsigma_{-i}, \sigma'_i \> \leq \mu_i\< \bsigma \>$.
\end{defi}

When the strategy $\bsigma$ is \emph{not} an NE, we call the deviations $\sigma'_i$ that do not observe the inequality above \emph{profitable} deviations.

\section{Parity games} \label{sec:NEparity}

Let us consider a system required to ensure a given condition, typically called a \emph{specification}, over the long run.
A concrete example would be a robot tasked with regularly delivering parcels arriving at a warehouse to an apartment block.

As a first abstraction, we assume that the robot has infinite carrying capacity.
At any given moment, it therefore has two available actions: either do nothing, or take all the parcels in the warehouse and deliver them to the apartment block.
A second abstraction consists in disregarding the delivery time, as long as every parcel that appears in the warehouse is eventually delivered.
A third abstraction, which is central to all our formalisms, is to assume that the robot operates indefinitely and can thus plan its strategy over an infinite horizon.

The company managing the robot and the warehouse may then choose to care about their customers only as long as they continue placing orders: if they stop ordering, delivering the remaining parcels is no longer necessary.
In this framework, the robot's objective can be expressed in a simple way: if parcels appear in the warehouse infinitely often, then the robot must infinitely often carry them to the apartment block.

Now, let us extend this setting by introducing a castle, which may also receive deliveries.
Since the castle owner has subscribed to a premium service, deliveries to the castle take precedence over those to the apartment block.
The robot's objective then becomes the following: if parcels for the castle appear in the warehouse infinitely often, the robot must infinitely often deliver them to the castle.
Otherwise, and only in that case, it must fulfill the previously defined objective.

We observe that the robot's objective is defined using \emph{priorities} among different possible situations.
A natural and elegant way to formalize such an objective is through a \emph{parity} condition, a concept widely used in automata theory.
Indeed, it is well known that $\omega$-regular languages are precisely those recognized by a non-deterministic automaton whose acceptance condition is a parity condition.

\subsection{Definition}

Given a graph $(V, E)$, a \emph{color mapping} on $(V, E)$ is a mapping $\kappa: V \to \NN$.

	\begin{defi}[Parity]
		The game $\Game$ is a \emph{parity game} if for each player $i$, there exists a color mapping $\kappa_i$ on the underlying graph of $\Game$ such that for every play $\pi$, we have $\mu_i(\pi) = 1$ if the color $\min_{v \in \Inf(\pi)} \kappa_i(v)$ is even, and $\mu_i(\pi) = 0$ if it is odd.
	\end{defi}

A \emph{Büchi game} is a parity game in which all colors are $0$ or $1$: then, each player's objective consists simply in visiting infinitely often vertices of color $0$.
Dually, a \emph{co-Büchi game} is a parity game in which all colors are $1$ and $2$: then, each player's objective consists in, eventually, avoiding vertices of color $1$.

\subsection{Results}

The constrained existence problem of Nash equilibria in parity games has already been studied by Michael Ummels.

\begin{thm}[\cite{DBLP:conf/fossacs/Ummels08}]\label{thm:NE_parity}
    The constrained existence problem of Nash equilibria in parity games is $\NP$-complete.
    Hardness still holds in co-Büchi games, and when there is no effective lower threshold, and only one effective upper threshold.
    In Büchi games, the problem is in $\PTime$.
\end{thm}

Actually, in the cited article, Michael Ummels shows hardness with one effective lower (and not upper) threshold, but the result given here can be quickly showed by adding one player to his reduction---and we will need that result under this form in \cref{chap:verif}.

\section{Mean-payoff games}

Imagine a system designed to generate a specific resource---an automated solar plant, for example, that produces green hydrogen. The plant's objective would not be to maintain continuous hydrogen production. At times, such as when there is no sunlight, production may halt, but that downtime could be used for maintenance, like cleaning the equipment.
If the plant is located outside Belgium, where long periods of sunlight are possible, maintenance would still be necessary, even if it temporarily reduces production. What matters, however, is the \emph{average} production over time---periods of peak output can compensate for necessary pauses.
Similarly, an initially long phase of low or no production may not be an issue if it ultimately leads to higher efficiency in the long run. The key concept here is the \emph{asymptotic average} production, which is precisely what \emph{mean-payoff objectives} capture.

\subsection{Definition}

Contrary to parity games, mean-payoff games are typically \emph{quantitative games}.
The players receive a reward at each turn, and they aim at maximizing their limit average reward.

Given a graph $(V, E)$, a \emph{reward function} on $(V, E)$ is a mapping $r: E \to \QQ$.

\begin{defi}[Mean-payoff] \label{def:mp}
	    In a graph $(V, E)$, we define for each reward mapping $r$ the \emph{mean-payoff function} $\MP_r: h_0 \dots h_n \mapsto \frac{1}{n} \sum_k r\left(h_k h_{k+1}\right)$.
	    Then, we write
	    $\MPi_r (\pi) = \liminf_n \MP_r(\pi_{\leq n}).$
		The game $\Game$ is a \emph{mean-payoff game} if there exists a tuple $(r_i)_{i \in \Pi}$ of reward mappings such that for each player $i$, we have $\mu_i = \MPi_{r_i}$.
	\end{defi}

When the context is clear, we write $\MP_i$ for $\MP_{r_i}$, and $\MPi_i$ for $\MPi_{r_i}$.
We then also write $\MP(\pi)$ for the tuple $(\MP_i(\pi))_{i \in \Pi}$.

\subsection{Results}

Again, the constrained existence problem of Nash equilibria in mean-payoff games has already been studied by Michael Ummels.

\begin{thm}[\cite{Ummels2011}]\label{thm:NE_MP}
    The constrained existence problem of Nash equilibria in mean-payoff games is $\NP$-complete.
    Hardness still holds when all rewards are $0$ and $1$, and when there is no effective lower threshold and only one effective upper threshold.
\end{thm}

Again, Michael Ummel's proved hardness with one lower threshold, but a quick modification of his proof gives us the desired result.

\section{Discounted-sum games}

Let us return to our example of an automated solar plant producing hydrogen, which we used to illustrate mean-payoff games.
The assumption that only the \emph{asymptotic} average reward matters may seem bold: in practice, as time passes, uncertainties grow, and one might prefer to secure reasonable short-term rewards rather than rely on promises of greater gains in the distant future: \emph{a bird in the hand is worth two in the bush.}
This idea is captured by \emph{discounted-sum} objectives.

\subsection{Definition}

	\begin{defi}[Discounted-sum]\label{def:ds}
	    In a graph $(V, E)$, we define for each reward mapping $r$ and each \emph{discount factor} $\beta \in (0, 1)$ the \emph{discounted-sum function} $\ds_r^\beta: h \mapsto \sum_k \beta^k r(h_kh_{k+1})$.
		Then, we write
	    $\ds^\beta_r (\pi) = \lim_n \ds^\beta_r(\pi_{\leq n}).$
		The game $\Game$ is a \emph{discounted-sum game} if there exists a discount factor $\beta \in (0, 1) \cap \QQ$ and a tuple $(r_i)_{i \in \Pi}$ of reward mappings such that for each $i$ and every $\pi$, we have $\mu_i(\pi) = \ds^\beta_{r_i}(\pi)$.
		When the context is clear, we write $\ds_i$ for $\ds_{r_i}^\beta$.
	\end{defi}

\subsection{The target discounted-sum problem}

Discounted-sum objectives are closely related to the following problem, which is as simple to state as it is difficult to analyze.

\begin{prob}[Target Discounted-Sum Problem]
Given four quantities $\beta, a, b, t \in \mathbb{Q}$ with $0 < \beta < 1$, is there a sequence $(u_n)_{n \in \mathbb{N}} \in \{a, b\}^\omega$ such that $\sum_{n \in \mathbb{N}} u_n \beta^n = t$?
\end{prob}

Although this problem naturally arises in various fields, the Target Discounted-Sum ($\TDS$) problem is surprisingly difficult to solve, and its decidability remains an open question.
For further details, the interested reader may refer to~\cite{DBLP:conf/lics/BokerHO15}.
We will now show that the problem we wish to study is at least as challenging.

\subsection{Hardness result}
	
	\begin{thm} \label{thm:ds_ne_hardness}
		The TDS problem reduces to the constrained existence problem of NEs in discounted-sum games.
        This hardness result still holds when there is no effective lower threshold, and only one effective upper threshold.
	\end{thm}
	
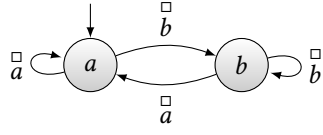
\begin{figure}
			\centering
			\begin{tikzpicture}[initial text={}]
				\node[vert, initial above] (a) at (0, 0) {$a$};
				\node[vert] (b) at (2, 0) {$b$};
				\path (a) edge[bend left=20] node[above] {$\stackrel{\Box}{b}$} (b);
				\path (b) edge[bend left=20] node[below] {$\stackrel{\Box}{a}$} (a);
				\path (b) edge[loop right] node[right] {$\stackrel{\Box}{b}$} (b);
				\path (a) edge[loop left] node[left] {$\stackrel{\Box}{a}$} (a);
			\end{tikzpicture}
			\caption{A game constructed from an instance of the TDS problem}
			\label{fig_reduction_tds1}
		\end{figure}

\begin{figure}
			\centering
			\begin{tikzpicture}[initial text={}]
                \newcommand{\deltay}{1.5}
				\node[vert, rectangle, initial left] (v0) at (0, 2*\deltay) {$v_0$};
				\node[vert, rectangle] (v1) at (0, 0*\deltay) {$v_1$};
				\node[vert, diamond] (v2) at (2, 2*\deltay) {$v_2$};
				\node[vert, diamond] (v3) at (2, 0*\deltay) {$v_3$};
				\node[vert] (a) at (4, 2*\deltay) {$a$};
				\node[vert] (b) at (4, 0*\deltay) {$b$};
				\path (a) edge[bend left=20] node[right] {$\stackrel{\circ}{-1}  \, \stackrel{\Box}{b}  \, \stackrel{\diamond}{-b}$} (b);
				\path (b) edge[bend left=20] node[left] {$\stackrel{\circ}{-1}  \, \stackrel{\Box}{a}  \, \stackrel{\diamond}{-a}$} (a);
				\path (b) edge[loop right] node[right] {$\stackrel{\circ}{-1}  \, \stackrel{\Box}{b}  \, \stackrel{\diamond}{-b}$} (b);
				\path (a) edge[loop right] node[right] {$\stackrel{\circ}{-1}  \, \stackrel{\Box}{a}  \, \stackrel{\diamond}{-a}$} (a);
				\path (v0) edge (v1);
				\path (v0) edge (v2);
				\path (v2) edge (v3);
				\path (v2) edge (a);
				\path (v1) edge[loop below] node[below] {$\stackrel{\Box}{t \lambda (1-\lambda)}$} (v1);
				\path (v3) edge[loop below] node[below] {$\stackrel{\diamond}{-t (1 - \lambda)}$} (v3);
			\end{tikzpicture}
			\caption{Another game constructed from the same instance of $\TDS$}
			\label{fig_reduction_tds2}
		\end{figure}
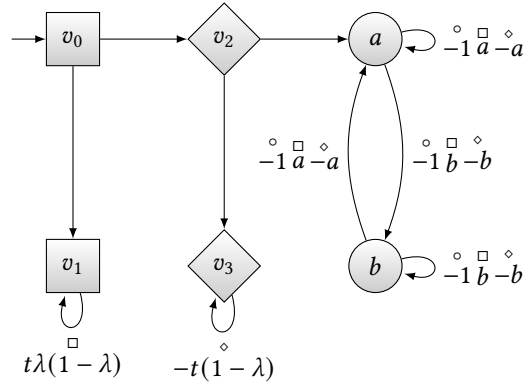

	\begin{proof}
	    Let $a, b, t \in \QQ$ and let $\beta \in (0, 1) \cap \QQ$ form an instance of the TDS problem.
        We construct from those inputs a discounted-sum game $\Game_{\|a}$, depicted by Figure~\ref{fig_reduction_tds1}, with discount factor $\beta$.
        Player $\Circle$'s rewards are zero on all edges.
		In that game, we claim that there exists an NE $\bsigma$ with $t \leq \mu_\Box\< \bsigma \> \leq t$, if and only if $a, b, t$, and $\beta$ form a positive instance of the TDS problem.

\begin{itemize}
        \item Indeed, if there exists an NE $\bsigma$ satisfying $t \leq \mu_\Box\< \bsigma \> \leq t$, i.e. $\mu_\Box\< \bsigma \> = t$, let $\pi = \< \bsigma \>$.
        Then, the sequence $(u_n)_n = (\pi_n)_n$ is such that $\sum_n u_n \beta^n = \mu_\Box(\pi) = t$.

        \item Conversely, let us note that player $\Circle$ is the only player who actually makes choices, and that her payoff is $0$ in every play.
        Consequently, every strategy profile in that game is an NE.
        Thus, if there exists a sequence $(u_n)_n \in \{a, b\}^\NN$ with $\sum_n u_n \beta^n = t$, then that sequence also defines a play $\pi$ with $\mu_\Box(\pi) = t$, and every strategy profile $\bsigma$ with $\< \bsigma \> = \pi$ is an NE.
\end{itemize}

        To establish hardness even with only one effective upper threshold, we can extend this reasoning to the slightly more involved construction depicted by \Cref{fig_reduction_tds2}.
	\end{proof}

\subsection{Co-recursive enumerability}

The previous theorem suggests that designing algorithms to solve these problems lies beyond the scope of this thesis.
However, we will now show that, like the TDS problem, our problem is co-recursively enumerable.
The key idea is as follows: a fundamental property of discounted-sum objectives is that when a play yields a payoff outside a given interval for some player, there exists a prefix of that play after which the gap becomes irrecoverable. Beyond this point, we can stop reading the play and confidently conclude that the payoff will fall outside the interval.
Therefore, although strategy profiles are generally infinite objects and exist in uncountably many variations, profitable deviations can be identified by analyzing their behavior over a finite (though unbounded) number of histories.

	\begin{thm} \label{thm:ds_ne_easiness}
		The constrained existence problem of NEs in discounted-sum games is co-recursively enumerable.
	\end{thm}

\begin{proof}
We present here a semi-algorithm that recognizes negative instances of our problem.
	
		Given a discounted-sum game $\Game_{\|v_0}$ with discount factor $\beta$, a player $i$ and two threshold vectors $\bx$ and $\by$, we give an algorithm that stops if and only if there exists no NE $\bsigma$ in $\Game_{\|v_0}$ with $\bx \leq \mu_i\< \bsigma \> \leq \by$.
		But first, let us give a preliminary result that will justify the correctness of our result.
		
		\paragraph{Preliminary result}
        
        We show here the following lemma.

        \begin{lem}\label{lm:ds_preliminary}
            For every play $\pi$, each player $j$ and every index $n$, we have:
				$$\mu_j(\pi) \in \left[ \ds_j\left(\pi_{\leq n}\right) - M \beta^n,    \ds_j\left(\pi_{\leq n}\right) + M \beta^n \right],$$
				where:
            $$M = \frac{1}{1 - \beta} \max_{uv \in E} ~ \max_{j \in \Pi} ~ |r_j(uv)|$$
            is a bound on the payoff (in absolute value) of every player.
        \end{lem}
        
        \begin{proof}
			Let us proceed by induction on $n$.

            \subparagraph{Base case.}
			When $n = 0$, for every play $\pi$ starting from $v_0$ and each player $j$, we have:
			\begin{align*}
				\mu_j(\pi) & = \sum_k r_j(\pi_k \pi_{k+1}) \beta^k \\
				& \leq \sum_k \beta^k \max_{uv \in E} \max_{j \in \Pi} |r_j(uv)| = M
			\end{align*}
			and symetrically $\mu_j(\pi) \geq -M$, which constitutes the desired interval since $\ds_i(\pi_0) = 0$.

            \subparagraph{Inductive case.}
			Now, if the desired property is true for $n \geq 0$, let us prove that it is true for $n + 1$.
			Let $v \pi$ be a play.
			By induction hypothesis, we have $\mu_j(\pi) \in \left[ \ds_j(\pi_{\leq n}) - M \beta^n, \ds_j(\pi_{\leq n}) + M \beta^n \right].$
			Hence:
			\begin{align*}
				\mu_j(v\pi) & = r_j(v\pi_0) + \sum_k r_j(\pi_k \pi_{k+1})\beta^{k+1} \\
				& = r_j(v\pi_0) + \beta \mu_j(\pi) \\
				& \geq r_j(v\pi_0) + \beta \ds_j(\pi_{\leq n}) - M \beta^{n+1} \\
				& = \ds_j(v_0\pi_{\leq n}) - M \beta^{n+1},
			\end{align*}
			and analogously for the upper bound.
            \end{proof}
			
			\paragraph{Algorithm}
			
			Let $\left(h^{(n)}\right)_{n \in \NN}$ be a recursive enumeration of the nonempty histories in $\Game_{\|v_0}$ by increasing order of lengths (we can, for example, order histories of the same length with the lexicographic order induced by some arbitrary order on vertices).
			
			Now, let $T$ be the infinite tree whose nodes of depth $n+1$ are all possible $n$-uples $\left(\bsigma(h^{(0)}), \dots, \bsigma(h^{(n)})\right)$, and where the children of the node $\left(\bsigma(h^{(0)}), \dots, \bsigma(h^{(n)})\right)$ are the nodes of the form:
            $$\left(\bsigma(h^{(0)}), \dots, \bsigma(h^{(n)}), \bsigma(h^{(n+1)})\right).$$
			Thus, every node partially defines a complete strategy profile $\bsigma$, and every infinite branch entirely defines it---and conversely, every complete strategy profile is defined by an infinite branch.
			
			A node $\left(\bsigma(h^{(0)}), \dots, \bsigma(h^{(n)})\right)$ is called \emph{Nash-irrational} if there exist two indices $\l, m \leq n$, with $|h^{(\l)}| = |h^{(m)}| = p$, an index $k \leq p$ and a player $i$ such that:
			\begin{itemize}
				\item we have $h^{(\l)}_{\leq k} = h^{(m)}_{\leq k}$;
				
				\item the history $h^{(\l)}$ is compatible with the strategy profile $\bsigma$ that is partially defined by the node;
				
				\item the history $h^{(m)}$ is compatible with the strategy profile $\bsigma_{-i}$;
				
				\item and finally, we have:
				$$\ds_i\left(h^{(m)}\right) - \ds_i\left(h^{(\l)}\right) > 2M \beta^{p-1}.$$
			\end{itemize}
			
			The same node is called \emph{off-topic} if for some player $i$ we have $\ds_i\left(h\right) > y_i + M \beta^{|h|-1}$, or $\ds_i\left(h\right) < x_i - M \beta^{|h|-1}$
   where $h$ is the longest history that is compatible with $\bsigma$ as partially defined by the node.
			
			Our algorithm consists in constructing the tree $T'$, obtained from the tree $T$ by cutting every branch after the first Nash-irrational or off-topic node, and declaring the instance to be negative once that construction is finished.

			\paragraph{Correctness}

            To show that our algorithm is correct, we must prove that the tree $T'$ will be finite if and only if we have a negative instance of the constrained existence problem.
            By K\H{o}nig's lemma, that will be the case if and only if every branch is finite, i.e., if and and only if every branch of the tree $T$ contains either an off-topic or a Nash-irrational node.
            Therefore, we will be done if we prove that, given a branch of $T$ and the corresponding strategy profile $\bsigma$, the branch contains no such node if and only if the strategy profile $\bsigma$ is an NE and satisfies $\bx \leq \mu\< \bsigma \> \leq \by$.
            
\subparagraph{Off-topic nodes.}            
			Using \cref{lm:ds_preliminary}, we know that we have $\bx \leq \mu_i\< \bsigma \> \leq \by$ if and only if the corresponding branch contains no off-topic node.

            \subparagraph{If the branch contains a Nash-irrational node, then the strategy profile $\bsigma$ is not an NE.}
			If the branch contains a Nash-irrational node $\left(\bsigma(h^{(0)}), \dots, \bsigma(h^{(n)})\right)$.
			Let us use the notations $k, \l, m,$ and $i$ from the definition of Nash-irrationality.
			Let us define $\pi = \< \bsigma \>$: note that the history $h^{(\l)}$ is the prefix of length $p$ of the play $\pi$.
			Similarly, let us extend the history $h^{(m)}$ into some play $\pi'$, compatible with the strategy profile $\bsigma_{-i}$.
			By \cref{lm:ds_preliminary}, the inequality:
			$$\ds_i\left(h^{(m)}\right) - \ds_i\left(h^{(\l)}\right) > 2 M \beta^{p-1}$$
			implies $\mu_j(\pi') > \mu_j(\pi)$, and therefore the strategy profile $\bsigma$ is not an NE.

            \subparagraph{If the strategy profile $\bsigma$ is not an NE, then the branch contains a Nash-irrational node.}
			If $\bsigma$ is not an NE, then there exists a player $i$ and a strategy $\sigma'_i$ such that we have:
			$$\mu_i\< \bsigma \> < \mu_i\< \bsigma_{-i}, \sigma'_i \>.$$
			
			Then, let $\pi = \< \bsigma \>$, and let $\pi' = \< \bsigma_{-i}, \sigma'_i \>$: since $\mu_i(\pi) < \mu_i(\pi')$, by \cref{lm:ds_preliminary}, there exists an index $k$ such that:
			$$\ds_i\left(\pi_{<k}\right) + M \beta^{k-1} < \ds_i\left(\pi'_{<k}\right) - M \beta^{k-1}.$$
			Let now $\l$ and $m$ be the indices such that $h^{(\l)} = \pi_{<k}$, and $h^{(m)} = \pi'_{<k}$.
			Then, we have $\ds_j\left(h^{(m)}\right) - \ds_j\left(h^{(\l)}\right) > 2M \beta^{k-1}$: along the branch corresponding to $\bsigma$, the node of depth $\max\{\l, m\}$ is  Nash-irrational.
			Which ends the proof.
\end{proof}

\section{Energy games}

Let us revisit the example of our green hydrogen plant, but now assume that it is connected to another system that consumes the hydrogen. For instance, the delivery robot from \cref{sec:NEparity}, which arrives whenever necessary to collect fuel from a tank where the station stores the hydrogen it produces.
The plant's goal is no longer to maximize hydrogen production, which would after all be a rather productivist objective.
Instead, its aim is to ensure that the robot will always find the required amount of fuel available---in other words, to avoid a situation in which the robot would attempt to take more fuel than what is stored in the tank.
Such scenarios are modeled with \emph{energy objectives}.

\subsection{Definition}

Like mean-payoff and discounted-sum objectives, energy objectives are based on reward functions.
However, energy objectives are Boolean.

\begin{defi}[Energy]\label{def:energy}
    In a graph $(V, E)$, we associate to each reward mapping $r$ the \emph{energy level function} $\EL_r$ that maps a finite path $h = h_0 \dots h_n$ to the element of $\NN \cup \{\bot\}$ defined as follows.
    \begin{itemize}
        \item If the path has length $1$, we define $\EL_r(h_0) = 0$.
        \item Otherwise, if we have $\EL_r(h_{\leq n}) \neq \bot$ and $\EL_r(h_{\leq n}) + r(h_nh_{n+1}) \geq 0$, we define $\EL_r(h_{\leq n+1}) = \EL_r(h_{\leq n}) + r(h_nh_{n+1})$.
        \item In every other case, we define $\EL_r(h_{\leq n+1}) = \bot$.
    \end{itemize}

	The game $\Game$ is an \emph{energy game} if there exists a tuple $(r_i)_{i \in \Pi}$ of reward mappings such that for each player $i$ and every play $\pi$, we have $\mu_i(\pi) = 0$ if we have $\EL_{r_i}(\pi_{\leq n}) = \bot$ for some $n$, and $\mu_i(\pi) = 1$ otherwise.
	When the context is clear, we write $\EL_i$ for $\EL_{r_i}$.
\end{defi}

\subsection{Two-counter machines}

We will see in \cref{ssec:ne_energy_results} that energy games can be used to encode \emph{counter machines}.
Let us first define that model.

\begin{defi}[Two-counter machine]
A \emph{two-counter machine} is a tuple
	$\Kount = \left(Q, q_0, q_\f, \Delta_{C_1^+}, \Delta_{C_2^+}, \Delta_{C_1^-}, \Delta_{C_2^-}\right)$, with:
    \begin{itemize}
        \item a finite set $Q$ of \emph{states};

        \item an \emph{initial state} $q_0 \in Q$;

        \item a \emph{final state} $q_\f$;

        \item two sets $\Delta_{C_1^+}, \Delta_{C_2^+} \in Q \times Q$ of \emph{incremental transitions};

        \item and two sets $\Delta_{C_1^-}, \Delta_{C_2^-} \subseteq Q \times Q \times Q$ of \emph{test transitions};
    \end{itemize}
	such that every state $q \in Q \setminus \{q_\f\}$ admits exactly one outgoing transition (either incremental or test), and such that the state $q_\f$ admits none.
    We call \emph{sequence of transitions} a finite or infinite sequence $\eta_0\eta_1\eta_2\dots$ such that we have $\eta_0 = q_0$ and that for each $k \in \NN$, there is a \emph{counter} $C \in \{C_1, C_2\}$ such that we have either $(\eta_k, \eta_{k+1}) \in \Delta_{C^+}$, $(\eta_k, \eta_{k+1}, \cdot) \in \Delta_{C^-}$, or $(\eta_k, \cdot, \eta_{k+1}) \in \Delta_{C^-}$.
    
	We define the \emph{interpretation} $\hC$ of each counter $C$ on sequences of transitions as follows:

\begin{itemize}
    \item on the one-state sequence of transitions, we define $\hC(q_0) = 0$;

    \item if we have $(q_{n-1}, q_n) \in \Delta_{C^+}$, we define $\hC(q_0 \dots q_n) = \hC(q_0 \dots q_{n-1}) + 1$;

    \item if we have $(q_{n-1}, q_n, \cdot) \in \Delta_{C^-}$, we define $\hC(q_0 \dots q_n) = \hC(q_0 \dots q_{n-1}) - 1$;

    \item if we have $(q_{n-1}, \cdot, q_n) \in \Delta_{C^-}$, we define $\hC(q_0 \dots q_n) = \hC(q_0 \dots q_{n-1})$.
\end{itemize}

	A \emph{run} of the machine $\Kount$ is a sequence of transition, either infinite or ending with $q_\f$, such that for every index $k$:
    \begin{itemize}
        \item if we have $(\eta_k, \eta_{k+1}, \cdot) \in \Delta_{C^-}$, then we have $\hC(\eta_0 \dots \eta_k) > 0$;

        \item and if we have $(\eta_k, \cdot, \eta_{k+1}) \in \Delta_{C^-}$, then we have $\hC(\eta_0 \dots \eta_k) = 0$.
    \end{itemize}
	Note that each two-counter machine admits exactly one run.
	A machine \emph{halts} if that unique run is finite.
\end{defi}
	
\begin{figure}
			\caption{A machine that does not halt on $(0, 0)$}
				\centering
				\begin{tikzpicture}[initial text={}]
					\node[vert, initial left] (q0) at (0, 0) {$q_0$};
					\node[vert] (q1) at (2, 0) {$q_1$};
					\node[vert, double] (qf) at (4, 0) {$q_\f$};
					
					\path[bend left] (q0) edge node[above] {$C_1^{+}$} (q1);
					\path[bend left] (q1) edge node[below] {$C_1^{-}$} (q0);
					\path (q1) edge node[above] {$C_1=0$} (qf);
				\end{tikzpicture}
				\label{fig_counter_machine2}	
		\end{figure}
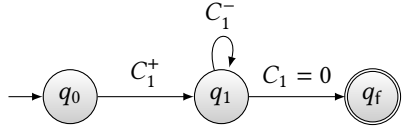

	\Cref{fig_counter_machine1,fig_counter_machine2} depict two examples of two-counter machines.
		The arrow with label $C_1^{+}$ from $q_0$ to $q_1$ indicates a transition $(q_0, q_1) \in \Delta_{C_1^{+}}$.
		The arrows with the label $C_1^{-}$ from $q_1$ to itself and with the label $C_1 = 0$ from $q_1$ to $q_\f$ indicate a transition $(q_1, q_1, q_\f) \in \Delta_{C_1^{-}}$.
		
		\begin{figure}
			\caption{A machine that halts on $(0, 0)$}
				\centering
				\begin{tikzpicture}[initial text={}]
					\node[vert, initial left] (q0) at (0, 0) {$q_0$};
					\node[vert] (q1) at (2, 0) {$q_1$};
					\node[vert, double] (qf) at (4, 0) {$q_\f$};
					
					\path (q0) edge node[above] {$C_1^{+}$} (q1);
					\path (q1) edge [loop above] node[above] {$C_1^{-}$} (q1);
					\path (q1) edge node[above] {$C_1=0$} (qf);
				\end{tikzpicture}
				\label{fig_counter_machine1}
        \end{figure}

We will use the following well-known result:

\begin{thm}[\cite{minsky1961recursive}]
    The problem of deciding whether a given two-counter machine halts is undecidable, and in particular not co-recursively enumerable.
\end{thm}

\subsection{Results}\label{ssec:ne_energy_results}

In energy games, our problem is undecidable, since a counter machine can be encoded in an energy game.
However, while we showed co-recursive enumerability in discounted-sum games, we will show that in energy games, it is recursively enumerable.

\begin{thm}\label{thm:NE_energy}
    The constrained existence problem of Nash equilibria in energy games is undecidable, even with no effective lower threshold and only one effective upper threshold, and recursively enumerable.
\end{thm}

\begin{proof}
    \paragraph{Undecidability.}

    We show undecidability by reduction from the halting problem of a two-counter machine.

\begin{figure}
    \centering
    \begin{subfigure}[b]{0.5\textwidth}
			\centering
			\begin{tikzpicture}
				\node (1) at (3, 0) {};
				\node[ownedVert=$C_1^\top$, initial left] (q0) at (0, 0) {$q_0^1$};
                \node[ownedVert=$C_2^\top$] (q0') at (2, 0) {$q_0^2$};
				\path (q0) edge (q0');
                \path (q0') edge (1);
				\path (q0) edge[loop above] (q0);
                \path (q0') edge[loop above] (q0');
			\end{tikzpicture}
			\caption{Initial state}
			\label{fig_gadget_initial_app}
		\end{subfigure}
		\begin{subfigure}[b]{0.4\textwidth}
			\centering
			\begin{tikzpicture}
				\node[vert, initial above] (qf) at (0, 0) {$q_\f$};
				\path (qf) edge[loop right] node[right] {$\stackrel{C_1^\bot}{-1}\,\stackrel{C_2^\bot}{-1}\,\stackrel{\Witness}{-1}$} (qf);
			\end{tikzpicture}
			\caption{Final state}
			\label{fig_gadget_final_app}
		\end{subfigure}
		\begin{subfigure}[b]{0.35\textwidth}
			\centering
			\begin{tikzpicture}
				\node[vert, initial left] (q) at (0, 0) {$q$};
				\node (2) at (1, 0) {};
				\path (q) edge node[above] {$\stackrel{C^\top}{1}\,\stackrel{C^\bot}{1}$} (2);
			\end{tikzpicture}
			\caption{Incrementations}
			\label{fig_gadget_incrementation_app}
		\end{subfigure}
	\begin{subfigure}[b]{0.55\textwidth}
		\centering
		\begin{tikzpicture}
            \newcommand{\deltay}{1}
			\node (1) at (-1, 0) {};
			\node[ownedVert=$C^\top$] (q) at (0, 0) {$q$};
			\node (2) at (2, 0*\deltay) {(if $C > 0$)};
			\node[ownedVert=$C^\bot$] (q') at (0, 2*\deltay) {$q'$};
			\node (3) at (2, 2*\deltay) {(if $C = 0$)};
			\node[vert] (s) at (-1.2, 2*\deltay) {$\triangle$};
			
			\path (1) edge (q);
			\path (q) edge node[above] {$\stackrel{C^\top}{-1}\,\stackrel{C^\bot}{-1}$} (2);
			\path (q) edge (q');
			\path (q') edge (3);
			\path (q') edge node[above] {$\stackrel{C^\bot}{-1}$} (s);
			\path (s) edge [loop left] (s);
		\end{tikzpicture}
		\caption{Tests}
		\label{fig_gadget_test_app}
	\end{subfigure}
    \caption{Gadgets}
    \label{fig_gadget_app}
\end{figure}

	Let $\Kount$ be a two-counter machine.
	We define an energy game $\Game_{\|q_0^1}$ with five players (players $C_1^\top$, $C_1^\bot$, $C_2^\top$, $C_2^\bot$, and $\Witness$, called \emph{Witness}) by assembling the gadgets presented in Figure~\ref{fig_gadget_app}.
    The rewards that are not presented are equal to $0$, and the players controlling relevant vertices are written in black boxes.
	For each state of $\Kount$, we define from one to two vertices, plus the additional vertex $\triangle$.
	Then, a play in $\Game_{\|q_0^1}$ that does not reach the vertex $\triangle$ simulates a sequence of transitions of $\Kount$, that is a valid run if and only if each time the test gadget is visited, the player $C^\top$ (with $C \in \{C_1, C_2\}$) controlling $q$ goes to $q'$ if and only if her energy level is $0$.
    Then, at each step, the counter $C_i$ is captured by the energy level of player $C_i^\top$, always equal to the energy level of player $C_i^\bot$.
	Let us now prove that the game $\Game_{\|q_0^1}$ admits an NE where Witness loses if and only if the machine $\Kount$ terminates.
	
	\subparagraph{If there is an NE where Witness loses, then the machine $\Kount$ terminates.}
		
		Let $\bsigma$ be an NE such that $\mu_\Witness\< \bsigma \> = 0$.
        Let $\pi = \< \bsigma \>$: that play is then lost by Witness.
        Since the only edge that makes Witness lose energy is the edge $q_\f q_\f$, we deduce that the play $\pi$ reaches the vertex $q_\f$, and therefore simulates a terminating sequence of transitions of $\Kount$ (note that one transition may be represented by several edges).
        We must now prove that this run is valid, i.e. that tests are simulated correctly.

        \begin{itemize}
            \item Let us assume that at some point along the play $\pi$, from a vertex $q$, player $C_i^\top$, with $i \in \{1, 2\}$, does not take the edge to the vertex $q'$ while her energy level is zero.
        Then, her energy level drops to $\bot$ and she loses.
        Therefore, she has a profitable deviation at the beginning of the play, by looping on the vertex $q_0^i$.

        \item Let us now assume that she goes to the vertex $q'$, while her energy is positive.
        Then, player $C_i^\bot$'s energy is also positive: he can go to the vertex $\triangle$ and win.
        That would be a profitable deviation, since, as the play $\pi$ reaches the vertex $q_\f$, it is lost by player $C_i^\bot$.
        \end{itemize}

        Therefore, the play $\pi$ does not fake any test, and simulates correctly the machine $\Kount$, which terminates.

		\subp{If the machine $\Kount$ terminates, then there is an NE where Witness loses}
		
		Then, let us define a strategy profile $\bsigma$ in $\Game_{\|q_0^1}$ as follows: in tests of counter $C$, player $C^\top$ goes to $q'$ if and only if her energy level is positive; from $q'$, player $C^\bot$ never goes to the vertex $\triangle$.
        Let $\pi = \< \bsigma \>$: since tests are simulated correctly, the play $\pi$ simulates the run of $\Kount$, and therefore reaches the vertex $q_\f$.
        It is therefore lost by Witness, as well as by the players $C_1^\bot$ and $C_2^\bot$.
  
		Witness has no profitable deviation, since he does not control any vertex.
        The only vertices controlled by each player $C^\bot$ are the vertices of the form $q'$.
        Those vertices are reached only when player $C^\top$'s, and therefore player $C^\bot$'s energy level is zero.
        Then, deviating and going to the vertex $\triangle$ is not profitable for player $C^\bot$, since it makes him immediately lose.
        The strategy profile $\bsigma$ is therefore an NE, lost by Witness.

    \subp{Conclusion}
    Every NE outcome in the game $\Game_{\|q_0^1}$ is won by Witness if and only if the machine $\Kount$ does not terminate.
    Therefore, the halting problem of two-counter machines reduces to the constrained existence problem of NEs in energy games, which is therefore undecidable.

    \paragraph{Recursive enumerability}
    
    Recursive enumerability will be a consequence of the following result.

\begin{lem} \label{lm:energy_ne_finitememory}
    For every NE $\bsigma$ in the energy game $\Game_{\|v_0}$, there exists a finite-memory NE $\bsigma^\star$ with $\mu(\bsigma) = \mu(\bsigma^\star)$.
\end{lem}
    
\begin{proof}
    Let $\bsigma$ be an NE in the game $\Game_{\|v_0}$.
    Let $\pi = \< \bsigma \>$.
    By Dickson's lemma, there exist two indices $m$ and $n$, with $m < n$, such that $\pi_m = \pi_n$ and such that for every player $i$ that does not lose the play $\pi$, we have $\EL_i(\pi_{\leq m}) \leq \EL_i(\pi_{\leq n})$.
    Moreover, we can chose $m$ great enough to have $\EL_j(\pi_{\leq m}) = \bot$ for every player $j$ that loses the play $\pi$.
    Thus, the players winning the play $\pi$ are exactly the players winning the play $\chi = \pi_{< m} \left(\pi_m \dots \pi_{n - 1}\right)^\omega$.

    Now, let $k \leq n$, and let $i$ be the player controlling the vertex $\pi_k$.
    If $i$ is a player who loses the play $\pi$, then since $\bsigma$ is an NE, any play of the form $\pi_{\leq k} \chi$ compatible with $\bsigma_{-i}$ is lost by player $i$; in other words, the strategy profile:
    $$\bsigma_{-i\|\pi_{\leq k}}: \begin{cases}
        \Hist_{-i} \Game_{\|\pi_k} &\to V \\
        h &\mapsto \bsigma_{-i}(\pi_{<k}h)
    \end{cases}$$
    is a strategy profile against which player $i$ cannot win.
    It is known (see for example~\cite[Lemma~10]{DBLP:conf/formats/BouyerFLMS08}) that stationary strategies are sufficient to falsify an energy objective.
    Therefore, let $\btau^k_{-i}$ be a stationary strategy profile, from the vertex $\pi_k$, against which player $i$ cannot win.
    Let $\tau^k_i$ be an arbitrary stationary strategy.
    If $i$ is not a player who loses the play $\pi$, then we define $\btau^k$ as an arbitrary stationary strategy profile.

\begin{figure}
			\centering
			\begin{tikzpicture}[node distance=3cm]
				\node[vert, rectangle, rounded corners, initial above] (q0) {$q_0$};
				\node[right of=q0] (d1) {\dots};
				\node[vert, rectangle, rounded corners, right of=d1] (qm) {$q_m$};
				\node[right of=qm] (d2) {\dots};
				\node[vert, rectangle, rounded corners, right of=d2] (qn-1) {$q_{n-1}$};
				\node[vert, rectangle, rounded corners, below of=qm] (qpm) {$q'_m$};
				\node[vert, rectangle, rounded corners, below of=qn-1] (qpn-1) {$q'_{n-1}$};

				\path (q0) edge node[above] {$\pi_0|\pi_1$} (d1);
                \path (d1) edge node[above] {$\pi_{m-1}|\pi_m$} (qm);
                \path (qm) edge node[above] {$\pi_m|\pi_{m+1}$}  (d2);
                \path (d2) edge node[above] {$\pi_{n-2}|\pi_{n-1}$} (qn-1);
                \path (qn-1) edge[bend right] node[above] {$\pi_{n-1}|\pi_m$} (qm);
                \path (qm) edge node[left] {$v \in V \setminus \{\pi_m\}|\btau^m(v)$} (qpm);
                \path (qn-1) edge node[left] {$v \in V \setminus \{\pi_{n-1}\}|\btau^{n-1}(v)$} (qpn-1);
                \path (qpm) edge[loop below] node[below] {$v|\btau^m(v)$} (qpm);
                \path (qpn-1) edge[loop below] node[below] {$v|\btau^n(v)$} (qpn-1);
			\end{tikzpicture}
			\caption{The memory structure $\Mem$} \label{fig:nash_energy_finite_memory}
		\end{figure} 

    Let $\Mem$ be the memory structure depicted by \Cref{fig:nash_energy_finite_memory}, and defined as follows: it has $2n$ states, namely $q_0, \dots, q_{n-1}$ and $q'_0, \dots, q'_{n-1}$.
    From each state $q_k$, the transition reading the vertex $\pi_k$ leads to the state $q_{k+1}$ (or $q_m$ if $k = n-1$), and outputs the vertex $\pi_{k+1}$ (or $\pi_m$ if $k = n-1$).
    The transition reading any other vertex $v$ (if $k \geq 1$) leads to the state $q'_k$ and outputs the vertex $\btau^{k-1}(v)$.
    From the state $q'_k$, the transition reading each vertex $v$ leads to $q'_k$, and outputs the vertex $\btau^{k-1}(v)$ (since $\btau^{k-1}$ is stationary).
    Thus, the strategy profile induced by the memory structure $\Mem$ is the strategy profile that follows the play $\chi$, and that punishes any player who deviates by following the stationary strategy profile $\btau^k$.
    That strategy profile is finite-memory, and generates the same payoff vector as $\bsigma$, as desired.
    \end{proof}

    Moreover, once we have a memory structure, there is an algorithm that says whether it induces an NE or not.

    \begin{lem}\label{lm:energy_ne_checking}
    Given an energy game $\Game_{\|v_0}$ and a multiplayer memory structure $\Mem$, deciding whether the strategy profile $\bsigma$ induced by $\Mem$ is an NE can be done in deterministic polynomial time.
\end{lem}

\begin{proof}
    Given a player $i$, that player has a profitable deviation from $\bsigma$ if and only if these two conditions are satisfied:
    \begin{itemize}
        \item the play $\< \bsigma \>$ is lost by player $i$;
        
        \item there exists a play $\pi$ compatible with $\bsigma_{-i}$ such that $\mu_i(\pi) = 1$.
    \end{itemize}
    
    The first condition can be checked in polynomial time, since the play $\< \bsigma \>$ is a lasso whose size is bounded by a polynomial function of $\lv \Mem \rv$.
    The second condition is satisfied if and only if there exists a play giving player $i$ the payoff $1$ in the one-player game:
    $$\Game' = \left( \{i\}, V \times Q, (V \times Q), E', \mu'\right),$$
    where $\mu': (\pi_0, q_0)(\pi_1, q_1) \dots \mapsto \mu_i(\pi)$, and $E'$ contains the transitions $(u, p)(v, q)$ such that $\delta(p, u) = q$ and either $\nu(p, u) = v$, or $u \in V_i$.
    Checking the existence of such a play can be done in polynomial time according to~\cite[Theorem~7]{DBLP:conf/formats/BouyerFLMS08}.
\end{proof}

    Thus, a semi-algorithm that recognizes the positive instances of the constrained existence problem consists in enumerating the deterministic multiplayer memory structures on $\Game_{\|v_0}$, and for each of them, to check (by diagonalization):
    \begin{itemize}
        \item whether the only strategy profile compatible with it is an NE: that problem is decidable (in polynomial time) by \cref{lm:energy_ne_checking};

        \item whether that strategy profile generates a payoff vector between $\bx$ and $\by$: that is recursively enumerable, by constructing step by step its outcome and computing the energy levels on the fly.
    \end{itemize}
    
    By \cref{lm:energy_ne_finitememory}, we have a positive instance of the constrained problem if and only if at least one memory structure satisfies those two conditions.
    The constrained existence problem is therefore recursively enumerable.
\end{proof}

%% file: 2aNego.tex
Let us imagine the following situation: in a parking lot, cars---which can be automated, semi-automated, or driven solely by a human---must enter and exit through the same entrance, which is too narrow to allow two cars to pass simultaneously.
When one car arrives and another one leaves, they (and their possibly existing drivers) must agree on who should pass first to avoid getting stuck.
Since endless negotiations are undesirable, they need a protocol to make this decision. In game-theoretic terms, such a protocol would be represented by a strategy profile.
For example, the protocol could be as follows: when two cars arrive at the same time, one direction (typically exiting) has priority over the other. When there are queues on both sides, cars alternate.

A more challenging issue arises when a car does not follow the protocol.
Ideally, such deviations should not allow the car to gain time in any way; otherwise, impatient drivers would always be tempted to deviate, making the protocol ineffective in practice. This is why it is desirable for the strategy profile representing the protocol to be a Nash equilibrium.

On the other hand, let us consider the following protocol, which does satisfy the conditions of a Nash equilibrium: first, cars adhere to the priority rules defined above. Then, if a car goes through the entrance when it is not supposed to, all other cars stop driving, resulting in a complete and permanent standstill.
The flaw in this approach is that the threat meant to prevent deviations is enforced by agents who have their own objectives (reaching their destinations within a reasonable time), which they will not simply disregard when a deviation occurs.

This situation highlights a fundamental limitation of Nash equilibria in sequential settings: a problem known as \emph{non-credible threats}.
In doing so, it also advocates for a more robust equilibrium concept, one that leads to interesting algorithmic challenges: \emph{subgame-perfect equilibria}.
That concept was first introduced by Reinhard Selten~\cite{selten}, and earned him the Nobel prize in economic sciences in 1994, shared with John Nash and John Harsanyi.

\section{Subgame-perfect equilibria}

Formally, subgame-perfect equilibria will be defined by the fact that players play rationally not only in the outcome of the strategy profile, but also in \emph{subgames}.

\begin{defi}[Subgame, substrategy]
	Let $hv$ be a history in the game $\Game$. The \emph{subgame} of the game $\Game$ after the history $hv$ is the initialized game $\left(\Pi, V, (V_i)_i, E, \mu_{\|hv}\right)_{\|v}$, where $\mu_{\|hv}$ maps each play to its payoff in the game $\Game$, assuming that the history $hv$ has already been played: formally, for every $\pi \in \Plays \Game_{\|hv}$, we have $\mu_{\|hv}(\pi) = \mu(h\pi)$.
    
    If $\sigma_i$ is a strategy in the game $\Game_{\|v_0}$, its \emph{substrategy} after $hv$ is the strategy $\sigma_{i\|hv}$ in the game $\Game_{\|hv}$, defined by $\sigma_{i\|hv}(h') = \sigma_i(hh')$ for every $h' \in \Hist_i \Game_{\|hv}$.
\end{defi}

\begin{rem}
    The initialized game $\Game_{\|v_0}$ is also the subgame of $\Game$ after the one-vertex history $v_0$.
\end{rem}

We can now define subgame-perfect equilibria.

\begin{defi}[Subgame-perfect equilibrium]
	Let $\Game_{\|v_0}$ be a game.
	The strategy profile $\bsigma$ is a \emph{subgame-perfect equilibrium}, or \emph{SPE} for short, in the game $\Game_{\|v_0}$, if and only if for every history $h$ in the game $\Game_{\|v_0}$, the strategy profile $\bsigma_{\|h}$ is a Nash equilibrium in the subgame $\Game_{\|h}$.
\end{defi}

\begin{figure}
    \centering
    \begin{tikzpicture}[node distance=3cm]
        \node[vert,rectangle,rounded corners=7pt,dotted,minimum size=2.5cm] (aa) {$\begin{matrix}~\\~\\~\\~\\\mu_\circ = \mu_\square = 0\end{matrix}$};
        \node[vert, initial] (a) at (0, 0) {$a$};
        \node[vert,rectangle,rounded corners=7pt,dotted,minimum size=2.5cm, right of=a] (bb) {$\begin{matrix}~\\~\\~\\~\\\mu_\circ = \mu_\square = 0\end{matrix}$};
        \node[vert, rectangle, right of=a] (b) {$b$};
        \node[vert,rectangle,rounded corners=7pt,dotted,minimum size=2.5cm, right of=b] (cc) {$\begin{matrix}~\\~\\~\\~\\\mu_\circ = \mu_\square = 1\end{matrix}$};
        \node[vert, right of=b] (c) {$c$};
        \path[thick, red] (a) edge[loop above] (a);
        \path[thick, red] (b) edge[loop above] (b);
        \path (c) edge[loop above] (c);
        \path[thick, blue] (a) edge (b);
        \path[thick, blue] (b) edge (c);
    \end{tikzpicture}
    \caption{Two Nash equilibria}
    \label{fig_ne_spe}
\end{figure}
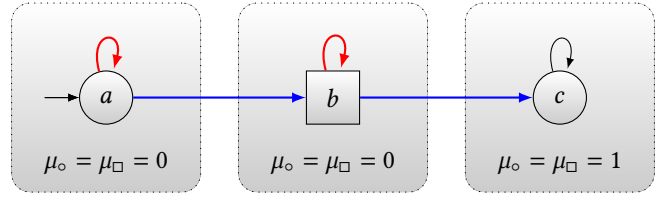

\begin{exa} \label{ex_ne_spe}
We have recalled in Figure~\ref{fig_ne_spe} the game that was already given in \cref{ex:first_game}.
In this game, there are actually two NEs: 
\begin{itemize}
    \item one, depicted in blue, where $\Circle$ goes to the vertex $b$ and then player $\square$ goes to $c$, and both get the payoff $1$;
    \item and one, depicted in red, where player $\Circle$ stays in $a$, and has no incentive to deviate to $b$ because player $\square$ plans to stay in $b$.
\end{itemize}
However, only the first one is an SPE, because in the subgame after the history $ab$, player $\square$ has a profitable deviation by going to $c$.
\end{exa}

While Nash equilibria are known to exist in all the game classes we study in this thesis, it is not the case of SPEs~\cite{SOLAN2003911,DBLP:conf/csl/BrihayeBMR15}.
We will therefore, sometimes, need a quantitative relaxation of SPEs, namely \emph{$\epsilon$-SPEs}.
That notion will be built on the notion of \emph{$\epsilon$-Nash equilibrium}: given a quantity $\epsilon \geq 0$, a strategy profile $\bsigma$ is an $\epsilon$-Nash equilibrium, or $\epsilon$-NE for short, if for each player $i$ and every deviation $\sigma'_i$ of player $i$ from $\bsigma$, the inequality $\mu_i\< \bsigma_{-i}, \sigma'_i \> \leq \mu_i\< \bsigma \> + \epsilon$ holds---i.e., the deviation $\sigma'_i$ is not profitable \emph{by more than $\epsilon$}.
We can then define $\epsilon$-SPEs.

\begin{defi}[$\epsilon$-Subgame-perfect equilibrium]
	Let $\Game_{\|v_0}$ be a game, and let $\epsilon \geq 0$.
	The strategy profile $\bsigma$ is an \emph{$\epsilon$-subgame-perfect equilibrium}, or \emph{$\epsilon$-SPE} for short, in the game $\Game_{\|v_0}$, if and only if for every history $h$ in the game $\Game_{\|v_0}$, the strategy profile $\bsigma_{\|h}$ is an $\epsilon$-NE in the subgame $\Game_{\|h}$.
\end{defi}

\begin{rem}
The definitions of $0$-NEs and $0$-SPEs coincide, respectively, with those of NEs and SPEs.
\end{rem}

This part of this thesis is dedicated to finding and proving the complexities of the constrained existence problem of SPEs, and $\epsilon$-SPEs when it is relevant, in the game classes we have introduced in \cref{part:nash}.

\section{Negotiation}
\subsection{Origin of the concept}

The idea behind the concept of negotiation seems to have emerged in two parallel ways, without a full formalization.

On the one hand, János Flesch and Arkadi Predtetchinski introduced in 2017 a characterization of SPEs in Borel games where the payoff functions have a finite range (such as Boolean games), based on an iterative process corresponding to what we call here the \emph{negotiation function}, defined from what we refer to as the \emph{abstract negotiation game}~\cite{DBLP:journals/mor/FleschP17}.
When the payoff functions have an infinite range (as in mean-payoff or discounted-sum games), their work characterizes a set of plays that is known to contain only outcomes of $\epsilon$-SPEs, for every $\epsilon > 0$, as well as all outcomes of SPEs.
We will show that it actually characterizes SPE outcomes in all games \emph{with steady negotiation}.
Since their work was primarily conducted with the objective of proving existence results, Flesch and Predtetchinski's approach does not provide a characterization that can be directly used for algorithmic purposes, as their iterative process may be infinite, and each iteration requires solving an infinite number of two-player zero-sum games, which can themselves be of infinite size.

On the other hand, in 2019, Marie van den Bogaard, Thomas Brihaye, Véronique Bruyère, Aline Goeminne, and Jean-François Raskin proved the $\PSpace$-completeness of the constrained existence problem of SPEs in a class of games that we do not study here: \emph{quantitative reachability games}, where players aim to reach a given set of vertices as quickly as possible~\cite{DBLP:conf/concur/BrihayeBGRB19}.
Their upper bound relied on an algorithm that labels each vertex with an integer, representing the number of steps within which the player controlling the vertex can, and therefore must, reach their target.
The algorithm then updates this labeling iteratively, leveraging the fact that a player may discover new ways to reach their target more quickly when other players, while trying to prevent them from doing so, are also attempting to satisfy their own requirements.
These updates, which correspond to one iteration of the negotiation function (in a case where Flesch and Predtetchinski's work did not provide a full characterization, since quantitative reachability payoff functions have infinite range), continue until a fixed point is reached, which characterizes SPEs.

In both cases, the negotiation function was not explicitly formalized as a function but was instead an implicit part of an iterative process.
The reader will observe in the following chapters that reasoning about these iterations as a function, whose fixed points characterize SPEs, is useful in some cases where simply iterating the function is not the most efficient way to solve our problem, or could even fail to terminate, as is the case in mean-payoff games.

		\subsection{Requirements}

In the method we will develop further, we will need to analyze the players' behaviors when they have some \emph{requirement} to satisfy.
Intuitively, one can see requirements as \emph{rationality constraints} for the players, that is, a threshold payoff value under which a player will not accept to follow a play, because they would be better off deviating from it.

\begin{defi}[Requirement]
	A \emph{requirement} on the game $\Game$ is a mapping $\lambda: V \to \RR \cup \{\pm\infty\}$.
\end{defi}

For a given vertex $v$, the quantity $\lambda(v)$ represents the minimal payoff that the player controlling $v$ will require in a play starting from $v$.

\begin{defi}[$\lambda$-consistency]\label{def:lambda_cons}
	Let $\lambda$ be a requirement on a game $\Game$.
 A play $\pi$ in the game $\Game$ is \emph{$\lambda$-consistent} if and only if, for every player $i \in \Pi$ and integer $n \in \NN$ with $\pi_n \in V_i$, we have $\mu_i(\pi_{\geq n})~\geq~\lambda(\pi_n)$.
	The set of $\lambda$-consistent plays from a vertex $v$ is denoted by $\lCons\Game_{\|v}$, or $\lCons(v)$ when the context is clear.
\end{defi}

Such a requirement also induces a notion of rationality for strategy profiles aiming at punishing one player.

\begin{defi}[$\lambda$-rationality] \label{def:lambda_rat}
	Let $\lambda$ be a requirement on a game $\Game$.
 Let $i \in \Pi$.
 The strategy profile $\bsigma_{-i}$ is \emph{$\lambda$-rational assuming} the strategy $\sigma_i$ if and only if for every history $hv$ compatible with $\bsigma_{-i}$, the play $\< \bsigma_{\|hv} \>$ is $\lambda$-consistent.
	A strategy profile is simply \emph{$\lambda$-rational} when it is $\lambda$-rational assuming some strategy.
	The set of $\lambda$-rational strategy profiles in the game $\Game_{\|v}$ is denoted by $\lRat_{-i}\Game_{\|v}$, or $\lRat(v)$ when the context is clear.	
\end{defi}
	
Note that $\lambda$-rationality is a property of a strategy profile for all the players but one, player $i$. Intuitively, their rationality is justified by the fact that they collectively assume that player $i$ will, eventually, play according to the strategy $\sigma_i$: if it is the case, then everyone gets their payoff satisfied.

Finally, let us define a particular requirement: the \emph{vacuous requirement}, that requires nothing, and with which every play is consistent.

\begin{defi}[Vacuous requirement]\label{def:lambda0}
	In any game, the \emph{vacuous requirement}, denoted by $\lambda_0$, is the requirement constantly equal to $-\infty$.
\end{defi}

		\subsection{Negotiation} \label{ss_def_nego}

We will show that SPEs in prefix-independent games are characterized by the fixed points of a function on requirements. That function captures a \emph{negotiation} process: when a player has a requirement to satisfy, another player can hope for a better payoff than what they can secure in general, and therefore update their own requirement.
Note that we always use the convention $\inf \emptyset = +\infty$.

\begin{defi}[Negotiation function, steady negotiation] \label{def_nego}
	Let $\Game$ be a game.
	The \emph{negotiation function} is the function that transforms each requirement $\lambda$ on $\Game$ into a requirement $\nego(\lambda)$ on $\Game$, such that for each player $i \in \Pi$ and every vertex $v \in V_i$, we have:
	$$\nego(\lambda)(v) = \inf_{\bsigma_{-i} \in \lRat(v)} \sup_{\sigma_i \in \Strat_i\Game_{\|v}} \mu_i\< \bsigma\>.$$
	If that infimum is realized for every requirement $\lambda$, player $i$ and vertex $v \in V_i$ such that $\lRat(v) \neq \emptyset$, then the game $\Game$ is called a game \emph{with steady negotiation}.
\end{defi}

The quantity $\nego(\lambda)(v)$ is then the worst case value that the player controlling $v$ can ensure, assuming that the other players play $\lambda$-rationally.

\begin{rem}
    The negotiation function satisfies the following properties.
    
    \begin{itemize}
        \item It is monotone: if we have $\lambda \leq \lambda'$, then we have $\nego(\lambda) \leq \nego(\lambda')$.
        
        \item It is also non-decreasing: for every $\lambda$, we have $\lambda \leq \nego(\lambda)$.
        
        \item There exists a $\lambda$-rational strategy profile from $v$ against the player controlling $v$ if and only if $\nego(\lambda)(v) \neq +\infty$.
    \end{itemize}
\end{rem}

\begin{exa}    
    Let us consider again the game depicted by Figure~\ref{fig_ne_spe}, and let us consider the vacuous requirement $\lambda_0: v \to -\infty$.
    Let us compute $\lambda_1 = \nego(\lambda_0)$.
    
    From the vertex $c$, the only payoff that can be obtained is $\lambda_1(c) = 1$.
    From the vertex $b$, player $\square$ can go to the vertex $c$ and get the payoff $\lambda_1(b) = 1$.
    From the vertex $a$, even if player $\Circle$ goes to $b$, player $\square$ can stay there, hence she cannot get more than $\lambda_1(a) = 0$.

    Let us now compute $\lambda_2 = \nego(\lambda_1)$.
    From $c$, the only payoff possible is still $\lambda_2(c) = 1$, and from $b$, it is still possible to get $\lambda_2(b) = 1$ and nothing more.
    But from the vertex $a$, a $\lambda$-rational strategy profile must necessarily plan to go to the vertex $c$ if the vertex $b$ is reached, hence $\lambda_2(a) = 1$.
    
    Finally, we have $\nego(\lambda_2) = \lambda_2$.
\end{exa}

	\section{Link between negotiation and equilibria}

        \subsection{Nash equilibria}

Although all our computational results about Nash equilibria have already been stated in \cref{chap:nash}, let us show how we can, using the negotiation function, rephrase a folklore characterization of Nash equilibria outcomes: a play is a Nash equilibrium outcome if and only if it gives to every player at least the payoff that they could enforce by deviating, if all the other players try to punish them---i.e., if for every vertex it traverses, it gives to the player controlling that vertex its adversarial value.

\begin{thm}\label{thm_ne}
	Let $\Game$ be a game with steady negotiation, and let $\pi$ be a play in the game $\Game$.
    Then, the play $\pi$ is an NE outcome if and only if $\pi$ is $\nego(\lambda_0)$-consistent.
\end{thm}

\begin{proof} \paragraph{If $\pi$ is an NE outcome, then it is $\nego(\lambda_0)$-consistent.}

Let $\bsigma$ be a Nash equilibrium in the game $\Game$, and let $\pi = \< \bsigma \>$: let us prove that the play $\pi$ is $\nego(\lambda_0)$-consistent.
	Let $k \in \NN$, let $i \in \Pi$ be such that $\pi_k \in V_i$, and let us prove that $\mu_i\left(\pi_{\geq k}\right) \geq \nego(\lambda_0)(\pi_k)$.
	For any deviation $\sigma'_i$ of player $i$ from the strategy profile $\bsigma_{\|\pi_{\leq k}}$, by definition of NEs, we have $\mu_i\< \bsigma_{-i\|\pi_{\leq k}}, \sigma'_i \> \leq \mu_i(\pi)$.
	Therefore, we have $\mu_i(\pi) \geq \sup_{\sigma'_i} \mu_i\< \bsigma_{-i\|\pi_{\leq k}}, \sigma'_i \>$, hence the inequality:
    $$\mu_i(\pi) \geq \inf_{\btau_{-i}} ~\sup_{\tau_i} ~ \mu_i\< \btau_{-i\|\pi_{\leq k}}, \tau_i \>,$$ i.e.	$\mu_i(\pi) \geq \nego(\lambda_0)(\pi_k)$: the play $\pi$ is $\nego(\lambda_0)$-consistent.

	\paragraph{If $\pi$ is $\nego(\lambda_0)$-consistent, then it is an NE outcome.}
    
    Let $\pi$ be a $\nego(\lambda_0)$-consistent play in the game $\Game$.
    Let us define a strategy profile $\bsigma$ generating $\pi$ as follows.
	
	\begin{itemize}
		\item First, we define $\< \bsigma \> = \pi$.
		
		\item Then, for each history of the form $\pi_{\leq k} v$ with $v \neq \pi_{k+1}$, let $i$ be the player controlling $\pi_k$.
		Since the game $\Game$ is with steady negotiation, the infimum:
		$$\inf_{\btau_{-i} \in \lambda_0\Rat(\pi_k)}~ \sup_{\tau_i}~ \mu_i\< \btau \> = \nego(\lambda_0)(v) \neq +\infty$$
		is a minimum.
		Let $\btau^k_{-i}$ be $\lambda_0$-rational strategy profile from $\pi_k$ realizing that minimum, and let $\tau^k_i$ be some strategy from $\pi_k$.
        Then, we define:
		$$\< \bsigma_{\|\pi_{\leq k} v} \> = \< \btau^k_{\|\pi_k v} \>.$$
		
		\item For every other history $h$, the vertex $\bsigma(h)$ is defined arbitrarily.
	\end{itemize}

	Let us prove that $\bsigma$ is an NE: let $\sigma'_i$ be a deviation of $\sigma_i$, and let $\pi' = \< \bsigma_{-i}, \sigma'_i \>$.
    Since the case $\pi' = \pi$ is trivial, we assume $\pi \neq \pi'$ and define $\pi_{\leq k}$ as the longest common prefix of $\pi$ and $\pi'$.
    Let $v = \pi'_{k+1} \neq \pi_{k+1}$.
	Then, the play $\pi'_{\geq k+1}$ is compatible with the strategy profile $\bsigma_{-i\|\pi_{\leq k}v} = \btau_{-i\|\pi_kv}$, hence the inequality:
	$$\mu_i(\pi') \leq \sup_{\tau_i^k} ~\mu_i\< \btau^k \> = \nego(\lambda_0)(\pi_k).$$
	On the other hand, since the play $\pi$ is $\lambda_0$-consistent, we have $\nego(\lambda_0)(\pi_k) \leq \mu_i(\pi)$, hence $\mu_i(\pi') \leq \mu_i(\pi)$: the deviation $\sigma'_i$ is not profitable, and the strategy profile $\bsigma$ is a Nash equilibrium.
\end{proof}

\subsection{Link with (\texorpdfstring{$\epsilon$}---)subgame-perfect equilibria}

The notion of negotiation will enable us to find the SPEs, but also more generally the $\epsilon$-SPEs, in a game. For that purpose, we need the notion of $\epsilon$-fixed points of a function.

\begin{defi}[$\epsilon$-fixed point]
	Let $\epsilon \geq 0$, let $D$ be a finite set and let $f: (\RR \cup \{\pm\infty\})^D \to (\RR \cup \{\pm\infty\})^D$ be a mapping. A tuple $\bx \in \RR^D$ is an \emph{$\epsilon$-fixed point} of $f$ if for each $d \in D$, for $\by = f(\bx)$, we have $y_d \in [x_d - \epsilon, x_d + \epsilon]$.
\end{defi}

\begin{rem}
    The definition of $0$-fixed points coincide with that of fixed points.
\end{rem}

We now have all the tools required to state and prove the following theorem.
We give it here in a form that applies only to prefix-independent games, even though the same result extends (at least when $\epsilon = 0$) to classes of games that do not observe that property, such as quantitative reachability games or discounted-sum games.
However, since we do not use this result in such classes, we favored the following proof, less general but simpler.

\begin{thm} \label{thm:nego_spe}
	Let $\Game$ be a game with steady negotiation, and let $\pi$ be a play in the game $\Game$.
    Let $\epsilon \geq 0$.
    Then, the play $\pi$ is an $\epsilon$-SPE outcome if and only if $\pi$ is $\lambda$-consistent, for some $\epsilon$-fixed point $\lambda$ of the negotiation function.
\end{thm}

\begin{proof}
        \paragraph{If $\pi$ is an $\epsilon$-SPE outcome, then it is $\lambda$-consistent for some $\epsilon$-fixed point $\lambda$.}

Let $\bsigma$ be an $\epsilon$-SPE in the game $\Game$ such that $\pi = \< \bsigma \>$.

        Let us define a requirement $\lambda$ by, for each $i \in \Pi$ and $v \in V_i$:
        $$\lambda(v) = \inf_{hv \in \Hist \Game_{\|v_0}} \mu_i\< \bsigma_{\|hv} \>.$$
        Then, for every history $hv$ starting in $\pi_0$, the play $\< \bsigma_{\|hv} \>$ is $\lambda$-consistent.
        And in particular, the play $\pi$ is.
        Let us now prove that the requirement $\lambda$ is an $\epsilon$-fixed point of $\nego$.

Let $i \in \Pi$, let $v \in V_i$, and let us show that we have $\nego(\lambda)(v) \leq \lambda(v) + \epsilon$.
Since $\bsigma$ is an $\epsilon$-SPE, for every history $hv \in \Hist\Game_{\|\pi_0}$, we have:
$$\sup_{\tau_i} ~ \mu_i\< \bsigma_{-i\|hv}, \tau_i \> \leq \mu_i\< \bsigma_{\|hv} \> + \epsilon.$$
By taking the infima over $hv$, we deduce:
$$\inf_{hv \in \Hist\Game_{\|\pi_0}} ~ \sup_{\tau_i} ~ \mu_i\< \bsigma_{-i\|hv}, \tau_i \> \leq \inf_{hv} \mu_i\< \bsigma_{\|hv} \> + \epsilon = \lambda(v) + \epsilon.$$
Let us now notice that all the strategy profiles of the form $\bsigma_{-i\|hv}$ are $\lambda$-rational.
Thus, we obtain:
$$\inf_{\btau_{-i} \in \lRat(v)} ~ \sup_{\tau_i} ~ \mu_i\< \btau \> \leq \lambda(v) + \epsilon,$$
that is, we obtain $\nego(\lambda)(v) \leq \lambda(v) + \epsilon$, as desired.

        \paragraph{If $\pi$ is $\lambda$-consistent for some fixed point $\lambda$, then $\pi$ is an $\epsilon$-SPE outcome.}
    
\subp{A particular case: if there exists $v$ accessible from $\pi_0$ such that $\lambda(v) = +\infty$}
	
	In that case, for each vertex $u$ such that $uv \in E$, if the player controlling $u$ chooses to go to $v$, no $\lambda$-consistent play can be proposed to them from there, hence there is no $\lambda$-rational strategy profile against that player from $u$, and $\nego(\lambda)(u) = +\infty$.
	Since $\epsilon$ is finite and since $\lambda$ is an $\epsilon$-fixed point of the negotiation function, it follows that $\lambda(u) = +\infty$. Since $v$ is accessible from $\pi_0$, we can repeat this argument and show that $\lambda(\pi_0) = +\infty$; in that case, there is no $\lambda$-consistent play $\pi$ from $u$, and then the proof is done.
	
	Therefore, for the rest of the proof, we assume that for all $v$, we have $\lambda(v) \neq +\infty$.
	As a consequence, since $\lambda$ is an $\epsilon$-fixed point of the function $\nego$, for each $v$ accessible from $\pi_0$, we have $\nego(\lambda)(v) \neq +\infty$; which implies that for each such $v$, there exists a $\lambda$-consistent strategy profile against the player controlling $v$, starting from $v$.

    The rest of the proof constructs the strategy profile $\bsigma$ and proves that it is an $\epsilon$-SPE.
    That construction is illustrated by Figure~\ref{fig_construction_bsigma}.
    
    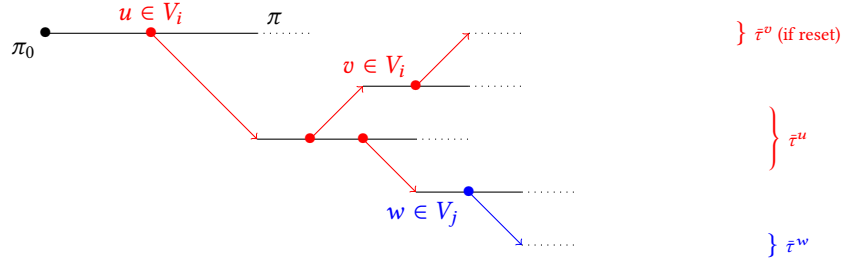
\begin{figure}
        \centering
        \begin{center}
		\begin{tikzpicture}[scale=0.7]
		\draw (0, 0) node {$\bullet$};
		\draw (0, 0) node[below left] {$\pi_0$};
		\draw (0, 0) -- (4, 0);
		\draw [dotted] (4, 0) -- (5, 0);
		\draw (4, 0) node[above right] {$\pi$};
		
		\draw[red] (2, 0) node {$\bullet$};
		\draw[red] (2, 0) node[above] {$u \in V_i$};
		\draw[red, ->] (2, 0) -- (4, -2);
		
		\draw (4, -2) -- (7, -2);
		\draw[dotted] (7, -2) -- (8, -2);
		
		\draw[red] (5, -2) node {$\bullet$};
		\draw[red, ->] (5, -2) -- (6, -1);
		\draw (6, -1) -- (8, -1);
		\draw[dotted] (8, -1) -- (9, -1);
		\draw[red] (6, -2) node {$\bullet$};
		\draw[red, ->] (6, -2) -- (7, -3);
		\draw (7, -3) -- (9, -3);
		\draw[dotted] (9, -3) -- (10, -3);
		
		\draw[red] (14, -2) node {$\left. \begin{matrix} \\ \\ \end{matrix} \right\}$ \scriptsize{$\btau^u$}};
		
		\draw[blue] (8, -3) node {$\bullet$};
		\draw[blue] (8, -3) node[below left] {$w \in V_j$};
		\draw[blue, ->] (8, -3) -- (9, -4);
		\draw[dotted] (9, -4) -- (10, -4);
		\draw[blue] (14, -4) node {$\left. \right\}$ \scriptsize{$\btau^w$}};
		
		\draw[red] (7, -1) node {$\bullet$};
		\draw[red] (7, -1) node[above left] {$v \in V_i$};
		\draw[red, ->] (7, -1) -- (8, 0);
		\draw[dotted] (8, 0) -- (9, 0);
		
		\draw[red] (14, 0) node {$\left. \right\}$ \scriptsize{$\btau^v$ (if reset)}};

		\end{tikzpicture}
	\end{center}
        \caption{The construction of $\bsigma$}
        \label{fig_construction_bsigma}
    \end{figure}

	\subp{Spare parts: the strategy profiles $\btau^{v*}$}
	
	Let us recall that since $\Game$ is a game with steady negotiation.
    Then, let $i \in \Pi$ and $v \in V_i$: since, by the previous point, we assume $\lambda\Rat(v) \neq \emptyset$ for each $v$, we know that there exists a strategy profile $\btau^v_{-i}$ from $v$ that is $\lambda$-rational assuming a strategy $\tau^v_i$ and that satisfies the inequality:
	$$\sup_{\tau_i} ~ \mu_i\< \btau^v_{-i}, \tau_i \> = \inf_{\btau_{-i} \in \lRat(v)} ~\sup_{\tau_i} ~\mu_i\< \btau \> = \nego(\lambda)(v).$$
	In other words, there exists a worst $\lambda$-rational strategy profile against player $i$ from the vertex $v$, with regards to player $i$'s payoff.
    
	Our goal in this part of the proof is to construct a strategy profile $\btau^{v*}_{-i}$, that is $\lambda$-rational assuming a strategy $\tau^{v*}_i$, and that will be used to punish player $i$ when they deviate from $\bsigma$ until another player deviates.
	The strategy profile $\btau^v_{-i}$ and the strategy $\tau^v_i$ are not sufficient for that purpose, because if some history $h$ compatible with $\btau^v_{-i}$ is such that $\mu_i\< \btau^v_{\|h} \> < \mu_i\< \btau^v \>$, then in the corresponding subgame, it may be possible for player $i$ to deviate and get a payoff that would be smaller than or equal to the quantity $\mu_i\< \btau^v \>$, but greater than $\mu_i\< \btau^v_{\|h} \>$.
	On the contrary, the construction of the strategy profile $\btau^{v*}_{-i}$ will ensure that each time player $i$ deviates, the other players punish them at least as harshly as they were planning to do before the deviation.
	
	Let us construct inductively the strategy profile $\btau^{v*}$.
	We define it only on histories that are compatible with $\btau^{v*}_{-i}$, since it can be defined arbitrarily on other histories.
	We proceed by assembling the strategy profiles of the form $\btau^w$ for various $w \in V_i$, and the histories after which we follow a new $\btau^w$ will be called the \emph{resets} of $\btau^{v*}$: they will be histories of the form $hw'$, where $h$ is empty or $\last(h) = w$.

    \begin{itemize}
	\item First, we set $\< \btau^{v*} \> = \< \btau^v \>$: the one-vertex history $v$ is then the first reset of $\btau^{v*}_{-i}$.
	
    \item Then, for every history $hww'$ from $v$ such that $h$ is compatible with $\btau^{v*}_{-i}$, that $w \in V_i$, and that $w' \neq \tau^{v*}_i(hw)$: let us decompose $hww' = h_1h_2$, so that the history $h_1 \first(h_2)$ is the longest reset of $\btau^{v*}_{-i}$ among the prefixes of $hw$.
    Or, in other words, so that the strategy profile $\btau^{v*}_{\|h_1\first(h_2)}$ has been defined as equal to $\btau^u$ over the prefixes of $h_2$ until $w$, where $u = v$ if $h_1$ is empty, or $u = \last(h_1)$ otherwise.
	By prefix-independence of $\Game$ and by definition of $\btau^u$ and $\btau^w$, we have:
	$$\inf_{\btau_{-i} \in \lRat(w')} \sup_{\tau_i} \mu_i\< \btau \> \leq \sup_{\tau_i} \mu_i\< \btau^w_{-i}, \tau_i \> = \nego(\lambda)(w).$$
	Let us now separate two cases.
		
		\begin{itemize}
			\item Suppose first that there is equality:
			$$\inf_{\btau_{-i} \in \lRat(w')} \sup_{\tau_i} \mu_i\< \btau \> = \nego(\lambda)(w).$$
			Then, we choose $\< \btau^{v*}_{\|hww'} \> = \< \btau^u_{\|uh_2} \>$: the coalition of players against player $i$ keeps following the same strategy profile.
			
			\item Suppose now that the inequality is strict:
			$$\inf_{\btau_{-i} \in \lRat(w')} \sup_{\tau_i} \mu_i\< \btau \> < \nego(\lambda)(w).$$
			Then, we choose $\< \btau^{v^*}_{\|hww'} \> = \< \btau^w_{\|ww'} \>$: player $i$ has done something that lowers the payoff they can ensure, and therefore the other players have to update their strategy profile in order to punish them more harshly.
			The history $hw$ is a reset of $\btau^{v*}_{-i}$.
		\end{itemize}
\end{itemize}
    
	Since there are finitely many histories of each length, this process completely defines the strategy profile $\btau^{v*}$.
    Moreover, all the plays constructed are $\lambda$-consistent, hence the strategy profile $\btau^{v*}_{-i}$ is $\lambda$-rational assuming the strategy $\tau^{v*}_i$, as desired.

	\subp{Construction of $\bsigma$}

	Let us now construct inductively the strategy profile $\bsigma$ itself: we will prove in the next part of the proof that it is an $\epsilon$-SPE.
    We proceed inductively, by defining all the plays $\< \bsigma_{\|hv} \>$, for $hv \in \Hist\Game_{\pi_0}$ with $v \neq \bsigma(h)$.
    We maintain the induction hypothesis that such a play is always $\lambda$-consistent.

	\begin{itemize}
		\item First, we choose $\< \bsigma \> = \pi$, which satisfies the induction hypothesis.

		\item Let now $huv$ be a history such that the strategy profile $\bsigma$ has been defined on all the prefixes of $hu$, which we now assume to be nonempty, but not on $huv$ itself, and such that $v \neq \bsigma(hu)$.
		Let $i$ be the player controlling the vertex $u$.
		
		Then, we define $\< \bsigma_{\|huv} \> = \< \btau^{u*}_{\|uv} \>$, and inductively, for every history $h'w$ starting from $v$ and compatible with $\bsigma_{-i\|huv}$, we define $\< \bsigma_{\|huh'w} \> = \< \btau^{u*}_{\|uh'w} \>$.
		The strategy profile $\bsigma_{\|huv}$ is then equal to $\btau^{v*}_{\|uv}$ on any history compatible with $\btau^{v*}_{-i}$.
	\end{itemize}

	Since there are finitely many histories of each length, this process completely defines $\bsigma$.

	\subp{The strategy profile $\bsigma$ is an $\epsilon$-SPE}
	
	Consider a history $h_0 w \in \Hist \Game_{\|\pi_0}$, a player $i \in \Pi$, and a deviation $\sigma'_i$ of $\sigma_i$.
    Let $\chi = h_0 \< \bsigma_{\|h_0w} \>$, and let $\chi' = h_0 \< \bsigma_{-i\|h_0w}, \sigma'_{i\|h_0w} \>$.
	We wish to prove the inequality $\mu_i(\chi') \leq \mu_i(\chi) + \epsilon$.
    The different notations in this proof are illustrated by \cref{fig_bsigma_SPE}.

        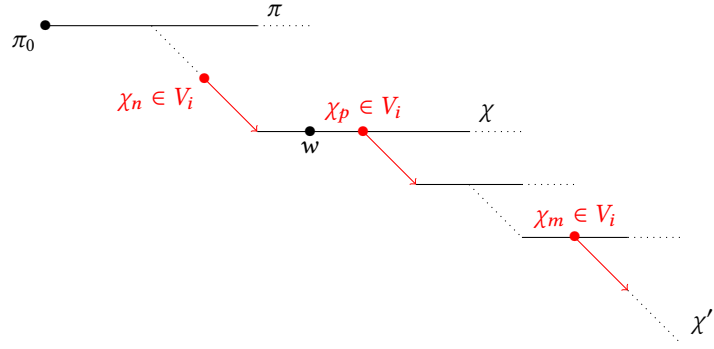
\begin{figure}
        \centering
        \begin{center}
		\begin{tikzpicture}[scale=0.7]
		\draw (0, 0) node {$\bullet$};
		\draw (0, 0) node[below left] {$\pi_0$};
		\draw (0, 0) -- (4, 0);
		\draw [dotted] (4, 0) -- (5, 0);
		\draw (4, 0) node[above right] {$\pi$};
		
		\draw[dotted] (2, 0) -- (3, -1);
        \draw[red] (3, -1) node {$\bullet$};
		\draw[red] (3, -1) node[below left] {$\chi_n \in V_i$};
        \draw[red, ->] (3, -1) -- (4, -2);
        
		\draw (4, -2) -- (8, -2);
		\draw[dotted] (8, -2) -- (9, -2);
        \draw (8, -2) node[above right] {$\chi$};
		
		\draw (5, -2) node {$\bullet$};
        \draw (5, -2) node[below] {$w$};
		\draw[red] (6, -2) node {$\bullet$};
        \draw[red] (6, -2) node[above] {$\chi_p \in V_i$};
        
		\draw[red, ->] (6, -2) -- (7, -3);
		\draw (7, -3) -- (9, -3);
		\draw[dotted] (9, -3) -- (10, -3);
		
		\draw[dotted] (8, -3) -- (9, -4);
		\draw (9, -4) -- (11, -4);
        \draw[dotted] (11, -4) -- (12, -4);
		
	    \draw[red] (10, -4) node {$\bullet$};
        \draw[red] (10, -4) node[above] {$\chi_m \in V_i$};
        \draw[red, ->] (10, -4) -- (11, -5);
		\draw[dotted] (11, -5) -- (12, -6);
		\draw (12, -6) node[above right] {$\chi'$};
		\end{tikzpicture}
	\end{center}
        \caption{The strategy profile $\bsigma$ is an SPE.}
        \label{fig_bsigma_SPE}
    \end{figure}
	
	First, if the play $\chi'$ is compatible with $\sigma_i$, then we have $\chi' = \chi$ and the proof is immediate.
	Now, if it is not, we let $n$ denote the least index such that $\chi'_n \in V_i$ and $\chi'_{n+1} \neq \sigma_i(\chi'_{\leq n})$, and such that the play $\chi'_{\geq n}$ is compatible with the strategy profile $\bsigma_{-i\|\chi'_{\leq n}}$. 
	Thus, the edge $\chi'_n\chi'_{n+1}$ marks the time when player $i$ begins to deviate unilaterally from $\sigma_i$.
 However, note that $\chi'_{\leq n}$ can be both longer or shorter than $h_0 w$: player $i$ may have already deviated in $h_0 w$, or may wait afterwards to effectively deviate.
	
	Be that as it may, the history $\chi'_{\leq n}$ is a common prefix of the plays $\chi$ and $\chi'$, and the substrategy profile $\bsigma_{\|\chi'_{\leq n+1}}$ has been defined during the construction of $\bsigma$ as equal to $\btau^{v*}_{\| \chi'_n\chi'_{n+1}}$, where $v = \chi'_n$, on any history compatible with $\bsigma_{-i\|\chi'_{\leq n+1}}$. 
	
	By construction of $\btau^{v*}$, the sequence:
    $$\left(\nego(\lambda)(\chi'_k)\right)_{k \geq n, \chi'_k \in V_i}$$
    is non-increasing.
	It is therefore ultimately constant (or is finite), because it can take only a finite number of values.
	Consequently, there is a finite number of resets along the play $\chi'_{\geq n}$.
	Let then $\chi'_n \dots \chi'_{m+1}$ be the last (longest) one.
	Afterwards, the play $\chi'_{\geq m+1}$ is compatible with the strategy profile $\btau^{\chi'_m}_{-i}$.
    By definition of that strategy profile, we have the inequality $\mu_i(\chi') \leq \nego(\lambda)(\chi'_m)$.
    We need now to prove the inequality $\nego(\lambda)(\chi'_m) \leq \mu_i(\chi) + \epsilon$.

	Let now $\chi_{\leq p} = \chi'_{\leq p}$ denote the longest common prefix of $\chi$ and $\chi'$.
	Note that, then, we have $n \leq p$ and $\chi_p \in V_i$.
    Moreover, we have $\chi_{\geq p} = \< \bsigma_{\|\chi_{\leq p}} \>$, which is $\lambda$-consistent.
	As a consequence, we have the inequality $\mu_i(\chi) \geq \lambda(\chi_p)$.
 
 Finally, since the sequence of the quantities $\nego(\lambda)(\chi'_k)$ with $\chi'_k \in V_i$ is non-increasing for $k \geq n$, we also have $\nego(\lambda)(\chi'_{m-1}) \leq \nego(\lambda)(\chi_p)$.
 Consequently, we have:
 $$\mu_i(\chi') \leq \nego(\lambda)(\chi'_m) \leq \nego(\lambda)(\chi_p) \leq \lambda(\chi_p) + \epsilon \leq \mu_i(\chi) + \epsilon.$$
	
	The strategy profile $\bsigma$ is an $\epsilon$-SPE.
\end{proof}

\begin{exa}
    Let us consider again the game of Figure~\ref{fig_ne_spe}.
    The two NE outcomes that we identified in Example~\ref{ex_ne_spe} are indeed the two plays from $a$ that are $\lambda_1$-consistent: the play $a^\omega$ and the play $abc^\omega$.
    However, only the latter is $\lambda_2$-consistent, and $\lambda_2$ is a fixed point of the negotiation function.
\end{exa}

This theorem will enable us to design efficient algorithms for the ($\epsilon$-)SPE constrained existence problems.
Indeed, given a game $\Game$ and two thresholds $\bx$ and $\by$, one can decide whether there exists an ($\epsilon$-)SPE generating a payoff vector between $\bx$ and $\by$ by:

\begin{itemize}
    \item guessing a requirement $\lambda$;

    \item checking that there exists a $\lambda$-consistent play in the game $\Game$ that generates a payoff vector between $\bx$ and $\by$;

    \item checking that $\lambda$ is an ($\epsilon$-)fixed point of the negotiation function.
\end{itemize}

The last point will usually be the one that induces more substantial work, since it requires a general method to compute the negotiation function.
Such a method will be provided by tools called \emph{negotiation games}.

\section{Negotiation games}

\subsection{The abstract negotiation game}

Given a requirement $\lambda$ and a vertex $v_0 \in V$, the quantity $\nego(\lambda)(v_0)$ can be characterized as the value of a \emph{negotiation game}, a two-player zero-sum game opposing the player \emph{Prover}, who simulates a $\lambda$-rational strategy profile and wants to minimize player $i$'s payoff (where $i$ is the player controlling $v_0$), and the player \emph{Challenger}, who simulates player $i$'s reaction by accepting or refusing Prover's proposals.
First defined by János Flesch and Arkadi Predtetchinski~\cite{DBLP:journals/mor/FleschP17} (without being linked to the concept of negotiation function), the \emph{abstract negotiation game}, written $\abs_{\lambda i}(\Game)_{\|v_0}$ unfolds as follows.

\begin{itemize}
    \item From the vertex $v_0$, Prover chooses a $\lambda$-consistent play $\pi$ and proposes it to Challenger.
    If Prover has no play to propose, the game is over and Challenger gets the payoff $+\infty$.
        
    \item Once a play $\pi$ has been proposed, Challenger can accept it.
    Or he can \emph{deviate}, and choose a prefix $\pi_{\leq k}$ with $\pi_k \in V_i$ and a new edge $\pi_k v \in E$.
        
    \item In the former case, the game is over.
    In the latter, it starts again from the vertex $v$.
\end{itemize}

In the play that Prover and Challenger construct together, 
Challenger's objective consists in maximizing player $i$'s payoff, and Prover's objective in minimizing it.
More formally, the abstract negotiation game $\abs_{\lambda i}(\Game)_{\|v_0}$ is a game with an uncountable vertex space, defined as follows.

\begin{defi}[Abstract negotiation game]\label{def:abstract}
	Let $\Game_{\|v_0}$ be a game, let $i \in \Pi$, and let $\lambda$ be a requirement on $\Game$. The \emph{abstract negotiation game} of $\Game_{\|v_0}$ for player $i$ with requirement $\lambda$ is the two-player zero-sum game:
	$$\abs_{\lambda i}(\Game)_{\|v_0} = \left( \{\Prover, \Challenger\}, V^\a, (V^\a_{\Prover}, V^\a_{\Challenger}), E^\a, \mu^\a\right)_{\|v_0},$$
	where:
	
	\begin{itemize}
		\item $\Prover$ denotes the player \emph{Prover} and $\Challenger$ the player \emph{Challenger}.

    \item Prover's vertices are the vertices of $\Game$, and two special vertices $\top$ and $\bot$.
    
    \item Challenger's vertices are the vertices of the form $[hv]$, where $hv$ can be any history in the game $\Game$, and the vertices of the form $[\pi]$, where $\pi$ is a $\lambda$-consistent play in the game $\Game$.

    \item From the vertex $\bot$, the only available edge is the self-loop to $\bot$, and similarly from $\top$.
    From a vertex of the form $v$, Prover can move to:
    \begin{itemize}
        \item any vertex $[\pi]$ where $\pi$ starts from $v$ (she proposes the play $\pi$);
        
        \item the vertex $\bot$ (she gives up, last option if she cannot propose any $\lambda$-consistent play).
    \end{itemize}

    \item From a vertex of the form $[\pi]$, Challenger can move to the vertex $\top$ (he accepts the play $\pi$), or to any vertex of the form $[\pi_{\leq k}v]$, where $k \in \NN$, $\pi_k \in V_i$ and $v \neq \pi_{k+1}$ (he deviates to $v$).
    From a vertex of the form $[hv]$, he can only go to the vertex $v$.

    \item In a play $\chi$ in this game, if Prover gives up and takes an edge to the vertex $\top$, Challenger's payoff is defined by $\mu^\a_\Challenger(\chi) = +\infty$.
    If Challenger finally accepts a proposal $\pi$, and takes the edge $[\pi]\top$, then his payoff is defined by $\mu^\a_\Challenger(\chi) = \mu_i(\pi)$.
If he deviates infinitely often, then Prover's proposals and his deviations construct a play $\dchi = \pi^{(0)}_{\leq k_0} \pi^{(1)}_{\leq k_1} \pi^{(2)}_{\leq k_2} \dots$.
Then, Challenger's payoff is defined by $\mu^\a_\Challenger(\chi) = \mu_i(\dchi)$.
In all those cases, Prover's payoff is the opposite of Challenger's one.
\end{itemize}
\end{defi}

The abstract negotiation game constitutes an alternative definition of the negotiation function.
Note that the following theorem does not use the notation $\val_\Challenger$, because the abstract negotiation game is not guaranteed to be Borel.

\begin{thm}\label{thm:abstract}
	Let $\Game_{\|v_0}$ be a game, let $\lambda$ be a requirement on $\Game$ and let $i \in \Pi$ be such that $v_0 \in V_i$.
	Then, we have: 
	$$\inf_{\tau_\Prover \in \Strat_\Prover \abs_{\lambda i}(\Game)_{\|v_0}} ~ \sup_{\tau_\Challenger \in \Strat_\Challenger \abs_{\lambda i}(\Game)_{\|v_0}} ~ \mu^\a_\Challenger\< \btau \> = \nego(\lambda)(v_0).$$
\end{thm}

\begin{proof}
Let $\alpha \in \RR$, and let us prove that the following statements are equivalent.

\begin{enumerate}
	\item \label{case1_pf_abs} There exists a strategy $\tau_\Prover$ such that for every strategy $\tau_\Challenger$, we have $\mu^\a_\Challenger\< \btau \> < \alpha$.
	
	\item \label{case2_pf_abs} There exists a $\lambda$-rational strategy profile $\bsigma_{-i}$ in the game $\Game_{\|v_0}$ such that for every strategy $\sigma_i$, we have $\mu_i\< \bsigma \> < \alpha$.
\end{enumerate}

\paragraph{Assertion~\ref{case1_pf_abs} implies Assertion~\ref{case2_pf_abs}.}
	
	Let $\tau_\Prover$ be such that for every strategy $\tau_\Challenger$, we have $\mu^\a_\Challenger\< \btau \> < \alpha$.
	
	In what follows, any history $h$ compatible with an already defined strategy profile $\bsigma_{-i}$ in the game $\Game_{\|v_0}$ will be decomposed in:
	$$h = v_0 h^{(0)} v_1 h^{(1)} \dots h^{(n-1)} v_n h^{(n)},$$
	so that there exist plays $\pi^{(0)}, \dots, \pi^{(n-1)}, \chi$ and a history:
	$$[v_0] \left[\pi^{(0)}\right] \left[v_1 h^{(1)} v_2\right] \dots \left[v_{n-1}h^{(n-1)}v_n\right] \left[v_nh^{(n)}\chi\right]$$
	in the game $\abs_{\lambda i}(\Game)_{v_0}$ compatible with $\tau_{\Prover}$: the existence and the unicity of that decomposition can be proved by induction. Intuitively, the history $h$ is cut in histories which are prefixes of plays that can be proposed by Prover.
	
	Then, let us define inductively the strategy profile $\bsigma_{-i}$ by $\bsigma_{-i}(h) = \chi_0$, with $\chi$ defined from $h$ as above, for every $h$ such that $\bsigma_{-i}$ has been defined on the prefixes of $h$, and such that the last vertex of $h$ is not controlled by player $i$.
    Let us prove that $\bsigma_{-i}$ is the desired strategy profile.
	
	\subp{The strategy profile $\bsigma_{-i}$ is $\lambda$-rational}
		
		Let us define $\sigma_i$ so that for every history $hv$ compatible with $\bsigma_{-i}$, the play $\< \bsigma_{\|hv} \>$ is $\lambda$-consistent.
		
		For each history:
		$$h = v_0 h^{(0)} v_1 h^{(1)} \dots h^{(n-1)} v_n h^{(n)}$$
		compatible with $\bsigma_{-i}$ and ending in $V_i$, let $\sigma_i(h) = \chi_0$ with $\chi$ corresponding to the decomposition of $h$, so that by induction:
		$$\< \bsigma_{\|v_0 h^{(0)} v_1 h^{(1)} \dots h^{(n-1)} v_n} \> = v_n h^{(n)} \chi.$$
		
		Let now $hv$ be a history in the game $\Game_{\|v_0}$, and let us show that the play $\< \bsigma_{\|hv} \>$ is $\lambda$-consistent. If we decompose:
		$$hv = v_0 h^{(0)} v_1 h^{(1)} \dots h^{(n-1)} v_n h^{(n)}$$
		with the same definition of $\chi$ (note that the vertex $v$ is now included in the decomposition), then we have $\< \bsigma_{\|hv} \> = v\chi$, and by definition of the abstract negotiation game, the play $v_n h^{(n)} \chi$ is $\lambda$-consistent, and therefore so is the play $v \chi$.

		\subp{The strategy profile $\bsigma_{-i}$ keeps player $i$'s payoff under the value $\alpha$}
		
		Let $\sigma_i$ be some strategy for player $i$, and let $\pi = \< \bsigma \>$. We want to prove that $\mu_i(\pi) < \alpha$.
		
		Let us define two finite or infinite sequences $\left( \pi^{(k)} \right)_{k \in K}$ and $\left( h^{(k)} v_k \right)_{k \in K}$, where $K = \{1, \dots, n\}$ or $K = \NN \setminus \{0\}$, by for every $k \in K$:
		
		$$\left[ \pi^{(k)} \right] = \tau_{\Prover} \left( [v_0] \left[ \pi^{(0)} \right] \dots \left[ \pi^{(k-1)} \right] \left[ h^{(k)} v_k \right] \right)$$
		and so that for every $k$, the history $h^{(k)} v_k$ is the shortest prefix of $\pi$ that is not a prefix of $h^{(1)} \dots h^{(k-1)} \pi^{(k-1)}$ (or equivalently, the history $h^{(k)}$ is the longest common prefix of $\pi$ and $h^{(1)} \dots h^{(k-1)} \pi^{(k-1)}$).
		
		Then, the length of the longest common prefix of $h^{(1)} \dots h^{(k-1)} \pi^{(k)}$ and $\pi$ increases with $k$, and the set $K$ is finite if and only if there exists $n$ such that $h^{(1)} \dots h^{(n-1)} \pi^{(n)} = \pi$.
		
		In the infinite case, let:
		$$\chi = [v_0] \left[ \pi^{(0)} \right] \left[ h^{(1)} v_1 \right] \dots \left[ \pi^{(k)} \right] \left[ h^{(k)} v_k \right] \dots.$$
		The play $\chi$ is compatible with $\tau_{\Prover}$, hence $\mu^\a_\Challenger(\chi) < \alpha$, that is to say:
		$$\mu_i\left(h^{(1)} h^{(2)} \dots\right) < \alpha,$$
		i.e. $\mu_i(\pi) < \alpha$.
		
		In the finite case, let:
		$$\chi = [v_0] \left[ \pi^{(0)} \right] \left[ h^{(1)} v_1 \right] \dots \left[ \pi^{(n)} \right] \top^\omega.$$
		For the same reason, we have $\mu^\a_\Challenger(\chi) < \alpha$, i.e., we have $\mu_i \left( h^{(1)} \dots h^{(n)} \pi^{(n)} \right) = \mu_i(\pi) < \alpha$.
	
	\paragraph{Assertion~\ref{case2_pf_abs} implies Assertion~\ref{case1_pf_abs}.}
	
	Let $\bsigma_{-i}$ be a strategy profile that is $\lambda$-rational assuming a strategy $\sigma_i$, and that maintains player $i$'s payoff below the quantity $\alpha$.
	Let us define a strategy $\tau_{\Prover}$ for Prover in the abstract negotiation game.
	
	Let $g = [v_0] \left[\pi^{(0)}\right] \left[h^{(1)}v_1\right] \left[\pi^{(1)}\right] \dots \left[h^{(n)} v_n\right]$ be a history in the abstract game, ending in $V^\a_\Prover$. Then, we define:
	$$\tau_{\Prover}(g) = \left[ \< \bsigma_{\|h^{(1)} \dots h^{(n)}v_n} \> \right].$$

	If $g$ is a history ending in $\top$, then we define $\tau_\Prover(g) = \top$, and similarly, if $g$ ends in $\bot$, we define $\tau_\Prover(g) = \bot$.

    Let us show that $\tau_{\Prover}$ is the strategy we were looking for. 
	Let $\chi$ be a play compatible with $\tau_{\Prover}$. Let us first note that the vertex $\bot$ cannot appear in $\chi$. Then, the play $\chi$ can only have two forms: a play in which Challenger eventually accepts Prover's proposal, or an infinite sequence of proposals and deviations.
	
	\subp{If Challenger eventually accepts Prover's proposal}
    If $\chi = [v_0] \left[\pi^{(0)}\right] \left[h^{(1)}v_1\right]  \dots \left[\pi^{(n)}\right] \top^\omega$, then we have:
		$$\pi^{(n)} = \< \bsigma_{\|h^{(1)} \dots h^{(n)} v_n} \>,$$
		and the history $h^{(1)} \dots h^{(n)} v_n$ in the game $\Game_{\|v_0}$ is compatible with $\bsigma_{-i}$. By hypothesis, we have:
		$$\mu_i\left(h^{(1)} \dots h^{(n)} \pi^{(n)}\right) < \alpha,$$
		hence $\mu^\a_\Challenger(\chi) < \alpha$.
		
		\subp{If Challenger deviates infinitely often}
        If $\chi = [v_0] \left[\pi^{(0)}\right]  \dots \left[h^{(n)} v_n\right] \left[\pi^{(n)}\right] \dots$, then the play $\pi = h^{(1)} h^{(2)} \dots$ is compatible with $\bsigma_{-i}$, and by hypothesis we have $\mu_i(\pi) < \alpha$, hence $\mu^\a_\Challenger(\chi) < \alpha$.
\end{proof}

Of course, the abstract negotiation game does not provide a straightforward way to compute the negotiation function, since its vertex space is infinite, and even uncountable, in the general case.
However, it gives a first intuition.

\begin{exa}
    Let us consider again the game depicted by Figure~\ref{fig_ne_spe}, and consider the requirement $\lambda_1$, defined by $\lambda_1(a) = 0$, and $\lambda_1(b) = \lambda_1(c) = 1$.
    From each of the three vertices, let us present a play of the abstract negotiation game where both Prover and Challenger play optimally.

    From the vertex $c$, Prover proposes the play $c^\omega$, and Challenger cannot deviate from it: he may deviate from any vertex controlled by player $\Circle$, which is the case of the only vertex traversed by the play, but he has no alternative edge to use.
    Therefore, he accepts the play, and gets the payoff $1$, hence $\nego(\lambda_2)(c) = 1$.

    From the vertex $b$, Prover is not allowed to propose the play $b^\omega$, which is not $\lambda_1$-consistent (player $\square$ should get the payoff $1$, and gets only the payoff $0$).
    Therefore, she proposes the play $bc^\omega$, and Challenger accepts, hence $\nego(\lambda_1)(b) = 1$.

    From the vertex $a$, Prover may propose the play $a^\omega$.
    But then, Challenger can deviate to the vertex $b$, and from there, again, Prover has to propose a play that eventually reaches the vertex $c$, providing player $\Circle$, and therefore Challenger, the payoff $1$, hence $\nego(\lambda_1)(a) = 1$.
\end{exa}

Thus, checking that a given requirement $\lambda$ is an ($\epsilon$-)fixed point of the negotiation function can be done by guessing a finite representation of a strategy for Prover in the abstract negotiation game, from each vertex $u$, that forces the player controlling the vertex $u$ to get, at most, the payoff $\lambda(u)$ (or $\lambda(u) + \epsilon$).
Sometimes (when $\epsilon$ is clear from the context), we will abuse language and call such a strategy \emph{winning}, assimilating the abstract negotiation game with a Boolean version in which Prover's goal is to keep Challenger's payoff under $\lambda(u)$ (or $\lambda(u) + \epsilon$)---\emph{proving}, then, that $\lambda$ is an ($\epsilon$-)fixed point of the function $\nego$.
To obtain effective algorithms from that idea, we need therefore to prove that if such a strategy exists, there exists one that is simple, i.e. that has a finite, and small, representation.

\subsection{The concrete negotiation game.}
In the general case, the use that can be made of the abstract negotiation game is limited by its infinite (and uncountable) vertex space.
However, under the hypothesis that the game $\Game$ is prefix-independent, it can be turned into a game on a finite graph if Prover does not propose plays as a whole, but edge by edge.
In the \emph{concrete negotiation game} $\conc_{\lambda i}(\Game)_{\|v_0^\con}$, the vertices controlled by Prover have the form $(v, M)$, where $M \subseteq V$ memorizes the vertices seen since the last time Challenger did deviate, in order to control that the play Prover is constructing from that point is $\lambda$-consistent: for each $u \in M$, Prover has to give to the player controlling $u$ at least the payoff $\lambda(u)$.
Similarly, the vertices controlled by Challenger are of the form $(vv', M)$, where $vv' \in E$ is an edge proposed by Prover.
The game unfolds as follows, initially from the vertex $v_0^\con = (v_0, \{v_0\})$.

\begin{itemize}
    \item From the vertex $(v, M)$, Prover chooses an edge $vv'$ and proposes it to Challenger.
    She therefore moves to Challenger's vertex $(vv', M)$.
        
    \item Once an edge $vv'$ has been proposed, Challenger can accept it, moving to Prover's vertex $(v', M \cup \{v'\})$.
    Or, if $v \in V_i$, he can \emph{deviate}, and choose a new edge $vw$, moving then to Prover's vertex $(w, \{w\})$.
        
    \item Then, the game starts again from that new vertex.
\end{itemize}

Again, those proposals and deviations draw a play in the game $\Game_{\|v_0}$, in which Challenger intends to maximize player $i$'s payoff, while Prover intends to minimize it.
We give a formal definition below.

\begin{defi}[Concrete negotiation game] \label{def:concrete}
	Let $\Game$ be a prefix-independent game played on a finite graph, let $i \in \Pi$ and $v_0 \in V_i$, and let $\lambda$ be a requirement on $\Game$.
	The \emph{concrete negotiation game} of $\Game_{\|v_0}$ is the two-player zero-sum game $\conc_{\lambda i}(\Game)_{\|v^\con_0} = \left( \{\Prover, \Challenger\}, V^\con, (V^\con_{\Prover}, V^\con_{\Challenger}), E^\con, \mu^\con\right)_{\|v^\con_0}$, defined as follows:

	\begin{itemize}
	    \item player $\Prover$ is called \emph{Prover}, and player $\Challenger$ is called \emph{Challenger}.
	
		\item The set of vertices controlled by Prover is $V^\con_{\Prover} = V \times 2^V$, where the vertex $v^\con = (v, M)$ contains the information of the current vertex $v$ on which Prover has to define the strategy profile, and the \emph{memory} $M$ of the vertices that have been traversed so far since the last deviation, defining the requirements Prover has to satisfy.
		The initial vertex is $v^\con_0 = (v_0, \{v_0\})$.
		
		\item The set of vertices controlled by Challenger is $V^\con_{\Challenger} = E \times 2^V$, where in the vertex $v^\con = (uv, M)$, the edge $uv$ is the edge proposed by Prover.
		
		\item The set $E^\con$ contains three types of edges: \emph{proposals}, \emph{acceptations} and \emph{deviations}.
		
		\begin{itemize}
			\item Proposals are edges in which Prover proposes an edge of the game $\Game$:
			$$\Prop = \left\{ (v, M) (vv', M) ~\left|~
			vv' \in E, M \in 2^V \right.\right\}.$$
			
			\item Acceptations are edges in which Challenger accepts to follow the edge proposed by Prover (it is in particular his only possibility when that edge begins on a vertex that is not controlled by player $i$):
			$$\Acc = \left\{ (vv', M)\left(v', M \cup \{v'\}\right) ~\left|~
			vv' \in E, M \in 2^V
			\right.\right\}.$$
            Note that the memory is updated.
			
			\item Deviations are edges in which Challenger refuses to follow the edge proposed by Prover, as he can if that edge begins in a vertex controlled by player $i$.
            The memory is then erased, and only the new vertex the deviating edge leads to is memorized:
			$$\Dev = \left\{ (vv', M) (w, \{w\})	~\left|~ 
			v \in V_i, w \neq v', vv', vw \in E, M \in 2^V
			\right.\right\}.$$
		\end{itemize}

		\item Let $g = (h_0, M_0) (h_0h'_0, M_0) \dots (h_nh'_n, M_n)$ be a history in $\conc_{\lambda i}(\Game)$: the \emph{projection} of the history $g$ is the history $\dg = h_0 \dots h_n$ in the game $\Game$.
	    That definition is naturally extended to plays.

		\item The payoff function $\mu^\con_\Challenger = -\mu^\con_\Prover$ measures player $i$'s payoff, with a winning condition if the constructed strategy profile is not $\lambda$-rational, that is to say if after finitely many player $i$'s deviations, it generates a play which is not $\lambda$-consistent:
		
		\begin{itemize}
		    \item we define $\mu^\con_\Challenger(\pi) = +\infty$ if after some index $n \in \NN$, the play $\pi_{\geq 2n}$ contains no deviation, and if the projection $\dpi_{\geq n}$ is not $\lambda$-consistent\footnote{When we combine the notations $\dpi$ and $\pi_{\geq n}$, the notation $\dpi$ is applied first; that is, the play $\dpi_{\geq n}$ is the projection of the play $\pi_{\geq 2n}$, not $\pi_{\geq n}$.};
		    
		    \item  and $\mu^\con_\Challenger(\pi) = \mu_i(\dpi)$ otherwise.
		\end{itemize}
	\end{itemize}
\end{defi}

Like in the abstract negotiation game, the goal of Challenger is to find a $\lambda$-rational strategy profile that forces the worst possible payoff for player $i$, and the goal of Prover is to find a possibly deviating strategy for player $i$ that gives them the highest possible payoff.

\begin{rem}
    The concrete negotiation game has the following properties.
    \begin{itemize}
        \item If the game $\Game$ is Borel, then the game $\conc_{\lambda i}(\Game)$ is Borel.
    
        \item When $\pi_{\geq 2n}$ contains no deviation, the memory of its vertices is increasing, and therefore eventually equal to the memory $M = \Occ(\dpi_{\geq n})$.
        If it is the longest such suffix of $\pi$, it means that the projection $\dpi_{\geq n}$ is $\lambda$-consistent if and only if for each player $j$ and each vertex $v \in V_j$, we have $\mu_j(\dpi) \geq \lambda(v)$.
    \end{itemize}
\end{rem}

We can now prove the equivalent of \cref{thm:abstract}.
For convenience, we prove it only when the game $\Game$ is Borel.

\begin{thm} \label{thm:concrete}
	Let $\Game$ be a Borel prefix-independent game played on a finite graph.
	Let $\lambda$ be a requirement, let $i$ be a player and let $v_0 \in V_i$.
	Then, we have the equality $\val_\Challenger\left(\conc_{\lambda i}(\Game)_{\|v^\con_0}\right) = \nego(\lambda)(v_0)$.
	Moreover, if for each player $i$ and every state $v_0 \in V_i$, Prover has an optimal strategy in $\conc_{\lambda i}(\Game)_{\|v_0^\con}$, then $\Game$ is a game with steady negotiation.
\end{thm}

\begin{proof} \paragraph{First direction: the inequality $\nego(\lambda)(v_0) \leq \val_\Challenger\left(\conc_{\lambda i}(\Game)_{\|v^\con_0}\right)$.}
	
	Let $\tau_\Prover$ be a strategy such that $\sup_{\tau_\Challenger} \mu^\con_\Challenger\< \btau \> \neq +\infty$, and let us define inductively the strategy profile $\bsigma$ as follows: for every history $h \in \Hist_{-i}\Game_{\|v_0}$ compatible with $\bsigma_{-i}$, there exists (by induction) exactly one history $g$ compatible with $\tau_\Prover$ such that $\dg = h$.
    Let then $\bsigma(\dg) = w$ for every history $g$ compatible with $\tau_\Prover$ with $\tau_\Prover(g) = (vw, M)$ for some $M$.
    The strategy profile $\bsigma$ is arbitrarily defined on other histories.
	We prove that the strategy profile $\bsigma_{-i}$ is $\lambda$-rational assuming the strategy $\sigma_i$, and that $\sup_{\sigma'_i} \mu_i\< \bsigma_{-i}, \sigma'_i \> \leq \sup_{\tau_\Challenger} \mu^\con_\Challenger\< \btau \>$.
	
	\subp{The strategy profile $\bsigma_{-i}$ is $\lambda$-rational, assuming the strategy $\sigma_i$}
    Indeed, let us assume it is not.
		Then, there exists a history $h = h_0 \dots h_n$ in the game $\Game_{\|v_0}$ compatible with $\bsigma_{-i}$ such that the play $\< \bsigma_{\|h} \>$ is not $\lambda$-consistent.
		Then, let:
		$$g v^\con = \left(h_0, M_0\right) \left(h_0\bsigma(h_0), M_0\right) \dots \left(h_n, M_n\right)$$
		be the only history in $\conc_{\lambda i}(\Game)_{\|v^\con_0}$ compatible with $\tau_\Prover$ such that $\dg = h$.
		Let $\tau_\Challenger$ be a strategy constructing the history $h$, defined by:
		$$\tau_\Challenger\left(g_0 \dots g_{2k-1}\right) = g_{2k}$$
		for every $k$, and:
		$$\tau_\Challenger\left(g' (vw, M)\right) = (w, M \cup \{w\})$$
		for any other history $g' (vw, M)$.
		Then, the play $\pi = \< \btau \>$ contains finitely many deviations (Challenger stops the deviations after having constructed the history $h$), and the projection $\dpi_{\geq n}$ is not $\lambda$-consistent. Therefore, we have $\mu^\con_\Challenger(\pi) = +\infty$, which is false by hypothesis.

		\subp{The inequality $\sup_{\sigma'_i} \mu_i\< \bsigma_{-i}, \sigma'_i \> \leq \sup_{\tau_\Challenger} \mu^\con_\Challenger\< \btau \>$ holds}
		Let $\sigma'_i$ be a strategy for player $i$, and let $\pi = \< \bsigma_{-i}, \sigma'_i \>$.
		Let $\tau_\Challenger$ be a strategy such that for every $k$:
		$$\tau_\Challenger\left((\pi_0, \cdot)(\pi_0 \cdot, \cdot) \dots (\pi_k \cdot, \cdot)\right) = (\pi_{k+1}, \cdot),$$
		i.e. a strategy forcing the play $\pi$ against $\tau_\Prover$.
		Then, since $\mu^\con_\Challenger\< \btau \> \neq +\infty$ by hypothesis on $\tau_\Prover$, we have $\mu_i(\pi) = \mu^\con_\Challenger\< \btau \>$, hence $\mu_i\< \bsigma_{-i}, \sigma'_i \> \leq \sup_{\tau_\Challenger} \mu^\con_\Challenger\< \btau \>$, hence the desired inequality.

	\subp{Steady negotiation}
	Moreover, if $\tau_\Prover$ is optimal, then the $\lambda$-rational strategy profile $\bsigma_{-i}$ realizes the infimum:
	$$\inf_{\bsigma_{-i} \in \lRat(v_0)} \sup_{\sigma'_i} \mu_i\< \bsigma_{-i}, \sigma'_i \>,$$
	hence if there exists such an optimal strategy for every vertex $v_0$, then the game $\Game$ is with steady negotiation.

	\paragraph{Second direction: the inequality $\val_\Challenger\left(\conc_{\lambda i}(\Game)_{\|v^\con_0}\right) \leq \nego(\lambda)(v_0)$.}
	
	Let $\bsigma_{-i}$ be a $\lambda$-rational strategy profile from $v_0$, assuming the strategy $\sigma_i$; let us define a strategy $\tau_\Prover$, by $\tau_\Prover(g(v, \cdot)) = \left(v\bsigma(\dg v), \cdot\right)$ for every history $g$ and for every $v \in V$.
	Let us prove the inequality $\sup_{\tau_\Challenger} ~\mu^\con_\Challenger\< \btau \> \leq \sup_{\sigma'_i} ~\mu_i\< \bsigma_{-i}, \sigma'_i \>.$
	
	Let $\tau_\Challenger$ be a strategy for Challenger, and let $\pi = \< \btau \>$.
	If $\mu^\con_\Challenger(\pi) = +\infty$, then there exists $n$ such that the play $\pi_{\geq 2n}$ contains no deviation, i.e. $\dpi_{\geq n} = \< \bsigma_{\|\dpi_{\leq n}} \>$, and that play is not $\lambda$-consistent, which is impossible. 
	Therefore, we have $\mu^\con_\Challenger(\pi) \neq + \infty$, and as a consequence $\mu^\con_\Challenger(\pi) = \mu_i(\dpi) = \mu_i\< \bsigma_{-i}, \sigma'_i \>$ for some strategy $\sigma'_i$, hence $\mu^\con_\Challenger(\pi) \leq \sup_{\sigma'_i} \mu_i\< \bsigma_{-i}, \sigma'_i \>$, hence the desired inequality.
\end{proof}

\begin{exa}
    Consider again the game of Figure~\ref{fig_ne_spe}.
    The arena of the concrete negotiation game from the vertex $a$ is depicted by Figure~\ref{fig_concrete}.
    The blue vertices belong to Prover, the orange ones to Challenger.
    The dashed arrows depict the deviations, and optimal strategies for both Prover and Challenger, for the requirement $\lambda_1$ defined by $\lambda_1(a) = 0$ and $\lambda_1(b) = \lambda_1(c) = 1$, are defined by the thick arrows.

    \begin{figure}
        \centering
        \begin{tikzpicture}[node distance=2.5cm]
            \node[vert, rectangle, rounded corners, proverVert, initial below] (a-a) {$a, \{a\}$};
            \node[vert, rectangle, rounded corners, challengerVert, left of=a-a] (ab-a) {$ab, \{a\}$};
            \node[vert, rectangle, rounded corners, challengerVert, right of=a-a] (aa-a) {$aa, \{a\}$};
            \node[vert, rectangle, rounded corners, proverVert, below of=ab-a] (b-ab) {$b, \{a, b\}$};
            \node[vert, rectangle, rounded corners, proverVert, below of=aa-a] (b-b) {$b, \{b\}$};
            \node[vert, rectangle, rounded corners, challengerVert, left of=b-ab] (bb-ab) {$bb, \{a,b\}$};
            \node[vert, rectangle, rounded corners, challengerVert, right of=b-b] (bb-b) {$bb, \{b\}$};
            \node[vert, rectangle, rounded corners, challengerVert, below of=b-ab] (bc-ab) {$bc, \{a,b\}$};
            \node[vert, rectangle, rounded corners, challengerVert, below of=b-b] (bc-b) {$bc, \{b\}$};
            \node[vert, rectangle, rounded corners, proverVert, below of=bc-ab] (c-abc) {$c, \{a, b, c\}$};
            \node[vert, rectangle, rounded corners, proverVert, below of=bc-b] (c-bc) {$c, \{b, c\}$};
            \node[vert, rectangle, rounded corners, challengerVert, left of=c-abc] (cc-abc) {$cc, \{a,b,c\}$};
            \node[vert, rectangle, rounded corners, challengerVert, right of=c-bc] (cc-bc) {$cc, \{b,c\}$};

            \path[very thick] (a-a) edge[bend left] (aa-a);
            \path (aa-a) edge[bend left] (a-a);
            \path (a-a) edge[bend left] (ab-a);
            \path (ab-a) edge[bend left, dashed] (a-a);
            \path (b-ab) edge[bend left] (bb-ab);
            \path[very thick] (bb-ab) edge[bend left] (b-ab);
            \path (b-b) edge[bend left] (bb-b);
            \path[very thick] (bb-b) edge[bend left] (b-b);
            \path[very thick] (c-abc) edge[bend left] (cc-abc);
            \path[very thick] (cc-abc) edge[bend left] (c-abc);
            \path[very thick] (c-bc) edge[bend left] (cc-bc);
            \path[very thick] (cc-bc) edge[bend left] (c-bc);
            \path[very thick] (ab-a) edge (b-ab);
            \path[very thick] (aa-a) edge[dashed] (b-b);
            \path[very thick] (b-ab) edge (bc-ab);
            \path[very thick] (b-b) edge (bc-b);
            \path[very thick] (bc-ab) edge (c-abc);
            \path[very thick] (bc-b) edge (c-bc);
        \end{tikzpicture}
        \caption{A concrete negotiation game}
        \label{fig_concrete}
    \end{figure}
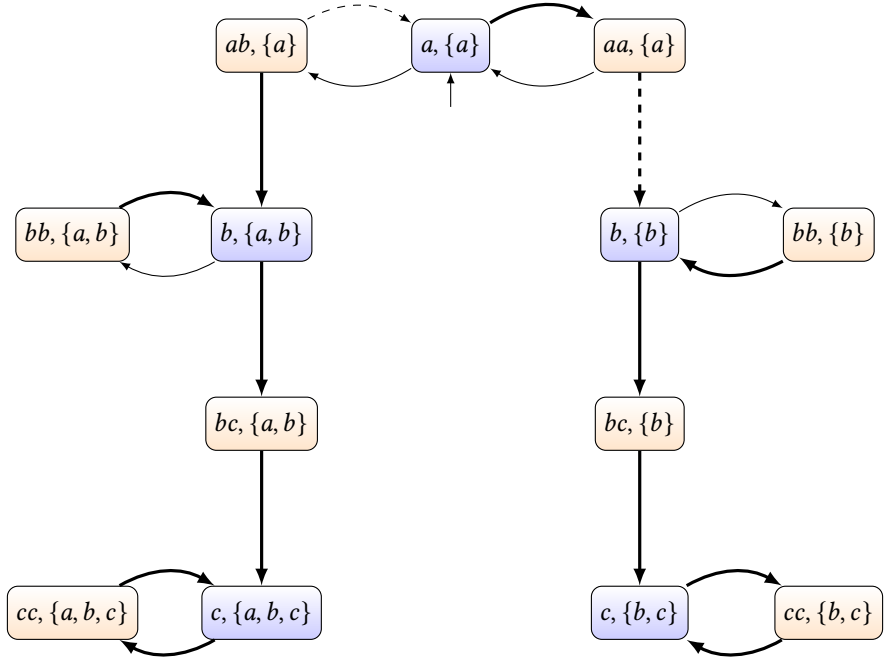
\end{exa}

Contrary to the abstract negotiation game, the concrete negotiation game has a finite vertex space.
However, the size of that vertex space is exponential in the size of the original game: constructing that game and solving it is therefore costly.
The concrete negotiation game will nevertheless be used to prove the fixed-parameter tractability of the SPE constrained existence problem in parity games (\cref{thm:spe_parity_fpt}), and to establish an intermediary result about the fixed points of the negotiation function in mean-payoff games (\cref{lm:mp_fixed_point}).

%% file: 2bParity.tex
In this chapter, we use the tools given in \cref{chap:nego} to prove $\NP$-completeness for the constrained existence problem of SPEs in parity games.
Note that, here, we are only interested in SPEs and not $\epsilon$-SPEs, since parity games are Boolean games, in which the notion of $\epsilon$-SPE has little relevance.

In order to design an efficient algorithm for that problem, we define an equivalence relation between histories and between plays; and, then, we show that in the abstract negotiation game, Prover can propose only plays that are simple representatives of their equivalence class, and propose always the same play from each vertex.

	\section{Reduced plays and reduced strategies}

\subsection{Definitions}
    
The equivalence relation that we use is based on the order in which vertices appear.
	
	\begin{defi}[Occurrence-equivalence]
		Two histories $h$ and $h'$ are \emph{occurrence-equivalent}, written $h \approx h'$, if and only if $\first(h) = \first(h')$, $\last(h) = \last(h')$ and $\Occ(h) = \Occ(h')$.

		Two plays $\pi$ and $\pi'$ are \emph{occurrence-equivalent}, written $\pi \approx \pi'$, if and only if the three following conditions are satisfied:
		\begin{itemize}
			\item we have $\Inf(\pi) = \Inf(\pi')$;
			
			\item for each history prefix of $\pi$, there exists a occurrence-equivalent history prefix of $\pi'$;
			
			\item for each history prefix of $\pi'$, there exists a occurrence-equivalent history prefix of $\pi$.
		\end{itemize}
	\end{defi}
	
\begin{exa}
    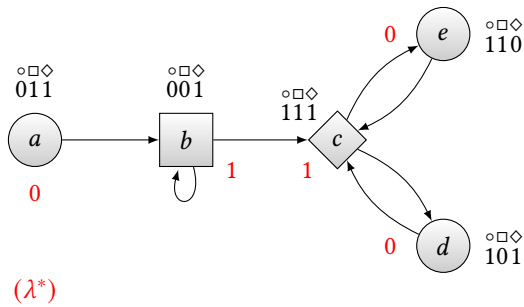
\begin{figure}
		    \centering
			\begin{tikzpicture}
			\node[vert] (a) at (-4, 0) {$a$};
			\node[vert, rectangle] (b) at (-2, 0) {$b$};
			\node[vert, diamond] (c) at (0, 0) {$c$};
			\node[vert] (d) at (1.4, -1.4) {$d$};
			\node[vert] (e) at (1.4, 1.4) {$e$};
			
			\path (a) edge (b);
			\path[loop below] (b) edge (b);
			\path (b) edge (c);
			\path[bend left = 20] (c) edge (d);
			\path[bend left = 20] (d) edge (c);
			\path[bend left = 20] (c) edge (e);
			\path[bend left = 20] (e) edge (c);
			
			\node (a') at (-4, 0.8) {$\stackrel{\circ}{0} \stackrel{\Box}{1} \stackrel{\Diamond}{1}$};
			\node (b') at (-2, 0.8) {$\stackrel{\circ}{0} \stackrel{\Box}{0} \stackrel{\Diamond}{1}$};
			\node (c') at (-0.5, 0.5) {$\stackrel{\circ}{1} \stackrel{\Box}{1} \stackrel{\Diamond}{1}$};
			\node (d') at (2.2, -1.4) {$\stackrel{\circ}{1} \stackrel{\Box}{0} \stackrel{\Diamond}{1}$};
			\node (e') at (2.2, 1.4) {$\stackrel{\circ}{1} \stackrel{\Box}{1} \stackrel{\Diamond}{0}$};
			
			\node[red] (l) at (-4, -2) {$(\lambda^*)$};
			\node[red] (a'') at (-4, -0.7) {$0$};
			\node[red] (b'') at (-1.4, -0.4) {$1$};
			\node[red] (c'') at (-0.4, -0.4) {$1$};
			\node[red] (d'') at (0.7, -1.4) {$0$};
			\node[red] (e'') at (0.7, 1.4) {$0$};
			\end{tikzpicture}
			\caption{A Büchi game.}
			\label{fig_buchi}
	\end{figure}

    Let us consider the game of Figure~\ref{fig_buchi}.
    In that game, the play $ab(cdce)^\omega$ is occurrence-equivalent to the play $abcd(cdce)^\omega$, but not to the play $ab(cecd)^\omega$.
    Indeed, the latter has the history $abce$ as a prefix, which is not occurrence-equivalent to any prefix of $ab(cdce)^\omega$, in which the vertex $e$ occurs only when the vertex $d$ has already occurred.
\end{exa}
	
	\begin{rem}
		The operators $\Occ$ and $\Inf$, and any parity payoff function, are stable by occurrence-equivalence.
	\end{rem}

\subsection{Representatives}\label{ssec:parity_representatives}

	The interest of that equivalence relation lies in the finite number of its equivalence classes, and by the existence of simple representatives for each of them.
	
	\begin{lem} \label{lm_lasso}
		Let $\pi$ be a play of $\Game$.
		There exists a lasso $h c^\omega \approx \pi$ where the history $h$ has length $|h| \leq n^3 + n^2$ and the cycle $c$ has length $|c| \leq n^2$, where $n = \card V$.
	\end{lem}

 \begin{proof}
		Let us write $W_0 \subset \dots \subset W_t$ for all the sets of the form $\Occ(\pi_{\leq k})$ with $k \in \NN$, without repetition.
		Note that for each index $s \in \{0, \dots, t-1\}$, the set $W_{s+1}$ contains the set $W_s$ plus one additional vertex.
		
		Let us construct the history $h$ and the cycle $c$ as follows, maintaining the hypothesis that for all $p$, the set $\Occ(h_{\leq p})$ is equal to $W_s$ for some $s$.
		
		\subp{Base case}
        First, we define $h_0 = \pi_0$, and $\{h_0\} = W_0$.
			
			\subp{Inductive case}
            Then, when the prefix $h_{\leq p}$ is constructed: let $s$ be such that $\Occ(h_{\leq p}) = W_s$, and let $k$ be the minimal integer such that $\Occ(\pi_{\leq k}) = W_s$.
			Let $U$ be the set of all the vertices $u$ such that there exists $\l$ with $\Occ(\pi_{\leq \l}) = W_s$ and $\pi_\l = u$: any such $\l$ is greater than or equal to $k$, and $U \subseteq W_s$. Then, there exists at least one path from $\pi_k = h_p$ that traverses all the vertices of $U$ and only them. Let $h_p \dots h_q$ be such a path with minimal length: it has at most length $n^2$.
			If $s < t$, let now $\l$ be the minimal index greater than $k$ such that $\Occ(\pi_{\leq \l}) = W_{s+1}$. Then, there exists a path from $h_q \in W_s$ to $\pi_\l$ that uses only vertices of $U$: let $h_q \dots h_r$ be such a path with minimal length. Then, it traverses all vertices at most once, and has therefore length at most $n$.
			If $s = t$, let $h_q \dots h_r$ be a path of minimal length from $h_q$ to a vertex $h_r \in \Inf(\pi)$: for the same reasons as above, such a path exists and has length at most $n$. Then, we can stop here the construction of $h$, and observe that the vertex $h_r$ belongs to the graph $(\Inf(\pi), E \cap \Inf(\pi)^2)$, which is strongly connected. We can therefore choose a cycle $c$ that traverses all its vertices and only them, and that has length at most $n^2$.

        \subp{Conclusion}
		By construction, the lasso $hc^\omega$ is occurrence-equivalent to $\pi$, and satisfies the desired size conditions.
	\end{proof}

	We call such lassos \emph{reduced plays}.
	For each play $\pi$, we write $\tpi$ for an arbitrary occurrence-equivalent reduced play.
	Then, given $\Game$ and a reduced play $\tpi$, operations such as computing the vector $\mu(\tpi)$, the sets $\Occ(\tpi)$ and $\Inf(\tpi)$, or checking whether the play $\tpi$ is $\lambda$-consistent, can be done in time $O(n^3)$.

\begin{exa}
    In the game of Figure~\ref{fig_buchi}, the play $ab(cdce)^\omega$ is reduced, but the occurrence-equivalent play $ab^{150}\left(cdce\right)^\omega$ is not.
\end{exa}

\begin{defi}[Reduced strategy]
    A strategy $\tau_\Prover$ for Prover in $\abs_{\lambda i}(\Game)$ is \emph{reduced} if and only if it is stationary, and for each vertex $v$, the play $\pi$ with $[\pi] = \tau_\Prover([v])$ is a reduced play.
\end{defi}

If we have $\pi \approx \pi'$, and if Challenger can deviate from the play $\pi$ after the history $hv$, then he can also deviate in $\pi'$ after some history $h'v$ that traverses the same vertices.
Thus, Prover can play optimally while proposing only reduced plays, and by proposing always the same play from each vertex; that is, by following a reduced strategy.
Note that since we will use the abstract negotiation game to check whether a given requirement is a fixed point of the negotiation function, we abuse language here by considering that Prover \emph{wins} if she maintains player $i$'s payoff below $\lambda(v_0)$, even though the abstract negotiation game is not exactly Boolean.

\begin{lem}\label{lm_parity_reduced_strategy}
	Prover has a winning strategy in the abstract negotiation game if and only if she has a reduced one.
\end{lem}

\begin{proof}
	Let us consider the \emph{reduced negotiation game}, i.e. the abstract negotiation game in which one would have removed all the vertices but:
	\begin{itemize}
        \item Prover's vertices, i.e., the set $V^\a_\Prover = V \cup \{\top, \bot\}$;
    
	    \item those of the form $[\tpi]$, where $\tpi$ is a reduced play;
	    
	    \item those of the form $[hv]$, where $h$ is a prefix of a $\lambda$-consistent reduced play $\tpi$, and has minimal length among the occurrence-equivalent prefixes of $\tpi$.
	\end{itemize}
	
	If the game $\Game$ has $n$ vertices, then this game has at most $n^{n^3 + 2n^2} (n^3 + 3n^2) + 1$ vertices.
	Let us now notice that it has another interesting property.

    \begin{slem}\label{slm:reduced_stationary}
        In the reduced negotiation game, either Prover or Challenger has a stationary winning strategy.
    \end{slem}

    \begin{proof}
	By Lemma~\ref{lm:concave} (and since the reduced game is Borel), that is the case if both Prover's and Challenger's objectives are convex.
 Let therefore $\pi$ and $\chi$ be two plays in the reduced negotiation game, and $\xi$ be a shuffling of $\pi$ and $\chi$.
 Then, we observe that the play $\dxi$ is also a shuffling of the plays $\dpi$ and $\dchi$.
        The convexity of Prover's and Challenger's objectives is then a consequence of the convexity of parity objectives: the minimal color seen infinitely often by player $i$ in $\dxi$ is the minimum of the minimal colors seen infinitely often in $\dpi$ and in $\dchi$.
\end{proof}
	
	We can now prove our lemma by using the equivalence between the abstract and the reduced negotiation game.

		\subp{If Prover has a winning strategy in the abstract negotiation game, she has one in the reduced negotiation game}
		    
			We proceed by contraposition: if Prover has no winning strategy in the reduced game, then, by \cref{slm:reduced_stationary}, Challenger has a stationary one: let us write it $\tau_\Challenger$.
			
			Now, let us extend $\tau_\Challenger$ into a stationary winning strategy $\tau_\Challenger^\a$ in the abstract negotiation game.
			
			Let $[\pi] \in V^\a_\Challenger$, and let $[hvw] = \tau_\Challenger([\tpi])$.
			By occurrence-equivalence, there exists $k \in \NN$ such that $\pi_k = v$, and $\Occ(\pi_{\leq k}) = \Occ(hv)$.
			We then set $\tau_\Challenger^\a([\pi]) = [\pi_{\leq k} w]$.
			Note that the vertices $\tau_\Challenger^\a([\pi]) = [\pi_{\leq k} w]$ and $\tau_\Challenger([\tpi]) = [hvw]$ can be different, but they both have as unique successor the vertex $[w]$. 
			
			Let us prove, now, that $\tau_\Challenger^\a$ is winning: let $\chi^\a$ be a play compatible with $\tau_\Challenger^\a$.
			When Prover proposes a play $\pi$, in the abstract game, against the strategy $\tau_\Challenger^\a$, and when she proposes the play $\tpi$ in the reduced game against the strategy $\tau_\Challenger$, the same thing happens in both cases: either Challenger accepts in both games, or he deviates, and Prover has to propose a new play from the same vertex $w$.
			Therefore, we can define from $\chi^\a$ a play $\chi$ compatible with $\tau_\Challenger$ in the reduced game, in which each Challenger's vertex $[\pi]$ is replaced by the vertex $[\tpi]$, and Prover's vertices are replaced accordingly.
			
			Since the play $\chi$ is compatible with $\tau_\Challenger$, it is winning for Challenger: let us then prove that so is $\chi^\a$.
			If $\chi^\a$ has the form $g [\pi] \top^\omega$, then $\chi$ has the form $g' [\tpi] \top^\omega$, hence $\tpi$ is winning for player $i$, and therefore $\pi$ is winning for player $i$ and $\chi^\a$ is winning for Challenger.
			If $\chi$ never reaches the vertex $\top$, then we have $\dchi = h^0 h^1 \dots$ and $\dchi^\a = h^{0\a} h^{1\a} \dots$ where, for every $k$, the histories $h^k$ and $h^{k\a}$ are possibly different, but contain exactly the same vertices.
			Then, the set of player $i$'s colors appearing infinitely often in $\dchi$ and $\dchi^\a$ are the same, and the play $\dchi^\a$ is winning for player $i$, i.e. the play $\chi^\a$ is winning for Challenger: the strategy $\tau^\a_\Challenger$ is winning.
		
			\subp{If Prover has a winning strategy in the reduced negotiation game, she has a reduced one in the abstract negotiation game}
			
			Indeed, let $\tau_\Prover$ be a winning strategy for Prover in the reduced negotiation game.
			As said above, we can assume that $\tau_\Prover$ is stationary.
			Since in the abstract game, the only vertices controlled by Prover where she has several possible choices are the ones of the form $[v]$, for $v \in V$, we can see $\tau_\Prover$ as a reduced strategy in the abstract game: let us write it $\tau_\Prover^\a$ in that case.
			We now have to prove that $\tau_\Prover^\a$ is also a winning strategy.
			
			Let $\chi^\a$ be a play in the abstract negotiation game compatible with $\tau_\Prover^\a$.
			For any sequence of vertices $[v] [\tpi] [hw] [w]$ that appears in $\chi^\a$, the history $h$ is occurrence-equivalent to some prefix $\th$ of $\tpi$ such that $[\th w]$ is a vertex of the reduced game.
			Therefore, we can transform the play $\chi^\a$ into a play $\chi$ of the reduced game, compatible with $\tau_\Prover$, where each vertex of the form $[hw]$ have been replaced by $[\th w]$.
			Since $\tau_\Prover$ is a winning strategy in the reduced negotiation game, the play $\chi$ is winning for Prover.
            Let us prove that so is $\chi^\a$.
			
			If $\chi^\a$ has the form $g [\tpi] \top^\omega$, then $\chi$ has the form $g' [\tpi] \top^\omega$, and since $\chi$ is winning for Prover, the play $\tpi$ is losing for player $i$, and therefore the play $\chi^\a$ is winning for Prover.
			If $\chi^\a$ has the form:
			$$\chi^\a = [v_0] [\tpi^0] [h^0v_1] [v_1] [\tpi^1] [h^1v_2] \dots$$
			then we have:
			$$\chi = [v_0] [\tpi^0] [\th^0v_1] [v_1] [\tpi^1] [\th^1v_2] \dots$$
			and since for each $k$, we have $\Occ(h^k) = \Occ(\th^k)$, we find $\Inf(h^0 h^1 \dots) = \Inf(\th^0 \th^1 \dots)$ and therefore, if the play $\chi$ is winning for Prover, so is the play $\chi^\a$.
			The strategy $\tau_\Prover^\a$ is a reduced winning strategy.
		
		Therefore, to conclude, if Prover has a winning strategy in the abstract negotiation game, she has a reduced one.
	\end{proof}

	\section{Checking that a reduced strategy is winning}
	
	We have established that Prover is winning the abstract negotiation game if and only if she has a reduced winning strategy.
	Such a strategy has polynomial size and can thus be guessed in nondeterministic polynomial time.
	We must now show that we can verify in deterministic polynomial time that a guessed strategy is winning.

	\begin{lem} \label{lm_parity_check_winning}
		Given a parity game $\Game$, a requirement $\lambda$, a player $i$, a vertex $v_0 \in V_i$, and a reduced strategy $\tau_\Prover$ in the abstract negotiation game $\abs_{\lambda i}(\Game)_{\|v_0}$, deciding whether $\tau_\Prover$ is a winning strategy can be done in polynomial time.
	\end{lem}
	
	\begin{proof}
Let us consider the induced game $\abs_{\lambda i}(\Game)_{\|v_0}[\tau_\Prover]$, as defined in \cref{sec:stationary_finite_memory}.
In that game, there remain only:

\begin{itemize}
    \item Prover's vertices, i.e., the set $V \cup \{\top, \bot\}$;

    \item Challenger's vertices of the form $[\tpi]$, where there exists $v \in V$ from which the strategy $\tau_\Prover$ proposes the reduced play $\tpi$;

    \item Challenger's vertices of the form $[hv]$ accessible from one vertex $\tpi$ as defined in the previous item, that have minimal length among occurrence-equivalent prefixes of $\tpi$;
\end{itemize}
and the edges binding those vertices.
It therefore contains at most $n + n + n (n^3 + 2n^2)$ vertices (at most $n$ of the form $v$, at most $n$ of the form $[\tpi]$, and at most $n (n^3 + 2n^2)$ of the form $[hv]$: assuming each play $\tpi$ has the form $hc^\omega$ with $|hc| \leq n^3 + 2n^2$, a minimal prefix is necessarily a prefix of $hc$).
Deciding whether Challenger can win against the strategy $\tau_\Prover$ amounts, then, to looking for a play in that game that is winning for him, i.e. either (assuming $\lambda(v_0) = 0$---otherwise, Challenger cannot win except if Prover gives up, which is excluded by the definition of a reduced strategy):
\begin{itemize}
    \item that reaches a vertex $[\tpi]$ where $\tpi$ is won by player $i$ and then goes to the vertex $\top$;

    \item that has the form $[v_0][\tpi^0][h^0v_1][v_1][\tpi^1]\dots$ where the play $h^0 h^1 \dots$ is won by player $i$.
\end{itemize}

In the first case, the existence of such a play can clearly be decided in polynomial time.
In the second case, looking for such a play amounts to looking for a play satisfying the parity condition defined by $\kappa([v]) = m$ for each vertex $v \in V$, by $\kappa([\tpi]) = m$ for each play $\tpi$, and by $\kappa([hv]) = \min_k \kappa_i(h_k)$, where $m$ is the maximal color in the game $\Game$.
That can also be done in polynomial time.
\end{proof}

We have then all the necessary ingredients for an algorithm solving the SPE constrained existence problem.
Indeed, given a game $\Game_{\|v_0}$, two thresholds $\bx, \by \in (\QQ \cup \{\pm \infty\})^\Pi$ and a requirement $\lambda$, if there exists a $\lambda$-consistent play that generates a payoff vector between $\bx$ and $\by$, then it is the case of every occurrence-equivalent play, and therefore, there exists a reduced one that satisfies those conditions.
Therefore, an algorithm solving the SPE constrained existence problem consists in:
    \begin{itemize}
        \item guessing a requirement $\lambda$ on $\Game$ with values $0$ or $1$;
        
        \item guessing a reduced strategy $\tau_\Prover^u$ in each abstract negotiation game $\abs_{\lambda i}(\Game)_{\|u}$ with $i \in \Pi$ and $u \in V_i$;

        \item guessing a reduced play $\tpi$ in $\Game_{\|v_0}$;

        \item checking that $\tpi$ is $\lambda$-consistent and that $\bx \leq \mu(\tpi) \leq \by$;

        \item checking that each $\tau_\Prover^v$ is winning (and thus that $\nego(\lambda) = \lambda$).
    \end{itemize}

By Theorems~\ref{thm:nego_spe} and Lemmas~\ref{lm_parity_reduced_strategy} and \ref{lm_parity_check_winning}, that algorithm is correct and runs in non-deterministic polynomial time.
Finally, it is already known, since the work of Erich Grädel and Michael Ummels~\cite{GU08}, that the constrained existence problem of SPEs in parity (and co-Büchi) games is $\NP$-hard, even with only one effective lower threshold.
Once again, a quick modification of their proof, by adding one player, entails hardness with only one effective upper threshold.
Hence the following.

\begin{thm} \label{thm:spe_parity_np}
    In parity games, the constrained existence problem of SPEs is $\NP$-complete.
    Hardness still holds in co-Büchi games, and when there is no effective lower threshold, and only one effective upper threshold.
\end{thm}

While the NE constrained existence problem in Büchi games has been proved to be decidable in polynomial time, the problem is, to the best of our knowledge, still open for the SPE constrained existence problem.

\begin{oprob}\label{op:spe_buchi}
    Is the SPE constrained existence problem in Büchi games decidable in polynomial time?
\end{oprob}

	\section{A deterministic upper bound}

We end this chapter by mentioning an additional complexity result on the constrained existence problem of SPEs in parity games: it is fixed-parameter tractable. 

\begin{thm} \label{thm:spe_parity_fpt}
    The SPE constrained existence problem on parity games is fixed-parameter tractable when the number of players and the number of colors are fixed.
    More precisely, there exists a deterministic algorithm that solves that problem in time $O(2^{2^{pm}} n^{12})$, where $n$ is the number of vertices, where $p$ is the number of players and where $m$ is the number of colors.
\end{thm}

\begin{proof}
Let $\Game$ be a parity game, let $i \in \Pi$, and let $u \in V_i$.
    Let us assume, without loss of generality, that $m$ is even, and that the vertices of $\Game$ or labeled by the colors $0, \dots, m-1$.
    Let $\lambda$ be a requirement.
    The value of $\nego(\lambda)(u)$ can be computed as the value of the corresponding concrete negotiation game.
    Let us recall that in that game, Challenger wins a play $\chi$ if and only if either:
    \begin{itemize}
        \item player $i$ wins the projection $\dchi$;
    
        \item or Challenger stops deviating, and the play proposed by Prover after the last deviation is not $\lambda$-consistent. 
    \end{itemize}

    Moreover, since parity games are Boolean, a vertex $v$ that is visited after the last deviation induces either no constraint (if $\lambda(v) \leq 0$), or the constraint that the player controlling $v$ must win (if $\lambda(v) > 0$).
    Thus, the game need only memorize the set of players $i$ such that a vertex $v \in V_i$ with $\lambda(v) > 0$ has been visited.
    We therefore slightly modify the definition of the concrete game, to use a version in which the vertices are of the form $(v, M)$ or $(uv, M)$ with $M \subseteq \Pi$, instead of $M \subseteq V$, which gives it size at most $(n^2 + n) 2^p$ instead of $(n^2 + n) 2^n$.

    Thus, that game can be seen as a \emph{multi-parity game}, where a color function is defined on edges (and not on vertices as it usually is) for each \emph{dimension} $d \in \Pi \cup \{\star\}$:
    \begin{itemize}
        \item if $d = \star$, then for each edge $(u, M)(uv, M)$ or $(uv, M)(w, M')$, we define:
        $$\hkappa_\star((u, M)(uv, M)) = \hkappa_\star((uv, M)(w, M')) = \kappa_i(u);$$

        \item if $d = j \in \Pi$, then for each edge of the form $(u, M)(uv, M)$ we define:
        $$\hkappa_j((u, M)(uv, M)) = m-1,$$
        and for each edge of the form $(uv, M)(w, M')$ we define $\hkappa_j((uv, M)(w, M')) = 1$ if $w \neq v$ (i.e., if the edge is a deviation) and $\hkappa_j((uv, M)(w, M')) = \kappa_j(uv) + 1$ if $w = v$ (i.e., if it is not a deviation);
    \end{itemize}
    and where Challenger's goal consists in satisfying \emph{at least one} of the corresponding parity conditions.

    It is a multi-parity game with colors on edges, but one can easily transform it into a game with colors on vertices, up to adding one vertex in the middle of each edge, i.e. decomposing each edge $u^\con v^\con$ into two edges $u^\con \delta_{u^\con v^\con}$ and $\delta_{u^\con v^\con} v^\con$.
    
    Then, that game can also be interpreted as a Boolean Büchi game in the sense of \cite{DBLP:conf/concur/BruyereHR18}, i.e. a two-player zero-sum game in which the objective of the first player (Prover) is to validate a Boolean formula whose atoms are Büchi conditions.
    Indeed, Prover's objective can be written:
    $$\bigwedge_{d \in \Pi \cup \{\star\}} \bigvee_{k = 0}^{\frac{m}{2}} \left( \B\{ \delta_{u^\con v^\con} ~|~ \kappa_d(u^\con v^\con) = 2k\} \wedge \neg \B \{\delta_{u^\con v^\con} ~|~ \kappa_d(u^\con v^\con) < 2k\} \right),$$
    where $\B(W)$ is the Büchi objective associated to the set $W$, i.e. the objective of visiting infinitely often at least one vertex of $W$.
    That formula has at most $dm$ atoms, and has size $dm$, in a game of size $O(n^2 2^p)$, hence by Proposition~5 from \cite{DBLP:conf/concur/BruyereHR18}, there exists a deterministic algorithm that decides which player has a winning strategy in time:
    $$O\left( 2^{2^{(p+1)m}} (p+1)m + \left( 2^{(p+1)m 2^{(p+1)m}} O(n^2 2^p) \right)^5 \right)
    = 2^{2^{O(pm)}} n^{10}.$$
    
    Therefore, given a requirement $\lambda$, by constructing the concrete negotiation game and applying that algorithm on each vertex, it is possible to compute the requirement $\nego(\lambda)$ in time $2^{2^{O(pm)}} n^{11}$.
    Thus, it is possible to compute the iterations of the negotiation function from the vacuous requirement $\lambda_0$: since the function $\nego$ is non-decreasing, its least fixed point $\lambda^*$ (whose existence is guaranteed by Tarski's fixed point theorem, since the set of requirements is a complete lattice) will be reached in at most $n$ steps, and will therefore be found in time $2^{2^{O(pm)}} n^{12}$.
    
    Once $\lambda^*$ has been computed, given two thresholds $\bx$ and $\by$, the SPE constrained existence problem can be solved by searching, for each tuple $\bz \in \{0, \dots, m-1\}^\Pi$ where $z_i$ is even whenever $x_i = 1$ and odd whenever $y_i = 0$, a play $\pi$ in $\Game$ that avoids the set:
    $$W_{\bz} = \{v \in V_i ~|~ z_i \in 2\ZZ \mathrm{~and~} \lambda^*(v) = 1\},$$
    and such that for each $i$, we have $\min \kappa_i(\Inf(\pi)) = z_i$.
    For a given tuple $\bz$, the existence of a play can be decided by removing all the vertices of $W_{\bz}$ and the vertices $v$ such that $\kappa_i(v) < z_i$ for some $i$, then looking for a strongly connected component that contains at least one vertex $v$ with $\kappa_i(v) = z_i$ for each $i$, and finally check whether the vertices of that strongly connected component are accessible from the initial vertex in $\Game$ without visiting the vertices of $W_{\bz}$.
    All those computations can be done in time $O(n)$.
    Thus, once $\lambda^*$ has be computed, the SPE constrained existence problem can be solved in time $O(m^p n)$.
    
    Given a parity game $\Game_{\|v_0}$ and two thresholds $\bx$ and $\by$, solving the SPE constrained existence problem can be done in time $2^{2^{O(pm)}} n^{12}$, and is therefore fixed-parameter tractable with parameters $m$ and $p$.
\end{proof}

%% file: 2cMP.tex
Let us now move to the study of mean-payoff games.
Here, a similar algorithm will be used but with additional steps, due to the fact that optimal strategies in the abstract negotiation game may not be stationary.
Moreover, we will not only consider the constrained existence of SPEs, but also of $\epsilon$-SPEs, where $\epsilon \geq 0$ is given with the instance.
We will see that this relaxation does not require specific additional tools, nor entail additionnal complexity.

\section{Hardness}

We first show a lower bound for our problem.

\begin{lem} \label{lm:spe_mp_hardness}
    The constrained existence problem of $\epsilon$-SPE in mean-payoff games is $\NP$-hard, even when $\epsilon$ is fixed equal to $0$, and when there is no effective lower threshold and only one effective upper threshold.
\end{lem}

\begin{proof}
    The structure of this proof is strongly inspired from several proofs from Michael Ummels, in articles that have already been cited above.
    We proceed by reduction from the $\NP$-complete problem $\SAT$.
	
	Let $\phi = \bigwedge_{i = 1}^n \bigvee_{j = 1}^m L_{ij}$ be a formula from propositional logic, written in conjunctive normal form, over the finite variable set $X$.
    We construct a mean-payoff game $\Game^\phi_{\|v_0}$ that admits an SPE where the player $\Witness$ gets the payoff $1$, if and only if $\phi$ is satisfiable.

\paragraph{Construction of the game $\Game^\phi$}
    
	First, we define the set of players $\Pi = \{\Solver\} \cup X$: every variable of $\phi$ is a player and there are two additional players $\Solver$ and $\Witness$, called \emph{Solver} and \emph{Witness}, respectively.
	
	Then, let us define the vertex space: for each clause $C_i$, with $i \in \ZZ/n\ZZ$, of $\phi$, we define a vertex $C_i$ that is controlled by Solver, and for each literal $L_{ij}$ of $C_i$ we define a vertex $(C_i, L_{ij})$, that is controlled by the player $x$ such that $L_{ij} = x$ or $\neg x$.
    Witness controls no vertex.
	We add an edge from $C_i$ to $(C_i, L_{ij})$, and another one from $(C_i, L_{ij})$ to $C_{i+1}$.
	Moreover, we add a sink vertex $\bot$, with an edge from it to itself, and edges from all the vertices of the form $(C, \neg x)$ to $\bot$.
	
	We define the reward function $r$ on this game as follows:
	\begin{itemize}
		\item $r_\Solver(\bot\bot) = 0$, and $r_\Solver(uv) = 1$ for any other edge $vw$;
        
		\item $r_\Witness(\bot\bot) = 1$, and $r_\Witness(uv) = 0$ for any other edge $vw$;
		
		\item for each player $x$, we have $r_x(uv) = 0$ for every edge leading to a vertex of the form $v = (C, x)$, and $r_x(uv) = 1$ for any other edge.
	\end{itemize}
	
	Note that Solver and Witness can only get the payoffs $0$ or $1$, and that in every play, exactly one of them gets the payoff $1$.
	Another player $x$ gets the payoff $1$ in a play that never visits (or finitely often, or infinitely often but with negligible frequence) a vertex of the form $(C, x)$.
	Otherwise, he may get any payoff between $0.5$ and $1$, depending on the frequence with which such a vertex is visited.
	Finally, we initialize that game in $v_0 = C_1$.
	
	\begin{exa}
    The game $\Game^\phi$, when $\phi$ is the tautology $(x_1 \vee \neg x_1) \wedge \dots \wedge (x_6 \vee \neg x_6)$, is represented by Figure~\ref{fig_Gphi}.
    The rewards that are not written are equal to $1$, or to $0$ in the case of Witness.
    
    \begin{figure}
        \centering
        \begin{tikzpicture}
    		\node[vert, initial right] (C1) at (0:5) {$C_1$};
    		\node[vert] (C11) at (30:6) {$x_1$};
    		\node[vert] (C12) at (30:4) {$\neg x_1$};
    		\node[vert] (C2) at (60:5) {$C_2$};
    		\node[vert] (C21) at (90:6) {$x_2$};
    		\node[vert] (C22) at (90:4) {$\neg x_2$};
    		\node[vert] (C3) at (120:5) {$C_3$};
    		\node[vert] (C31) at (150:6) {$x_3$};
    		\node[vert] (C32) at (150:4) {$\neg x_3$};
    		\node[vert] (C4) at (180:5) {$C_4$};
    		\node[vert] (C41) at (210:6) {$x_4$};
    		\node[vert] (C42) at (210:4) {$\neg x_4$};
    		\node[vert] (C5) at (240:5) {$C_5$};
    		\node[vert] (C51) at (270:6) {$x_5$};
    		\node[vert] (C52) at (270:4) {$\neg x_5$};
    		\node[vert] (C6) at (300:5) {$C_6$};
    		\node[vert] (C61) at (330:6) {$x_6$};
    		\node[vert] (C62) at (330:4) {$\neg x_6$};
    		\node[vert] (b) at (0,0) {$\bot$};
    		
    		\path[->] (C1) edge node[right] {$\stackrel{x_1}{0}$} (C11);
    		\path[->] (C1) edge (C12);
            \path[->] (C11) edge (C2);
            \path[->] (C12) edge (C2);
            \path[->] (C12) edge (b);
            \path[->] (C2) edge node[above right] {$\stackrel{x_2}{0}$} (C21);
    		\path[->] (C2) edge (C22);
            \path[->] (C21) edge (C3);
            \path[->] (C22) edge (C3);
            \path[->] (C22) edge (b);
            \path[->] (C3) edge node[above left] {$\stackrel{x_3}{0}$} (C31);
    		\path[->] (C3) edge (C32);
            \path[->] (C31) edge (C4);
            \path[->] (C32) edge (C4);
            \path[->] (C32) edge (b);
            \path[->] (C4) edge node[left] {$\stackrel{x_4}{0}$} (C41);
    		\path[->] (C4) edge (C42);
            \path[->] (C41) edge (C5);
            \path[->] (C42) edge (C5);
            \path[->] (C42) edge (b);
            \path[->] (C5) edge node[below left] {$\stackrel{x_5}{0}$} (C51);
    		\path[->] (C5) edge (C52);
            \path[->] (C51) edge (C6);
            \path[->] (C52) edge (C6);
            \path[->] (C52) edge (b);
            \path[->] (C6) edge node[below right] {$\stackrel{x_6}{0}$} (C61);
    		\path[->] (C6) edge (C62);
            \path[->] (C61) edge (C1);
            \path[->] (C62) edge (C1);
            \path[->] (C62) edge (b);
            \path (b) edge[loop left] node[left] {$\stackrel{\Solver}{0}\stackrel{\Witness}{1}$} (b);
        \end{tikzpicture}
        \caption{The game $\Game^\phi$}
        \label{fig_Gphi}
    \end{figure}
\end{exa}

	Now, let us prove that there is an SPE in $\Game^\phi_{\|C_1}$ in which Solver gets the payoff $1$, if and only if the formula $\phi$ is satisfiable, that is, if there exists a valuation $\nu: X \to \{0, 1\}$ that satisfies it.
	
	\paragraph{If there is an SPE in $\Game^\phi_{\|C_1}$ in which Witness gets the payoff $0$, then $\phi$ is satisfiable.}
    
    Let us write $\bsigma$ for such an SPE, and let $\pi = \< \bsigma \>$.
		Since $\mu_\Witness(\pi) = 0$, the sink vertex $\bot$ is never visited.
		Let us define a valuation $\nu$ on $X$ as follows: for each variable $x$, we have $\nu(x) = 1$ if and only if $\mu_x(\pi) < 1$.
		
		Now, let $C$ be a clause of $\phi$: since $C$, as a vertex, is necessarily visited infinitely often and with a fixed frequence in the play $\pi$ (because no player ever go to the sink vertex $\bot$), one of its successors, say $(C, L)$, is visited with a non-negligible frequence (more formally, the time between two occurrences of $(C, L)$ is bounded).
		If $L$ is a positive literal, say $x$, then by definition of $\nu$, we have $\nu(x) = 1$ and the clause $C$ is satisfied.
		
		If $L$ has the form $\neg x$, then each time the vertex $(C, \neg x)$ is traversed, player $x$ has the possibility to deviate and to go to the sink vertex $\bot$, where he is sure to get the payoff $1$.
		Since $\bsigma$ is an SPE, it means that he already gets the payoff $1$ in the play $\pi$.
		By definition of $\nu$, we then have $\nu(x) = 0$, hence the literal $\neg x$ is satisfied, hence so is the clause $C$.
		
		The valuation $\nu$ satisfies all the clauses of $\phi$, and therefore satisfies the formula $\phi$ itself.

		\paragraph{If $\phi$ is satisfiable, then there is an SPE in which Witness gets payoff $0$.}
        
        Let $\nu$ be a valuation satisfying $\phi$, and let us define a strategy profile $\bsigma$ by:
		\begin{itemize}
			\item $\sigma_\Solver(hC) = (C, L)$ for each history $h C$ where $C$ is a clause of $\phi$, where $L$ is a literal of $C$ that is satisfied in the valuation $\nu$;
			
			\item and $\sigma_x(h(C, \neg x)) = \bot$ if and only if $\nu(x) = 1$ for each history $h(C, \neg x)$ where $C$ is a clause of $\phi$ and $x$ is a variable.
		\end{itemize}
	
		Any other vertex has only one successor, hence we now have completely defined a strategy profile.
		Now, let us prove it is an SPE, in which Witness gets the payoff $0$.
		
		Let $hC$ be a history, where $C$ is a clause of $\phi$.
		We want to prove that $\bsigma_{\|hC}$ is a Nash equilibrium, in which Witness gets the payoff $0$.
		Let $\pi = \< \bsigma_{\|hC} \>$.
		If $\mu_\Witness(\pi) = 1$, i.e. if $\pi$ is of the form $h D (D, \neg x) \bot^\omega$, then by definition of $\bsigma$ we have $\nu(x) = 0$.
		But then, we cannot have $\sigma_\Solver(D) = (D, \neg x)$: contradiction.
		The play $\pi$ never reaches the vertex $\bot$, and Witness gets the payoff $0$.
        Since Witness controls no vertex, he has no profitable deviation.
        On her side, Solver gets the payoff $1$, and as a consequence she does not have any profitable deviation.
		
		Now, if another player $x$ has a profitable deviation, it means that he does not get the payoff $1$ in $\pi$, and therefore that some vertex of the form $(D, x)$ is visited infinitely often.
		But then, if Solver chooses to go to the vertex $(D, x)$, it means that the literal $x$ is satisfied in $\nu$, i.e. that $\nu(x) = 1$.
		In that case, if some clause $D'$ contains the literal $\neg x$, it is not a literal satisfied by $\nu$, and therefore the strategy $\sigma_\Solver$, as we defined it, never chooses the edge to the vertex $(D', \neg x)$, where player $x$ could have the possibility to deviate from his strategy.
		Contradiction.

		Finally, after a history of the form $h(C, L)$, either:
		\begin{itemize}
			\item we have $L = \neg x$ with $\nu(x) = 1$, and in that case, we have $\< \bsigma_{\|h(C, L)} \> = (C, L) \bot^\omega$, player $x$ gets the payoff $1$, and no player has a profitable deviation;
			
			\item or $L$ is a positive literal, and then there exists only one edge from the vertex $(C, L)$ to another clause $D$, and we go back to the previous case;
			
			\item or we have $L = \neg x$ with $\nu(x) = 0$, and in that case, we have $\sigma_x(C, L) = D$ where $D$ is the following clause, and by the first case the strategy profile $\bsigma_{\|h(C, \neg x)D}$ is a Nash equilibrium.
			Moreover, since the literal $\neg x$ is not satisfied in $\nu$, the play $\< \bsigma_{\|h(C, \neg x)D} \>$ does never traverse again any vertex of the form $(D', \neg x)$, hence player $x$ wins, and therefore has no profitable deviation: the strategy profile $\bsigma_{\|h(C, \neg x)}$ is a Nash equilibrium.
		\end{itemize}
	
	The constrained existence problem of SPEs, and therefore of $\epsilon$-SPEs, is $\NP$-hard in mean-payoff games.
\end{proof}

The rest of this chapter is dedicated to prove that the same problem is $\NP$-easy, and therefore $\NP$-complete.

\section{Manipulating sets of payoff vectors}

This section provides us with a toolbox to compute and manipulate the sets of payoff vectors that can be achieved in a mean-payoff game.

\subsection{Achievable payoff vectors in a mean-payoff game} \label{ssec:achievable_payoffs}

A first important result that we need is the characterization of the set of possible payoff vectors in a mean-payoff game, which has been introduced in \cite{DBLP:conf/concur/ChatterjeeDEHR10}.
Let us recall that given a graph $(V, E)$, we denote by $\SC(V, E)$ the set of simple cycles it contains.
Given a finite set $D$ of dimensions and a set $X \subseteq \RR^D$, we write $\Conv X$ for the convex hull of $X$, i.e., the set of vectors $\by$ such that there exists a family of positive real numbers $(\alpha_{\bx})_{\bx \in X}$ that satisfies the equalities $\sum_{\bx \in X} \alpha_{\bx} = 1$ and $\sum_{\bx \in X} \alpha_{\bx} \bx = \by$.
We will often use the subscript notation $\Conv_{x \in X} f(x)$ for the set $\Conv f(X)$.

\begin{defi}[Downward sealing] \label{def_dseal}
    Given a set $Y \subseteq \RR^D$, the \emph{downward sealing} of $Y$ is the set:
    $$\dseal Y = \left\{\left. \left( \min_{\bz \in Z} z_d \right)_{d \in D} ~\right|~ Z \mathrm{~is~a~finite~subset~of~} Y \right\}.$$
\end{defi}

\begin{lem}[\cite{DBLP:conf/concur/ChatterjeeDEHR10}]\label{lm:dseal}
    Let $\Game$ be a mean-payoff game, whose underlying graph is strongly connected.
    Then, we have the equality:
    $$\mu\left(\Plays\Game\right) = \dseal\left( \underset{c \in \SC(V, E)}{\Conv} \MP(c) \right).$$
\end{lem}

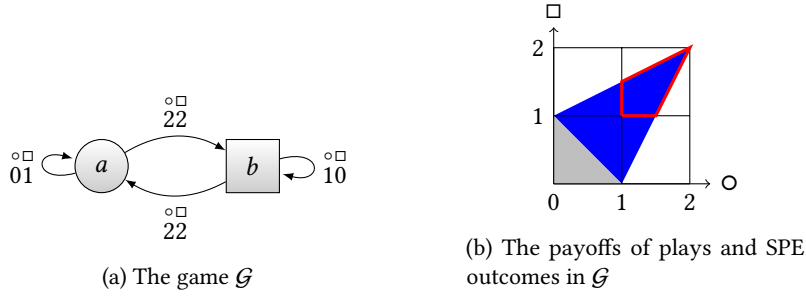
\begin{figure}
	\begin{subfigure}[b]{0.5\textwidth}
	    \centering
		\begin{tikzpicture}
		\node[vert] (a) at (0, 0) {$a$};
		\node[vert, rectangle] (b) at (2, 0) {$b$};
		\path (a) edge[bend left] node[above, rectangle] {$\stackrel{\circ}{2}\stackrel{\Box}{2}$} (b);
		\path (b) edge[bend left] node[below, rectangle] {$\stackrel{\circ}{2}\stackrel{\Box}{2}$} (a);
		\path (a) edge [loop left] node {$\stackrel{\circ}{0}\stackrel{\Box}{1}$} (a);
		\path (b) edge [loop right] node {$\stackrel{\circ}{1}\stackrel{\Box}{0}$} (b);
		\end{tikzpicture}
		\caption{The game $\Game$}
		\label{fig:inf_spe}
	\end{subfigure}
	\begin{subfigure}[b]{0.3\textwidth}
	    \centering
		\begin{tikzpicture}[scale=0.9]
		\draw [->] (0,0) -- (2.3,0);
		\draw (2.3,0) node[right] {${\Circle}$};
		\draw [->] (0,0) -- (0,2.3);
		\draw (0,2.3) node[above] {${\Box}$};
		\fill [blue] (0, 1) -- (1, 0) -- (2, 2);
		\fill [gray!50] (0, 1) -- (1, 0) -- (0, 0);
		\draw (0, 0) grid (2, 2);
		\foreach \x in {1,2} \draw(0,\x)node[left]{\x};
		\foreach \x in {0,1,2} \draw(\x,0)node[below]{\x};
		
		\draw [red, very thick] (1, 1) -- (1, 1.5) -- (2, 2) -- (1.5, 1) -- (1, 1);
		\end{tikzpicture}
	    \caption{The payoffs of plays and SPE outcomes in $\Game$} \label{fig:inf_spe_payoffs}
	\end{subfigure}
	
	\caption{An example of mean-payoff game}
\end{figure}

\begin{exa}
    If the set $Y$ is depicted by the blue area in Figure~\ref{fig:inf_spe_payoffs}, then the set $\dseal Y$ is obtained by adding the gray area.
    As a consequence, if $\Game$ is the game depicted by Figure~\ref{fig:inf_spe}, then the blue and the gray area in Figure~\ref{fig:inf_spe_payoffs} form the set of achievable payoffs in $\Game$.
    
	Indeed, following exclusively one of the three simple cycles $a$, $ab$ and $b$ of the game graph during a play yields the payoffs $01, 10$ and $22$, respectively.
	By combining those cycles with well chosen frequencies, one can obtain any payoff in the convex hull of those three points.
	It is also possible to obtain the point $00$ by using the properties of the limit inferior: it is for instance the payoff vector of the play $a^2 b^4 a^{16} b^{256} \dots a^{2^{2^n}} b^{2^{2^{n+1}}} \dots$.
	Then, by combining that play with the simple cycles, one can construct a play that yields any payoff in the convex hull of the four points $(0, 0), (0, 1), (1, 0)$, and $(2, 2)$.
\end{exa}

\subsection{Equations and inequations}

As the experienced reader might have guessed after reading the previous subsection, our algorithms for mean-payoff games will require cautious manipulation of sets of real numbers.
The proof of Theorem~\ref{lm:size_lambda} below, in particular, requires manipulation of \emph{polytopes}, e.g. downward sealings of convex hulls (from Lemma~\ref{lm:dseal}), expressed as solution sets of \emph{systems of linear inequations}.

\begin{defi}[Linear equations, inequations, systems]
    Let $D$ be a finite set.
    A \emph{linear equation} in $\RR^D$ is a pair $(\ba, b) \in \left(\RR^D \setminus \left\{\bzero\right\}\right) \times \RR$.
    The \emph{solution set} of the equation $(\ba, b)$ is the set $\Sol_=(\ba, b) = \{ \bx \in \RR^D ~|~ \ba \cdot \bx = b\}$, where $\cdot$ denotes the canonical scalar product on the euclidian space $\RR^D$.
    A set $X \subseteq \RR^D$ is a \emph{hyperplane} of $\RR^D$ if it is the solution set of some linear equation.
    A \emph{system of linear equations} is a finite set $\Sigma$ of linear equations.
    The \emph{solution set} of the system $\Sigma$ is the set $\Sol_= \Sigma = \bigcap_{(\ba, b) \in \Sigma} \Sol_=(\ba, b)$.
    A set $X \subseteq \RR^D$ is a \emph{linear subspace} of $\RR^D$ if it is the solution set of some system of linear equations.
    
    A \emph{linear inequation} in $\RR^D$ is also a pair $(\ba, b) \in \left(\RR^D \setminus \left\{\bzero\right\}\right) \times \RR$.
    The \emph{solution set} of the inequation $(\ba, b)$ is the set $\Sol_\geq(\ba, b) = \{ \bx \in \RR^D ~|~ \ba \cdot \bx \geq b\}$.
    A set $X \subseteq \RR^D$ is a \emph{half-space} of $\RR^D$ if it is the solution set of some linear inequation.
    A \emph{system of linear inequations} is a finite set $\Sigma$ of linear inequations.
    The \emph{solution set} of the system $\Sigma$ is the set $\Sol_\geq \Sigma = \bigcap_{(\ba, b) \in \Sigma} \Sol_\geq(\ba, b)$.
    A set $X \subseteq \RR^D$ is a \emph{polyhedron} of $\RR^D$ if it is the solution set of some system of linear inequations $\Sigma$.
    A \emph{vertex} of $X$ is a point $\bx \in \RR^D$ such that $\{\bx\} = \Sol_=(\Sigma')$ for some subset $\Sigma' \subseteq \Sigma$.
    A \emph{polytope} is a bounded polyhedron.
\end{defi}

\begin{rem}
\begin{itemize}
    \item Polyhedra are closed sets.

    \item The polytopes of $\RR^D$ are exactly the sets of the form $\Conv(S)$, where $S$ is a finite subset of $\RR^D$.

    \item A same pair $(\ba, b)$ may alternatively be seen as a linear equation or a linear inequation.
\end{itemize}    
\end{rem}

\subsection{Lemmas}

A first important result is the following: when the set $X$ is a polyhedron, so is $\dseal X$.
Given a system of inequations defining the polyhedron $X$, it is therefore possible to find a system of inequations defining $\dseal X$.
An exponential blowup might appear in the cardinality of that system, but not in the size of individual inequations.
Let us recall that what we mean with the \emph{size} of a number, or of a set, has been defined precisely in \cref{sec:conventions}.

\begin{lem}[\cite{DBLP:conf/concur/ChatterjeeDEHR10}] \label{lm:dseal_size}
    Let $\Sigma$ be a system of inequations, and let $X = \Sol_{\geq}(\Sigma)$.
    The set $\dseal X$ is itself a polyhedron, and there exists a system of inequations $\Sigma'$ such that $\dseal X = \Sol_\geq (\Sigma')$ and that for every $(\ba', b') \in \Sigma'$, there exists $(\ba, b) \in \Sigma$ with $\lv (\ba', b') \rv \leq \lv (\ba, b) \rv$.
\end{lem}

The previous lemma will be completed by the following, which bounds polynomially the size of the vertices of a polyhedron given the system of inequations that defines it.

\begin{lem}[\cite{DBLP:conf/cav/BrenguierR15}, Theorem~1] \label{lm:system_eq_to_point}
    There exists a polynomial $P_1$ such that, for every system of equations $\Sigma$, there exists a point $\bx \in \Sol_= \Sigma$, such that $\lv \bx \rv \leq P_1\left(\max_{(\ba, b) \in \Sigma} \lv (\ba, b) \rv\right)$.
\end{lem}

\begin{cor} \label{cor_size_vertices}
    For every system of inequations $\Sigma$, each vertex $\bx$ of the polyhedron $\Sol_{\geq}(\Sigma)$ has size:
    $$\lv x \rv \leq P_1\left(\max_{(\ba, b) \in \Sigma} \lv (\ba, b) \rv\right).$$
\end{cor}

Note that in Lemma~\ref{lm:dseal_size}, in Lemma~\ref{lm:system_eq_to_point} and in Corollary~\ref{cor_size_vertices}, the number of equations or inequations has no influence.
In the reverse direction, the size of the inequations defining a polyhedron can be polynomially bounded given the size of its vertices.

\begin{lem}\label{lm:conv_to_system}
    There exists a polynomial $P_2$ such that, for each finite set $D$ and every finite subset $X \subseteq \RR^D$, there exists a system of linear inequations $\Sigma$, such that $\Sol_\geq(\Sigma) = \Conv(X)$ and $\lv (\ba, b) \rv \leq P_2(\lv X \rv)$ for every $(\ba, b) \in \Sigma$.
\end{lem}

\begin{proof}
    First, let us recall the notion of \emph{facet}: a facet of a polytope $P$ is a subset of $P$ of dimension $\dim P - 1$ and of the form $P \cap H$, where $H$ is a hyperplane defined by an equation $(\ba, b)$, such that $P \subseteq \Sol_\geq(\ba, b)$.

    Each facet of the polytope $\Conv(X)$ is of the form $\Conv\{\bx_1, \dots, \bx_n\}$, where $\bx_1, \dots, \bx_n$ are vertices of $X$ and $n \geq d = \card D$.
    Let $\Phi$ be the set of those facets.
    We can then write (see for example~\cite{brondsted1983convex} for a proof):
    $$\Conv(X) = \Sol_{\geq} \left\{ (\ba_F, b_F) ~|~ F \in \Phi\right\},$$
    where for each $F$, the equation $(\ba_F, b_F)$ defines the hyperplane to which the facet $F$ belongs.
    Let us now study the complexity of each of those equations (or inequations).
    
    For a given facet $F = \Conv\{\bx_1, \dots, \bx_n\}$, let us choose $d$ points $\by_1, \dots, \by_d \in \{\bx_1, \dots, \bx_n\}$ that are linearly independent.
    The equation $(\ba_F, b_F)$ has the points $\by_1, \dots, \by_d$ among its solutions, i.e. it satisfies:
    $$\forall i \in \{1, \dots, d\}, \sum_{j=1}^d a_{Fj} y_{ij} - b_F = 0.$$
    Those $d$ equalities can themselve be understood as equations that are satisfied by the pair $(\ba_F, b_F)$.
    Let us add a $(d+1)$-th equation: for some dimension $i_0$, we have $a_{Fi_0} \neq 0$, and by multiplying if needed by a nonzero factor, we can assume $a_{Fi_0} = 1$.
    By Lemma~\ref{lm:system_eq_to_point}, there exists a pair $(\ba, b)$ satisfying that system of $d+1$ equations and the inequality $\lv (\ba, b) \rv \leq P_1 \left( 2d + 5 + \max_i \left( \sum_j \lv x_{ij} \rv \right) \right) \leq P_1 (\lv X \rv)$.
    That pair is an equation of the hyperplane containing $F$.
    
    As a consequence, the polynomial $P_2 = P_1$ satisfies the desired inequalities.
\end{proof}

    \section{About negotiation in mean-payoff games}

We now give some observations about the behavior of the negotiation function in mean-payoff games.
We first show a useful result: in the concrete negotiation game, stationary strategies are sufficient for Challenger (\cref{lm:concrete_stationary}).
We infer from that result that mean-payoff games are games with steady negotiation (\cref{lm:mp_steady_nego}), a result that is not as trivial as it was for parity games, and that is still necessary to apply \cref{thm:nego_spe}, and use the negotiation function as a tool to characterize SPEs and $\epsilon$-SPEs.
We end the section with two important examples, which should help the reader forge intuition on how the negotiation function can be used on mean-payoff games, before moving toward an algorithm.

\subsection{Stationary optimal strategies for Challenger}\label{ssec:mp_stat_optimal}

We first give a useful result for the sequel: in the concrete negotiation game, Challenger has a stationary optimal strategy.

\begin{lem} \label{lm:concrete_stationary}
	Let $\Game_{\|v_0}$ be a mean-payoff game, let $i$ be a player, and let $\lambda$ be a requirement.
	In the corresponding concrete negotiation game, there exists a stationary strategy $\tau_\Challenger$ that is optimal for Challenger.
\end{lem}

\begin{proof}
    The structure of that proof is inspired from the proof of Lemma~14 in \cite{DBLP:journals/iandc/VelnerC0HRR15}.
    
    Let $\nu_\Challenger$ be the payoff function defined by:
    \begin{itemize}
        \item $\nu_\Challenger(\pi) = +\infty$ if there exists $n$ such that $\pi_{\geq 2n}$ contains no deviation, and such that the play $\dpi_{\geq n}$ is not $\lambda$-consistent.
        
        \item $\nu_\Challenger(\pi) = \limsup_n \MP_i(\dpi_{\leq n})$ otherwise.
    \end{itemize}
    
    The payoff function $\nu_\Challenger$ is then defined as $\mu^\con_\Challenger$, but with a limit superior instead of inferior.
    We will first prove that Challenger has a stationary optimal strategy to maximize $\nu_\Challenger$, and then show that such a strategy is also optimal to maximize $\mu^\con_\Challenger$.
    
    \paragraph{The payoff function $\nu_\Challenger$ is concave.}
    Indeed, let $\pi$ and $\chi$ be two plays in $\conc_{\lambda i}(\Game)_{\|v_0}$, and let $\xi$ be a shuffling of those plays.
    Let us check that $\nu_\Challenger(\xi) \leq \max\{\nu_\Challenger(\pi), \nu_\Challenger(\chi)\}$.
    
    If either $\nu_\Challenger(\pi) = +\infty$ or $\nu_\Challenger(\chi) = +\infty$, it is immediate.
    Otherwise, we also have $\nu_\Challenger(\xi) \neq +\infty$: if either $\pi$ or $\chi$ contains infinitely many deviations, then so does $\xi$.
    If both contain finitely many deviations, then so does $\xi$: the vertices of $\xi$ have therefore eventually the same memory $M$, which is also the memory of, eventually, the vertices of both $\pi$ and $\chi$.
    Now, since $\nu_\Challenger(\pi), \nu_\Challenger(\chi) \neq +\infty$, we have $\mu_j(\dpi), \mu_j(\dchi) \geq \lambda(v)$ for each player $j$ and every $v \in M \cap V_j$.
    Since mean-payoff functions are convex, it is also the case for the play $\dxi$, which is a shuffling of $\dpi$ and $\dchi$.
    Hence $\nu_\Challenger(\xi) \neq +\infty$.
    
    Therefore, we have $\nu_\Challenger(\xi) = \limsup_n \MP_i(\dxi_{\leq n})$, as well as $\nu_\Challenger(\pi) = \limsup_n \MP_i(\dpi_{\leq n})$ and $\nu_\Challenger(\dchi) = \limsup_n \MP_i(\dchi_{\leq n})$.
    Since, as shown in \cite{DBLP:journals/iandc/VelnerC0HRR15}, mean-payoff functions defined with a limit superior are concave, it implies $\nu_\Challenger(\xi) \leq \max\{ \nu_\Challenger(\pi), \nu_\Challenger(\chi)\}$: the payoff function $\nu_\Challenger$ is concave.
    
    Therefore, by Lemma~\ref{lm:concave} Challenger has a stationary strategy that is optimal with regards to the payoff function $\nu_\Challenger$: let us write it $\tau_\Challenger$.

    \paragraph{The stationary strategy $\tau_\Challenger$ is also optimal with regards to $\mu^\con_\Challenger$.}
    
    Note that for every play $\pi$, we have $\mu^\con_\Challenger(\pi) \leq \nu_\Challenger(\pi)$, and therefore $\val_\Challenger\left( \conc_{\lambda i}(\Game)_{\|v^\con_0} \right) \leq \alpha$, where $\alpha$ is the value of the game $\conc_{\lambda i}(\Game)_{\|v^\con_0}$ with the payoff function $\nu_\Challenger$ instead of $\mu^\con_\Challenger$.
    Therefore, we have proven that $\tau_\Challenger$ is optimal with regards to $\mu^\con_\Challenger$ if we prove that $\inf_{\tau_\P} \mu^\con_\Challenger\< \btau \> \geq \alpha$.
    
    Let $\pi$ be a play compatible with $\tau_\Challenger$, i.e., a play in the game $\conc_{\lambda i}(\Game)_{\|v^\con_0}[\tau_\Challenger]$.
    If $\mu^\con_\Challenger(\pi) = +\infty$, then clearly $\mu^\con_\Challenger(\pi) \geq \alpha$.
    Otherwise, we have $\mu^\con_\Challenger(\pi) = \mu_i(\dpi)$, and by Lemma~\ref{lm:dseal}, we have:
    $$\mu_i(\dpi) \geq \min\left\{\MP_i(\dc) \mid c \in \SC(\conc_{\lambda i}(\Game)_{\|v^\con_0}[\tau_\Challenger])\right\}.$$
    Now, for each cycle $c \in \SC(\conc_{\lambda i}(\Game)_{\|v^\con_0}[\tau_\Challenger])$, there exists a history $g$ such that the play $g c^\omega$ is compatible with the strategy $\tau_\Challenger$, and therefore satisfies $\nu_\Challenger(gc^\omega) \geq \alpha$, and consequently $\MP_i(\dc) \geq \alpha$.
    Therefore, we have $\mu_i(\dpi) \geq \alpha$, and the strategy $\tau_\Challenger$ is optimal with regards to the payoff function $\mu^\con_\Challenger$, which concludes the proof.
\end{proof}

\subsection{Steady negotiation}

Using this lemma, computing $\nego(\lambda)$ for any given $\lambda$ amounts to looking for an optimal play for Prover in the game $\conc_{\lambda i}(\Game)_{\|v_0^\con}[\tau_\Challenger]$.
We show that there exists an optimal such play; and that this proves that mean-payoff games are games with steady negotiation, which will enable us to apply \cref{thm:nego_spe}.

\begin{lem} \label{lm:mp_steady_nego}
    Mean-payoff games are games with steady negotiation.
\end{lem}

\begin{proof}
    According to Theorem \ref{thm:concrete}, to prove that mean-payoff games are games with steady negotiation, it suffices to prove that Prover always has an optimal strategy in every concrete negotiation game constructed from a mean-payoff game.
    We have already noted that mean-payoff games are Borel, and that when a game is Borel, every concrete negotiation game constructed from it is Borel.
    By \cref{lm_borel_determinacy}, it is then equivalent to prove that, in every concrete negotiation game $\conc_{\lambda i}(\Game)_{\|v_0^\con}$ built from a mean-payoff game $\Game$, against every strategy $\tau_\Challenger$ of Challenger, Prover has a strategy to obtain at least the payoff $\val_\Prover(\conc_{\lambda i}(\Game)_{\|v_0^\con})$, i.e. a strategy so that Challenger obtains at most the payoff $\val_\Challenger(\conc_{\lambda i}(\Game)_{\|v_0^\con})$.
    Finally, by \cref{lm:concrete_stationary}, it is then sufficient to prove that result only for stationary strategies of Challenger.

    Let therefore $\tau_\Challenger$ be a stationary strategy of Prover.
    By definition of the adversarial value, we have:
    $$\inf \left\{ \mu^\con_\Challenger(\pi) \mid \pi \in \Plays\left(\conc_{\lambda i}(\Game)_{\|v_0^\con}[\tau_\Challenger]\right)\right\} \leq \val_\Challenger(\conc_{\lambda i}(\Game)_{\|v_0^\con}).$$
    We must now prove that this infimum is a minimum, i.e. that Prover can choose a worst play for Challenger in the game $\conc_{\lambda i}(\Game)_{\|v_0^\con}[\tau_\Challenger]$.
    
    For any play $\pi$ in that graph, there exists a strongly connected component $K$ of $\conc_{\lambda i}(\Game)_{\|v_0^\con}[\tau_\Challenger]$ such that after a finite number of steps, the play $\pi$ remains in $K$.
    Since there are finitely many strongly connected components in a graph, there is a \emph{worst} such strongly connected component, i.e. a strongly connected component $K$ such that we have:
$$\inf \left\{ \mu^\con_\Challenger(\pi) \mid \pi \in \Plays\left(\conc_{\lambda i}(\Game)_{\|v_0^\con}[\tau_\Challenger]\right)\right\} = \inf \left\{ \mu^\con_\Challenger(\pi) \mid \pi \in \Plays(K)\right\}.$$
We must now prove that there is a worst play in $K$.
Let us distinguish two cases.
    
    \subp{If there is at least one deviation in $K$}
    	
    	Then, for every play $\pi$ in $K$, it is possible to transform the play $\pi$ into a play $\pi'$ with $\mu(\dpi') = \mu(\dpi)$, which contains infinitely many deviations: it suffices to add round trips to a deviation, endlessly, but less and less often.
    	Therefore, the outcomes $\mu^\con_\Challenger(\pi)$ of plays in $K$ are exactly the mean-payoffs $\mu_i(\dpi)$ of plays in $K$ (plus possibly $+\infty$); and in particular, the lowest payoff that can be given to Challenger in $K$ is the quantity (using \cref{lm:dseal}):
    	$$\min_{c \in \SC(K)} \MP_i(\dc),$$
    	which is achieved by the play $c^\omega$, where $c$ is a simple cycle realizing that minimum.

    	\subp{If there is no deviation in $K$}
        Then, let us notice that deviations are the only edges that go from a vertex of the form $(\cdot, M)$ to a vertex of the form $(\cdot, M')$ with $M' \subset M$.
        Therefore, necessarily, there exists a set $M \subseteq V$ such that all vertices in $K$ have the form $(\cdot, M)$.
        
    	By Lemma~\ref{lm:dseal}, the set of possible values of $\mu(\dpi)$ for all plays $\pi$ in $K$ is exactly the set:
    	$$X = \dseal \left( \underset{c \in \SC(K)}{\Conv} \mu(\dc^\omega) \right).$$
    	
    	Since all plays in $K$ contain finitely many deviations (actually none), for every play $\pi$ in $K$, we have $\mu^\con_\Challenger(\pi) = +\infty$ if and only if there exists $j \in \Pi$ and $u \in V_j \cap M$ such that $\mu_j(\dpi) < \lambda(u)$.
    	Then, the lowest payoff that can be given to Challenger in $K$ is the quantity:
    	$$\inf \left\{ x_i ~|~ \bx \in X, \forall j \in \Pi, \forall u \in V_j \cap M, x_j \geq \lambda(u) \right\},$$
    	and since that set is closed, this infimum is a minimum, which concludes the proof.
\end{proof}

Note that the proof is constructive, suggesting an algorithm to compute the negotiation function.
We will however not use the concrete negotiation game for that purpose, because of its prohibitive size: it will only be used for intermediary results.

        \subsection{First application: an example with no SPE} \label{ssec:mp_no_spe}

It is a well-known fact that there are mean-payoff games that do not contain SPEs, while they are guaranteed to exist in Borel Boolean games such as parity or energy games~\cite{GU08}, and while NEs are guaranteed to exist in mean-payoff games~\cite{DBLP:conf/lfcs/BrihayePS13}.
We present here a classical example of such a game, along with a proof of the non-existence of SPEs based on the negotiation function.

\begin{thm}[\cite{SOLAN2003911}]\label{thm:no_spe}
    There exists a mean-payoff game with no SPE.
\end{thm}

\begin{proof}
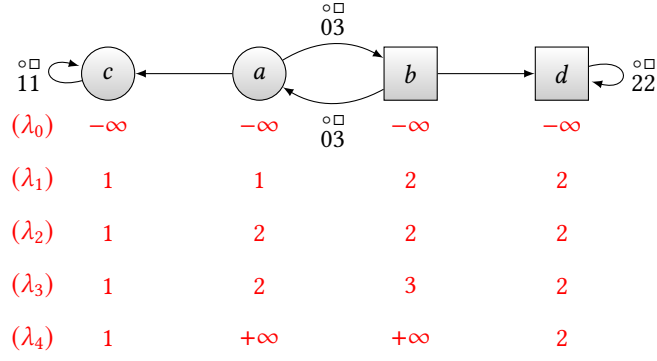
\begin{figure}
        \centering
		\begin{tikzpicture}
		\node[vert] (a) at (0, 0) {$a$};
		\node[vert] (c) at (-2, 0) {$c$};
		\node[vert, rectangle] (b) at (2, 0) {$b$};
		\node[vert, rectangle] (d) at (4, 0) {$d$};
		\path[->] (a) edge (c);
		\path (a) edge[bend left] node[above] {$\stackrel{\circ}{0} \stackrel{\Box}{3}$} (b);
		\path (b) edge[bend left] node[below] {$\stackrel{\circ}{0} \stackrel{\Box}{3}$} (a);
		\path[->] (b) edge (d);
		\path (d) edge [loop right] node {$\stackrel{\circ}{2} \stackrel{\Box}{2}$} (d);
		\path (c) edge [loop left] node {$\stackrel{\circ}{1} \stackrel{\Box}{1}$} (c);
		
		\node[red] (l0) at (-3, -0.7) {$(\lambda_0)$};
		\node[red] (l0a) at (0, -0.7) {$-\infty$};
		\node[red] (l0b) at (2, -0.7) {$-\infty$};
		\node[red] (l0c) at (-2, -0.7) {$-\infty$};
		\node[red] (l0d) at (4, -0.7) {$-\infty$};
		
		\node[red] (l1) at (-3, -1.4) {$(\lambda_1)$};
		\node[red] (l1a) at (0, -1.4) {$1$};
		\node[red] (l1b) at (2, -1.4) {$2$};
		\node[red] (l1c) at (-2, -1.4) {$1$};
		\node[red] (l1d) at (4, -1.4) {$2$};
		
		\node[red] (l2) at (-3, -2.1) {$(\lambda_2)$};
		\node[red] (l2a) at (0, -2.1) {$2$};
		\node[red] (l2b) at (2, -2.1) {$2$};
		\node[red] (l2c) at (-2, -2.1) {$1$};
		\node[red] (l2d) at (4, -2.1) {$2$};
		
		\node[red] (l3) at (-3, -2.8) {$(\lambda_3)$};
		\node[red] (l3a) at (0, -2.8) {$2$};
		\node[red] (l3b) at (2, -2.8) {$3$};
		\node[red] (l3c) at (-2, -2.8) {$1$};
		\node[red] (l3d) at (4, -2.8) {$2$};
		
		\node[red] (l4) at (-3, -3.5) {$(\lambda_4)$};
		\node[red] (l4a) at (0, -3.5) {$+\infty$};
		\node[red] (l4b) at (2, -3.5) {$+\infty$};
		\node[red] (l4c) at (-2, -3.5) {$1$};
		\node[red] (l4d) at (4, -3.5) {$2$};
		\end{tikzpicture}
		\caption{A game without SPE} 
		\label{fig:sans_spe}
\end{figure}

    Let us consider the game depicted by Figure~\ref{fig:sans_spe}.
    On the two first lines below the vertices, we present the requirements $\lambda_0$ and $\lambda_1 = \nego(\lambda_0)$.
    The latter is easy to compute since for each $v$, we have the equality $\lambda_1(v) = \val(v)$.
    Let us now compute the following iterations.

    \paragraph{Computation of $\lambda_2 = \nego(\lambda_1)$}
	From the vertex $c$, there exists exactly one $\lambda_1$-rational strategy profile $\bsigma_{-\circ} = \sigma_\Box$, which is the empty strategy since player $\Box$ never has to choose anything.
    Against that strategy, the best and the only payoff player $\Circle$ can get is $1$, hence $\lambda_2(c) = 1$.
	For the same reasons, we have $\lambda_2(d) = 2$.
	
	From the vertex $b$, player $\Circle$ can force $\Box$ to get the payoff $2$ or less, with the strategy (profile) $\sigma_\circ: h \mapsto c$. Such a strategy (profile) is $\lambda_1$-rational, assuming the strategy $\sigma_\Box: h \mapsto d$.
    Therefore, we have $\lambda_2(b) = 2$.
		
	Finally, from the vertex $a$, player $\Box$ can force $\Circle$ to get the payoff $2$ or less, with the strategy profile $\sigma_\Box: h \mapsto d$.
    Such a strategy is $\lambda_1$-rational, assuming the strategy $\sigma_\circ: h \mapsto c$.
    However, he cannot force her to get less than the payoff $2$, because she can force the access to the vertex $b$, and the only $\lambda_1$-consistent plays from $b$ are the plays with the form $(ba)^k b d^\omega$.
    Therefore, we have $\lambda_2(a) = 2$.

    \paragraph{Computation of $\lambda_3 = \nego(\lambda_2)$}
    The computation of $\lambda_3(c) = 1$ and $\lambda_3(d) = 2$ is immediate.

From the vertex $a$, player $\Circle$ cannot obtain more than the payoff $2$ against the strategy $\sigma_\square: h \mapsto d$.
And the strategy profile composed of that strategy is still $\lambda_2$-rational, assuming the strategy $\sigma_\circ: h \mapsto c$.
We therefore have $\lambda_3(a) = 2$.

From the vertex $b$, on the other hand, the only $\lambda_2$-rational strategy profile for player $\Circle$ is composed of the strategy $\sigma_\circ: ha \mapsto b$, which is $\lambda_2$-rational assuming the strategy $\sigma_\Box: h \mapsto d$: taking the edge from $a$ to $c$ is now unacceptable for player $\Circle$.
Consequently, using the strategy $\sigma'_\Box: hb \mapsto a$, player $\Box$ can force the payoff $3$, hence $\lambda_3(b) = 3$.

\paragraph{Computation of $\lambda_4 = \nego(\lambda_3)$ and conclusion}
Finally, let us observe that there is no $\lambda_3$-consistent play, and therefore no $\lambda_3$-rational strategy profile, from the vertices $a$ and $b$, hence $\lambda_4(a) = \lambda_4(b) = +\infty$.
The requirement $\lambda_4$ is then a fixed point of the negotiation function.

By a quick induction, we can show that every fixed point of the negotiation function is always greater than or equal to all the iterated $\nego^k(\lambda_0)$, and therefore every SPE outcome from the vertices $a$ or $b$ is necessarily $\lambda_4$-consistent.
Since there is no such play, there is no SPE that starts from one of those vertices.
\end{proof}

        \subsection{The impossibility of an algorithm based on iterations} \label{ssec:mp_not_stationary}

The two previous subsections suggest an algorithm based on the computation of the \emph{negotiation sequence}, i.e. the iterations of the negotiation function on the vacuous requirement $\lambda_0$, until a fixed point is reached.
Such an algorithm would be analogous to the one described in the proof of \cref{thm:spe_parity_fpt}.
However, such an approach cannot be used on mean-payoff games, because there are mean-payoff games in which the negotiation sequence does never reach any fixed point.

\begin{thm} \label{thm:not_stationary}
    There exists a mean-payoff game on which the negotiation sequence is not ultimately constant.
\end{thm}

\begin{proof}
    Let $\Game$ be the game of Figure~\ref{fig:not_stationary}, and let us compute the negotiation sequence $(\lambda_n)_{n \in \NN}$.
    Since all player $\Diamond$'s rewards are equal to $0$, for all $n > 0$, we have $\lambda_n(c) = \lambda_n(d) = \lambda_n(e) = \lambda_n(f) = 0$.
    Moreover, by symmetry of the game, we always have $\lambda_n(a) = \lambda_n(b)$. Therefore, to compute the negotiation sequence, it suffices to compute $\lambda_{n+1}(a)$ as a function of $\lambda_n(b)$, knowing that $\lambda_1(a) = \lambda_1(b) = 1$, and therefore that for all $n > 0$, we have $\lambda_n(a) = \lambda_n(b) \geq 1$.
    For convenience, we compute here the negotiation function using the abstract negotiation game.
    
    \begin{figure} 
    \begin{center}
    	\begin{tikzpicture}
    	\node[vert, diamond] (c) at (-1.7, -1) {$c$};
    	\node[vert, diamond] (d) at (-1.7, 1) {$d$};
    	\node[vert] (a) at (0, 0) {$a$};
    	\node[vert, rectangle] (b) at (2, 0) {$b$};
    	\node[vert, diamond] (e) at (3.7, 1) {$e$};
    	\node[vert, diamond] (f) at (3.7, -1) {$f$};
    	
    	\path[->] (a) edge[bend left] node[above] {$\stackrel{\circ}{2} \stackrel{\Box}{2} \stackrel{\Diamond}{0}$} (b);
        \path[->] (b) edge[bend left] node[below] {$\stackrel{\circ}{2} \stackrel{\Box}{2} \stackrel{\Diamond}{0}$} (a);
    	\path[->] (a) edge (c);
    	\path[->] (c) edge[bend left] node[left] {$\stackrel{\circ}{0} \stackrel{\Box}{1} \stackrel{\Diamond}{0}$} (d);
    	\path[->] (d) edge[bend left] node[above right] {$\stackrel{\circ}{0} \stackrel{\Box}{1} \stackrel{\Diamond}{0}$} (c);
    	\path (d) edge [loop above] node {$\stackrel{\circ}{2} \stackrel{\Box}{2} \stackrel{\Diamond}{0}$} (d);
    	\path[->] (b) edge (e);
    	\path[->] (e) edge[bend left] node[right] {$\stackrel{\circ}{1} \stackrel{\Box}{0} \stackrel{\Diamond}{0}$} (f);
    	\path[->] (f) edge[bend left] node[below left] {$\stackrel{\circ}{1} \stackrel{\Box}{0} \stackrel{\Diamond}{0}$} (e);
    	\path (f) edge [loop below] node {$\stackrel{\circ}{2} \stackrel{\Box}{2} \stackrel{\Diamond}{0}$} (f);
    	\end{tikzpicture}
    \end{center}
    \caption{A game where the negotiation sequence is not ultimately constant} \label{fig:not_stationary}
    \end{figure}
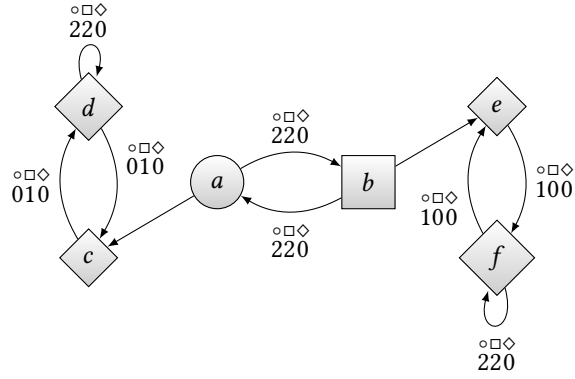

    From $a$, the worst play that Prover can propose would be a combination of the cycles $cd$ and $d$ giving her exactly $1$.
    But then, Challenger would deviate to go to $b$, from which, if Prover always proposes plays in the strongly connected component containing $c$ and $d$, Challenger will always deviate and generate the play $(ab)^\omega$, and then get the payoff $2$.
    
    Then, in order to give player $\Circle$ a payoff lower than $2$, Prover has to go to the vertex $e$.
    Since player $\Circle$ does not control any vertex in that strongly connected component, the play that Prover will propose will be accepted: she will, then, propose the worst possible combination of the cycles $ef$ and $f$ for player $\Circle$, such that player $\Box$ gets at least his requirement $\lambda_n(b)$.
    The payoff $\lambda_{n+1}(a)$ is then the minimal solution of the system:
    $$\left\{\begin{matrix}
    \lambda_{n+1}(a) = x + 2(1-x) \\
    2(1-x) \geq \lambda_n(b) \\
    0 \leq x \leq 1
    \end{matrix}\right.$$
    that is to say $\lambda_{n+1}(a) = 1 + \frac{\lambda_n(b)}{2} = 1 + \frac{\lambda_n(a)}{2}$. By induction, we obtain for all $n > 0$:
    $$\lambda_n(a) = \lambda_n(b) = 2 - \frac{1}{2^{n-1}}$$
    which converges to $2$ but never reaches it.
\end{proof}

Along with the prohibitive size of the concrete negotiation game, this theorem suggests that, if the constrained existence problem of SPEs in mean-payoff games is fixed-parameter tractable, a proof of that result must use a completely different approach.
We leave that problem open.

\begin{oprob}\label{op:spe_mp_fpt}
    Is the constrained existence problem of ($\epsilon$-)SPEs in mean-payoff games fixed-parameter tractable, when the number of players is fixed?
\end{oprob}

    \section{Size of the least \texorpdfstring{$\epsilon$}---fixed point} \label{sec:mp_fixed_point}

Now that the necessary conceptual tools are given, we present a technical but important result about mean-payoff games: the size of the least $\epsilon$-fixed point of the negotiation function in a mean-payoff game $\Game$ is bounded by a polynomial of the size of $\Game$ and $\epsilon$.
The first piece of the witnesses identifying positive instances of the $\epsilon$-SPE constrained existence problem will then be an $\epsilon$-fixed point of the negotiation function of polynomial size.

\subsection{Existence}

Before proving anything about its size, we must first prove that the negotiation function always has a least $\epsilon$-fixed point.

\begin{lem} \label{lm:mp_fixed_point}
	Let $\epsilon \geq 0$.
	On every game, the function $\nego$ has a least $\epsilon$-fixed point.
\end{lem}

\begin{proof}
    The following proof is a generalization of a classical proof of Tarski's fixed point theorem.
    Let $\Lambda_\epsilon$ be the set of $\epsilon$-fixed points of the negotiation function.
    The set $\Lambda_\epsilon$ is not empty, since it contains at least the requirement $v \mapsto +\infty$.
    Let $\lambda_\epsilon$ be the requirement defined by:
    $$\lambda_\epsilon: v \mapsto \inf_{\lambda \in \Lambda_\epsilon} \lambda(v).$$
    
    For every $\epsilon$-fixed point $\lambda$ of the negotiation function, and for each vertex $v$, we have then $\lambda_\epsilon(v) \leq \lambda(v)$, and therefore $\nego(\lambda_\epsilon)(v) \leq \nego(\lambda)(v)$ since $\nego$ is monotone; and therefore, we have $\nego(\lambda_\epsilon)(v) \leq \lambda(v) + \epsilon$.
    As a consequence, we have:
    $$\nego(\lambda_\epsilon)(v) \leq \inf_{\lambda \in \Lambda_\epsilon} \lambda(v) + \epsilon = \lambda_\epsilon(v) + \epsilon.$$
    The requirement $\lambda_\epsilon$ is an $\epsilon$-fixed point of the negotiation function, and is therefore the least of them.
\end{proof}

\subsection{Size}

\begin{lem}\label{lm:size_lambda}
    There exists a polynomial $P_3$ such that for every mean-payoff game $\Game$, the least $\epsilon$-fixed point $\lambda_\epsilon$ of the negotiation function has size $\lv \lambda_\epsilon \rv \leq P_3(\lv \Game \rv + \lv \epsilon \rv)$.
\end{lem}

\begin{proof}
    This proof will use a characterization of $\lambda_\epsilon$ as a vertex (or, more exactly, a well-chosen projection of a vertex) in a union of polyhedra, to which we will apply \cref{cor_size_vertices}.
    Those polyhedra will be defined using the concrete negotiation games, and the fact that, by \cref{lm:concrete_stationary}, stationary strategies are optimal for Challenger in those games.

    \paragraph{Reminders about the concrete negotiation game}

    Without loss of generality, we assume that all the rewards $r_i(uv)$ for $i \in \Pi$ and $uv \in E$ are integers---otherwise, we can multiply all the rewards by the least common denominator, compute the least $\epsilon$-fixed point in the resulting game, and finally divide each of its values by the least common denominator.
    
    Let now $i \in \Pi$ and $v \in V_i$, let $\lambda$ be a requirement, and let us consider the corresponding concrete negotiation game $\conc_{\lambda i}(\Game)_{\|(v, \{v\})}$.
    By \cref{lm:concrete_stationary}, we know that Challenger has a stationary optimal strategy in that game.
    Let then $\tau_\Challenger$ be a stationary strategy for Challenger in $\conc_{\lambda i}(\Game)_{\|(v, \{v\})}$ and let us consider the game $\conc_{\lambda i}(\Game)_{(v, \{v\})}[\tau_\Challenger]$.
    Facing the strategy $\tau_\Challenger$, Prover must choose an optimal play in that game, i.e., since the concrete negotiation game is prefix-independent, a strongly connected component $K$ of its underlying graph, and a good combination of cycles in $K$.
    There are, then, two possibilities: either Challenger does not deviate anywhere in $K$, and then, Prover must find a play that minimizes player $i$'s payoff while remaining $\lambda$-consistent.
    Or, there is such a deviation, and it is then possible for Prover to visit that deviation infinitely often (if necessary, with negligible frequency), and thus to ignore the constraints defined by $\lambda$.
    
    \paragraph{Size of mean-payoffs of cycles}
    
    Let therefore $K$ be a strongly connected component of $\conc_{\lambda i}(\Game)_{(v, \{v\})}[\tau_\Challenger]$. 
    Let $c = c_1 \dots c_n$ be a simple cycle of $K$: necessarily, we have $n \leq \card V^\con = \card V 2^{\card V}$.
    Therefore, we have, for each player $j$:
    \begin{align*}
        \lv \MP_j(\dc) \rv &= \left\lv \frac{1}{n} \sum_{k \in \ZZ/n\ZZ} r_j(\dc_k \dc_{k+1}) \right\rv \\
        &= 1 + \left\lceil \log_2\left( \left\vert \sum_{k \in \ZZ/n\ZZ} r_j(\dc_k \dc_{k+1}) \right\vert + 1 \right) \right\rceil + \lceil \log_2( n ) \rceil \\
        &\leq 1 + \left\lceil \log_2\left( \left\vert 2^{\card V} \sum_{e \in E} r_j(e) \right\vert + 1 \right) \right\rceil + \lceil \card V \log_2(\card V) \rceil \\
        &\leq 1 + \card V + \sum_{e \in E} \left\lceil \log_2\left( | r_j(e) | + 1 \right) \right\rceil + (\card V)^2 \\
        &\leq (\card E)^3 + \sum_{e \in E} \left\lceil \log_2\left( | r_j(e) | + 1 \right) \right\rceil \\
        &\leq \lv r_j \rv^3 \leq \lv \Game \rv^3.
    \end{align*}

    \paragraph{Feasible payoff vectors in a strongly connected component}
    
    Let us now define the polytope:
    $$F_K = \dseal \left(\underset{c \in \SC(K)}{\Conv} \MP(\dc) \right).$$
    By Lemma~\ref{lm:dseal}, the polytope $F_K$ is exactly the set of the payoff vectors of the projections of plays in the strongly connected component $K$.
    By Lemma~\ref{lm:conv_to_system}, the polytope:
    $$\underset{c \in \SC(K)}{\Conv} \MP(\dc)$$
    is the solution set of a system of inequations whose sizes are bounded by a polynomial function of $\lv \{\MP(\dc) ~|~ c \in \SC(K) \} \rv$, i.e., according to the previous point, of $\lv \Game \rv$.
    By Lemma~\ref{lm:dseal_size}, this is also the case for the polytope $F_K$.

    \paragraph{Consistent payoff vectors in a strongly connected component}
    
    If the strongly connected component $K$ contains no deviation, then all the vertices of $K$ share the same memory: there is a set $M_K \subseteq V$ such that all the vertices of $K$ have the form $(\cdot, M_K)$.
    If $K$ contains a deviation, it might not be the case: then, we define $M_K = \emptyset$.
    Thus, in both cases, the set $M_K$ defines the set of constraints a play must satisfy so that Challenger does not get the payoff $+\infty$: the vertices memorized so far if there is no deviation, and nothing if there is a deviation.
    
    Given a requirement $\lambda$, we can now define the polytope of \emph{consistent} payoff vectors:
    $$C_{K \lambda} = \left\{\bx \in F_K ~|~ \forall j, \forall u \in V_i \cap M_K, x_i \geq \lambda(u)\right\}.$$
    That polytope is the set of tuples $\bx$ such that there exists a play $\pi$ in $K$ realizing $\mu(\dpi) = \bx$, \emph{and} $\mu^\con_\Challenger(\pi) \neq +\infty$.
    We can then write:
    $$\nego(\lambda)(v) = \sup_{\tau_\Challenger} \inf_K \inf_{\bx \in C_{K\lambda}} x_i.$$
    The inequations defining $C_{K\lambda}$ are the same as those defining $F_K$, plus the inequations of the form $x_i \geq \lambda(u)$.

    \paragraph{Union, intersection and product}
    
    Let us now work in the space $\RR^{V \times \Pi}$.
    Given a requirement $\lambda$, we define the set:
    $$X_\lambda = \prod_{\substack{i \in \Pi\\v \in V_i}} \bigcap_{\tau_\Challenger \in \Stat\left( \conc_{\lambda i}(\Game)_{\|(v, \{v\})}\right)} \bigcup_K C_{K\lambda}^\uparrow,$$
    where $Y \mapsto Y^\uparrow$ denotes the upward closure operation, i.e. $Y^\uparrow = \{z ~|~ \exists y \in Y, z \geq y\}$.
    Then, for every tuple of tuples $\bbx \in \RR^{V \times \Pi}$, we have $\bbx \in X_\lambda$ if and only if for each $i$ and $v \in V_i$, for every stationary strategy of Challenger, there exists a play $\pi$ compatible with $\tau_\Challenger$ satisfying $\mu^\con_\Challenger(\pi) \neq +\infty$ and $\mu(\dpi) \leq \bx_v$.
    Therefore, for each $i$ and $v \in V_i$, we have $\nego(\lambda)(v) = \inf\left\{x_{vi} ~\left|~ \bbx \in X_\lambda\right.\right\}$.
    
    In terms of inequations, the set $X_\lambda$ is a union of polyhedra which are all defined by inequations that are inequations defining some $C_{K\lambda}$, padded with $0$ to fit with the dimension change.

    \paragraph{$\epsilon$-fixed points}
    
    To each tuple of tuples $\bbx$, we associate the requirement $\lambda_{\bbx}$ defined, for each $i \in \Pi$ and $v \in V_i$, by $\lambda_{\bbx}(v) = x_{vi} - \epsilon$.
    Now, let us consider the diagonal set:
    $$X = \left\{ \bbx \in \RR^{V \times \Pi} ~\left|~ \bbx \in X_{\lambda_{\bbx}} \right.\right\}.$$
    
    In terms of inequations, the set $X$ is a union of polyhedra defined by the same inequations as $X_\lambda$, but where those of the form $x_{ui} \geq \lambda(v)$ are replaced by equations of the form $x_{ui} \geq x_{vi}$, each of size $2 + 4\card V \card\Pi$.

    We can now link the set $X$ to $\epsilon$-fixed points of the negotiation function.

    \begin{slem}
    Let $\lambda$ be a requirement.
    Then, it is an $\epsilon$-fixed point of the negotiation function if and only if we have $\lambda = \lambda_{\bbx}$ for some $\bbx \in X$.
    \end{slem}

    \begin{proof}
    \begin{itemize}
        \item If there exists $\bbx \in X$ such that $\lambda = \lambda_{\bbx}$, then we have $\bbx \in X_{\lambda_{\bbx}}$, and by the previous point, for each $i$ and $v \in V_i$, we have $\nego(\lambda_{\bbx})(v) \leq x_{vi} = \lambda_{\bbx}(v) + \epsilon$.
        Therefore, $\lambda$ is an $\epsilon$-fixed point of the negotiation function.
        
        \item Conversely, if for each $i$ and $v \in V_i$, we have $\nego(\lambda)(v) \leq \lambda(v) + \epsilon$, then according to the previous point and since the set $X_{\lambda}$ is closed, there exists a tuple of tuples $\bbx^{(v)} \in X_{\lambda}$ such that $x^{(v)}_{vi} = \lambda(v) + \epsilon$.
        Then, since $X_{\lambda}$ is defined as a cartesian product over $v$, the tuple of tuples $\bbx = \left(\bx^{(v)}_v\right)_v$ does also belong to $X_{\lambda}$, and satisfies $\lambda = \lambda_{\bbx}$.
    \end{itemize}
\end{proof}
    
    Then, in particular, the least $\epsilon$-fixed point $\lambda_\epsilon$ is the (unique) minimal element in the set $\left\{\lambda_{\bbx} ~\left|~ \bbx \in X\right.\right\}$.
    The set $X$ is itself a union of polyhedra: consequently, the linear mapping $\bbx \mapsto \sum_v \lambda_{\bbx}(v)$ finds its minimum over $X$ on some vertex $\bbx$ of one of those polyhedra, which is therefore such that $\lambda_\epsilon = \lambda_{\bbx}$.
    By Corollary~\ref{cor_size_vertices}, such a vertex has size bounded by a polynomial function of the maximal size of the inequations defining $X$, and therefore a polynomial function of $\lv \Game \rv + \lv \epsilon \rv$.
\end{proof}

With this proof, we are now done with our utilization of the concrete negotiation game.
Like in the case of parity games, the negotiation function will rather be computed using simple strategies in the abstract negotiation game.

    \section{Constrained existence of a \texorpdfstring{$\lambda$}---consistent play}

We claim that a non-deterministic algorithm can recognize the positive instances of the $\epsilon$-SPE constrained existence problem by guessing an $\epsilon$-fixed point $\lambda$ of the negotiation function, as first piece of a three-piece witness.
Once $\lambda$ has been guessed, following Theorem~\ref{thm:nego_spe}, two assertions must be checked: on the one hand, that there exists a $\lambda$-consistent play between the two desired thresholds, and on the other hand, that $\lambda$ is actually an $\epsilon$-fixed point of the negotiation function.
The latter will be handled later through a new notion of reduced strategy.
Now, we tackle the former, and provide the second piece of our notion of witness: to prove the existence of a $\lambda$-consistent play $\pi$ with $\bx \leq \mu(\pi) \leq \by$, we need to guess the sets $W = \Inf(\pi)$ and $W' = \Occ(\pi)$, and a tuple of tuples $\bbalpha \in [0, 1]^{\Pi \times \SC(W)}$ indicating how $\pi$ combines the cycles of $W$ (we liberally assimilate the set $W$ to the induced subgraph $(W, E \cap (W \times W))$ to simplify notations), i.e. such that we have:
$$\mu(\pi) = \left( \min_{j \in \Pi} \sum_{c \in \SC(W)} \alpha_{jc} \MP_i(c) \right)_i.$$

\begin{lem} \label{lm:constrained_existence_lambda_cons}
    There exists a polynomial $P_4$ such that for every mean-payoff game $\Game_{\|v_0}$, for every $\bx, \by \in (\RR \cup \{\pm\infty\})^\Pi$, and for every requirement $\lambda$ on $\Game$, there exists a $\lambda$-consistent play $\pi$ in $\Game_{\|v_0}$ satisfying $\bx \leq \mu(\pi) \leq \by$ if and only if there exist two sets $W \subseteq W' \subseteq V$ and a tuple of tuples $\bbalpha \in [0, 1]^{\Pi \times \SC(W)}$ such that:
    \begin{itemize}
        \item the set $W$ is strongly connected in $(V, E)$, and accessible from the vertex $v_0$ using only vertices of $W'$;
        
        \item for each player $i$, we have $\sum_c \alpha_{ic} = 1$, and:
        $$x_i \leq \min_{j \in \Pi} \sum_{c \in \SC(W)} \alpha_{jc} \MP_i(c) \leq y_i;$$
        
        \item for each player $i$ and $v \in W \cap V_i$, we have:
        $$\min_{j \in \Pi} \sum_{c \in \SC(W)} \alpha_{jc} \MP_i(c) \geq \lambda(v);$$
        
        \item and finally, we have $\lv \bbalpha \rv \leq P_4(\lv \Game, \bx, \by, \lambda \rv)$.
    \end{itemize}
\end{lem}

\begin{proof}
    Let us first notice that given a set $X \subseteq \RR^\Pi$, the elements of the set $\dseal (\Conv X)$ are exactly the tuples of the form:
    $$\left( \min_{j \in \Pi} \sum_{x \in X} \alpha_{jx} x_i \right)_{i \in \Pi}$$
    for some tuple $\bbalpha \in \RR^{\Pi \times X}$ satisfying $\sum_x \alpha_{ix} = 1$ for each $i$.

    \subp{Now, let us assume that the sets $W$, $W'$ and the tuple $\bbalpha$ exist}
    Then, there exists a play $\chi$ with $\Occ(\chi) = \Inf(\chi) = W$ with payoff vector:
    $$\mu(\chi) = \left( \min_{j \in \Pi} \sum_{c \in \SC(W)} \alpha_{jx} \MP_i(c) \right)_{i \in \Pi}.$$
    Moreover, since $W$ is accessible from $v_0$ using only vertices of $W'$, there exists a history $h \chi_0$ from $v_0$ to $\chi_0$ with $\Occ(h) = W'$.
    Then, the play $\pi = h \chi$ is $\lambda$-consistent and satisfies $\bx \leq \mu(\pi) \leq \by$.
    
    \subp{Conversely, let us assume that the play $\pi$ exists}
    Let $W = \Inf(\pi)$ and $W' = \Occ(\pi)$.
    The polytope:
    \begin{align*}
    Z &= \left\{\mu(\chi) ~\left|~ \begin{matrix}
        \chi \in \lCons\Game_{\|v_0}, \\
        \Inf(\chi) = W, \\
        \Occ(\chi) = W', \\
        \mathrm{and~} \bx \leq \mu(\chi) \leq \by
    \end{matrix} \right. \right\} \\
    &= \left\{ \bz \in \dseal \left( \underset{c \in \SC(W)}{\Conv} \MP(c) \right) ~\left|~ \begin{matrix}
        \bx \leq \bz \leq \by, \mathrm{~and} \\
        \forall i, \forall v \in W' \cap V_i, z_i \geq \lambda(v)
    \end{matrix} \right. \right\}
    \end{align*}
    (the equality holds by Lemma~\ref{lm:dseal}) is nonempty (it contains at least the vector $\mu(\pi)$).
    By Lemma~\ref{lm:conv_to_system}, the set $\Conv_{c \in \SC(W)} \MP(c)$ is defined by a system of inequations which all have size:
    $$\lv (\ba, b) \rv \leq P_2\left(\max_c \lv \MP(c) \rv\right).$$
    Since by Lemma~\ref{lm:dseal_size}, the inequations defining the polytope $\dseal \left( \Conv_{c \in \SC(W)} \MP(c)\right)$ are not larger, there exists a polynomial $P_6$, independent of $\Game, \bx, \by$ and $\lambda$, such that the polytope $Z$ is defined by a system of inequations $\Sigma$ such that for every $(\ba, b) \in \Sigma$, we have $\lv (\ba, b) \rv \leq P_6(\lv (\Game, \bx, \by, \lambda) \rv)$. 
    Therefore, by Corollary~\ref{cor_size_vertices}, the polytope $Z$ admits a vertex $\bz$ of size $\lv \bz \rv \leq P_1(P_6(\lv (\Game, \bx, \by, \lambda) \rv))$.
    
    Then, since we have $\bz \in \dseal \left( \underset{c \in \SC(W)}{\Conv} \MP(c) \right)$, that vertex is, according to Definition~\ref{def_dseal}, of the form:
    $$\bz = \left( \min_j \sum_c \alpha_{jc} \MP_i(c) \right)_i$$
    for some tuple of tuples $\bbalpha \in [0, 1]^{\Pi \times \SC(W)}$ with $\sum_c \alpha_{ic} = 1$ and having, by Corollary~\ref{cor_size_vertices} again, size:
    $$\lv \bbalpha \rv \leq P_1\left( \max_{i \in \Pi} \sum_{c \in \SC(W)} \lv \MP_i(c) \rv + \lv z_i \rv\right),$$
    i.e. $\lv \bbalpha \rv \leq P_4(\lv (\Game, \bx, \by, \lambda) \rv)$ for some polynomial $P_4$ independent of $\Game, \bx, \by$ and $\lambda$.
\end{proof}

Now, we need the third and last piece of our witness, which will be the evidence of the fact that the requirement $\lambda$ is an $\epsilon$-fixed point of the negotiation function.

\section{Computing the negotiation function}

        \subsection{A disturbing example} \label{ssec:mp_disturbing_example}

Let us study more deeply the game $\Game$ depicted by Figure~\ref{fig:inf_spe}: consider the requirement $\lambda$, which maps both $a$ and $b$ to the value $1$.
We would like to know whether $\lambda$ is a fixed point of the negotiation function; or in other words, using the symmetry of the game, whether there exists a strategy for Prover, in the abstract negotiation game from vertex $a$, that forces player $\Circle$ to get at most the payoff $1$.
Prover can propose, for example, the play $(aaabbb)^\omega$, that is $\lambda$-consistent, and in which player $\Circle$ gets exactly the payoff $1$.
But then, Challenger can make player $\Circle$ deviate and go directly to $b$.
That strategy will give player $\Circle$ a payoff greater than $1$ against every finite-memory strategy of Prover.

Now, consider the following strategy of Prover: propose the play $(aaabbb)^\omega$.
Then, whenever Challenger deviates, and reaches prematurely the vertex $b$, propose the play $b^{|h|^2} (aaabbb)^\omega$, where $h$ is the history in $\Game_{\|a}$ that has already been drawn by Prover and Challenger's previous interactions.
Such a play is $\lambda$-consistent, and even if Challenger deviates to go to the vertex $b$ whenever he has the opportunity to do so, the play that will be generated is $abab^9ab^{169}ab^{33489}\dots$, in which player $\Circle$'s payoff is still $1$.
The requirement $\lambda$ is therefore a fixed point of the negotiation function (hence the set of SPE payoff vectors is the red-circled area in \cref{fig:inf_spe_payoffs}), but Prover requires infinite memory to maintain it.
The algorithm that we used for parity games in \cref{chap:parity} can therefore not be used as such.

However, we can see in this example that the plays proposed by Prover are still very similar: only the number of repetitions of the loop $b$ does increase.
More generally, one observes that Prover can play optimally while always proposing a play of the form $h c^n \pi$, where the history $h$, the cycle $c$ and the play $\pi$ are constant, and only the number $n$ increases, quadratically with time---so that Challenger's payoff is dominated by the mean-payoff $\MP_i(c)$ if he deviates infinitely often.
This will define a new notion of reduced strategy, inspired with what we did in \cref{chap:parity} but more sophisticated.

    \subsection{Reduced strategies for mean-payoff games} \label{ssec:mp_reduced_strategy}

Let us first define the notion of punishment family, the brick from which our reduced strategies will be built.

\begin{defi}[Punishment family]
    A \emph{punishment family} is a set of plays of the form:
    $$\left\{\left. h c^n \pi ~\right|~ n > 0, \mu(\pi) = \bx, \Occ(\pi) = W \right\}$$
    where $h$ is a simple history, where $c$ is a (nonempty) simple cycle, and where $W \subseteq V$ and $\bx \in \RR^\Pi$.
    The cycle $c$ is called its \emph{punishing cycle}.
    For every $\beta \in \NN$, a \emph{$\beta$-punishment family} is a punishment family with $\lv \bx \rv \leq \beta$.
    A $\beta$-punishment family is represented by the data $h$, $c$, $\bx$ and $W$, and that representation has a size smaller than or equal to the quantity $3 \card V \lceil \log_2(\card V + 1) \rceil + \beta$.
\end{defi}

We write $h c^\infty \pi$ for the punishment family $\{ h c^n \pi' ~|~ n > 0, \mu(\pi') = \mu(\pi), \Occ(\pi') = \Occ(\pi)\}$.
Beware that the play $\pi$ matters only for its payoff vector and the vertices it traverses: if we have $\Occ(\pi) = \Occ(\pi')$ and $\mu(\pi) = \mu(\pi')$, then we have $h c^\infty \pi = h c^\infty \pi'$.
We write $\mu(h c^\infty \pi)$ for the common payoff vector of all elements of the set $h c^\infty \pi$, and we will say that $h c^\infty \pi$ is $\lambda$-consistent if all its elements are (which is the case as soon as one of its elements is).
Furthermore, let us clarify that a punishment family is not an equivalence class: for example, in the game of Figure~\ref{fig:inf_spe}, the play $ab^\omega$ belongs to both sets $a^\infty b^\omega$ and $ab^\infty b^\omega$, which are distinct.

We can now define the \emph{reduced negotiation game}, where Prover proposes $\beta$-punishment families instead of plays.

\begin{defi}[Reduced negotiation game] \label{def_reduced_game}
    Let $\Game$ be a mean-payoff game, let $\lambda$ be a requirement, let $i$ be a player, let $v_0 \in V_i$ and let $\beta$ be a natural integer.
    The corresponding \emph{reduced negotiation game} is the game $\red^\beta_{\lambda i}(\Game)_{\|v_0} = (\{\Prover, \Challenger\}, V^\r, (V^\r_\Prover, V^\r_\Challenger), E^\r, \mu^\r)_{\|v_0}$, where:
    \begin{itemize}        
        \item the vertices controlled by Prover are the vertices of $\Game$, i.e. the set $V^\r_\Prover = V$;
        
        \item the vertices controlled by Challenger are the vertices:
        \begin{itemize}
            \item of the form $[h c^\infty \pi]$, where $h c^\infty \pi$ is a $\lambda$-consistent $\beta$-punishment family;

            \item of the form $(c, u)$, where $h c^\infty \pi$ is a $\lambda$-consistent $\beta$-punishment family, and where there exists a vertex $\pi_k \in V_i$ along the play $\pi$ such that $\pi_k u \in E$;

            \item of the form $[h'v]$, where $h c^\infty \pi$ is a $\lambda$-consistent $\beta$-punishment family, and the history $h'v$ is such that $h'$ is a prefix of the history $hc$, and $\last(h') \in V_i$;

            \item $\top$ and $\bot$;
        \end{itemize}
        
        \item with the same notations, the set $E^\r$ contains the edges of the form:
        \begin{itemize}
            \item $v [h c^\infty \pi]$ (Prover proposes a punishment family);
            
            \item $v \bot$ (Prover gives up);
            
            \item $[h c^\infty \pi] \top$ (Challenger accepts Prover's proposal);
            
            \item $[h c^\infty \pi] [h'v]$ (Challenger deviates before the punishing cycle---\emph{pre-cycle deviation});
            
            \item $[h c^\infty \pi] (c, u)$ (Challenger deviates after the punishing cycle---\emph{post-cycle deviation});
            
            \item $[h' v] v$ and $(c, u) u$ (Prover has now to propose a new play);
            
            \item $\top\top$ and $\bot\bot$ (the play is over);
        \end{itemize}
        
        \item given a history $g = g_0 \dots g_n \in \Hist\,\red^\beta_{\lambda i}(\Game)$ that does not reach the vertex $\bot$, we denote by $\dg = h^{(1)} \dots h^{(n)}$ the history or play in $\Game$ defined by, for each $k$:
        \begin{itemize}
            \item if $g_{k-1}g_k = v [h c^\infty \pi]$, then $h^{(k)}$ is empty;
            
            \item if $g_{k-1}g_k = [h c^\infty \pi] \top$, then $h^{(k)} \dots h^{(n)} = h c^{\left|h^{(1)} \dots h^{(k-1)} h\right|^2} \pi$ (the number of times the cycle $c$ is repeated depends quadratically on time);
            
            \item if $g_{k-1}g_k = [h c^\infty \pi] [h'v]$, then $h^{(k)} = h'$;
            
            \item if $g_{k-1}g_k = [h c^\infty \pi] (c, v)$, then $h^{(k)} = h c^{\left|h^{(1)} \dots h^{(k-1)} h\right|^2} h'$, where $h'$ is one of the shortest histories such that $\Occ(h') \subseteq \Occ(\pi)$,  $\last(h') \in V_i$ and $\last(h') v \in E$;
            
            \item if $g_{k-1}g_k = [h' v] v$ or $(c, v) v$, then $h^{(k)}$ is empty;
        \end{itemize}
        and that definition is naturally extended to plays: for example, if $\Game$ is the game of Figure~\ref{fig:inf_spe} and if $\pi = a [ab^\infty a^\omega] [aa] a [ab^\infty a^\omega] (b, b) b [b^\infty a^\omega] \top^\omega$, then we write $\dpi = a \cdot ab^{2^2} \cdot a b^{7^2} a^\omega = a^2b^4ab^{49}a^\omega$;
        
        \item the payoff function $\mu^\r$ is defined, for each play $\pi$, by $\mu^\r_\Challenger(\pi) = -\mu^\r_\Prover(\pi) = +\infty$ if $\pi$ reaches the vertex $\bot$, and by $\mu^\r_\Challenger(\pi) = -\mu^\r_\Prover(\pi) = \mu_i(\dpi)$ otherwise.
    \end{itemize}
\end{defi}

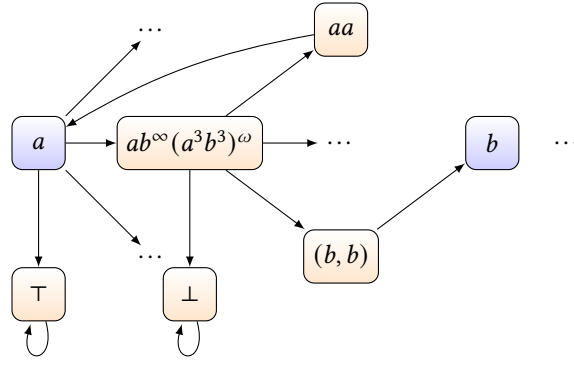
\begin{figure}
    \centering
	\begin{tikzpicture}
	\node[proverVert] (a) at (0, 0) {$a$};
	\node[challengerVert] (red) at (2, 0) {$a b^\infty (a^3b^3)^\omega$};
	\node[challengerVert] (ac) at (4, 1.5) {$aa$};
	\node[challengerVert] (abc) at (4, -1.5) {$(b, b)$};
	\node[proverVert] (c) at (6, 0) {$b$};
	\node (dots1) at (1.5, 1.5) {\dots};
	\node (dots2) at (1.5, -1.5) {\dots};
	\node (dots3) at (4, 0) {\dots};
	\node (dots4) at (7, 0) {\dots};
	\node[challengerVert] (bot) at (0, -2) {$\top$};
	\node[challengerVert] (top) at (2, -2) {$\bot$};
	
	\path (a) edge (red);
	\path (red) edge (ac);
	\path (red) edge (abc);
	\path[bend right=10] (ac) edge (a);
	\path (abc) edge (c);
	\path (a) edge (dots1);
	\path (a) edge (dots2);
	\path (red) edge (dots3);
	\path (a) edge (bot);
	\path (bot) edge[loop below] (bot);
	\path (red) edge (top);
	\path (top) edge[loop below] (top);
	\end{tikzpicture}
\caption{The reduced negotiation game} \label{fig:reduced}
\end{figure}

\begin{exa}
    Figure~\ref{fig:reduced} illustrates a (small) part of the game $\red_{\lambda \circ}^2(\Game)_{\|a}$, where $\Game$ is the game of Figure~\ref{fig:inf_spe}, and $\lambda(a) = \lambda(b) = 1$.
    Blue vertices are owned by Prover, orange ones by Challenger.
    When Prover proposes the punishment family $ab^\infty (a^3b^3)^\omega$, the function $\mu^\r_\Challenger$ interprets it as the play $ab^{|h|^2} (a^3b^3)^\omega$, where $h$ is the history that has already been constructed so far.
\end{exa}

\begin{rem}
    Reduced negotiation games are Borel, and are played on a finite graph.
\end{rem}

We will now prove that the reduced negotiation game captures the negotiation function, as do the abstract and concrete ones.
For that purpose, we first need the following key result.

\begin{lem} \label{lm:reduced_stationary}
    In a reduced negotiation game, Prover has a stationary optimal strategy.
\end{lem}

\begin{proof}
    This lemma is a consequence of Lemma~\ref{lm:concave}: the payoff function $\mu^\r_\Challenger$ is concave.
    Indeed, let $\xi$ be a shuffling of two plays $\pi$ and $\chi$.
    If either $\pi$ or $\chi$ reaches the vertex $\bot$ (in which case both do), then we immediately have $\mu^\r_\Challenger(\dxi) \leq \max\{\mu^\r_\Challenger(\dpi), \mu^\r_\Challenger(\dchi)\} = +\infty$.
    Otherwise, the play $\dxi$ is a shuffling of $\dpi$ and $\dchi$, and since mean-payoff objectives defined with a limit inferior are convex, we have $\mu^\r_\Challenger(\dxi) \leq \max\{\mu^\r_\Challenger(\dpi), \mu^\r_\Challenger(\dchi)\}$.
\end{proof}

This lemma enables us to prove that the reduced negotiation game is equivalent to the other negotiation games.

\begin{lem} \label{lm:reduced_game}
    There exists a polynomial $P_5$ such that for every mean-payoff game $\Game$, every requirement $\lambda$ with rational values, each player $i$ and each $v_0 \in V_i$, for every $\beta \geq P_5(\lv \Game \rv + \lv \lambda \rv)$, we have $\nego(\lambda)(v_0) = \val_\Challenger \left( \red^\beta_{\lambda i}(\Game)_{\|v_0} \right).$
\end{lem}

\begin{proof}
    For every mean-payoff game $\Game$ and every requirement $\lambda$, we assume:
    $$\beta \geq P_1 \circ P_2\left(\lv \left\{ \MP(c) ~|~ c \in \SC(\Game)\right\} \rv\right) $$
    and for each $v \in V$:
    $$\beta \geq \lv \lambda(v) \rv + 3,$$
    which are indeed quantities that are bounded by a polynomial of $\lv \Game \rv +  \lv \lambda \rv$.

    Contrary to what we did with parity games in the proof of \cref{lm_parity_reduced_strategy}, here, the equivalence between the reduced and the abstract game cannot be established directly (the proof would be too technical): we need to get back to the definition of the negotiation function.
    
    \paragraph{First direction: $\nego(\lambda)(v_0) \geq \val_\Challenger \left( \red^\beta_{\lambda i}(\Game)_{\|v_0} \right).$}
        
        Let $\bsigma_{-i}$ be a strategy profile in $\Game$ that is $\lambda$-rational assuming a strategy $\sigma_i$, and let $x = \sup_{\sigma'_i} \mu_i\< \bsigma_{-i}, \sigma'_i \>$.
        We wish to prove that there exists a strategy $\tau_\Prover$ in the reduced negotiation game such that $\sup_{\tau_\Challenger} \mu^\r_\Challenger\< \btau \> \leq x$.
        Thus, we will have proved that the quantity $\val_\Challenger \left( \red^\beta_{\lambda i}(\Game)_{\|v_0} \right)$ is smaller than or equal to every such $x$, and therefore smaller than or equal to $\nego(\lambda)(v_0)$.
        
        Let us define simultaneously the strategy $\tau_\Prover$ and a mapping $\phi: \Hist_\Prover\left(\red^\beta_{\lambda i}(\Game)_{\|v_0}\right) \to \Hist \Game_{\|v_0}$, such that for each history $g$, the punishment family $\tau_\Prover(g)$ will be defined from the play $\< \bsigma_{\|\phi(g)} \>$.
        We guarantee inductively that if $g \in \Hist_\Prover\left(\red^\beta_{\lambda i}(\Game)_{\|v_0}\right)$ is compatible with $\tau_\Prover$, then $\phi(g) \in \Hist \Game_{\|v_0}$ is compatible with $\bsigma_{-i}$.
        First, let us define $\phi(v_0) = v_0$.
        
        Let $g \in \Hist_\Prover\left(\red^\beta_{\lambda i}(\Game)_{\|v_0}\right)$ be a history compatible with $\tau_\Prover$ as it has been defined so far, and such that $\phi(g)$ has already been defined.
        Let $\chi^0 = \< \bsigma_{\|\phi(g)} \>$.
        By induction hypothesis, the history $\phi(g)$ is compatible with $\bsigma_{-i}$, hence the play $\chi^0$ is $\lambda$-consistent, and satisfies $\mu_i(\chi^0) \leq x$.
        
        Let $\chi^0_{\leq \l}$ be the shortest prefix of $\chi^0$ that is not simple, i.e. such that there exists $k < \l$ with $\chi^0_k = \chi^0_\l$.
        If we have $\MP_i(\chi^0_{k+1} \dots \chi^0_\l) \leq x$, then we define:
        $$\tau_\Prover(g) = \left[\chi^0_{\leq k} \left( \chi^0_{k+1} \dots \chi^0_\l \right)^\infty \pi \right],$$
        where $\pi$ is a play such that $\Occ(\pi) = \Occ\left(\chi^0_{>\l}\right)$, that $\mu_i(\pi) \leq x$, and that $\lv \mu(\pi) \rv \leq \beta$.
        Such a play exists, because the polytope:
        $$Z = \left\{ \mu(\pi) ~\left|~ \begin{matrix}
            \forall j, \forall v \in V_j \cap \Occ(\chi^0), \mu_j(\pi) \geq \lambda(v), \\
            \mathrm{and~} \Occ(\pi) = \Occ\left(\chi^0_{>\l}\right)
        \end{matrix} \right. \right\}$$
        is nonempty (it contains the vector $\mu(\chi^0_{>\l})$), and has at least one vertex $\bz$ with $z_i \leq x$ (because we have $\mu_i(\chi^0_{>\l}) \leq x$), which by Lemma~\ref{lm:conv_to_system} and Corollary~\ref{cor_size_vertices} has size $\lv\bz \rv \leq \beta$.
        
        Otherwise, if we have $\MP_i(\chi^0_{k+1} \dots \chi^0_\l) > x$, we define $\chi^1 = \chi^0_{\leq k} \chi^0_{>\l}$, and we iterate the process, which does necessarily terminate since we have $\mu_i(\chi^0) \leq x$.
        As a consequence, it effectively defines the proposal:
        $$\tau_\Prover(g) = \left[\chi^n_{\leq k} \left( \chi^n_{k+1} \dots \chi^n_\l \right)^\infty \pi \right],$$
        for some $n$.
        Then, for each prefix $hv$, we define:
        $$\phi\left(g \left[\chi^n_{\leq k} \left( \chi^n_{k+1} \dots \chi^n_\l \right)^\infty \pi \right] [hv] v\right) = \phi(g) \chi^0_{\leq m},$$
        where $\chi^0_{\leq m}$ is the prefix of $\chi^0$ from which $n$ simple cycles have been pulled out to obtain the prefix $h$ of $\chi^n$; and similarly, for each pair $(c, v)$, we define:
        $$\phi\left(g \left[\chi^n_{\leq k} \left( \chi^n_{k+1} \dots \chi^n_\l \right)^\infty \pi \right] (c, v) v\right) = \phi(g) \chi^0_{\leq p},$$
        where $\chi^0_{\leq p}$ is the prefix of $\chi^0$ from which $n$ simple cycles have been pulled out to obtain the shortest prefix $\chi^n_{\leq p}$ of $\chi^n$ such that $\chi^n_p \in V_i$ and $\chi^n_p v \in E$.
        
        Thus, the mapping $\phi$ is defined on every history compatible with $\tau_\Prover$, and the image of such a history is always a history compatible with $\bsigma_{-i}$.
        We define it arbitrarily on other histories.
        Note that for each history $g$, the history $\dg$ can be obtained from $\phi(g)$ by pulling out cycles $c$ satisfying $\MP_i(c) > x$, and adding cycles $d$ with $\MP_i(d) \leq x$.
        As a consequence, if we have $\MP_i(\phi(g)) \leq x$, then we have $\MP_i(\dg) \leq x$---and the same result is true when we naturally extend the mapping $\phi$ to plays.
        
        Let us now prove that we have $\sup_{\tau_\Challenger} \mu^\r_\Challenger\< \btau \> \leq \sup_{\sigma'_i} \mu_i\< \bsigma_{-i}, \sigma'_i \>$.
        Let $\xi$ be a play compatible with $\tau_\Prover$:
        \begin{itemize}
            \item the vertex $\bot$ does not appear in $\xi$, because Prover's strategy does never use a edge to it.
            
            \item If the play $\xi$ has the form $\xi = g [h c^\infty \pi] \top^\omega$: then, we have $\mu^\r_\Challenger(\xi) = \mu_i(\pi) \leq x$.
            
            \item If the play $\xi$ is made of infinitely many deviations: the play $\phi(\xi)$ is compatible with $\bsigma_{-i}$, hence $\mu_i(\phi(\xi)) \leq x$; which implies $\mu_i(\dchi) \leq x$, i.e. $\mu^\r_\Challenger(\xi) \leq x$.
        \end{itemize}
        
        \paragraph{Second direction: $\nego(\lambda)(v_0) \leq \val_\Challenger \left( \red^\beta_{\lambda i}(\Game)_{\|v_0} \right).$}
        
        Let $\tau_\Prover$ be a stationary strategy for Prover in the reduced negotiation game, and let $y = \sup_{\tau_\Challenger} \mu^\r_\Challenger\< \btau \>$.
        We want to show that $\nego(\lambda)(v_0) \leq y$: by Lemma~\ref{lm:reduced_stationary}, it will be enough to conclude.
        If we have $y = +\infty$, it is clear.
        Let us assume $y \neq +\infty$.
        Then, we will define a strategy profile $\bsigma$, where $\bsigma_{-i}$ is $\lambda$-rational assuming $\sigma_i$, such that $\sup_{\sigma'_i} \mu_i\< \bsigma_{-i}, \sigma'_i \> \leq y$.
        We proceed inductively by defining the play $\< \bsigma_{\|hv} \>$ for each history $hv$ compatible with $\bsigma_{-i}$ such that $h$ is empty, or $\last(h) \in V_i$ and $v \neq \sigma_i(h)$.
        Such a history is called a \emph{bud history}.
        After other histories, the strategy profile can be defined arbitrarily.
        To that end, we construct a mapping $\psi$ which maps each bud history to a history $\psi(hv) \in \Hist_\Prover \left(\red_{\lambda i}^\beta(\Game)_{\|v_0}\right)$ that is compatible with $\tau_\Prover$.
        This mapping will induce a definition of $\bsigma$: since $y \neq +\infty$, we have $\tau_\Prover(\psi(hv)) \neq \bot$: let then $[h' c^\infty \pi] = \tau_\Prover(\psi(hv))$.
        We then define $\< \bsigma_{\|hv} \> = h' c^{|hh'|^2} \pi$, which is a $\lambda$-consistent play since $h'c^\infty \pi$ is a $\lambda$-consistent punishment family, by definition of the reduced negotiation game.
        
        Let now $h_0 v$ be a bud history: we assume that $\bsigma$ has been defined on every prefix of $h_0$, but not on $h_0 v$ itself.
        If $h_0$ is empty, that is if $h_0v = v_0$, then we define $\psi(h_0v) = \psi(v_0) = v_0$.
        Otherwise, let us write $h_0 = h_1 w h_2$, where $h_1 w$ is the longest prefix of $h_0$ that is a bud history---that is, its longest prefix such that $\psi(h_1 w)$ has been defined, or its shortest prefix such that $wh_2$ is compatible with $\bsigma_{\|h_1 w}$.
        Let $g = \psi(h_1 w)$, and let $[h c^\infty \pi] = \tau_\Prover(g)$.
        We have defined $\< \bsigma_{\|h_1 w} \> = h c^{|h_1 h|^2} \pi$, and consequently, the history $w h_2$ is a prefix of that play.
        If it is a prefix of the history $hc$, then we define $\psi(h_0 v) = g [h c^\infty \pi] [w h_2 v] v$.
        Otherwise, we define $\psi(h_0 v) = g [h c^\infty \pi] (c, v) v$.
        
        Now, the strategy profile $\bsigma$ has been defined, and since all the punishment families proposed by Prover are $\lambda$-consistent, the strategy profile $\bsigma_{-i}$ is $\lambda$-rational assuming $\sigma_i$.
        Let $\chi$ be a play compatible with $\bsigma_{-i}$, and let us prove that $\mu_i(\chi) \leq y$.
        If $\chi$ has finitely many prefixes that are bud histories, then let $\chi_{\leq n}$ be the longest one: we have $\chi_{\geq n} = h c^{n+|h|} \pi$, where $[h c^\infty \pi] = \tau_\Prover(\psi(\chi_{\leq n}))$.
        Then, we have $\mu_i(\chi) = \mu_i(\pi) \leq y$.
        
        Now, if $\chi$ has infinitely many such prefixes, then there exists a unique play $\pi$ in the reduced negotiation game such that for any prefix $\chi_{\leq n}$ of $\chi$ that is a bud history, the history $\psi(\chi_{\leq n})$ is a prefix of $\pi$.
        Then, if $\pi$ contains finitely many post-cycle deviations, then there exist two indices $m$ and $n$ such that $\chi_{\geq m} = \dpi_{\geq n}$, hence $\mu_i(\chi) = \mu_i(\dpi) \leq y$.
        
        Finally, if $\pi$ contains infinitely many post-cycle deviations, i.e. infinitely many occurrences of a vertex of the form $(c, v)$, then let us choose such vertex that minimizes the quantity $\MP_i(c)$.
        The play $\chi$ has the form:
        $$\chi = h_0 c^{k_0^2} h_1 c^{k_1^2} h_2 \dots,$$
        where for each $n$, we have $k_n = \left|h_0 c^{k_0^2} \dots c^{k_{n-1}^2} h_n\right|$.
        Then, if we write $M = \max r_i$, we have:
        $$\MP_i\left(h_0 c^{k_0^2} \dots h_n c^{k_n^2}\right) \leq \frac{1}{k_n + k_n^2 |c| - 1} \left( k_n M + \left(k_n^2 |c| - 1\right) \MP_i(c) \right),$$
        which converges to $\MP_i(c)$ when $n$ tends to $+\infty$, hence $\mu_i(\chi) \leq \MP_i(c)$.
        Now, since $\tau_\Prover$ is stationary, there exists a play of the form $g d^\omega$ that is compatible with it, and such that $(c, v) \in \Occ(d) \subseteq \Inf(\pi)$; and by definition of $y$, we have $\MP_i(\dd) = \mu^\r_\Challenger(gd^\omega) \leq y$.
        By minimality of $\MP_i(c)$, we have $\MP_i(\dd) = \MP_i(c)$, hence $\MP_i(c) \leq y$, and therefore $\mu_i(\chi) \leq y$.
\end{proof}

Thus, a given requirement $\lambda$ is an $\epsilon$-fixed point of the negotiation function if and only if for each $i$ and $v \in V_i$, there exists a stationary strategy $\tau_\Prover$ in the game $\red_{\lambda i}^\beta$, with $\beta = P_5(\lv \Game \rv + \lv \lambda \rv)$, such that $\sup_{\tau_\Challenger} \mu^\r_\Challenger\< \btau \> \leq \lambda(v) + \epsilon$.
The reduced negotiation game has an exponential size, but it contains only $\card V$ vertices that are controlled by Prover: stationary strategies for Prover are therefore objects of polynomial size.
Such strategies will be called \emph{reduced} strategies, and can alternatively be seen as simple strategies in the asbtract negotiation game, as suggested in the end of \cref{ssec:mp_disturbing_example}.
They constitute the third and last piece of our notion of witness.

        \section{Algorithm and complexity}

We are now in a position to define formally our notion of witness.

\begin{defi}[Witness]
    Let $I = (\Game_{\|v_0}, \bx, \by, \epsilon)$ be an instance of the $\epsilon$-SPE constrained existence problem.
    A \emph{witness} for $I$ is a tuple $\left(W, W', \bbalpha, \lambda, (\tau^v_\Prover)_v\right)$, where:
    \begin{itemize}
        \item $W \subseteq W' \subseteq V$;

        \item $\bbalpha \in [0,1]^{\Pi \times \SC(W)}$;

        \item $\lambda$ is a requirement;

        \item and each $\tau^v_\Prover$ is a stationary strategy in the game $\red^\beta_{\lambda i}(\Game)_{\|v}$, where $\beta = P_5(\lv \Game \rv + \lv \lambda \rv)$.
    \end{itemize}    
    A witness is \emph{valid} if:
    \begin{itemize}
        \item each strategy $\tau^v_\Prover$ satisfies the inequality $\sup_{\tau_\Challenger} \mu^\r_\Challenger\< \tau^v_\Prover, \tau_\Challenger\> \leq \lambda(v) + \epsilon$;
        
        \item the sets $W$ and $W'$ and the tuple of tuples $\bbalpha$ satisfy the hypotheses of Theorem~\ref{lm:constrained_existence_lambda_cons}.
    \end{itemize}
\end{defi}

The $\epsilon$-SPE constrained existence problem will be $\NP$-easy if we show, first, that there exists a valid witness of polynomial size if and only if the instance is positive, and second, that the validity of a witness can be decided in polynomial time.
The former is a consequence of \cref{thm:nego_spe,lm:size_lambda,lm:constrained_existence_lambda_cons,lm:reduced_stationary,lm:reduced_game}.

\begin{lem} \label{lm:witness_existence}
    There exists a polynomial $P_6$ such that an instance $I$ of the $\epsilon$-SPE constrained existence problem admits a valid witness of size at most $P_6(\lv I \rv)$ if and only if it is a positive instance.
\end{lem}

Let us now tackle the latter.

\begin{lem} \label{lm:mp_check_polynomial}
    Given an instance of the $\epsilon$-SPE constrained existence problem and a witness for it, deciding whether that witness is valid is $\PTime$-easy.
\end{lem}

\begin{proof}
    The validity of a witness has been defined by two conditions.
    As regards the second one, all the hypotheses of Lemma~\ref{lm:constrained_existence_lambda_cons} can be checked in polynomial time with classical algorithms.
    Let us now show how the first condition can also be checked in polynomial time.

    Let us recall that in our algorithm for parity games, at the same point, we constructed the game induced by each strategy guessed for Prover, and searched for a strategy for Challenger, i.e. for a play in that game, that would contradict the claim that Prover's strategy is winning.
    Such a play could be seen as a winning path for either a reachability or a parity objective.
    Here, the idea will be similar, but we need to consider Prover's strategy in the reduced negotiation game, so that it indeed induces a finite graph; and then, the search of a path for Challenger will be slightly more complicated.
        
    Let $n = \card V$.
    Given a stationary strategy $\tau^v_\Prover$ of Prover in a reduced negotiation game, one can construct in a time polynomial in $\lv \tau^v_\Prover \rv$ the game $\red_{\lambda i}^\beta(\Game)_{\|v}[\tau^v_\Prover]$.
    The underlying graph of that game has indeed a polynomial size, because it is composed only of:
    \begin{itemize}
        \item at most $n$ vertices of the form $w \in V$;
        
        \item at most $n$ vertices of the form $\tau_\Prover(w)$ (either equal to $\bot$ or of the form $[h c^\infty \pi]$);
        
        \item at most $n^2$ vertices of the form $(c, w)$;
        
        \item at most $2n^2$ vertices of the form $[h'w']$, where $h'$ is a prefix of the history $hc$ for some punishment family $[hc^\infty\pi] = \tau_\Prover(w)$;
        
        \item possibly the vertices $\top$ and $\bot$.
    \end{itemize}
    
    We call this connected graph the \emph{deviation graph}.
    Note that if among those vertices, there is the vertex $\bot$, then since the vertices that are not accessible have been removed, we have $\sup_{\tau_\Challenger} \mu^\r_\Challenger\< \btau \> = +\infty$ and the problem can be solved immediately.
    In what follows, we assume that it is not the case, i.e. that for each $w$, the vertex $\tau_\Prover(w)$ has the form $[h c^\infty \pi]$.
    Deciding whether $\sup_{\tau_\Challenger} \mu^\r_\Challenger\< \btau \> \leq \alpha$ is then equivalent to deciding whether there exists a path $\pi$, in that graph, such that we have $\mu^\r_\Challenger(\pi) > \alpha$.
    Such a play can have three forms.
    
    \begin{itemize}
        \item It can end in the vertex $\top$, i.e. with Challenger accepting Prover's proposal.
        The existence of such a play can be decided immediately, by checking whether in the deviation graph, there exists a vertex of the form $[h c^\infty \pi]$ with $\mu_i(\pi) > \alpha$.
        
        \item It can avoid the vertex $\top$, and comprise finitely many post-cycle deviations.
        This is the case if and only if there exists a cycle $d$ in the deviation graph, without post-cycle deviations, such that we have $\MP_i(\dd) > \alpha$.
        The existence of such a cycle can be decided in polynomial time with Karp's algorithm~\cite{DBLP:journals/dm/Karp78}.
        
        \item It can avoid the vertex $\top$, and comprise infinitely many post-cycle deviations.
        In that case, we have $\mu^\r_\Challenger(\pi) \leq \MP_i(c)$ for each vertex of the form $(c, w)$ appearing infinitely often along $\pi$; then, there exists a cycle $d$ in the deviation graph, such that every vertex of the form $(c, w)$ along $d$ satisfies $\MP_i(c) > \alpha$.
        Conversely, if such a cycle exists, then $\pi$ exists.
        The existence of such a cycle can be decided in polynomial time with Karp's algorithm.
    \end{itemize}
    
    Therefore, the existence of such a play is decidable in polynomial time.
\end{proof}

Thus, given an instance of the $\epsilon$-SPE constrained existence problem, a valid witness can be guessed and checked in polynomial time.
Applying \cref{lm:spe_mp_hardness}, we finally obtain the following theorem.

\begin{thm}\label{thm:spe_mp_np_complete}
    The $\epsilon$-SPE constrained existence problem in mean-payoff games is $\NP$-complete.
    Hardness still holds when there is only one effective upper threshold.
\end{thm}

%% file: 2dEnergy_DS.tex
We now tackle subgame-perfect equilibria in energy and in discounted-sum games.
Here, the negotiation function will not be used, and our results and proofs will mostly be extensions of those presented in \cref{part:nash}.

\section{Discounted-sum games}\label{sec:spe_ds}

\subsection{Hardness}

We showed in \cref{part:nash} that the constrained existence problem of NEs in discounted-sum games was at least as hard as the target discounted-sum problem, whose decidability remains an open problem.
That result also holds when we replace NEs with SPEs.

	\begin{thm} \label{thm:ds_spe_hardness}
		The TDS problem reduces to the constrained existence problem of SPEs in discounted-sum games.
        This hardness result still holds when there is no effective lower threshold, and only one effective upper threshold.
	\end{thm}
	
	\begin{proof}
	    The proof is analogous to that of \cref{thm:ds_ne_hardness}.
	\end{proof}

\subsection{Co-recursive enumerability}

Since, by \cref{thm:ds_spe_hardness}, we know that finding an algorithm for the SPE constrained existence problem would imply solving a long-standing open problem, we only present a semi-algorithm that recognizes its negative instances, leaving decidability as an open question.

	\begin{thm}\label{thm:ds_spe_easiness}
		In discounted-sum games, the constrained existence problem of SPEs is co-recursively enumerable.
	\end{thm}
	
	\begin{proof}
	    We present here a modified version of the semi-algorithm presented in the proof of \cref{thm:ds_ne_easiness}, so that it recognizes the negative instances of the SPE constrained existence problem.
	    
		\paragraph{Algorithm}
			
			Let $\left(h^{(n)}\right)_{n \in \NN}$ be a recursive enumeration of the nonempty histories in $\Game_{\|v_0}$ by increasing order of lengths (we can, for example, order histories of the same length with the lexicographic order induced by some arbitrary order on vertices).
			
			Now, let $T$ be the infinite tree whose nodes of depth $n+1$ are all possible $n$-uples $\left(\bsigma(h^{(0)}), \dots, \bsigma(h^{(n)})\right)$, and where the children of the node $\left(\bsigma(h^{(0)}), \dots, \bsigma(h^{(n)})\right)$ are the nodes of the form:
            $$\left(\bsigma(h^{(0)}), \dots, \bsigma(h^{(n)}), \bsigma(h^{(n+1)})\right).$$
			Thus, every node partially defines a complete strategy profile $\bsigma$, and every infinite branch entirely defines it---and every complete strategy profile is defined by an infinite branch.
			
			A node $\left(\bsigma(h^{(0)}), \dots, \bsigma(h^{(n)})\right)$ is called \emph{subgame-irrational} if there exist two indices $\l, m \leq n$, with $|h^{(\l)}| = |h^{(m)}| = p$, an index $k \leq p$ and a player $i$ such that:
			\begin{itemize}
				\item we have $h^{(\l)}_{\leq k} = h^{(m)}_{\leq k}$;
				
				\item the history $h^{(\l)}_{\geq k}$ is compatible with the strategy profile $\bsigma_{\|h^{(\l)}_{\leq k}}$, as partially defined by the node;
				
				\item the history $h^{(m)}_{\geq k}$ is compatible with the strategy profile $\bsigma_{-i\|h^{(m)}_{\leq k}}$;
				
				\item and finally, we have:
				$$\ds_i\left(h^{(m)}\right) - \ds_i\left(h^{(\l)}\right) > 2M \beta^{p-1}.$$
			\end{itemize}
			
			The same node is called \emph{off-topic} if for some player $i$ we have $\ds_i\left(h\right) > y_i - M \beta^{|h|-1}$, or $\ds_i(h) < x_i + M \beta^{|h|-1}$, where $h$ is the longest history that is compatible with $\bsigma$ as defined so far.
			
			Our algorithm consists in constructing the tree $T'$, obtained from the tree $T$ by cutting every branch after the first subgame-irrational or off-topic node, and terminating once that construction is finished.

			\paragraph{Correctness}

Our correctness proof will use \cref{lm:ds_preliminary}, that was proved inside the proof of \cref{thm:ds_ne_easiness}.
            
To show that our algorithm is correct, we must prove that the tree $T'$ will be finite if and only if we have a negative instance of the constrained existence problem.
            By K\H{o}nig's lemma, that will be the case if and only if every branch is finite, i.e., if and and only if every branch of the tree $T$ contains either an off-topic or a Nash-irrational node.
            Therefore, we will be done if we prove that, given a branch of $T$ and the corresponding strategy profile $\bsigma$, the branch contains no such node if and only if the strategy profile $\bsigma$ is an SPE and satisfies $\bx \leq \mu\< \bsigma \> \leq \by$.
            
\subparagraph{Off-topic nodes.}            
			Using \cref{lm:ds_preliminary}, we know that we have $\bx \leq \mu_i\< \bsigma \> \leq \by$ if and only if the corresponding branch contains no off-topic node.

            \subparagraph{If the branch contains a subgame-irrational node, then the strategy profile $\bsigma$ is not an SPE.}
			
			Let us assume that some branch contains a subgame-irrational node $\left(\bsigma(h^{(0)}), \dots, \bsigma(h^{(n)})\right)$.
			Let us use the notations $k, \l, m, p$ and $i$ from the definition of subgame-irrationality.
			Let us define $\pi = \< \bsigma_{\|h^{(\l)}_{\leq k}} \>$: note that the history $h^{(\l)}_{\geq k}$ is a prefix of length $p-k$ of the play $\pi$.
			Similarly, let us extend the history $h^{(m)}_{\geq k}$ into some play $\pi'$, compatible with the strategy profile $\bsigma_{-i\|h^{(m)}_{\leq k}}$.
			By \cref{lm:ds_preliminary}, the inequality:
			$$\ds_i\left(h^{(m)}\right) - \ds_i\left(h^{(\l)}\right) > 2 M \beta^{p-1}$$
			implies $\mu_i(h^{(\l)}_{<k} \pi') > \mu_i(h^{(\l)}_{<k} \pi)$, and therefore the strategy profile $\bsigma$ is not an SPE.
			
			Conversely, if $\bsigma$ is not an SPE, then there exists a history $hv$, a player $i$ and a strategy $\sigma'_i$ such that we have:
			$$\mu_i(h \< \bsigma_{\|hv} \>) < \mu_i(h \< \bsigma_{-i\|hv}, \sigma'_{i\|hv} \>).$$
			
			Then, let $\pi = \< \bsigma_{\|hv} \>$, and let $\pi' = \< \bsigma_{-i\|hv}, \sigma'_{i\|hv} \>$: since we have $\mu_i(h\pi) < \mu_i(h\pi')$, by \cref{lm:ds_preliminary}, there exists an index $q$ such that:
			$$\ds_i\left(h\pi_{\leq q}\right) + M \beta^{|h|+q-1} < \ds_i\left(\pi'_{\leq q}\right) - M \beta^{|h|+q-1}.$$
			Let now $\l$ and $m$ be the indices such that $h^{(\l)} = h\pi_{\leq q}$, and $h^{(m)} = h\pi'_{\leq q}$.
			Let $p = |h^{(\l)}| = |h^{(m)}|$, and let $k = |h|$.
			Then, we have $\ds_i\left(h^{(m)}_{\geq k}\right) - \ds_i\left(h^{(\l)}_{\geq k}\right) > 2M \beta^{p-1}$: along the branch corresponding to $\bsigma$, the node of depth $\max\{\l, m\}$ is subgame-irrational.
			Which ends the proof.
	\end{proof}

\section{Energy games}

\subsection{Undecidability}

We proved in \cref{part:nash} that the constrained existence problem of Nash equilibria in energy games is undecidable, since counter machines can be encoded as energy games.
It should therefore not come as a surprise that the same problem for SPEs is also undecidable.
We prove however a slightly stronger result: undecidability holds even when the game contains only two players.

\begin{thm}\label{thm:energy_spe_undec}
		In energy games, the SPE constrained existence problem is undecidable, even on games with only two players, and even with no effective lower threshold and only one effective upper threshold.
	\end{thm}

\begin{proof}
	We proceed by reduction from the halting problem of a two-counter machine.
	
	\begin{figure}
    \centering
        \begin{subfigure}[b]{0.3\textwidth}
			\centering
			\begin{tikzpicture}
				\node (1) at (2, 0) {};
				\node[vert, initial left] (q0) at (0, 0) {$q_0$};
				\path (q0) edge (1);
				\path (q0) edge[loop above] (q0);
			\end{tikzpicture}
			\caption{Gadget for the initial state}
			\label{fig_gadget_initial_sp}
		\end{subfigure}
		\begin{subfigure}[b]{0.2\textwidth}
			\centering
			\begin{tikzpicture}
				\node (1) at (-1, 0) {};
				\node[vert] (qf) at (0, 0) {$q_\f$};
				\path (1) edge (qf);
				\path (qf) edge[loop right] node[right] {$\stackrel{C_2}{-1}$} (qf);
			\end{tikzpicture}
			\caption{Gadget for the final state}
			\label{fig_gadget_final_sp}
		\end{subfigure}
		\begin{subfigure}[b]{0.3\textwidth}
			\centering
			\begin{tikzpicture}
				\node[vert] (q) at (0, 0) {$q$};
				\node (1) at (-1, 0) {};
				\node (2) at (2, 0) {};
				\path (1) edge (q);
				\path (q) edge node[above] {$\stackrel{C}{1}$} (2);
			\end{tikzpicture}
			\caption{Gadget for incrementations of counter $C \in \{C_1, C_2\}$}
			\label{fig_gadget_incrementation_sp}
		\end{subfigure}
	\begin{subfigure}[b]{0.4\textwidth}
		\centering
		\begin{tikzpicture}
			\node (1) at (-1, 0) {};
			\node[vert] (q) at (0, 0) {$q$};
			\node (2) at (4, -1) {(if $C_1 > 0$)};
			\node[vert, rectangle] (q') at (2, 1) {$q'$};
			\node (3) at (4, 1) {(if $C_1 = 0$)};
			\node[vert] (q'') at (2, 3) {$a$};
			\node[vert] (top) at (1, 5) {$\triangle$};
			\node[vert] (bot) at (3, 5) {$\triangledown$};
			
			\path (1) edge (q);
			\path (q) edge node[below left] {$\stackrel{C_1}{-1}$} (2);
			\path (q) edge (q');
			\path (q') edge (3);
			\path (q') edge (q'');
			\path (q'') edge node[below left] {$\stackrel{C_1}{-1}$} (top);
			\path (q'') edge (bot);
			\path (top) edge [loop above] (top);
			\path (bot) edge [loop above] node[above] {$\stackrel{C_1}{-1} \, \stackrel{C_2}{-1}$} (bot);
		\end{tikzpicture}
		\caption{Gadget for tests of counter $C_1$}
		\label{fig_gadget_test_C1_sp}
	\end{subfigure}
	\begin{subfigure}[b]{0.4\textwidth}
				\centering
				\begin{tikzpicture}
					\node (1) at (-1, 0) {};
					\node[vert] (q) at (0, 0) {$q$};
					\node[vert, rectangle] (q') at (2, 1) {$q''$};
					\node (2) at (4, 1) {(if $C_2 = 0$)};
					\node[vert] (s1) at (2, 3) {$\triangle$};
					\node[vert, rectangle] (q'') at (2, -1) {$q'''$};
					\node (3) at (4, -1) {(if $C_2 > 0$)};
					\node[vert] (q''') at (2, -3) {$b$};
					\node[vert, rectangle] (q'''') at (1, -5) {$c$};
					\node[vert] (s2) at (0, -7) {$\triangle$};
					\node[vert] (mc) at (2, -7) {$\triangleone$};
					\node[vert] (m) at (3, -5) {$\triangledown$};
					
					\path (1) edge (q);
					\path (q) edge (q');
					\path (q) edge (q'');
					\path (q') edge node[left] {$\stackrel{C_2}{-1}$} (s1);
					\path (s1) edge[loop above] (s1);
					\path (q') edge (2);
					\path (q) edge (q'');
					\path (q'') edge node[above] {$\stackrel{C_2}{-1}$} (3);
					\path (q'') edge (q''');
					\path (q''') edge (m);
					\path (q''') edge (q'''');
					\path (q'''') edge (s2);
					\path (q'''') edge node[above right] {$\stackrel{C_2}{-1}$} (mc);
					\path (m) edge [loop below] node[below] {$\stackrel{C_1}{-1} \, \stackrel{C_2}{-1}$} (m);
					\path (s2) edge [loop below] (s2);
					\path (mc) edge [loop below] node[below] {$\stackrel{C_1}{-1}$} (mc);
				\end{tikzpicture}
				\caption{Gadget for tests of counter $C_2$}
				\label{fig_gadget_test_C2_sp}
			\end{subfigure}
    \caption{Gadgets}
\end{figure}

	Let $\Kount$ be a two-counter machine.
	We define an energy game $\Game_{\|q_0}$ with two players, player $C_1$ and player $C_2$, by assembling the gadgets presented in \Cref{fig_gadget_initial_sp,fig_gadget_final_sp,fig_gadget_test_C1_sp,fig_gadget_test_C2_sp,fig_gadget_incrementation_sp}.
    The rewards that are not presented are equal to $0$, the round vertices are those controlled by player $C_1$, and the square ones are those controlled by player $C_2$.
	For each state of the machine $\Kount$, we define from one to three vertices, plus six additional vertices, written $a$, $b$, $c$, $\triangle$, $\triangledown$, and $\triangleone$.
	
	Then, a play in $\Game_{\|q_0}$ that does not reach one of the sink vertices $\triangle$, $\triangledown$, or $\triangleone$ simulates a sequence of transitions of the machine $\Kount$, that can be a run or not: at each step, the counter $C$ is captured by the energy level of player $C$.
	Let us now prove that the game $\Game_{\|q_0}$ admits an SPE where player $C_2$ loses if and only if the machine $\Kount$ terminates.
	
	\paragraph{If such an SPE exists, then the machine $\Kount$ terminates.}
		
		Let us write $\bsigma$ for such an SPE, and let $\pi = \< \bsigma \>$.
  Let us show that the play $\pi$ simulates a run of $\Kount$.
		Since $\pi$ is lost by player $C_2$, there are \emph{a priori} three possibilities.
		
		\subp{The play $\pi$ reaches the vertex $\triangledown$}
			Then, player $C_1$ loses, and has therefore a profitable deviation by looping on $\pi_0 = q_0$, which is impossible.
			
			\subp{Player $C_1$ makes player $C_2$ lose by going to a vertex of the form $q'''$, when his energy level is zero}
            Thus, the play $\pi$ simulates a spurious run of $\Kount$, since such an action amounts to faking a test of $C_2$ above zero.
			But then, player $C_2$ has a profitable deviation by going to $b$: there, player $C_1$ cannot go to the vertex $\triangledown$, because it would make her lose, while she can win by going to the vertex $c$; indeed, from there, player $C_2$ cannot go to the vertex $\triangleone$ because it would make him lose, since he has no more energy, while he can go to the vertex $\triangle$ and win---and let player $C_1$ win.
			Therefore, this case is also impossible.
			
			\subp{The play $\pi$ reaches the vertex $q_\f$}
			Then, it simulates a correct run of the machine $\Kount$, that reaches the state $q_\f$.
			Indeed, we have already shown that $\pi$ cannot fake a test of $C_2$ above $0$.
			It cannot fake a test of $C_1$ above $0$ either, because then, player $C_1$ would lose, while she can win by looping on $q_0$.
			It cannot fake a test of $C_2$ to $0$, because then, from the vertex $q''$, if player $C_2$'s energy is greater than $0$, he has a profitable deviation by going to the vertex $\triangle$.
			Finally, it cannot fake a test of $C_1$ to $0$, because then, from the vertex $q'$, if player $C_1$'s energy is greater than $0$, player $C_2$ has a profitable deviation by going to $a$, from where player $C_1$ cannot go to the vertex $\triangledown$, because it would make her lose while she can win by going to $\triangle$.
   Therefore, the machine $\Kount$ terminates.

		\paragraph{If the machine $\Kount$ terminates, then there is an SPE where player $C_2$ loses.}
        
		If the machine $\Kount$ terminates, let us define a strategy profile $\bsigma$ in $\Game_{\|q_0}$ as follows.
		\begin{itemize}
			\item In tests of counter $C_1$, player $C_1$ goes to $q'$ if and only if her energy level is zero or $\bot$; from $q'$, player $C_2$ goes to the vertex $a$ if and only if player $C_1$'s energy level is positive.
			
			\item In tests of counter $C_2$, player $C_1$ goes to $q''$ if and only if the energy level of player $C_2$ is zero or $\bot$; from $q''$, player $C_2$ goes to the vertex $\triangle$ if and only if his energy level is positive, and from $q'''$, he goes to $b$ if and only if it is zero or $\bot$.
			
			\item From the vertex $a$, player $C_1$ goes to the vertex $\triangledown$ if and only if her energy level is zero or $\bot$.
			
			\item From the vertex $b$, player $C_1$ goes to the vertex $\triangledown$ if and only if the energy level of player $C_2$ is positive.
			
			\item From the vertex $c$, player $C_2$ goes to the vertex $\triangleone$ if and only if his energy level is positive.
		\end{itemize}
		
		Let us show that the strategy profile $\bsigma$ is an SPE.
		Let $hv$ be a history from the vertex $q_0$: let us prove that $\bsigma_{\|hv}$ is an NE.
		
		\subp{If $v$ is a vertex of the form $q$}
			
			Then, the play $\< \bsigma_{\|hv} \>$ simulates the correct run of the machine $\Kount$ from $q$ when the counters are initialized to $\EL_{C_1}(hv)$ and $\EL_{C_2}(hv)$ (if one of those energy levels is $\bot$, then it simulates the correct run of $\Kount$ where that counter is locked to $0$).
			Therefore, player $C_1$ wins (or has already lost), hence she cannot have a profitable deviation.
			As for player $C_2$, he cannot have a profitable deviation from a vertex of the form $q'$: when such a vertex is reached, player $C_1$'s energy is zero or $\bot$, hence if player $C_2$ chooses to go to $a$, player $C_1$ will go to $\triangledown$ and he will lose.
			He cannot have a profitable deviation from a vertex of the form $q''$ either: when such a vertex is reached, he has a zero or $\bot$ energy level, hence if he chooses to go to the vertex $\triangle$, he loses.
			Finally, he cannot have a profitable deviation from a vertex of the form $q'''$: when such a vertex is reached, he has a positive energy level, hence if he goes to the vertex $b$, player $C_1$ will go to the vertex $\triangledown$ and he will lose.
			
			\subp{If $v$ is a vertex of the form $q'$}
			
			Then, if player $C_1$'s energy level is positive, player $C_2$ goes to $a$, then player $C_1$ goes to $\triangle$, and both player $C_1$ and $C_2$ win---and have therefore no profitable deviation.
			Otherwise, the substrategy profile $\bsigma_{\|hv}$ simulates a correct run of $\Kount$, and we can use the same arguments as in the previous point.
			
			\subp{If $v$ is a vertex of the form $q''$}
			
			Then, if player $C_2$'s energy level is positive, he goes to $\triangle$ and wins---and no player has a profitable deviation.
			Otherwise, the substrategy profile $\bsigma_{\|hv}$ simulates a correct run of $\Kount$, and we can use the same arguments than in the first point.
			
			\subp{If $v$ is a vertex of the form $q'''$}
			
			Then, if player $C_2$'s energy level is zero or $\bot$, player $C_2$ goes to $b$, then player $C_1$ goes to $c$, and finally player $C_2$ goes to $\triangle$, and both player $C_1$ and $C_2$ win---they have therefore no profitable deviation.
			Otherwise, the substrategy profile $\bsigma_{\|hv}$ simulates a correct run of $\Kount$, and we can use the same arguments than in the first point.
			
			\subp{If $v = a$}
			
			Then, either player $C_1$ has a positive energy level, and then she goes to $\triangle$ and wins, or she has a zero or $\bot$ energy level, and then she cannot win.
			
			\subp{If $v = b$}
			
			Then, either player $C_2$ has a zero or $\bot$ energy level, and then the play ends in $\triangle$ and both players win (or have already lost), or he has a positive energy level, and then player $C_1$ cannot win, since player $C_2$ plans to go to $\triangleone$.
			
			\subp{If $v = c$}
			
			Then, either $C_2$ has a positive energy level, and he goes to $\triangleone$ and wins, or he has a zero energy level, and he goes to $\triangle$ and wins, or he has already lost.
			
			\subp{If $v = \triangledown$, $\triangleone$, $q_\f$ or $\triangle$} Then, the proof is immediate.

		Therefore, the strategy profile $\bsigma$ is an SPE, that simulates the run of the machine $\Kount$.
  That run terminates, hence the play $\< \bsigma \>$ reaches the vertex $q_\f$, and is lost by player $C_2$. 
Which concludes the proof.
\end{proof}

\subsection{About recursive enumerability}

Like for NEs, this proof shows that, in particular, the SPE constrained existence problem is not co-recursively enumerable in energy games.
It might still be the case that it is recursively enumerable.
That would in particular be the case if finite memory was sufficient for an SPE to achieve a given payoff vector, when that is possible, as in the case of NEs.
Unfortunately, one cannot follow this approach, because that statement is false: in order to be able to punish some player, without making another player lose, an SPE may have to memorize their energy levels, and therefore require infinite memory.

	\begin{thm} \label{thm:energy_infinite_memory}
        There exists an energy game in which there exists an SPE such that no finite-memory SPE generates the same payoff vector.
	\end{thm}

\begin{proof}
		In the energy game presented in Figure~\ref{fig_energy_infinite_memory}, let us show that there exists an SPE that makes player $\Box$ lose, but that no finite-memory SPE can achieve that result.

		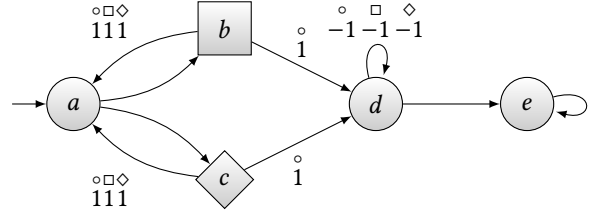
\begin{figure}
			\centering
			\begin{tikzpicture}
				\node[vert, initial left] (a) at (0, 0) {$a$};
				\node[vert, rectangle] (b) at (2, 1) {$b$};
				\node[vert, diamond] (c) at (2, -1) {$c$};
				\node[vert] (d) at (4, 0) {$d$};
				\node[vert] (e) at (6, 0) {$e$};
				\path (a) edge[bend right=20] (b);
				\path (b) edge[bend right=20] node[above left] {$\stackrel{\circ}{1}\stackrel{\Box}{1} \stackrel{\Diamond}{1}$} (a);
				\path (a) edge[bend left=20] (c);
				\path (c) edge[bend left=20] node[below left] {$\stackrel{\circ}{1}\stackrel{\Box}{1} \stackrel{\Diamond}{1}$} (a);
				\path (b) edge node[above] {$\stackrel{\circ}{1}$} (d);
				\path (c) edge node[below] {$\stackrel{\circ}{1}$} (d);
				\path (d) edge [loop above] node {$\stackrel{\circ}{-1} \, \stackrel{\Box}{-1} \, \stackrel{\Diamond}{-1}$} (d);
				\path (d) edge (e);
				\path (e) edge [loop right] (e);
			\end{tikzpicture}
			\caption{A game where infinite memory is necessary to make player $\Box$ lose}
			\label{fig_energy_infinite_memory}
		\end{figure}

        \paragraph{There is an SPE in which player $\Box$ loses.}
Consider the strategy profile $\bsigma$ defined by $\bsigma(a) = c$, by $\bsigma(hca) = b$ for every history $hca$, by $\bsigma(hb) = d$ for every history $hb$, by $\bsigma(hba) = c$ for every history $hba$, by $\bsigma(hc) = d$ for every history $hc$, and finally by $\< \bsigma_{\|hd} \> = d^{\EL_\circ(hd)+1} e^\omega$ for every history $hd$.

    Intuitively: in every subgame, player $\Circle$ makes player $\Box$ and player $\Diamond$ lose.
    But to do so, she needs one of those players to cooperate with her: for example, in the main subgame $\Game_{\|a}$, she wants to make player $\Box$ lose, and to do so, she traverses the vertex $c$ to go to the vertex $d$.
    But then, player $\Diamond$ may deviate, and go back to the vertex $a$.
    Then, player $\Diamond$ has to be punished: and to do so, player $\Circle$ must go to the vertex $d$ through the vertex $b$, with player $\Box$'s cooperation\dots and so on.
    Once the vertex $d$ is reached (which will eventually be the case in every subgame), player $\Circle$'s energy level is equal to player $\Box$'s and player $\Diamond$'s one, plus 1.
    Thus, to make those players lose without losing herself, player $\Circle$ loops on the vertex $d$ exactly the right number of times, before going to the vertex $e$.

    Thus, the strategy profile $\bsigma$ is an SPE in which player $\Box$ loses---but it is not a finite-memory SPE.

    \paragraph{There is no finite-memory SPE in which player $\Box$ loses.}
    Let $\btau$ be a finite-memory SPE in $\Game_{\|a}$, compatible with a memory structure $\Mem$, and let us assume toward contradiction that $\mu_\Box\< \btau \> = 0$.
    Since we have $\mu_\Box\< \btau \> = 0$ (and therefore $\mu_\diamond\< \btau \> = 0$, since players $\Box$ and $\Diamond$ always receive the same rewards), we also have, by induction and because $\btau$ is an SPE, the equality $\mu_\Box\< \btau_{\|ha} \> = \mu_\diamond\< \btau_{\|h a} \> = 0$ for every history $h$ that is compatible with $\sigma_\circ$.
    
    Let now $n$ be the number of states of the memory structure $\Mem$, and let us consider a history $ha$ with $|h| = 2n$ (and therefore $\EL_\circ(ha) = \EL_\Box(ha) = \EL_\diamond(ha) = n$).
    By the previous proposition, we know that $\mu_\Box\< \btau_{\|ha} \> = \mu_\diamond\< \btau_{\|h a} \> = 0$, i.e. that the play $\< \btau_{\|ha} \>$ reaches the vertex $d$, and takes the edge $dd$ more than $n$ times.
    Since the memory structure $\Mem$ has only $n$ states, it means that that play actually loops on the vertex $d$ infinitely often, and is therefore also lost by player $\Circle$.
    Then, player $\Circle$ has a profitable deviation in that subgame, by going to the vertex $e$: contradiction.

    There exists an SPE that makes player $\Box$ lose in that game, but no finite-memory one.
\end{proof}
        
We leave therefore the question open.

\begin{oprob}\label{op:spe_energy_re}
    Is the constrained existence problem of SPEs in energy games recursively enumerable?
\end{oprob}

%% file: 2eVerif.tex
We have now established all our results concerning the constrained existence problem of NEs and SPEs.
Before turning to more specialized notions of equilibria, as we will do in \cref{part:other_equilibria}, we use this chapter to discuss a possible application of our results: \emph{rational verification}.
Our observations will lead us to define a new decision problem, for which we also provide tight complexity bounds.

\section{Rational verification}

\emph{Verification}, without further precision, usually refers to the study of computational systems (typically abstracted as collections of automata or machine models) with the goal of verifying whether they satisfy some property, called a \emph{specification}.

A particular case of verification is \emph{reactive verification}, already described in \cref{sec:intro_2players}, where the computational model under study interacts with an \emph{environment} that may behave unpredictably.
In such a case, one seeks to check whether the system satisfies the specification against every possible behavior of the environment.
Illustrations of this concept often involve coffee machines; for the sake of detoxification, let us instead consider an elevator.
Its specification could be that, for every $i$, if at some point in time a user presses button $i$, then the elevator must eventually reach floor $i$ (or reach it within a specified duration).
If the elevator is located in a place where its use is essential, and where a malfunction could have serious consequences---such as in a hospital---it is reasonable to require that the elevator always guarantees this specification, even if a mischievous child persistently presses button 1.
Such an example motivates the study of two-player zero-sum games, which serve as a relevant metaphor for reactive verification: if the system (the elevator) and the environment (the child) are two players, we can define the system's objective as the specification, and we then want to verify whether the system's strategy is winning.

Now, let us imagine that our system is a self-driving car.
It would have an obvious Boolean specification: it must not cause the death of any human, whether the passenger or any other person nearby.
This specification can be refined into a quantitative objective by defining priorities: for example, we may require that it does not kill any human, and if that condition is satisfied, we may further wish for it to reach a given destination from location $A$ to location $B$, if possible within a reasonable duration.
In such a framework, it might seem excessively cautious to demand that the system satisfies its specification under \emph{every} possible behavior of the environment---or equivalently, to treat the environment as a purely adversarial entity whose sole aim is to cause the specification to fail.
If we assume, for instance, that all drivers in the world suddenly conspire to assassinate the passenger, then the most rational strategy, given the defined priorities, would be to keep the car safely locked in a garage.
This example illustrates that, in some cases, it might be useful to employ the game metaphor without modeling the environment as a purely adversarial player, but rather as a player or coalition of players with their own objectives, acting rationally according to some notion of \emph{rationality}.

This is the motivation of \emph{rational verification}.

        \section{Definitions}

Let us consider a game $\Game_{\|v_0}$ with one specific player, called \emph{Leader}, and denoted by $\Leader \in \Pi$.
Our decision problem will take as input a strategy $\sigma_\Leader$ for Leader (given as a memory structure), and ask whether that strategy ensures a given specification (encoded by Leader's payoff function and a threshold) against all \emph{rational} responses of the other players.

We must therefore define an adapted notion of rationality, that does not qualify a strategy profile for all players, but for all players except Leader.

\begin{defi}[$\Leader$-fixed NE, $\Leader$-fixed SPE]
    A strategy profile $\bsigma$ is a \emph{$\Leader$-fixed Nash equilibrium}, or \emph{$\Leader$-fixed NE}, if for each player $i \neq \Leader$ and every strategy $\sigma'_i$, we have $\mu_i\< \bsigma_{-i}, \sigma'_i \> \leq \mu_i\< \bsigma \>$.
    It is a \emph{$\Leader$-fixed subgame-perfect equilibrium}, or \emph{$\Leader$-fixed SPE}, if for every history $h$, the strategy profile $\bsigma_{\|h}$ is a $\Leader$-fixed NE.
\end{defi}

We can then define the rational verification problem, for a given class of game $\Class$ and a given \emph{rationality concept} $\rho \in \{\nash, \subgamep\}$.
A ($\Leader$-fixed) \emph{$\rho$-equilibrium} is a Nash equilibrium if $\rho = \nash$, and an SPE if $\rho = \subgamep$.
Let us recall that memory structures have been defined in \cref{def:memory_structure}, and that a non-deterministic memory structure $\Mem$ for a player $i$, from a given vertex $v_0$, induces a set $\Ind_{\|v_0}(\Mem)$ of strategies for player $i$.

\begin{prob}[(Deterministic) $\rho$-rational verification problem in the class $\Class$]
	Given a game $\Game_{\|v_0} \in \Class$, a threshold $t \in \QQ$ and a (deterministic) memory structure $\Mem$ for Leader on $\Game$, does every $\Leader$-fixed $\rho$-equilibrium $\bsigma$ with $\sigma_\Leader \in \Ind_{\|v_0}(\Mem)$ satisfy the inequality $\mu_\Leader\< \bsigma \> > t$?
\end{prob}

\section{Link with the constrained existence problem and complexities}

Rational verification problems are phrased in a way that makes them particularly annoying to treat directly, since their instances contain two graph structures: the game $\Game$, and the memory structure $\Mem$.
However, the experienced reader may have noticed that this annoyance can easily be removed, by considering the product of those two structures, or equivalently, by incorporating the memory structure $\Mem$ in the game, so that Leader's moves are constrained by the game structure itself.
To capture non-determinism, a fictional player must be added, who will choose which strategy $\sigma_\Leader \in \Ind_{\|v_0}(\Mem)$ Leader will follow, without having any interest in that choice (his payoff function will be constantly zero).
Since that player may choose the very strategy that does not guarantee the specification, if there is one, we call him \emph{Demon}.
We shall see that such a construction links the rational verification problem to the constrained existence problem, as studied in the previous chapters.

	\begin{defi}[Product game] \label{defi_product_game}
		Let $\Game_{\|v_0}$ be a game, and let $\Mem$ be a memory structure for Leader in $\Game$.
		Their \emph{product game} is the game $\Game_{\|v_0} \otimes \Mem = (\Pi \cup \{\Demon\}, V^\times, (V^\times_i)_i, E^\times, \mu^\times)_{\|(v_0, q_0)}$ where the player $\Demon$, called \emph{Demon}, chooses how the memory structure $\Mem$ will run.
		Formally:
		\begin{itemize}
			\item the vertex set is $V^\times = (V \times Q) \cup (V \times Q \times Q)$;
			
			\item Leader controls the set $V^\times_\Leader = \emptyset$, each player $i \neq \Leader, \Demon$ controls the set $V^\times_i = V_i \times Q \times Q$, and Demon controls the set $V^\times_\Demon = (V \times Q) \cup (V_\Leader \times Q \times Q)$;
			
			\item the set $E^\times$ contains:
			\begin{itemize}
				\item the edge $(u, p)(u, p, q)$ for each transition $(p, u, q) \in \Delta$ (if $u \not\in V_\Leader$), or $(p, u, q, v) \in \Delta$ (if $u \in V_\Leader$);
				
				\item the edge $(u, p, q)(v, q)$ for each transition $(p, u, q, v) \in \Delta$ (if $u \in V_\Leader$);
				
				\item the edge $(u, p, q)(v, q)$ for each transition $(p, u, q) \in \Delta$, and each edge $uv \in E$ (if $u \not\in V_\Leader$);
			\end{itemize}
			
			\item each payoff function $\mu^\times_i$ maps every play $(\pi_0, q_0) (\pi_0, q_0, q_1) (\pi_1, q_1) \dots$ to the payoff $\mu_i(\pi_0\pi_1 \dots)$ if $i \neq \Demon$, and to the payoff $0$ if $i = \Demon$.
		\end{itemize}
	\end{defi}

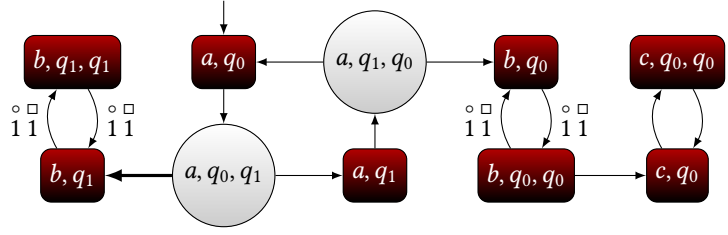
\begin{figure}
\centering
\begin{tikzpicture}
\newcommand{\deltay}{0.75}

   
	\node[demonVert] (bq1) at (-2, 2*\deltay) {$b, q_1$};
	\node[vert] (aq0q1) at (0, 2*\deltay) {$a, q_0, q_1$};
    \node[demonVert, initial above] (aq0) at (0, 4*\deltay) {$a, q_0$};
    \node[demonVert] (bq1q1) at (-2, 4*\deltay) {$b, q_1, q_1$};
    \node[demonVert] (aq1) at (2, 2*\deltay) {$a, q_1$};
    \node[vert] (aq1q0) at (2, 4*\deltay) {$a, q_1, q_0$};
    \node[demonVert] (bq0q0) at (4, 2*\deltay) {$b, q_0, q_0$};
    \node[demonVert] (bq0) at (4, 4*\deltay) {$b, q_0$};
    \node[demonVert] (cq0) at (6, 2*\deltay) {$c, q_0$};
    \node[demonVert] (cq0q0) at (6, 4*\deltay) {$c, q_0, q_0$};
				
				\path (aq0) edge (aq0q1);
				\path[very thick] (aq0q1) edge (bq1);
				\path (bq1) edge[bend left] node[left] {$\stackrel{\circ}{1}\, \stackrel{\Box}{1}$} (bq1q1);
				\path (bq1q1) edge[bend left] node[right] {$\stackrel{\circ}{1}\, \stackrel{\Box}{1}$} (bq1);
				\path (aq0q1) edge (aq1);
				\path (aq1) edge (aq1q0);
				\path (aq1q0) edge (aq0);
				\path (aq1q0) edge (bq0);
				\path (bq0) edge[bend left] node[right] {$\stackrel{\circ}{1}\, \stackrel{\Box}{1}$} (bq0q0);
				\path (bq0q0) edge[bend left] node[left] {$\stackrel{\circ}{1}\, \stackrel{\Box}{1}$} (bq0);
				\path (bq0q0) edge (cq0);
				\path (cq0) edge[bend left] (cq0q0);
				\path (cq0q0) edge[bend left] (cq0);
			\end{tikzpicture}
			\caption{A product game}
			\label{fig:product_game}
		\end{figure}

\begin{exa}
		Figure~\ref{fig:product_game} depicts the game $\Game_{\|v_0} \otimes \Mem$, when $\Game_{\|v_0}$ is the game of Figure~\ref{fig:first_example_game} (\cref{sec:graph_games}) and $\Mem$ the memory structure of Figure~\ref{fig:ex_1player_machine} (\cref{sec:stationary_finite_memory}).
		Leader is then assimilated to player $\Box$, and Demon's vertices are represented by dark red boxes.
The unreachable vertices have been omitted, and we have given only the non-zero rewards.
Since, from the vertex $(a, q_0, q_1)$, player $\Circle$ has always the possibility to go to the vertex $(b, q_1)$ and to get the payoff $1$, it can be shown that every NE and every SPE in that game gives player $\Box$ the payoff $1$.
As we will see now, that means that the strategies induced by the memory structure $\Mem$ guarantee the payoff $1$ to player $\Box$ against Nash-rational or subgame-perfect rational responses, i.e., that the game $\Game_{\|v_0}$, the memory structure $\Mem$, and the quantity $1-\epsilon$ for every $\epsilon > 0$, form a positive instance of the Nash and subgame-perfect rational verification problems.
\end{exa}

	\begin{thm}\label{thm:product_game}
		Let $\rho \in \{\nash, \subgamep\}$.
		Let $\Game_{\|v_0}$ be a game, let $\Mem$ be a memory structure for Leader in $\Game$, and let $t \in \QQ$.
		Then, every $\Leader$-fixed $\rho$-equilibrium $\bsigma$ with $\sigma_\Leader \in \Ind_{\|v_0}(\Mem)$ satisfies $\mu_\Leader\< \bsigma \> > t$ if and only if every $\rho$-equilibrium $\btau$ in the game $\Game_{\|v_0} \otimes \Mem$ satisfies $\mu^\times_\Leader\< \btau \> > t$.
	\end{thm}

\begin{proof}
    We present the proof when $\rho = \subgamep$; the proof for $\rho = \nash$ is analogous.
	    
		\paragraph{If every $\Leader$-fixed SPE $\bsigma$ with $\sigma_\Leader \in \Ind_{\|v_0}(\Mem)$ satisfies $\mu_\Leader\< \bsigma \> > t$, then every SPE $\btau$ in $\Game_{\|v_0} \otimes \Mem$ satisfies $\mu^\times_\Leader\< \btau \> > t$.}
			
			Let $\btau$ be an SPE in the game $\Game_{\|v_0} \otimes \Mem$.
			Let us define a strategy profile $\bsigma$ in $\Game_{\|v_0}$ as follows: for every history $h = h_0 \dots h_n$ in $\Game_{\|v_0}$, let $g = (h_0, q_0)(h_0, q_0, q_1) \dots (h_n, q_{n-1}, q_n)$ be the unique history of that form in $\Game_{\|v_0} \otimes \Mem$ such that we have $(h_k, q_k, q_{k+1}) = \tau_\Demon((h_0, q_0) \dots (h_k, q_k))$ for each $k$, and let $(v, q_n) = \btau(g)$.
			Then, we define $\bsigma(h) = v$.
			
			Since the only edges available for Demon in the game $\Game_{\|v_0} \otimes \Mem$ are those that are induced by $\Mem$, we have $\sigma_\Leader \in \Ind_{\|v_0}(\Mem)$.
			Moreover, the strategy profile $\bsigma$ is a $\Leader$-fixed SPE: let $h = h_0 \dots h_n$ be a history from $v_0$ that is compatible with the strategy $\sigma_\Leader$, let $i \in \Pi \setminus \{\Leader\}$ and let $\sigma'_i$ be a deviation of $\sigma_i$.
   We want to prove the inequality $\mu_i\< \bsigma_{-i\|h}, \sigma'_{i\|h} \> \leq \mu_i\< \bsigma_{\|h} \>$.
			Then, let us define the history $g$ as above, and let $\tau'_i$ be the strategy that simulates $\sigma'_i$ in the game $\Game_{\|v_0} \otimes \Mem$, i.e., that maps each history of the form $g (v_1, q_n) (v_1, q_n, q_{n+1}) \dots (v_k, q_{n+k-1}, q_{n+k})$ to the vertex $(\sigma'_i(hv_1 \dots v_k), q_{n+k})$.
			Since the strategy profile $\btau$ is an SPE, we have:
            $$\mu^\times_i(g_{< 2n+1} \< \btau_{-i\|g}, \tau'_{i\|g} \>) \leq \mu^\times_i(g_{< 2n+1} \< \btau_{\|g} \>),$$
            and therefore:
            $$\mu_i(h_{<n} \< \bsigma_{-i\|h}, \sigma'_{i\|h} \>) \leq \mu_i(h_{<n} \< \bsigma_{\|h} \>).$$
			
			Thus, the strategy $\bsigma$ is a $\Leader$-fixed SPE that satisfies $\sigma_\Leader \in \Ind_{\|v_0}(\Mem)$.
			By hypothesis, it comes that we have $\mu_\Leader\< \bsigma \> > t$, and therefore $\mu^\times_\Leader\< \btau \> > t$.

			\paragraph{If every SPE $\btau$ in $\Game_{\|v_0} \otimes \Mem$ satisfies $\mu^\times_\Leader\< \btau \> > t$, then every $\Leader$-fixed SPE with $\sigma_\Leader \in \Ind_{\|v_0}(\Mem)$ satisfies $\mu_\Leader\< \bsigma \> > t$.}
			
			Indeed, let $\bsigma$ be a $\Leader$-fixed SPE in $\Game_{\|v_0}$ with $\sigma_\Leader \in \Ind_{\|v_0}(\Mem)$.
			We write $h \mapsto q_h$ for the mapping establishing the fact that the strategy $\sigma_\Leader$ is induced by the memory structure $\Mem$, as defined in \cref{def:memory_structure}.
			Let us define a strategy profile $\btau$ in $\Game_{\|v_0} \otimes \Mem$ as follows: for every history of the form $g = (h_0, q_0) (h_0, q_0, q_1) \dots (h_n, q_{n-1})$, we define $\btau(g) = (h_n, q_{n-1}, q_{h_0 \dots h_n})$.
			For every history of the form $g = (h_0, q_0) \dots (h_n, q_{n-1}, q_n)$, we define $\btau(g) = (\bsigma(h_0 \dots h_n), q_n)$.
			
			Then, the strategy profile $\btau$ is an SPE: let $g = g_0 \dots g_m$ be a history in $\Game_{\|v_0} \otimes \Mem$, let $i$ be a player and let $\tau'_i$ be a deviation of $\tau_i$.
			If $i = \Leader$, then we have:
            $$\mu^\times_i(g_{<m} \< \btau_{-i\|g}, \tau'_{i\|g} \>) \leq \mu^\times_i(g_{<m} \< \btau_{\|g} \>),$$
            because Leader does not control any vertex in $\Game_{\|v_0} \otimes \Mem$, hence actually $\tau'_i = \tau_i$.
			Likewise if $i = \Demon$, because Demon gets the payoff $0$ in every play.
			Now, if $i \neq \Demon, \Leader$, let us consider without loss of generality that $g$ has the form $g = (h_0, q_0) (h_0, q_0, q_1) \dots (h_n, q_n)$ (if the last vertex is controlled by Demon, it can be removed).
			Let:
            $$(\pi_0, q_n) (\pi_0, q_n, q_{n+1}) (\pi_1, q_{n+1}) \dots = \< \btau_{-i\|g}, \tau'_{i\|g} \>.$$
			The play $\pi = \pi_0 \pi_1 \dots$ is compatible with the strategy profile $\bsigma_{-i\|h}$.
			Therefore, since the strategy profile $\bsigma$ is a $\Leader$-fixed SPE, we have the inequality $\mu_i(h_{<n} \pi) \leq \mu_i(h_{<n} \< \bsigma_{\|h} \>)$, i.e., the inequality:
            $$\mu^\times_i(g_{<2n+1} \< \btau_{-i\|g}, \tau'_{i\|g} \>) \leq \mu^\times_i(g_{<2n+1} \< \btau_{\|g} \>).$$
			
			Thus, the strategy profile $\btau$ is an SPE. Then, by hypothesis, we have $\mu^\times_\Leader\< \btau \> > t$, and therefore $\mu_\Leader\< \bsigma \> > t$.
	\end{proof}

As we will show, this result entails that the $\rho$-rational verification problem in the class $\Class$ is computationally equivalent to the following problem.

\begin{prob}[$\rho$-universal threshold problem in the class $\Class$]
		Given a game $\Game_{\|v_0} \in \Class$, a player $i \in \Pi$, and a threshold $t \in \QQ$, is every $\rho$-equilibrium $\bsigma$ in $\Game_{\|v_0}$ such that $\mu_i\< \bsigma \> > t$?
\end{prob}

	\begin{cor} \label{cor:reductions}
		Let $\Class$ be a game class among parity games, mean-payoff games, discounted-sum games, and energy games.
    Then, in the class $\Class$, for a given rationality concept $\rho \in \{\nash, \subgamep\}$, the $\rho$-universal threshold problem, the $\rho$-rational verification problem, and the deterministic $\rho$-rational verification problem are reducible to each other in polynomial time.
	\end{cor}

\begin{proof}
\paragraph{The deterministic $\rho$-rational verification problem reduces to the $\rho$-rational verification problem.}
	This result is true because a deterministic memory structure is a memory structure.
    
\paragraph{The $\rho$-universal threshold problem reduces to the deterministic $\rho$-rational verification problem.}
			Let the game $\Game_{\|v_0}$, the player $i$ and the threshold $t$ form an instance of the $\rho$-universal threshold problem.
			We define the game $\Game'_{\|v_0}$ as equal to the game $\Game_{\|v_0}$, where Leader has been added to the player set, but controls no vertex.
			We define $\mu_\Leader = \mu_i$.
			If $\Game$ belongs to the class $\Class$, so does $\Game'$.
			Let $\Mem$ be the one-state deterministic memory structure on $\Game'$ that never outputs anything.
			Then, a strategy profile $\bsigma$ in $\Game'_{\|v_0}$ is a $\Leader$-fixed $\rho$-equilibrium, if and only if it is a $\Leader$-fixed $\rho$-equilibrium with $\sigma_\Leader \in \Ind_{\|v_0}(\Mem)$, if and only if the strategy profile $\bsigma_{-\Leader}$ is a $\rho$-equilibrium in the game $\Game_{\|v_0}$.
			As a consequence the game $\Game_{\|v_0}$, the player $i$, and the threshold $t$ form a positive instance of the $\rho$-universal threshold problem, if and only if the game $\Game'_{\|v_0}$, the memory structure $\Mem$, and the threshold $t$ form a positive instance of the deterministic $\rho$-rational verification problem.
            Moreover, the latter can be constructed from the former in polynomial time.
            
	\paragraph{The $\rho$-rational verification problem reduces to the $\rho$-universal threshold problem.}
            This result is a consequence of Theorem~\ref{thm:product_game}, since the product game $\Game_{\|v_0} \otimes \Mem$ can be constructed from $\Game_{\|v_0}$ and $\Mem$ in polynomial time, and since for each of the four game classes considered, if the game $\Game$ belongs to the class $\Class$, so does the game $\Game_{\|v_0} \otimes \Mem$.
\end{proof}

It is now time to note that this problem is the complement of a subproblem of the constrained existence problem of $\rho$-equilibria in the class $\Class$, where the threshold vectors $\bx$ and $\by$ are such that $x_j = -\infty$ for all $j$, and $y_j = +\infty$ for all $j$ except one.
This strong connection between the two problems explains that, in the literature, \emph{rational verification} is sometimes the name given to what we call here the constrained existence problem (with possibly different types of constraints)---for a recent example, see~\cite{DBLP:journals/amai/GutierrezNPW23}.

Therefore, the complexity of Nash and subgame-perfect rational verification in parity, mean-payoff, discounted-sum or energy games can immediately be infered from \cref{thm:NE_parity,thm:NE_MP,thm:ds_ne_hardness,thm:ds_ne_easiness,thm:NE_energy,thm:spe_parity_np,thm:spe_parity_fpt,thm:spe_mp_np_complete,thm:ds_spe_easiness,thm:ds_spe_hardness,thm:energy_spe_undec}.

\begin{thm}\label{thm:verif}
    The Nash rational verification problem is:
    \begin{itemize}
        \item $\coNP$-complete in parity games;
        \item $\coNP$-complete in mean-payoff games;
        \item $\coTDS$-hard and recursively enumerable in discounted-sum games;
        \item undecidable and co-recursively enumerable in energy games.
    \end{itemize}

    The subgame-perfect rational verification problem is:
    \begin{itemize}
        \item $\coNP$-complete and fixed-parameter tractable (with the number of players and the number of colors as parameters) in parity games;
        \item $\coNP$-complete in mean-payoff games;
        \item $\coTDS$-hard and recursively enumerable in discounted-sum games;
        \item undecidable, and in particular not recursively enumerable, in energy games.
    \end{itemize}
\end{thm}

\section{A potential limit: the temptation of chaos}

It is now worth noting that the definition we gave of rational verification entails, in the case of mean-payoff games, results that may be considered as counter-intuitive.

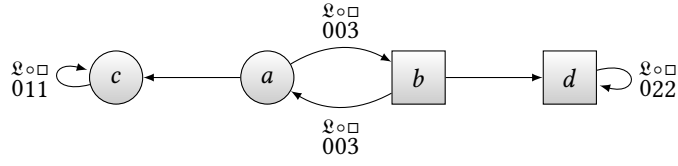
\begin{figure}
			\centering
			\begin{tikzpicture}
				\node[vert] (a) at (0, 0) {$a$};
				\node[vert] (c) at (-2, 0) {$c$};
				\node[vert, rectangle] (b) at (2, 0) {$b$};
				\node[vert, rectangle] (d) at (4, 0) {$d$};
				\path (a) edge (c);
				\path (a) edge[bend left] node[above] {$\stackrel{\Leader}{0}\stackrel{\circ}{0} \stackrel{\Box}{3}$} (b);
				\path (b) edge[bend left] node[below] {$\stackrel{\Leader}{0}\stackrel{\circ}{0} \stackrel{\Box}{3}$} (a);
				\path (b) edge (d);
				\path (d) edge [loop right] node {$\stackrel{\Leader}{0}\stackrel{\circ}{2} \stackrel{\Box}{2}$} (d);
				\path (c) edge [loop left] node {$\stackrel{\Leader}{0}\stackrel{\circ}{1} \stackrel{\Box}{1}$} (c);
			\end{tikzpicture}
			\caption{The temptation of chaos: an illustration}
			\label{fig:chaos}
		\end{figure}

\begin{exa}
Consider the mean-payoff game depicted by Figure~\ref{fig:chaos}, where Leader owns no vertex, and consider the only (vacuous) strategy available for Leader.
Does that strategy guarantee a payoff greater than $1$?
		Intuitively, it does not, since Leader always receives the payoff $0$.
		But still, that strategy, that game, and that threshold form a positive instance of the subgame-perfect rational verification problem, because no $\Leader$-fixed SPE exists in that game (see the proof of \cref{thm:no_spe}).
\end{exa}

More generally, the definition we give of rational verification considers that a \emph{good} strategy for Leader is a strategy such that for every response of the environment that is rational, the generated outcome observes some specification.
But a strategy is then good, in that sense, if \emph{no} rational response of the environment exists: that is the phenomenon that we can call \emph{temptation of chaos}.
While that case does never occur in all the other settings we have been studying, because rational responses are always guaranteed to exist (as we will see below), it must be considered specifically for subgame-perfect rational verification in mean-payoff games.
	
\section{Achaotic rational verification}

\subsection{Definitions}

	To avoid such phenomena, we generalize the approach used in a two-player context in~\cite{filiot_et_al:LIPIcs:2020:12534}, where \emph{best responses}, when they are not guaranteed to exist, are replaced by \emph{$\epsilon$-best responses}---i.e. best responses up to $\epsilon$, where $\epsilon \geq 0$ is as small as possible.

    We therefore introduce an alternative definition of rational verification, \emph{achaotic rational verification}: a good strategy for Leader is a strategy that guarantees the given threshold against every response that is \emph{as rational as possible}.
	A quantitative relaxation of our rationality concepts is therefore necessary: one is offered by the notions of $\epsilon$-NEs and $\epsilon$-SPEs, already defined in \cref{chap:nego}.
    For convenience, we write $\epsilon\SPE_\Leader\Game_{\|v_0}$ for the set of $\Leader$-fixed $\epsilon$-SPEs in the game $\Game_{\|v_0}$.
    Later, we will also write $\epsilon\SPE\Game_{\|v_0}$ for the set of $\epsilon$-SPEs in the same game.
    Finally, given a strategy $\sigma_\Leader$ in the game $\Game_{\|v_0}$, we denote by $\epsilon\SPR(\sigma_\Leader)$ the set of \emph{$\epsilon$-best responses to $\sigma_\Leader$}, i.e., the set of strategy profiles $\bsigma_{-\Leader}$ such that the strategy profile $\bsigma$ is a $\Leader$-fixed $\epsilon$-SPE.

\begin{prob}[Achaotic (deterministic) subgame-perfect rational verification in the class $\Class$]
	Given a game $\Game_{\|v_0} \in \Class$, a threshold $t \in \QQ$, and a memory structure (resp. a deterministic memory structure) $\Mem$ on $\Game$, do we have:
    $$\inf_{\sigma_\Leader \in \Ind_{\|v_0}(\Mem)}~
    \sup_{\substack{
        \epsilon \geq 0 \\
        \SPR(\sigma_\Leader) \neq \emptyset
    }}~
    \inf_{
        \bsigma_{-\Leader} \in \epsilon\SPR(\sigma_\Leader)
    }~
    \mu_\Leader\< \bsigma \> > t?$$
\end{prob}

In contrast with classical rational verification, the supremum and the infimum above are taken on sets that are never empty: if one considers $\epsilon$ large enough (for example $\epsilon = \max_{i \in \Pi} \max_{uv, u'v' \in E} |r_i(uv) - r_i(u'v')|$ in a mean-payoff game), there always exists an $\epsilon$-SPE in every game with a bounded set of payoff vectors, which is the case of all the game classes we considered so far.

The definition of \emph{achaotic (deterministic) Nash rational verification} is analogous.

\begin{exa}
    Let us consider again the example depicted by Figure~\ref{fig:chaos}.
    In that game, there exist ($\Leader$-fixed) $\epsilon$-SPEs for every $\epsilon \geq 1$ (for example, the stationary strategy profile $\bsigma$ defined by $\sigma(a) = c$ and $\sigma(b) = d$), but not for $\epsilon < 1$ (for the same reasons as there is no SPEs).
    Achaotic subgame-perfect rational verification amounts then to wondering whether all $\Leader$-fixed $1$-SPEs $\bsigma$ satisfy the inequality $\mu_\Leader\< \bsigma \> > t$, which is the case if and only if we have $t < 0$.
\end{exa}

In the general case, a proof analogous to that of \cref{thm:product_game,cor:reductions} can show that for each $\rho \in \{\nash, \subgamep\}$ and every game class $\Class$ among parity, mean-payoff, discounted-sum and energy games, the achaotic (deterministic or not) $\rho$-rational verification problem reduces in polynomial time to the following problem, and that a reverse reduction in polynomial time also exists.

\begin{prob}[Achaotic $\rho$-universal threshold problem in the class $\Class$]
    Given a game $\Game_{\|v_0} \in \Class$, a player $i \in \Pi$, and a threshold $t \in \QQ$, do we have:
    $$\sup_{\substack{
        \epsilon \geq 0 \\
        \epsilon\SPE\Game_{\|v_0} \neq \emptyset
    }}~
    \inf_{
        \bsigma \in \epsilon\SPE\Game_{\|v_0}
    }~
    \mu_\Leader\< \bsigma \> > t?$$
\end{prob}

\subsection{Coincidence with classical rational verification}

Among the problems we study here, this new definition is relevant in only one case: subgame-perfect rational verification in mean-payoff games.
In all other cases, the rational verification problems are equivalent to their achaotic versions, because Nash and subgame-perfect responses are guaranteed to exist.

\begin{thm}\label{thm:ach_verif}
    Let $\Class$ be a class of games, among the classes of parity, energy games and discounted-sum games.
	Let $\rho \in \{\nash, \subgamep\}$.
	Then, the positive instances of the achaotic $\rho$-universal threshold problem in $\Class$ are exactly the positive instances of the $\rho$-universal threshold problem in $\Class$.
\end{thm}

\begin{proof}
    \paragraph{Parity, energy, and discounted-sum games}
    The result is a consequence of the fact that in those three classes, SPEs are always guaranteed to exist---and therefore so are NEs.
        Indeed, it has been shown in~\cite{GU08} that every game with Borel Boolean winning condition contains an SPE.
        This covers parity, but also energy games.
        Thus, if $\Class$ is the class of parity or of energy games, then the game $\Game_{\|v_0} \otimes \Mem \in \Class$ contains an SPE.
        
        If $\Class$ is the class of discounted-sum games, then the game $\Game_{\|v_0}$, and therefore the game $\Game_{\|v_0} \otimes \Mem$, is a game with payoff functions that are continuous for the canonical distance on infinite words.
	    By~\cite{DBLP:conf/csl/BrihayeBMR15}, every such game contains an SPE.

        \paragraph{Nash-universal threshold problem in mean-payoff games}
        
        In this second case, SPEs may not exist in mean-payoff games, but NEs still always do---see for example~\cite{DBLP:conf/lfcs/BrihayePS13}.
\end{proof}

\subsection{The complexity of achaotic rational verification in mean-payoff games}

We now focus on the complexity of the achaotic subgame-perfect-universal threshold problem in mean-payoff games, knowing that it will also give us the complexity of the achaotic subgame-perfect rational verification problem.

An optimal algorithm for this problem requires the following lemma: in each game, there exists a least $\epsilon$ such that $\epsilon$-SPEs exist.
Moreover, that quantity can be written with a polynomially bounded number of bits.

	\begin{lem} \label{lm:epsilon}
		There exists a polynomial $P_1$ such that in every mean-payoff game $\Game_{\|v_0}$, there exists $\epsilon_{\min}$ with $\lv \epsilon_{\min} \rv \leq P_1(\lv \Game \rv)$ such that $\epsilon_{\min}$-SPEs exist in $\Game_{\|v_0}$, and $\epsilon$-SPEs, for every $\epsilon < \epsilon_{\min}$, do not.
	\end{lem}

\begin{proof}
    This proof will rely on some new technical considerations on the negotiation function in mean-payoff games.

\paragraph{A preliminary result}

We will need to prove that the negotiation function is \emph{continuous from below}.
Let us first recall a definition of that notion.

\begin{defi}[Continuity from below]
    A mapping $f: \RR^D \to \RR^D$ is \emph{continuous from below} if for every non-decreasing sequence $(\bx_n)$, we have $f(\sup_n \bx_n) = \sup_n f(\bx_n)$.
\end{defi}

\begin{slem}\label{slm:nego_continuous}
    In mean-payoff games, the negotiation function is continuous from below
\end{slem}

\begin{proof}
Let $(\lambda_n)_n$ be a non-decreasing sequence of requirements on a mean-payoff game $\Game$, and let $\lambda = \sup_n \lambda_n$. We want to prove the equality $\nego(\lambda) = \sup_n \nego(\lambda_n)$.
Since the negotiation function is monotone, we already have $\nego(\lambda) \geq \sup_n \nego(\lambda_n)$. Let us prove that $\nego(\lambda) \leq \sup_n \nego(\lambda_n)$.
Let $\delta > 0$: we want to find $n$ such that $\nego(\lambda_n)(v) \geq \nego(\lambda)(v) - \delta$ for each $v \in V$.

We will use for that purpose, for a given player $i$ and a given vertex $v \in V_i$, the concrete negotiation games:
$$\conc_{\lambda i}(\Game)_{\|v^\con} = \left(\{\Prover, \Challenger\}, V^\con, (V^\con_\Prover, V^\con_\Challenger), E^\con, \mu^\con\right)_{\|v^\con}$$
and:
$$\conc_{\lambda_n i}(\Game)_{\|v^\con} = \left(\{\Prover, \Challenger\}, V^\con, (V^\con_\Prover, V^\con_\Challenger), E^\con, \mu^n\right)_{\|v^\con}$$
for some requirement $\lambda_n$.
Let us note that both games have the same underlying graph, and that the only difference is between the payoff functions $\mu^\con$ and $\mu^n$.

Let us define:
$$\gamma_n = \sup_{w \in V} (\lambda(w) - \lambda_n(w)).$$
Then, the sequence $(\gamma_n)_n$ is non-increasing and converges to $0$.
Let $\tau_\Challenger$ be a stationary strategy for Challenger in the game $\conc_{\lambda i}(\Game)_{\|s_0}$; it can also be considered as a stationary strategy in the game $\conc_{\lambda_n i}(\Game)_{\|s_0}$.
Now, let $K$ be a strongly connected component of the underlying graph of the game $\conc_{\lambda i}(\Game)[\tau_\Challenger]$, accessible from the vertex $v^\con_0$.
Against the strategy $\tau_\Challenger$, Prover may go to the strongly connected component $K$, and give to Challenger the payoff that he gets in any play in that strongly connected component.

\subp{If the strongly connected component $K$ contains some deviation}
Then, the least payoff that Challenger can obtain in $K$ is the same when the payoff function is $\mu^\con$ and when it is $\mu^n$: it corresponds to the least payoff of the form $\mu_i(\dc^\omega)$ where $c$ is a simple cycle of $K$.

\subp{If the strongly connected component $K$ contains no deviation}
Then, the vertices of $K$ share a common memory $M$.
Let $R = \{\bx \in \RR^\Pi ~|~ \forall i \in \Pi, \forall v \in M \cap V_i, x_i \geq \lambda(v)\}$, and similarly, let $R_n = \{\bx \in \RR^\Pi ~|~ \forall i \in \Pi, \forall v \in M \cap V_i, x_i \geq \lambda_n(v)\}$.

If we have $(\Conv_{c \in \SC(K)} \mu(\dc^\omega)) \cap R = \emptyset$, since the sets $\Conv_{c \in \SC(K)} \mu(\dc^\omega)$ and $R$ are closed, then by taking the quantity $\gamma_n$ small enough, we also have $(\Conv_{c \in \SC(K)} \mu(\dc^\omega))\cap R_n = \emptyset$.
In this case, whatever Prover does in the strongly connected component $K$, Challenger will always get the payoff $+\infty$.
	
If that is not the case, the polytope $(\Conv_{c \in \SC(K)} \mu(\dc^\omega)) \cap R$ is included in the polytope $(\Conv_{c \in \SC(K)} \mu(\dc^\omega)) \cap R_n$, and the difference between the infima of those polytopes' projections on the dimension $i$ can be bounded by taking the biggest slope of those polytopes' edges, which are in finite number.
Formally, we have:
\begin{align*}
    &\inf\left\{ x_i ~\left|~ \bx \in \left(\underset{c \in \SC(K)}{\Conv} \mu(\dc^\omega)\right) \cap R_n \right.\right\} \\
    & \geq \inf\left\{ x_i ~\left|~ \bx \in \left(\underset{c \in \SC(K)}{\Conv} \mu(\dc^\omega)\right) \cap R \right.\right\}
    - \gamma_n \max_{
		c, d \in \SC(K)
		}
	\sum_{\substack{
		j \in \Pi, \\
		\mu_j(\dc^\omega) >
		\mu_j(\dd^\omega)
		}}
    \frac{\mu_i(\dc^\omega) - \mu_i(\dd^\omega)}{\mu_j(\dc^\omega) - \mu_j(\dd^\omega)}
\end{align*}
	and if $\gamma_n$ is small enough, the right-hand subtrahend can be made smaller than $\delta$.

\subp{Conclusion}
In both cases, we find that there exists $\gamma_n$ small enough, i.e. $n$ large enough, to ensure that the least payoff that Prover can give to Challenger in $K$ varies by less than the quantity $\delta$ when the requirement varies between $\lambda$ and $\lambda_n$.

We can find such $n$ for each strongly connected component $K$, and there exists a finite number of such components: we can therefore find some $n$ that satisfies the same property for all $K$.
Similarly, we can find some $n$ that also ensures it for all $K$ and for all stationary strategies $\tau_\Challenger$.
Since by \cref{lm:concrete_stationary} stationary strategies are optimal in the concrete negotiation game, we infer that there exists $n \in \NN$ such that:
$$\nego(\lambda_n)(v) \geq \nego(\lambda)(v) - \delta.$$

Finally, since there are finitely many states $v \in V$, we can conclude to the existence of $n \in \NN$ such that for each $v \in V$, we have:
$$\nego(\lambda_n)(v) \geq \nego(\lambda)(v) - \delta,$$
which concludes the proof.
\end{proof}

		\paragraph{Existence of $\epsilon_{\min}$}

Let us recall that, by \cref{lm:mp_fixed_point}, for every $\epsilon \geq 0$, the negotiation function has a least $\epsilon$-fixed point, which we will write $\lambda_\epsilon$.
Let us also recall that by \cref{thm:nego_spe}, a play in the game $\Game_{\|v_0}$ is an $\epsilon$-SPE outcome if and only if it is $\lambda_\epsilon$-consistent.
        
		Let us define:
        $$\epsilon = \inf\{\delta \geq 0 ~|~ \delta\SPE\Game_{\|v_0} \neq \emptyset\}.$$
		We only need to show that $\epsilon$ is also such that $\epsilon$-SPEs exist, or in other words, that $\lambda_\epsilon(v_0) \neq +\infty$.
		Let us note that for every $\delta, \delta'$ with $\delta \leq \delta'$, the requirement $\lambda_\delta$ is itself a $\delta'$-fixed point of the negotiation function, and therefore satisfies $\lambda_\delta \geq \lambda_{\delta'}$.
		
		Let us now define the requirement $\lambda: v \mapsto \sup_{\delta > \epsilon} \lambda_\delta(v)$.
		For each $v$ and every $\delta > \epsilon$, we have:
        $$\nego(\lambda_\delta)(v) \leq \lambda_\delta(v) - \delta.$$
        Using \cref{slm:nego_continuous}, we obtain:
        $$\nego(\lambda)(v) \leq \lambda(v) - \epsilon,$$
        i.e., the requirement $\lambda$ is an $\epsilon$-fixed point of the negotiation function, and therefore $\lambda = \lambda_\epsilon$.
		Since $\lambda(v_0) = \sup_\delta \lambda_\delta(v_0) \neq +\infty$, we obtain that $\epsilon$-SPEs exist in $\Game_{\|v_0}$, and therefore that $\epsilon_{\min} = \epsilon$ exists.

\paragraph{Size}
        
		Now, let us recall that we showed in the proof of \cref{lm:mp_fixed_point} that one can define, for every $\lambda$, a finite union of polyhedra $X_\lambda \subseteq \RR^{V \times \Pi}$, where each of those polyhedra is defined by a set of inequations either of the form $x_{vi} \geq \lambda(w)$ or of size bounded by a polynomial function of $\lv \Game \rv$, and such that for each $i \in \Pi$ and $v \in V_i$, we have $\nego(\lambda)(v) = \min\left\{x_{vi} ~\left|~ \bbx \in X_\lambda\right.\right\}$.
		
		Let us now define the set:
        $$Y = \left\{\left(\lambda, \bbx\right) ~\left|~ \bbx \in X_\lambda\right.\right\} \subseteq \RR^V \times \RR^{V \times \Pi}.$$
		This set is itself a finite union of polyhedra defined by the same inequations than $X_\lambda$, or by inequations of bounded size (those of the form $x_{vi} \geq \lambda(w)$, where $\lambda(w)$ is no longer a constant but a coordinate of $\left(\lambda, \bbx\right)$).
		Let us also define the mapping:
		$$f: \left\{ \begin{matrix}
			Y & \to & \RR \\
			\left(\lambda, \bbx\right) & \mapsto & \underset{i \in  \Pi, v \in V_i}{\max} \left(x_{vi} - \lambda(v)\right)
		\end{matrix} \right..$$
		Then, we have:
		$$\epsilon_{\min} = \min_{\left(\lambda, \bbx\right)} f\left(\lambda, \bbx\right),$$
		and since $f$ is a piecewise linear mapping on a finite union of polyhedra that has a minimum, that minimum is reached on a vertex $\left(\lambda, \bbx\right)$ of one of those polyhedra.
		By \cref{cor_size_vertices}, such a vertex has coordinates bounded by a polynomial of the maximal size of the inequations defining the polyhedron to which it belongs, i.e., by a polynomial of $\lv \Game \rv$.
		Therefore, that is also the case of $f(\lambda, \bbx) = \epsilon_{\min}$.
	\end{proof}

We are now equipped to prove the following theorem.
 
	\begin{thm}\label{thm:mp_ach_verif}
		In the class of mean-payoff games, the achaotic subgame-perfect-universal threshold problem, deterministic or not, is $\P^\NP$-complete.
	\end{thm}
	
\begin{proof}		
		\paragraph{Easiness}
			
			By \cref{thm:spe_mp_np_complete} (taking $\bx = (-\infty)_{i \in \Pi}$ and $\by = (+\infty)_{i \in \Pi}$), there is an $\NP$ algorithm that decides, given $\epsilon$ and $\Game_{\|v_0}$, whether there exists an $\epsilon$-SPE in $\Game_{\|v_0}$, i.e. whether $\epsilon \geq \epsilon_{\min}$.
			Using Lemma~\ref{lm:epsilon}, a dichotomous search can therefore compute $\epsilon_{\min}$ by a polynomial number of calls to that algorithm.
			Then, one last call to that same algorithm can decide whether there exists an $\epsilon_{\min}$-SPE $\bsigma$ such that $\mu_i\< \bsigma \> \leq t$.

			\paragraph{Hardness}

			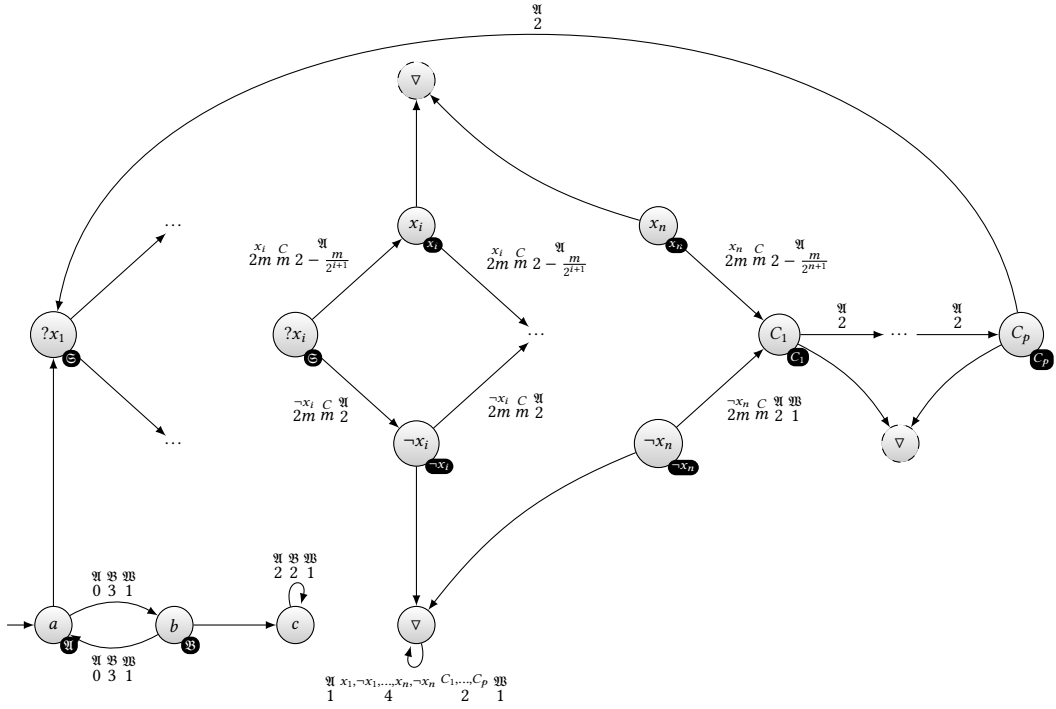
\begin{figure*}
				\centering
                \newcommand{\deltay}{1.2}
				\begin{tikzpicture}[scale=0.8, every node/.style={scale=0.7}, initial text={}]
					\node[ownedVert=$\Solver$] (?x1) at (0, 0*\deltay) {$?x_1$};
					\node (x1) at (2, 1.5*\deltay) {\dots};
					\node (nx1) at (2, -1.5*\deltay) {\dots};
					\node[ownedVert=$\Solver$] (?xi) at (4, 0*\deltay) {$?x_i$};
					\node[ownedVert=$x_i$] (xi) at (6, 1.5*\deltay) {$x_i$};
					\node[ownedVert=$\neg x_i$] (nxi) at (6, -1.5*\deltay) {$\neg x_i$};
					\node (?xi+1) at (8, 0*\deltay) {\dots};
					\node[ownedVert=$x_n$] (xn) at (10, 1.5*\deltay) {$x_n$};
					\node[ownedVert=$\neg x_n$] (nxn) at (10, -1.5*\deltay) {$\neg x_n$};
					\node[ownedVert=$C_1$] (C1) at (12, 0*\deltay) {$C_1$};
					\node (C2) at (14, 0*\deltay) {\dots};
					\node[ownedVert=$C_p$] (Cp) at (16, 0*\deltay) {$C_p$};
					\node[ownedVert=$\adel$, initial left] (a) at (0, -4*\deltay) {$a$};
					\node[ownedVert=$\bart$] (b) at (2, -4*\deltay) {$b$};
					\node[vert] (c) at (4, -4*\deltay) {$c$};
					\node[vert] (m1) at (6, -4*\deltay) {$\triangledown$};
					\node[vert, dashed] (m2) at (6, 3.5*\deltay) {$\triangledown$};
					\node[vert, dashed] (m3) at (14, -1.5*\deltay) {$\triangledown$};

					\path (a) edge[bend left] node[above] {$\stackrel{\adel}{0}~\stackrel{\bart}{3}~\stackrel{\Witness}{1}$} (b);
					\path (b) edge[bend left] node[below] {$\stackrel{\adel}{0}~\stackrel{\bart}{3}~\stackrel{\Witness}{1}$} (a);
					\path (b) edge (c);
					\path (c) edge[loop above] node[above] {$\stackrel{\adel}{2}~\stackrel{\bart}{2}~\stackrel{\Witness}{1}$} (c);
					\path (a) edge (?x1);
					\path (?x1) edge (x1);
					\path (?x1) edge (nx1);
					\path (?xi) edge node[above left] {$\stackrel{x_i}{2m}~\stackrel{C}{m}~\stackrel{\adel}{2 - \frac{m}{2^{i+1}}}$} (xi);
					\path (?xi) edge node[below left] {$\stackrel{\neg x_i}{2m}~\stackrel{C}{m}~\stackrel{\adel}{2}$} (nxi);
					\path (xi) edge node[above right] {$\stackrel{x_i}{2m}~\stackrel{C}{m}~\stackrel{\adel}{2 - \frac{m}{2^{i+1}}}$} (?xi+1);
					\path (nxi) edge node[below right] {$\stackrel{\neg x_i}{2m}~\stackrel{C}{m}~\stackrel{\adel}{2}$} (?xi+1);
					\path (xn) edge node[above right] {$\stackrel{x_n}{2m}~\stackrel{C}{m}~\stackrel{\adel}{2 - \frac{m}{2^{n+1}}}$} (C1);
					\path (nxn) edge node[below right] {$\stackrel{\neg x_n}{2m}~\stackrel{C}{m}~\stackrel{\adel}{2}~\stackrel{\Witness}{1}$} (C1);
					\path (C1) edge node[above] {$\stackrel{\adel}{2}$} (C2);
					\path (C2) edge node[above] {$\stackrel{\adel}{2}$} (Cp);
					\path (Cp) edge[bend right=80] node[above] {$\stackrel{\adel}{2}$} (?x1);
					\path (nxi) edge (m1);
					\path (nxn) edge[bend right=15] (m1);
					\path (xi) edge (m2);
					\path (xn) edge[bend left=15] (m2);
					\path (C1) edge[bend left=15] (m3);
					\path (Cp) edge[bend right=15] (m3);
					\path (m1) edge[loop below] node[below] {$\stackrel{\adel}{1}~\stackrel{x_1, \neg x_1, \dots, x_n, \neg x_n}{4}~ \stackrel{C_1, \dots, C_p}{2}~\stackrel{\Witness}{1}$} (m1);
				\end{tikzpicture}
				\caption{The game $\Game_{\|a}$}
				\label{fig:G_pnp}
			\end{figure*}
            
			We proceed by reducing, to the complement of our problem, the following $\P^\NP$-complete problem.

            \begin{prob}[Lexicographic minimum problem]
                Given a Boolean formula $\phi$ in conjunctive normal form over the ordered variables $x_1, \dots, x_n$, is the lexicographically first valuation $\nu_{\min}$ satisfying $\phi$ such that $\nu_{\min}(x_n) = 1$? (and in particular, does such a valuation exist?)
            \end{prob}

            \subp{Construction}
			Let us write $\phi = \bigwedge_{j=1}^p C_j$.
			We construct a game $\Game_{\|a}$, with a player called \emph{Witness} and written $\Witness$, in which there exists an $\epsilon_{\min}$-SPE $\bsigma$ such that $\mu_\Witness\< \bsigma\> \leq 0$ if and only if $\phi$ is satisfiable and is such that $\nu_{\min}(x_n) = 1$.
			That game, depicted in Figure~\ref{fig:G_pnp}, has $2n + p + 4$ players: the literal players $x_1, \neg x_1, \dots, x_n, \neg x_n$; the clause players $C_1, \dots, C_p$; the player \emph{Solver}, written $\Solver$; the player \emph{Witness}, written $\Witness$; the player \emph{Adélaïde}, written $\adel$; and the player \emph{Barthélémy}, written $\bart$.
			It contains $3n + p + 4$ vertices:
			\begin{itemize}
				\item the initial vertex $v_0 = a$, controlled by Adélaïde;
				
				\item two vertices $b$ and $c$, controlled by Barthélémy;
				
				\item for each variable $x_i$, a vertex $?x_i \in V_\Solver$ , a vertex $x_i \in V_{x_i}$, and a vertex $\neg x_i \in V_{\neg x_i}$;
				
				\item for each clause $C_j$, a vertex $C_j \in V_{C_j}$;
				
				\item a sink vertex $\triangledown$ (drawn three times in Figure~\ref{fig:G_pnp} for convenience).
			\end{itemize}
			
			Those vertices are connected by the following edges (unmentioned rewards are equal to $0$, and we write $m = 2n+p$):
			\begin{itemize}
				\item from the vertex $a$ to the vertex $b$ and from the vertex $b$ to the vertex $a$, two edges that give Adélaïde the reward $0$, Barthélémy the reward $3$, and Witness the reward $1$;
				
				\item from the vertex $a$ to the vertex $?x_1$ and from $b$ to $c$, an edge;
				
				\item from the vertex $c$ to itself, an edge giving both Adélaïde and Barthélémy the reward $2$, and giving Witness the reward $1$;
				
				\item from each vertex $?x_i$ to the vertex $\neg x_i$ and from each vertex $\neg x_i$ to the vertex $?x_{i+1}$ (or to the vertex $C_1$ if $i = n$), an edge giving:
				\begin{itemize}
					\item the reward $2m$ to player $\neg x_i$,
					\item the reward $m$ to every player $C_j$ such that the clause $C_j$ contains the literal $\neg x_i$,
					\item the reward $2$ to Adélaïde;
					\item and if $i = n$, the reward $1$ to Witness;
				\end{itemize}
				
				\item from each vertex $?x_i$ to the vertex $x_i$ and from each vertex $x_i$ to the vertex $?x_{i+1}$ (or to the vertex $C_1$ if $i = n$), an edge giving:
				\begin{itemize}
					\item the reward $2m$ to player $x_i$,
					\item the reward $m$ to every player $C_j$ such that the clause $C_j$ contains the literal $x_i$,
					\item and the reward $2 - \frac{m}{2^{i+1}}$ to Adélaïde;
				\end{itemize}
				
				\item from each vertex $C_j$ to the vertex $C_{j+1}$ (or the vertex $?x_1$ if $j = p$), an edge giving the reward $2$ to Adélaïde;
				
				\item from the sink vertex $\triangledown$ to itself, an edge giving the reward $1$ to Adélaïde, the reward $2$ to each clause player, the reward $4$ to each literal player, and the reward $1$ to Witness.
			\end{itemize}
			
We will show that in this game, when $\nu_{\min}$ exists, it defines a binary encoding for the quantity $\epsilon_{\min}$.
To do so, let us present a more general correspondence between $\epsilon$-SPEs and valuations satisfying $\phi$.

\subp{The strategy profile $\bsigma_\nu$}

    Let $\nu$ be a valuation satisfying $\phi$.
    We define the stationary strategy profile $\bsigma_\nu$ as follows: from the vertex $a$, Adélaïde always goes to ${?x_1}$; from the vertex $b$, Barthélémy always goes to $c$; from each vertex ${?x_i}$, Solver always goes to $x_i$ if $\nu(x_i) = 1$, and to $\neg x_i$ otherwise; from their vertices, the clause players and the literal players that are satisfied by $\nu$ do never go to the vertex $\triangledown$; and the literal players that are not satisfied by $\nu$ do whenever they have the opportunity.

\begin{slem}\label{slm:sigma_nu}
    The strategy profile $\bsigma_\nu$ is an $\epsilon$-SPE, where $\epsilon = \sum_{i=1}^n \frac{\nu(x_i)}{2^i}$, and is not a $\delta$-SPE for any $\delta < \epsilon$.
\end{slem}
                    
\begin{proof}
	Solver's payoff is constant, and Witness does not control any vertex, hence they have no profitable deviation.
	In every subgame where they have actions to choose, all the clause players get at least the payoff $2$, since at least one of their literals is satisfied (and therefore the corresponding vertex is visited at each turn).
	Similarly, from the vertex he controls, each literal player gets the payoff $4$, either because he is satisfied by $\nu$, or by going to the vertex $\triangledown$; they have therefore no profitable deviation.
	As for Barthélémy, in every subgame where he has an action to choose, he gets the payoff $2$, and cannot get a better one, since Adélaïde always plans to go to $?x_1$.
	Let us finally focus on Adélaïde: from the vertex $a$, the only one that she controls, she could get the payoff $2$ by going to $b$.
	By going to the vertex $?x_1$, she gets the payoff:
	\begin{gather*}
	\frac{1}{2n+p} \left(\sum_{i=1}^n 2\left( 2 - (2n+p) \frac{\nu(x_i)}{2^{i+1}} \right) + 2p \right) \\
	= 2 - \sum_{i=1}^n \frac{\nu(x_i)}{2^i} = 2 - \epsilon,
	\end{gather*}
	hence $\bsigma_\nu$ is an $\epsilon$-SPE, and is not a $\delta$-SPE for any $\delta < \epsilon$.
\end{proof}

    \subp{The valuation $\nu_{\bsigma}$}

    Let $\bsigma$ be a $1$-SPE such that the play $\pi = \< \bsigma \>$ traverses the vertex $?x_1$, but does never reach the vertex $\triangledown$.
    We define the valuation $\nu_{\bsigma}$ by $\nu_{\bsigma}(x_i) = 1$ if and only if $\pi$ traverses the vertex $x_i$.
    
	Let us note that for every vertex $x_i$ that is traversed by $\pi$, the player $x_i$ can deviate and go to the vertex $\triangledown$, where he can get the payoff $4$.
	Since $\bsigma$ is a $1$-SPE, we have therefore $\mu_{x_i}(\pi) \geq 3$, and therefore $\mu_{\neg x_i}(\pi) \leq 1$.
    As a consequence, the vertices $x_i$ and $\neg x_i$ cannot both be traversed; and if the vertex $\neg x_i$ is traversed, then we have $\nu_{\bsigma}(x_i) = 0$.

\begin{slem}\label{slm:nu_sigma}
    The valuation $\nu_{\bsigma}$ satisfies $\phi$, and we have:
    $$\mu_\adel(\pi) = 2 - \sum_{i=1}^n \frac{\nu_{\bsigma}(x_i)}{2^i}.$$
\end{slem}

\begin{proof}
	Let us consider a clause $C_j$ of $\phi$.
	Since the vertex $C_j$ is visited (infinitely often), there is a deviation available for player $C_j$ that grants her the payoff $2$.
	Since $\bsigma$ is a $1$-SPE, we have $\mu_{C_j}(\pi) \geq 1$.
	Therefore, there is at least one literal $\l$ of $C_j$ such that the vertex $\l$ is visited by $\pi$, i.e. such that $\nu_{\bsigma}$ satisfies $\l$.
    Consequently, the valuation $\nu_{\bsigma}$ satisfies $C_j$, and therefore $\phi$.

    Moreover, since the play $\pi$ does never traverse both $x_i$ and $\neg x_i$ for any $i$, it grants Adélaïde the payoff:
    \begin{gather*}
	\frac{1}{2n+p} \left(\sum_{i=1}^n 2\left( 2 - (2n+p) \frac{\nu_{\bsigma}(x_i)}{2^{i+1}} \right) + 2p \right) \\
	= 2 - \sum_{i=1}^n \frac{\nu_{\bsigma}(x_i)}{2^i}.
	\end{gather*}
\end{proof}

\subp{Consequence}
We can then prove the following proposition.

\begin{prop}
    if $\nu_{\min}$ is the least valuation satisfying $\phi$ (and in particular if such a valuation exists), then we have $\epsilon_{\min} = \sum_{i=1}^n \frac{\nu_{\min}(x_i)}{2^i}$.
\end{prop}

\begin{proof}
Let $\delta = \sum_{i=1}^n \frac{\nu_{\min}(x_i)}{2^i} < 1$.

By \cref{slm:sigma_nu}, the strategy profile $\bsigma_{\nu_{\min}}$ is a $\delta$-SPE.
Let us prove that there is no $\epsilon$-SPE with $\epsilon < \delta$.
Let $\epsilon < 1$ be such that there exists an $\epsilon$-SPE $\bsigma$ in $\Game_{\|a}$, and let us prove that $\epsilon \geq \delta$.

If $\bsigma$ is an $\epsilon$-SPE with $\epsilon < 1$, then necessarily, there exist infinitely many integers $k$ such that $\sigma_\adel((ab)^ka) = {?x_1}$, because otherwise each play $\< \bsigma_{\|(ab)^ka} \>$ would either end in the vertex $c$ (and then Barthélémy would have a deviation profitable by more than $1$ by refusing to go to $c$), or would be equal to $(ab)^\omega$ (and then Adélaïde would have a deviation profitable by more than $1$ by going to $?x_1$).
Then, for each such $k \neq 0$, we have $\sigma_\bart((ab)^k) = c$ (if Barthélémy does not go to $c$, then he gets only the payoff $1$, while by going to $c$, he gets the payoff $2$).
Therefore, if we choose some $k$ such that $\sigma_\adel((ab)^ka) = {?x_1}$, we also have $\mu_\adel\< \bsigma_{\|(ab)^kab} \> = 2$.
Since $\bsigma$ is an $\epsilon$-SPE with $\epsilon < 1$, we have therefore $\mu_\adel\< \bsigma_{\|(ab)^ka} \> > 1$, which means that the play $\< \bsigma_{\|(ab)^ka} \>$ does never reach the vertex $\triangledown$.
Then, the strategy profile $\bsigma_{\|(ab)^ka}$ is a $1$-SPE such that the play $\< \bsigma_{\|(ab)^ka} \>$ traverses the vertex $?x_1$, but does never reach the vertex $\triangledown$: the valuation $\nu_{\bsigma_{\|(ab)^ka}}$ is defined, and by \cref{slm:nu_sigma}, it satisfies the formula $\phi$, and is such that:
$$\mu_\adel\< \bsigma_{\|(ab)^ka} \> = 2 - \sum_{i=1}^n \frac{\nu_{\bsigma_{\|(ab)^ka}}(x_i)}{2^i}.$$
By going to the vertex $b$ instead of $?x_1$, Adélaïde has therefore a deviation that is profitable by $\sum_{i=1}^n \frac{\nu_{\bsigma_{\|(ab)^ka}}(x_i)}{2^i}$.
Moreover, since the valuation $\nu_{\bsigma_{\|(ab)^ka}}$ satisfies $\phi$, it is lexicographically greater than or equal to $\nu_{\min}$, hence the inequality:
$$\sum_{i=1}^n \frac{\nu_{\bsigma_{\|(ab)^ka}}(x_i)}{2^i} \geq \sum_{i=1}^n \frac{\nu_{\min}(x_i)}{2^i} = \delta,$$
and therefore $\epsilon \geq \delta$, as desired.
\end{proof}

\subp{Conclusion}
We can now conclude that there exists an $\epsilon_{\min}$-SPE $\bsigma$ in this game such that $\mu_\Witness\< \bsigma \> \leq 0$ if and only if $\phi$ is satisfiable and is such that $\nu_{\min}(x_n) = 1$.

\begin{itemize}
    \item If there exists such an $\epsilon_{\min}$-SPE, then necessarily the play $\< \bsigma \>$ visits infinitely often the vertex $x_n$ (every other play gives Witness a positive payoff).
    Therefore, the valuation $\nu_{\min} = \nu_{\bsigma}$ is such that $\nu_{\min}(x_n) = 1$.

    \item Conversely, if $\nu_{\min}(x_n) = 1$, then the $\epsilon_{\min}$-SPE $\bsigma_{\nu_{\min}}$ traverses the vertex $?x_1$ and does never traverse the vertex $\neg x_n$, nor reach the vertex $\triangledown$; it grants therefore Witness the payoff $0$. \hspace{1em plus 1fill}\qedhere
\end{itemize}
\end{proof}

\section{Rational synthesis}

Let us close this chapter with some considerations on the natural continuation of rational verification, where the strategy $\sigma_\Leader$ is not given: \emph{rational synthesis}, a concept due to Giuseppe Perelli, Orna Kupferman, and Moshe Vardi~\cite{DBLP:journals/amai/KupfermanPV16}.
Such a problem can be defined as follows, for a given game class $\Class$ and rationality concept $\rho$.

\begin{prob}[$\rho$-rational synthesis problem in the class $\Class$]
	Given a game $\Game_{\|v_0} \in \Class$ and a threshold $t \in \QQ$, does there exist a strategy $\sigma_\Leader$ such that every $\Leader$-fixed $\rho$-equilibrium $\btau$ with $\tau_\Leader = \sigma_\Leader$ satisfies the inequality $\mu_\Leader\< \btau \> > t$?
\end{prob}

Another common presentation of this problem in the literature (used, for example, in~\cite{filiot_et_al:LIPIcs:2020:12534}) is in terms of the computation of the \emph{Stackelberg value}.
Given a game $\Game_{|v_0}$, the Stackelberg value can be defined as the supremum of all thresholds $t$ for which the pair $(\Game_{|v_0}, t)$ is a positive instance of the decision problem described above.

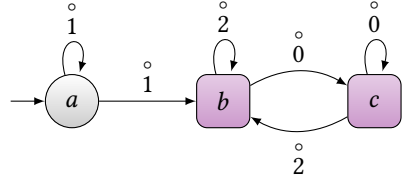
\begin{figure}
			\centering
			\begin{tikzpicture}
				\node[vert, initial left] (a) at (0, 0) {$a$};
				\node[vert, leaderVert] (b) at (2, 0) {$b$};
                \node[vert, leaderVert] (c) at (4, 0) {$c$};
                \path (a) edge[loop above] node {$\stackrel{\circ}{1}$} (a);
				\path (a) edge node[above] {$\stackrel{\circ}{1}$} (b);
                \path (b) edge[loop above] node {$\stackrel{\circ}{2}$} (b);
				\path (b) edge[bend left] node[above] {$\stackrel{\circ}{0}$} (c);
                \path (c) edge[loop above] node {$\stackrel{\circ}{0}$} (c);
				\path (c) edge[bend left] node[below] {$\stackrel{\circ}{2}$} (b);
			\end{tikzpicture}
			\caption{The temptation of chaos in a discounted-sum game}
			\label{fig:chaos2}
		\end{figure}

Variants of this problem can be obtained by restricting Leader's strategies to finite-memory ones (which makes sense if Leader's strategy is an abstraction for a program), or by considering its achaotic version.
Both variants might be relevant given the fact that if $\sigma_\Leader$ can use infinite memory, then the temptation of chaos can now be observed in all classes of quantitative games where payoff functions can have infinite range.
The game depicted by \Cref{fig:chaos2}, where violet vertices belong to Leader, illustrates that fact.
If Leader is allowed to use infinite memory, then she can design her strategy so that player $\Circle$ has an incentive to eventually go to the vertex $b$, but \emph{as late as possible}.
That will be the case for example with the strategy $\sigma_\Leader$ defined by:
$$\< \bsigma_{\|a^k b} \> = \left( b^k c \right)^\omega,$$
for every $k \in \NN$ and independently of $\sigma_\circ$.
Then, every strategy of player $\Circle$ admits a profitable deviation, by postponing the moment where she takes the edge $ab$.
By contrast, let us recall that we showed in the proof of \cref{thm:ach_verif} that in discounted-sum games, there is always a subgame-perfect, and \emph{a fortiori} a Nash response to every finite-memory strategy.

This document does not bring any significant result about rational verification.
Let us simply mention that the case of parity games, Nash rational synthesis has been proved to be $\ExpTime$-easy and $\PSpace$-hard, and $\PSpace$-easy, $\NP$-hard, and $\coNP$-hard when the number of players is fixed \cite{DBLP:conf/icalp/ConduracheFGR16}.
As for subgame-perfect rational synthesis, it has recently been proved to be $\NP$-hard and $\coNP$-hard, $2\ExpTime$-easy in general, and $\ExpTime$-easy when the number of players is fixed \cite{DBLP:journals/corr/abs-2412-08547}.

It is also worth noting here that we have the following general result:

\begin{lem}
    For each $\rho \in \{\subgamep, \nash\}$ and every game class $\Class$ among parity games, mean-payoff games, discounted-sum games and energy games, the $\rho$-universal threshold problem in the class $\Class$ reduces in polynomial time to the $\rho$-rational synthesis problem in $\Class$.
\end{lem}

The reduction consists simply in adding a player Leader who does not control any vertex.
Consequently, using \cref{thm:verif}, Nash and subgame-perfect rational synthesis are at least as hard as the target discounted-sum problem in discounted-sum games, and undecidable in energy games.

Finally, to our knowledge, little is known about rational synthesis in mean-payoff games with arbitrary many players in the environment (the case where the environment is made of one player has been studied in~\cite{filiot_et_al:LIPIcs:2020:12534}).
When Leader is restricted to finite-memory strategies, \cref{thm:NE_MP,thm:spe_mp_np_complete} entail recursive enumerability.
Similarly, one could obtain an analogous result for achaotic rational synthesis by proving that finite-memory strategies are optimal for Leader, which seems to be a reasonable conjecture (while it is clearly not true for classical rational verification, as examplified by the game depicted by \Cref{fig:chaos2}).
But all those problems might still be undecidable.

\begin{oprob}\label{op:rat_syn_mp}
    Are (achaotic) Nash and subgame-perfect rational synthesis decidable in mean-payoff games?
    Are they decidable when Leader is restricted to finite-memory strategies?
\end{oprob}

%% file: 3aSSE.tex
In the previous sections, our examples have primarily dealt with systems interacting with an unpredictable or partially predictable environment.
As we have argued, the most classical game-theoretic framework for modeling such situations is that of two-player zero-sum games.
However, when the assumption of full adversity is too restrictive, multiplayer settings can be more appropriate, incorporating relevant notions of collective rationality, typically captured by \emph{equilibria} concepts such as Nash equilibria or subgame-perfect equilibria. In this context, the environment---potentially consisting of multiple agents---may be better represented as acting rationally according to its own objectives.

Here, we focus on another class of computer-science-related problems that can be modeled using multiplayer games: \emph{protocol design}, where a set of agents interact according to a predefined protocol that must remain robust against deviations.
We argue that in such cases, a particularly relevant notion is that of \emph{strong secure equilibria}.
That notion combines the well-established concepts of \emph{strong equilibria} (resilient to deviations by coalitions) and \emph{secure equilibria} (where no player can deviate in a way that harms another player without also harming themself). However, to the best of our knowledge, it has not been studied as a standalone concept before this work.

\section{Motivating example and definition}\label{sec:motivatingExample}

The aim of designing a protocol is to establish a set of behavior rules that yield a specific outcome if thoroughly followed, and that are self-enforceable: every participating agent must adhere to these rules, lest they do not get anything out of the interactions.

\begin{figure}
    \centering
    \begin{tikzpicture}[auto,node distance=4cm]
        \node[abstractstate] (init) at (0,0) {$\begin{array}{rl}%
            \adel:&\itm_1 \\
            \bart:&\itm_2
        \end{array}$};
        \node[abstractstate] at (6,2) (aliceOwnsAll) {$\begin{array}{rl}%
            \adel:&\itm_1,\itm_2 \\
            \bart:&\emptyset
        \end{array}$};
        \node[abstractstate] at (6,-2) (bobOwnsAll) {$\begin{array}{rl}%
            \adel:& \emptyset\\
            \bart:&\itm_1,\itm_2
        \end{array}$};
        \node[abstractstate] at (12,0) (end) {$\begin{array}{rl}%
            \adel:&\itm_2 \\
            \bart:&\itm_1
        \end{array}$};

        \path[->,aliceMove] (init) edge node[swap,sloped] {$\adel$ sends $\itm_1$} (bobOwnsAll);
        \path[->,bobMove] (init) edge node[sloped] {$\bart$ sends $\itm_2$} (aliceOwnsAll);
        \path[->,aliceMove] (aliceOwnsAll) edge node[sloped] {$\adel$ sends $\itm_1$} (end);
        \path[->,bobMove] (bobOwnsAll) edge node[swap,sloped] {$\bart$ sends $\itm_2$} (end);
    \end{tikzpicture}
    \caption{Exchange between Adélaïde and Barthélémy only.}
    \label{fig:ongoingExampleTwoPlayers}
\end{figure}
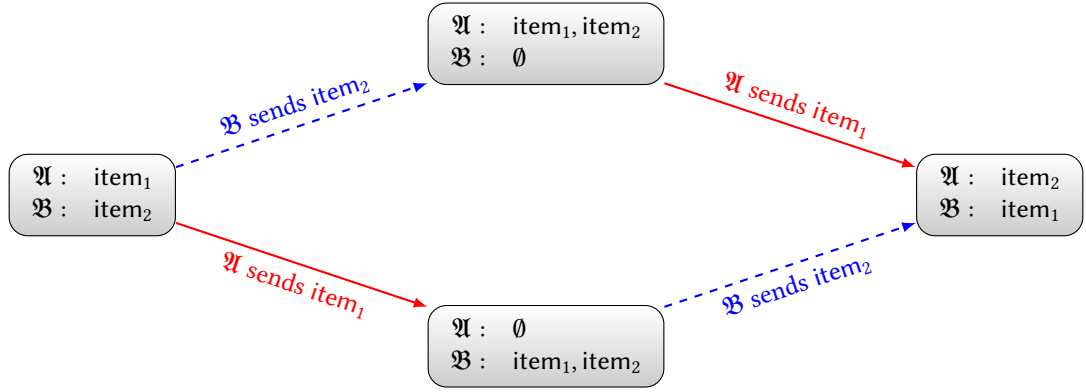

Before moving on the more formal definitions and results of the paper, let us run through the obstacles and possible solutions to the design of a good fair exchange protocol. 
Towards this, let us consider the simple setting of only two agents, Adélaïde and Barthélémy, that wish to exchange items online. (For instance, Adélaïde could want to purchase digital art from Barthélémy.)
A first naive approach to designing such a protocol could be ``let both send their part to the other one'': see \Cref{fig:ongoingExampleTwoPlayers} for a depiction of the possible sequences of actions following this basic guideline.
If both of them do this, even asynchronously, then the exchange is performed. 
Of course, there could be a malicious third party involved here, that could derail the execution of the protocol for any reason---an \emph{external} attacker.
Another source of trouble could be the failure of the communication channels. 
Here, we assume that the channels are \emph{resilient}: they deliver every message perfectly, albeit maybe with an arbitrary but finite delay (every message is \emph{eventually} delivered).

However, another critical aspect to take into account is what assumptions we make about the motivations of the agents. 
Indeed, one can argue that there is no reason for an agent to trust the other one. 
In fact, both may even prefer an outcome where they receive something from the other as intended, but do not send their share.
On the other hand, the agents wish in priority to avoid being \emph{wronged}, i.e. sending their share without receiving anything.
Depending on such assumptions, the solutions for designing a fair protocol may vary: clearly, the naive approach sketched above fails as soon as one agent is not fully trustworthy.

Let us briefly describe two approaches to ensure robustness against these possibly untrustworthy agents.

\subsection{With a trusted third party}\label{sec:motivatingExampleWithTTP}

\begin{figure}
    \centering
    \begin{tikzpicture}[auto,node distance=2.5cm, scale=0.9]
        \node[abstractstate] (init) at (1,0) {$\begin{array}{rl}%
            \adel:&\itm_1 \\
            \bart:&\itm_2
        \end{array}$};
        \node[abstractstate] at (5,3) (aliceOwnsAll) {$\begin{array}{rl}%
            \adel:&\itm_1,\itm_2 \\
            \bart:&\emptyset
        \end{array}$};
        \node[abstractstate] at (5,-3) (bobOwnsAll) {$\begin{array}{rl}%
            \adel:& \emptyset\\
            \bart:&\itm_1,\itm_2
        \end{array}$};
        \node[abstractstate] at (14,0) (end) {$\begin{array}{rl}%
            \adel:&\itm_2 \\
            \bart:&\itm_1
        \end{array}$};
        \node[abstractstate] at (12,3) (aliceOwnsAllTtpAlerted) {$\begin{array}{rl}%
            \adel:&\itm_1,\itm_2 \\
            \bart:&\emptyset \\
            & \ttp \text{ alerted}
        \end{array}$};
        \node[abstractstate] at (12,-3) (bobOwnsAllTtpAlerted) {$\begin{array}{rl}%
            \adel:& \emptyset\\
            \bart:&\itm_1,\itm_2\\
            & \ttp \text{ alerted}
        \end{array}$};

        \path[->,aliceMove] (init) edge node[sloped,pos=0.55] {$\adel$ sends $\itm_1$} (bobOwnsAll);
        \path[->,bobMove] (init) edge node[swap,sloped,pos=0.55] {$\bart$ sends $\itm_2$} (aliceOwnsAll);
        \path[->,aliceMove] (aliceOwnsAll) edge node[swap,sloped] {$\adel$ sends $\itm_1$} (end);
        \path[->,bobMove] (bobOwnsAll) edge node[sloped] {$\bart$ sends $\itm_2$} (end);
        \path[->,aliceMove] (bobOwnsAll) edge node[swap,sloped] {$\adel$ alerts $\ttp$} (bobOwnsAllTtpAlerted);
        \path[->,bobMove] (aliceOwnsAll) edge node[sloped] {$\bart$ alerts $\ttp$} (aliceOwnsAllTtpAlerted);
        \path[->,ttpMove] (aliceOwnsAllTtpAlerted) edge node[text width=2.5cm,pos=0.8] {$\ttp$ forces $\adel$ to send $\itm_1$} (end);
        \path[->,ttpMove] (bobOwnsAllTtpAlerted) edge node[swap,text width=2.5cm,pos=0.8] {$\ttp$ forces $\bart$ to send $\itm_2$} (end);
    \end{tikzpicture}
    \caption{Exchange between Adélaïde and Barthélémy with a trusted third party.}
    \label{fig:ongoingExampleWithTTP}
\end{figure}
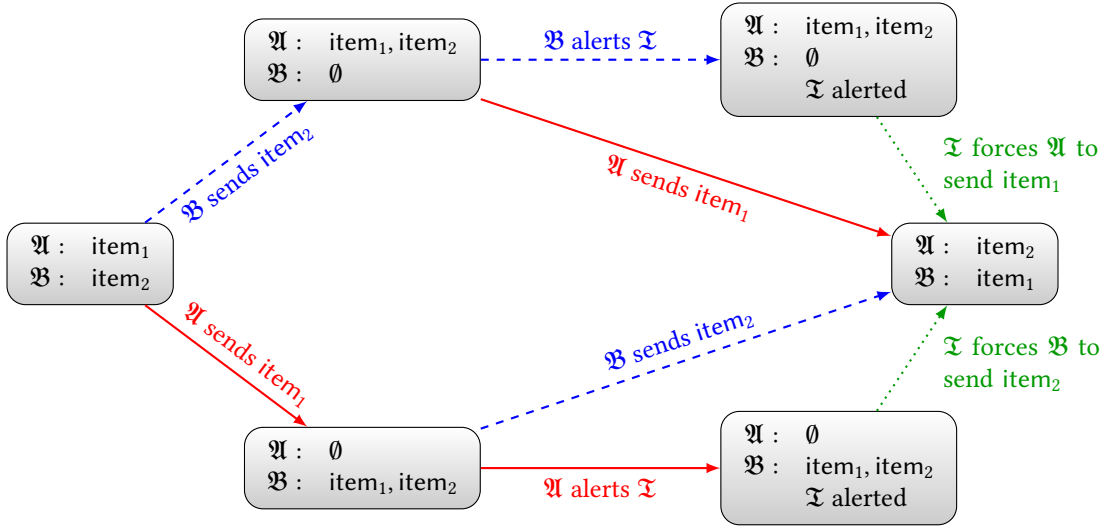

A first possible solution to ensure fairness while making no assumption on the trustworthiness of the participants is to involve a \emph{trusted third party} (TTP) in the protocol.
This particular agent is assumed to be trustworthy, and to have no other interest than the completion of the exchange. 
Furthermore, it is supposed to have sufficient leverage on the agents to enforce the completion of the exchange, if one of them tries to deflect from the intended course of actions.
The TTP is thus considered to be an agent with absolute \emph{authority}.
One could consider performing the exchange only via the TTP, sending both agents' parts to the TTP that would, upon full reception, redistribute them appropriately. This, however foolproof, is extremely costly, as the TTP is used in every instance of the protocol.
Usually, this is avoided by amending the naive protocol to add the rule ``if an agent does not receive anything while they have sent their share, they contact and alert the TTP, who then makes sure the other agent complies''. 
This is what is called an \emph{optimistic} version of a protocol involving a TTP. It is assumed that the participating agents desire the intended outcome of the protocol, and thus that in most cases, the TTP will not need to intervene at all.
A depiction of a such a TTP moderated protocol exchange can be found in \Cref{fig:ongoingExampleWithTTP}.
This is also the approach of the Zhou-Gollmann protocol~\cite{ZhouGollmann97} that uses a TTP only in case of a (suspected or real) cheating attempt (and is encompassed by our framework).

\subsection{Beyond trust: without a trusted third party}\label{sec:motivatingExampleInfinite}

The TTP approach is well-known and widely accepted as a necessity (as well-justified by the impossibility result of~\cite{pagnia1999impossibility}).
What can be done when this approach is not applicable?
Indeed, there are many cases where the trustworthiness of a third party is not clear enough to reasonably rely upon it.
For those cases, can we design protocols which rules are self-enforceable and that eliminate the need of a TTP?
In other terms, what if the incentives to complete the protocol properly were stronger than the ones to deviate from the intended behaviors?
An alternative, and the approach we choose in this work, is to rely, not on the trustworthiness of agents, but instead on their \emph{rationality}.
While not amenable to every possible context where fair exchange is needed, it can provide a viable solution in relevant situations.
We will therefore consider that a protocol is safe if no alliance of agents can wrong another agent \emph{without wronging a member of the alliance itself}. 
The security of the protocol stems from the assumption that all agents, even though they might be malicious, would nonetheless behave rationally.

For instance, in peer-to-peer settings, no one can be absolutely trusted in theory, but all agents have an incentive to be honest, since keeping a good reputation enable them to perform further exchanges.
As another example, a ride-share application can be used as a trusted third party between a driver and a traveler without being trusted \emph{per se}, but because both the driver and the traveler know that if the application had dishonest behaviors in some situations, then no one would trust it anymore, and the company that holds it would stop their profits.

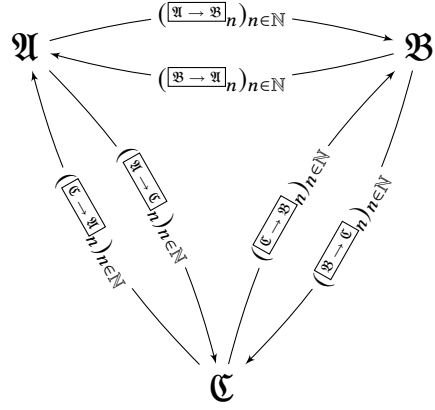
\begin{figure}
            \centering
            \begin{tikzpicture}
                \tikzstyle{toexchange}=[->,>=latex',draw,bend left=15]
                \tikzstyle{exlab}=[sloped,fill=white,inner sep=2pt]
                \node[font=\Large] (a) at (150:3cm) {$\adel$};
                \node[font=\Large] (b) at (30:3cm) {$\bart$};
                \node[font=\Large] (c) at (-90:3cm) {$\capu$};

                \path (a) edge[toexchange] node[exlab] {$(\iab_n)_{n\in\NN}$} (b);
                \path (b) edge[toexchange] node[exlab] {$(\iba_n)_{n\in\NN}$} (a);
                \path (a) edge[toexchange] node[exlab] {$(\iac_n)_{n\in\NN}$} (c);
                \path (c) edge[toexchange] node[exlab] {$(\ica_n)_{n\in\NN}$} (a);
                \path (b) edge[toexchange] node[exlab] {$(\ibc_n)_{n\in\NN}$} (c);
                \path (c) edge[toexchange] node[exlab] {$(\icb_n)_{n\in\NN}$} (b);
            \end{tikzpicture}
            \caption{Expected exchanged items between Adélaïde, Barthélémy, and Capucine.}
            \label{fig:noTTPcircularExchange}
        \end{figure}

In both these examples, important common features are their \emph{multi-agent} nature and the \emph{repeatability} of the exchange, albeit with different agents. 
To illustrate further and start going towards a more abstract model, consider the following setting, depicted by Figure~\ref{fig:noTTPcircularExchange}.
Adélaïde, Barthélémy and Capucine are three agents who wish to exchange an infinity of items: each agent has infinitely many items to send to both other agents.
Again, as shown in~\cite{pagnia1999impossibility}, there is no protocol that could satisfy the usual definitions of fairness. 
Indeed, there is an agent (say, for instance, Adélaïde) that can \emph{scam} another agent (say Barthélémy), i.e. that can send only finitely many items to Barthélémy and receive at least one more item from him.
That is the case if the agents exchange their items directly, but also if they use the third agent (Capucine) as an intermediary, because Capucine is not a TTP and can therefore choose to help Adélaïde, forming a (malicious) \emph{coalition}, to scam Barthélémy (or the reverse).

However, since the process is repeating for further exchanges, if Capucine does so, Barthélémy may react by stopping exchanges not only with Adélaïde, but also with Capucine.
Adélaïde would then be satisfied (she actually scammed Barthélémy), but Capucine would be worse off (she did not scam anyone herself, and can no longer exchange with Barthélémy).
So if Capucine behaves rationally, she has no incentive to help another agent scam the third.
In practice, therefore, such a protocol can be considered as reasonably \emph{safe}: Adélaïde, Barthélémy, and Capucine all take, alternatively, the role of the TTP and enable an exchange of items between the two other agents; and if an agent deviates from the protocol, the other ones stop all exchanges with them.

    \subsection{Strong secure equilibria}

At this point, we hope the reader can be convinced that a relevant equilibrium notion for capturing protocols that ensure a reasonable level of safety is the following.

\begin{defi}[Strong secure equilibrium]
    Let $\Game_{\|v_0}$ be a game.
    Let $\bsigma$ be a strategy profile in $\Game_{\|v_0}$: a \emph{harmful deviation} of the \emph{coalition} $C \subseteq \Pi$ from the strategy profile $\bsigma$ is a strategy profile $\bsigma'_C \in \Strat_C\Game_{\|v_0}$ such that for every player $i \in C$, we have $\mu_i\< \bsigma_{-C}, \bsigma'_C \> \geq \mu_i\< \bsigma \>$, and for some player $i \not\in C$, we have $\mu_i\< \bsigma_{-C}, \bsigma'_C \> < \mu_i\< \bsigma \>$.
    A \emph{strong secure equilibrium} in $\Game_{\|v_0}$, or \emph{SSE} for short, is a strategy profile from which there is no harmful deviation.
\end{defi}

Before moving to computational aspects, let us make two remarks to give an insight on how expressive a model based on SSEs is.
Indeed, one may believe that SSEs provide a very specific type of potential attackers---those that may participate in an attack as long as they are not wronged in the process.
But such a model can actually capture more classical types of agents, such as fully hostile ones, or on the contrary trusted ones.

\begin{itemize}
    \item A fully untrusted agent, on which no rationality assumptions can be made, can be modeled as an agent that is impervious to harm---i.e., one that always wins.

    \item Conversely, in a protocol that includes an agent who is fully trusted and will never participate in an attack, such an agent can be modeled as a player who is wronged as soon as any other player is wronged.
    As a result, if the protocol contains an attack, captured by the harmful deviation of a coalition, the trusted agent cannot be part of that coalition.
\end{itemize}

We provide a more detailed discussion in~\cite{DBLP:journals/corr/abs-2405-18958} regarding the protocol and attack models that can be captured by this notion, as well as its relevance to classical protocols such as the Zhou-Gollmann optimistic protocol~\cite{ZhouGollmann97}.
Since these security aspects diverge significantly from the game-theoretic focus of this document, we choose to concentrate here on algorithmic aspects.

\subsection{Problem}

Modeling secure protocols with strong secure equilibria is particularly relevant in Boolean games, where each player wins if and only if they are not wronged by other players.
In this setting, determining whether a given strategy profile (where every player wins) is an SSE reduces to checking whether there exists a coalition that can deviate and wrong some player without harming any of its own members.
Since such wronging conditions are typically represented by $\omega$-regular conditions (e.g., "at some point, Adélaïde sends the item $\itm_1$, but Barthélémy never sends the item $\itm_2$"), we restrict our study to parity games, as defined in \cref{chap:nash}, and, more generally, to \emph{$\omega$-regular games}, a broader class defined as follows.

\begin{defi}[Parity automaton]
    A {\em parity automaton} over alphabet $X$ is a tuple $\auto=(X,Q,q_0,\delta,\kappa)$ where $Q$ is a set of states, where $\delta:Q\times X \to Q$ is a deterministic transition function, and where $\kappa: Q \to \NN$ is a color mapping.
    Given a word $w=w_0w_1\cdots\in X^\omega$, the \emph{run} of $\auto$ on $w$ is $\rho_w = q_0 q_1 \cdots$ where for $i\in \NN$, $q_{i+1}=\delta(q_i,w_i)$.
    The set of infinite words accepted by $\auto$ is $\Lang(\auto)= \left\{w\in X^\omega \mid \min\kappa(\Inf(\rho_w)) \text{ is even}\right\}$.
\end{defi}

\begin{defi}[$\omega$-regular game]
    An \emph{$\omega$-regular game} is a multi-player game $\Game_{\|v_0}$ such that there exists, for each player $i$, a parity automaton $\auto_{i}$ over alphabet $V$, such that  for each play $\pi$, we have $\mu_i(\pi) = 1$ if and only if $\pi \in \Lang(\auto_{i})$.
\end{defi}

Moreover, since the primary objective in this framework is to design a protocol in which no player is wronged, we do not focus on the constrained existence problem of SSEs.
Instead, we study a variant of this problem, the \emph{fixed-payoff SSE problem}, where we seek a strategy profile that not only produces a given payoff vector but also satisfies a specified \emph{correctness condition}. This condition is of the same type as the players' objectives: a color function on the vertices of the game in the case of a parity game, or a parity automaton in the case of an $\omega$-regular game.
In both cases, we denote by $\pi \vDash \phi$ the fact that the play $\pi$ satisfies the condition $\phi$.
It is worth noting that, when considering Nash equilibrium or subgame-perfect equilibrium, such a correctness condition can be encoded as the objective of a fictional player. However, this approach is not applicable here, as it may introduce harmful deviations against that player.

\begin{prob}[Fixed-payoff SSE problem]
    Given a game $\Game_{\|v_0}$, a payoff vector $\bx \in \QQ^\Pi$, and a condition $\phi$, does there exist an SSE $\bsigma$ in $\Game_{\|v_0}$ such that $\mu\< \bsigma \> = \bx$ and $\< \bsigma \> \vDash \phi$?
\end{prob}

Algorithms for this problem can also be used to solve the constrained existence problem, at least in Boolean games (or more generally games with payoff functions ranging in a finite set), by checking each possible payoff vector $\bx$ between the two threshold vectors (with $\phi = \top$).
Of course, in the general case, the number of such calls grows exponentially with the number of players. However, since the lowest complexity we will obtain in this chapter is $\PSpace$, this additional cost will not be prohibitive, and our complexity results for the fixed-payoff SSE problem will immediately entail results for the constrained existence problem of SSEs.

The remainder of this chapter is therefore dedicated to establishing tight complexity bounds for the fixed-payoff SSE problem in parity games and $\omega$-regular games.
In particular, we will pay close attention to the complexities that arise when the number of players and the number of colors are small, as this is often the case in practical applications.

        \section{A tool: the deviator game}

The main tool we use to solve our problem takes the form of a new game structure, in which one player, \emph{Prover}, tries to prove that an SSE generating the desired payoff exists, while another one, \emph{Challenger}, tries to prove that the strategy profile she constructs is actually not an SSE.
That game is the \emph{deviator game}, very similar to a construction with the same name proposed in~\cite{DBLP:conf/fossacs/Brenguier16}, in a slightly different context.

\begin{defi}[Deviator game] \label{def:deviator_game}
    Let $\Game_{\|v_0}$ be a game, let $\bx$ be a payoff vector, and let $\phi$ a correctness condition.
    The \emph{deviator game} is the initialized game:
    \[\dev_{\bx\phi} (\Game)_{\|v_0^\cd} = \left( \{\Prover, \Challenger\}, V^\cd, E^\cd, \left(V^\cd_\Prover, V^\cd_\Challenger\right), \mu^\cd\right)_{\|v_0^\cd},\]
    where:

     \begin{itemize}
         \item the player $\Prover$ is called \emph{Prover}, and the player $\Challenger$ \emph{Challenger}.

         \item Prover controls the set $V^\cd_\Prover = V \times 2^\Pi$, and Challenger the set $V^\cd_\Challenger = E \times 2^\Pi$.

         \item The initial vertex is $v_0^\cd = (v_0, \emptyset)$.

         \item The edge set $E^\cd$ is defined as follows: from the vertex $(u, D)$, Prover can go to every vertex of the form $(uv, D) \in E$ (she \emph{proposes} the edge $uv$).
         From the vertex $(uv, D)$, Challenger can go to the vertex $(v, D)$ (he \emph{accepts} the edge $uv$), or to every vertex $(w, D \cup \{i\})$, with $w \neq v$ and $uw \in E$, and where $i$ is the player controlling $u$ (he \emph{deviates} from Prover's proposal, and player $i$ is added to the set of deviators).

         \item Given a play $\chi$ is this game, we write $\dchi$ for the play in $\Game$ constructed by the actions of Prover and Challenger: if $\chi = (u_0, D_0)(u_0v_0, D_0)(u_1, D_1) (u_1v_1, D_1) \dots$, then we define $\dchi = u_0 u_1 \dots$.
         We also define $\D(\chi) = \bigcup_k D_k$, the set of players who deviated along the play $\chi$.

         \item Then, the Boolean payoff function $\mu^\cd$ is defined as follows. A play $\chi$ is won by Challenger if and only if either:
         \begin{itemize}
             \item we have $\D(\chi) = \emptyset$ and $\mu(\dchi) \neq \bx$;
             \item we have $\D(\chi) = \emptyset$ and $\dchi \not\vDash \phi$;
             \item or for every player $i \in \D(\chi)$, we have $\mu_i(\dchi) \geq x_i$, and there exists a player $j \in \Pi$ such that $\mu_j(\dchi) < x_j$.
         \end{itemize}
     \end{itemize}
\end{defi}

\begin{thm} \label{thm_condev}
    Prover has a winning strategy in the game $\dev_{\bx\phi} (\Game)_{\|v_0^\cd}$ if and only if there exists an SSE $\bsigma$ in $\Game_{\|v_0}$ with $\mu\< \bsigma \> = \bx$ and $\< \bsigma \>\vDash \phi$.
\end{thm}

A proof of this theorem can easily be obtained by adapting the one presented in~\cite{DBLP:conf/fossacs/Brenguier16}.

        \section{Parity games}

Let us now consider the case of parity games: then, similarly to what was done in~\cite{DBLP:conf/fossacs/Brenguier16}, a cautious way to solve the deviator game leads to an algorithm that uses only polynomial space.
For convenience, we now write $\Parity(\kappa)$ for the set of plays satisfying the parity condition associated to each color function $\kappa$, and $\overline{\Parity}(\kappa)$ for its complement.
Similarly, we write $\B(v)$ for the set of plays that visit infinitely often the vertex $v$, and $\overline{\B}(v)$ for its complement.
The condition $\phi$ is defined by the color mapping $\kappa_\phi$.

\begin{thm}\label{thm:sseExistencePspaceEasy}
    In parity games, the fixed-payoff SSE problem is $\PSpace$-complete.
    Hardness still holds in co-Büchi games and with correctness condition $\phi = \top$.
    With two players, the problem is at least as hard as solving a two-player zero-sum parity game.
    If both the number of colors and the number of players are fixed, then the problem is fixed-parameter tractable.
\end{thm}

\begin{proof} Given a parity game $\Game_{\|v_0}$, we write $n$ for the number of players and $k$ for the highest color that appears in the automata $\auto_i$ and $\auto_\phi$.

\paragraph{$\PSpace$-easiness}

    By Theorem~\ref{thm_condev}, deciding the fixed-payoff SSE existence problem amounts to deciding which player, among Prover and Challenger, has a winning strategy in the corresponding deviator game.
    That game has an exponential size, but has a specific shape that makes it possible to solve it region by region, without using exponential space.
    Indeed, let us observe that along a play $\chi = (u_0, D_0)(u_0v_0, D_0)(u_1, D_1)(u_1v_1, D_1) \dots$, the sequence of sets $(D_\l)_{\l \in \NN}$ is non-decreasing.
    
    Let us therefore recursively define an algorithm that decides, given the game $\Game_{\|v_0}$, the vector $\bx$ and a vertex $(u_0, D)$ of the game $\dev_{\bx \phi} \Game$, whether Prover has a winning strategy from that vertex.
    Let $W = \{i\in\Pi ~|~ x_i=1\}$ be the set of players that win according to the payoff vector $\bx$.

    \subp{Definition of the game $\Gamebis^D_{\|(u_0, D)}$}
    First, construct the game $\Gamebis^D_{\|(u_0, D)}$ as follows: construct the region of the game $\dev_{\bx \phi} \Game$ that is made of vertices of the form $(u, D)$ or $(uv, D)$.
    Define in that region the color mappings so that player $i$'s payoff, in that region, is always equal to their payoff in the corresponding play in $\Game$: i.e., for all $i \in \Pi$, define $\kappa'_{i}(uv, D) = \kappa'_{i}(u, D) = \kappa_{i}(u)$ (and similarly for the correctness condition, define $\kappa'_{\phi}(uv, D) = \kappa'_{\phi}(u, D) = \kappa_{i}(u)$).
    
    For each edge $(uv, D)(w, D \cup \{i\})$ that leaves that region, add to the constructed game the edge $(uv, D)(w, D \cup \{i\})$ and the loop $(w, D \cup \{i\})(w, D \cup \{i\})$, and recursively decide whether Prover has a winning strategy from the vertex $(w, D \cup \{i\})$.
    If she does, call the vertex $(w, D \cup \{i\})$ a \emph{Prover's leaf}, and define the color mappings so that only players in $W$ would win on that loop: for each $i' \in W$, define $\kappa'_{i'}(w, D \cup \{i\}) = 0$; for $i'\notin W$, define $\kappa'_{i'}(w, D \cup \{i\}) = 1$; and finally, define $\kappa'_{\phi}(w, D \cup \{i\}) = 0$.
    If she does not, call the vertex $(w, D \cup \{i\})$ a \emph{Challenger's leaf}, and define the color mappings so that only players in $D \cup \{i\}$ would win on that loop: for each $i' \in D \cup \{i\}$, define $\kappa'_{i'}(w, D \cup \{i\}) = 0$; for $i \notin D \cup \{i\}$, define $\kappa'_{i'}(w, D \cup \{i\}) = 1$; and finally, define $\kappa'_{\phi}(w, D \cup \{i\}) = 1$.
    
    Now, define the payoff functions as follows.
    If $D \neq \emptyset$, then a play $\pi$ is won by Challenger if and only if we have:
    \[\pi \in \left( \bigcup_{i \in W} \overline{\Parity}(\kappa'_{i}) \right) \cap \left( \bigcap_{i \in D\cap W} \Parity(\kappa'_{i}) \right)\]
    (i.e., if the play $\pi$ gives to at least one player $i$ a payoff worse than $x_i$, and to every player who deviated at least the same payoff as in $\bx$).
    If $D = \emptyset$, then a play $\pi$ is won by Challenger if and only if we have:
    \[\pi \in \bigcup_{i \in W} \overline{\Parity}(\kappa'_{i}) \cup \bigcup_{i \notin W} \Parity(\kappa'_{i})\cup \overline{\Parity}(\kappa_\phi')\]
    (i.e., if the payoff vector generated by Prover, without any deviation, is not equal to $\bx$, or if the correctness constraint is not met).
    Note that in both cases, Challenger wins if he reaches a Challenger's leaf, and loses if he reaches a Prover's leaf.

    \subp{Equivalence with the game $\dev_{\bx \phi} \Game$}
    Prover has a winning strategy in the game $\Gamebis^D_{\|(u_0, D)}$ if and only if she has a winning strategy from the vertex $(u_0, D)$ in the game $\dev_{\bx \phi} \Game$.
    Indeed, if Prover has a winning strategy in $\Gamebis^D_{\|(u_0, D)}$, then she can follow that strategy in $\dev_{\bx \phi} \Game$.
    Then, she either stays in the region where the deviating coalition is $D$, or she reaches a vertex of the form $(w, D \cup \{i\})$ from which she has a winning strategy, that she can then follow.
    Conversely, if she has a winning strategy from $(u_0, D)$ in $\dev_{\bx \phi} \Game$, she can follow it in $\Gamebis^D_{\|(u_0, D)}$: she will then either stay in the region where the deviating coalition is $D$, or reach a Prover's leaf (since the strategy she is following afterwards in $\dev_{\bx \phi} \Game$ is still winning), which makes her win.

\subp{Resolution}
    This game can be seen as an Emerson-Lei game~\cite{EmersonLei87}: Prover's and Challenger's winning conditions are Boolean combinations of Büchi conditions.
    Indeed, we have the equality:
    \[\Parity(\kappa'_{i}) = \bigcup_{2m < k} \left(\bigcup_{\kappa'_{i}(v^\cd) = 2m} \B(v^\cd) \cap \neg \bigcup_{\l < 2m} \bigcup_{\kappa'_{i}(w^\cd) = \l} \B(w^\cd)\right).\]
    It was shown in~\cite{HunterDawar05} that such games can be solved using a space polynomial in the size of the game and of the objectives.
    Here, both are themselves polynomial in the size of $\Game$.
    Thus, the last step of our recursive algorithm consists in solving $\Gamebis^D_{\|(u_0, D)}$.

    Then, applying this recursive algorithm from the vertex $(v_0, \emptyset)$ decides our problem: there is an SSE $\bsigma$ in $\Game_{\|v_0}$ such that $\mu\< \bsigma \> = \bx$ and $\< \bsigma \> \vDash \phi$ if and only if Prover has a winning strategy from that vertex.
    Moreover, that algorithm uses polynomial space: each recursive call does, and the recursion stack has size at most $\card\Pi$, i.e. polynomial.
    Therefore, the fixed-payoff SSE existence problem in parity games is $\PSpace$-easy.

        \paragraph{Fixed-parameter tractability}

        In~\cite{BruyereHautemRaskin18}, it has been shown that an Emerson-Lei game $\Game$ with vertex space $S$ and objective $\psi$ can be solved in time:
        \[O\left( 2^{2^{|\psi|}} |\psi| + \left(2^{|\psi| 2^{|\psi|}} \card V^\cd \right)^5\right).\]
        In our case, the size of the objective $\psi$ depends only on $k$ and $n$, while $\card V^\d = 2^n \card V$: our problem is therefore fixed-parameter tractable with $k$ and $n$ as parameters.

    \paragraph{$\PSpace$-hardness (in co-Büchi games)}

        We reduce the quantified satisfiability problem ($\QSAT$) to the SSE fixed-payoff existence problem (with $\phi = \top$) in co-Büchi games (or more precisely to its complement).
        This will prove that our problem cannot be fixed-parameter tractable with $k$ as a single parameter, unless we have $\P = \PSpace$.

        \subp{Reduction}
    Let: \[\varphi=\exists x_1 \forall x_2  \cdots \exists x_{n-1} \forall x_n \bigwedge_{i=1}^p \bigvee_{j=1}^{m_i} \l_{ij}\] be an instance of $\QSAT$ in conjunctive normal form where for all $i,j$, we define $\l_{ij}$ as a literal of the form $x_k$ or $\neg x_k$ for some $k\in\{1,\dots,n\}$.
    We write $\psi'(x_1,\dots,x_n)=\bigwedge_{i=1}^p \bigvee_{j=1}^{m_i} \l_{ij}$ for the non-quantified part of the formula.

    We construct a game with $2n+2$ players: one for each literal player and two additional players \emph{Solver} (written $\Solver$) and \emph{Opponent} (written $\opponent$) as depicted in \figurename~\ref{fig:qsat2SSEGame}.
    Round vertices belong to Solver, hexagonal vertices belong to Opponent, rectangular vertices belong to the literal indicated in the bottom right corner (if any: some vertices only have a single outgoing edge and therefore could belong to anyone without loss of generality).
    Red clouds indicate the list of players for which a given vertex is in the co-B\"uchi set: infinitely many visits to these vertices means losing (payoff $0$) while only finitely many visits means winning (payoff $1$).
    
    The game is composed of three parts: in the first part, the \emph{setting} module, a valuation can be chosen by visiting the vertex $x_i^\s$ (setting $x_i$ to true) or the vertex $\neg x_i^\s$ (setting $x_i$ to false), the choice belonging to Solver if $x_i$ is existentially quantified and to Opponent if it is universally quantified.
    When the value of a variable is chosen, the corresponding literal player can choose to go to a sink vertex $\triangledown$ (duplicated in the figure for clarity) or continue the game.
    
    If all choose to continue, then the game moves into the second part, the \emph{checking} module, where the formula is ``checked'': each disjunctive clause $C_i$ has a vertex owned by Solver who must choose a literal $\l_{ij}$ from the clause.
    Vertex $\l_{ij}$ is in the co-B\"uchi set of player $\bar{\l}_{ij}$, who loses if said vertex was visited infinitely often.
    When this ``checking'' is done, the last part, called \emph{punishing} module, visits the co-B\"uchi set of one literal per variable, the choice of which being left to Solver.

    The game continues back to the checking module.
    As the last vertex of these modules is in the co-B\"uchi set of Opponent, Opponent will lose if the game enters these modules.
    Remark that there are no co-B\"uchi vertices for Solver who therefore always wins.

    We will now show that the formula $\varphi$ holds if and only if there is no SSE with payoff vector $(1)_{i \in \Pi}$.
    The intuition is that a coalition made of Solver and all true literals can deviate to make Opponent and all false literals lose while not penalizing the players in the coalition.

    \subp{If the formula $\psi$ does not hold, then there exists an SSE generating the payoff vector $(1)_{i \in \Pi}$} Let us assume $\varphi$ does not hold.
    Let $\bsigma$ be the following profile of strategies: each literal goes to $\triangledown$ whenever possible in the setting module, Solver plays any strategy, and Opponent plays the optimal strategy in response to Solver's choice of valuation so that $\psi'(x_1,\dots,x_n)$ is false.
    The outcome of this profile is either path $?x_1\cdot x_1^\s \cdot\triangledown^\omega$ or $?x_1\cdot\neg x_1^\s\cdot\triangledown^\omega$, both with payoff vector $(1)_{i \in \Pi}$.
    
    Let us prove that it is an SSE.
    Let $C$ be a coalition of players and assume that there exists a harmful deviation for this coalition.
    As Solver always wins, she can be assumed to be in the coalition.
    And since no one loses in the setting module or $\btd$, any harmful deviation must reach the checking module.
    Because that means that Opponent would then lose, Opponent cannot be in the coalition.
    For each variable $x_i$, players $x_i$ and $\neg x_i$ cannot be both in the coalition as (at least) one of them loses through the infinite visits in the punishing module.
    To actually reach the checking module, all literal players whose vertex was visited in the setting module must have deviated from $\bsigma$, and therefore are in the coalition.
    As a result the coalition is made of Solver and exactly one literal per variable, defining a valuation.
    Because Opponent played optimally, that valuation does not satisfy $\psi'(x_1,\dots,x_n)$, so there is a clause $C_i$ where all literals $\l_{i1}, \dots, \l_{im_i}$ are false, meaning that the player corresponding to their negations are all in the coalition.
    The choice by Solver of any of these literals therefore enforces a visit to the co-B\"uchi set of (at least) one player in the coalition.
    As this choice is infinitely repeated, that makes this player lose and therefore it should not be part of the coalition, which is a contradiction.
    So $\bsigma$ is an SSE.
    
    \subp{If the formula $\psi$ holds, then there is no such SSE} Let us assume that $\varphi$ holds.
    Let $\bsigma$ be a strategy profile with payoff $(1)_{i \in \Pi}$.
    As noted above, that payoff requires the outcome of $\bsigma$ to end in the vertex $\triangledown$.
    Consider the following strategy for Solver: in the setting module, choose the literal that ensures satisfaction of the formula $\psi'(x_1,\dots,x_n)$, based on the already known choices of Opponent.
    This is possible since the formula $\varphi$ holds.
    Let $C$ be the coalition made of Solver and all literal players corresponding to vertices $x_i^\s$ or $\neg x_i^\s$ visited in the setting module.
    These players deviate by reaching the vertex $?x_{i+1}$ (or $C_1$ if $i=n$) instead of going to $\triangledown$.
    Since the checking module is reached, Opponent (at least) will lose and be harmed by the deviation.
    The valuation thus built ensures that $\psi'(x_1,\dots,x_n)$ is satisfied, so in the checking module, for every clause there is a true literal.
    The strategy of Solver consists in consistently choosing the vertex for these literals.
    Note that this means visiting a co-B\"uchi set for a player that is not in the coalition.
    In the punishing module, Solver visits vertices corresponding to players in the coalition (i.e. the exact same one that were visited in the setting module: if $x_i^\s$ was visited, consistently choose $x_i^\p$, if $\neg x_i^\s$ was visited, consistently choose $\neg x_i^\p$.
    This ensures that the co-B\"uchi set of no player in the coalition $C$ is ever visited, hence all players in this coalition still win, hence the deviation is harmful and $\bsigma$ is not an SSE.

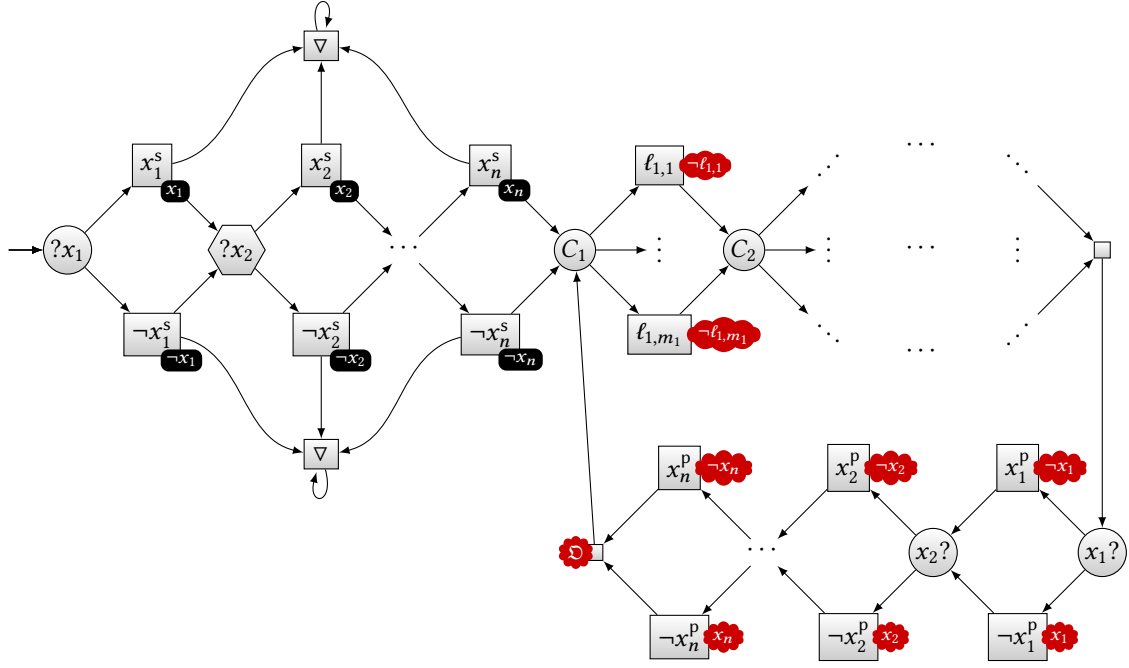
\begin{figure}
    \centering
    \begin{tikzpicture}[node distance=45pt]
        \node[solverState,initial] at (0,0) (q1) {$?x_1$};
        \node[literalState=$x_1$,above right of=q1] (p1) {$x_1^\s$};
        \node[literalState=$\neg x_1$,below right of=q1] (n1) {$\neg x_1^\s$};
        \node[opponentState,below right of=p1] (q2) {$?x_2$};
        \node[literalState=$x_2$,above right of=q2] (p2) {$x_2^\s$};
        \node[literalState=$\neg x_2$,below right of=q2] (n2) {$\neg x_2^\s$};
        \node[below right of=p2] (q3) {$\cdots$};
        \node[literalState=$x_n$,above right of=q3] (pn) {$x_n^\s$};
        \node[literalState=$\neg x_n$,below right of=q3] (nn) {$\neg x_n^\s$};

        \node[solverState,below right of=pn] (c1) {$C_1$};
        \node[nobodyState,above right of=c1] (l11) {$\l_{1,1}$}; \cobuchiLabel[inner sep=0pt,cloud puff arc=95]{l11}{$\neg \l_{1,1}$}
        \node[nobodyState,below right of=c1] (l1m1) {$\l_{1,m_1}$};  \cobuchiLabel[inner sep=0pt,cloud puff arc=95]{l1m1}{$\neg \l_{1,m_1}$}
        \node (c1dots) at (barycentric cs:l11=1,l1m1=1) {$\myvdots$};
        \node[solverState,below right of=l11] (c2) {$C_2$};
        \node[above right of=c2] (l21) {$\myiddots$};
        \node[below right of=c2] (l2m2) {$\myddots$};
        \node (c2dots) at (barycentric cs:l21=1,l2m2=1) {$\myvdots$};
        \node[right of=l21,node distance=2.5cm] (lp1) {$\myddots$};
        \node[right of=l2m2,node distance=2.5cm] (lpmp) {$\myiddots$};
        \node (cpdots) at (barycentric cs:lp1=1,lpmp=1) {$\myvdots$};
        \node[yshift=7pt] (ldots1) at (barycentric cs:l21=1,lp1=1) {$\cdots$};
        \node[yshift=-7pt] (ldotsm) at (barycentric cs:l2m2=1,lpmp=1) {$\cdots$};
        \node (ldots) at (barycentric cs:ldots1=1,ldotsm=1) {$\cdots$};
        \node[nobodyState,below right of=lp1] (ph) {};

        \node[solverState,below of=ph,node distance=4cm] (v1) {$x_1?$};
        \node[nobodyState,above left of=v1] (vp1) {$x_1^\p$}; \cobuchiLabel{vp1}{$\neg x_1$}
        \node[nobodyState,below left of=v1] (vn1) {$\neg x_1^\p$}; \cobuchiLabel{vn1}{$x_1$}
        \node[solverState,below left of=vp1] (v2) {$x_2?$};
        \node[nobodyState,above left of=v2] (vp2) {$x_2^\p$}; \cobuchiLabel{vp2}{$\neg x_2$}
        \node[nobodyState,below left of=v2] (vn2) {$\neg x_2^\p$}; \cobuchiLabel{vn2}{$x_2$}
        \node[below left of=vp2] (vn) {$\cdots$};
        \node[nobodyState,above left of=vn] (vpn) {$x_n^\p$};  \cobuchiLabel{vpn}{$\neg x_n$}
        \node[nobodyState,below left of=vn] (vnn) {$\neg x_n^\p$}; \cobuchiLabel{vnn}{$x_n$}
        \node[nobodyState,below left of=vpn] (ph2) {}; \cobuchiLabelLeft{ph2}{$\opponent$}

        \node[nobodyState,above of=p2] (sinkTop) {$\triangledown$}; 
        \node[nobodyState,below of=n2] (sinkBot) {$\triangledown$}; 

        \begin{pgfonlayer}{arrows}
            \path[->] (q1) edge (p1) edge (n1);
            \path[->] (p1) edge (q2) edge[bend left,out=-30] (sinkTop);
            \path[->] (n1) edge (q2) edge[bend right,out=30] (sinkBot);
            \path[->] (q2) edge (p2) edge (n2);
            \path[->] (p2) edge (q3) edge (sinkTop);
            \path[->] (n2) edge (q3) edge (sinkBot);
            \path[->] (q3) edge (pn) edge (nn);
            \path[->] (pn) edge (c1) edge[bend right,out=30] (sinkTop);
            \path[->] (nn) edge (c1) edge[bend left,out=-30] (sinkBot);
    
            \path[->] (c1) edge (l11) edge (c1dots) edge (l1m1);
            \path (l1m1) edge (c2);
            \path (l11) edge (c2);
            \path[->] (c2) edge (l21) edge (c2dots) edge (l2m2);
            \path (lpmp) edge (ph);
            \path (lp1) edge (ph);

            \path[->] (ph) edge (v1);

            \path[->] (v1) edge (vp1);
            \path (v1) edge (vn1);
            \path (vn1) edge (v2);
            \path (vp1) edge (v2);
            \path (v2) edge (vp2);
            \path (v2) edge (vn2);
            \path (vn2) edge (vn);
            \path (vp2) edge (vn);
            \path[->] (vn) edge (vpn) edge (vnn);
            \path (vnn) edge (ph2);
            \path (vpn) edge (ph2);

            \path[->] (ph2) edge (c1);
    
            \path[->] (sinkTop) edge[loop above] (sinkTop);
            \path[->] (sinkBot) edge[loop below] (sinkBot);
            \end{pgfonlayer}
    \end{tikzpicture}
    \caption{Game to encode $\QSAT$  instance $\exists x_1 \forall x_2  \cdots \exists x_{n-1} \forall x_n \bigwedge_{i=1}^p \bigvee_{j=1}^{m_i} \l_{ij}$ into the existence of an SSE with payoff $(1,\dots,1,1,1)$. Round vertices belong to Solver, hexagonal vertices belong to Opponent, rectangular vertices belong to the literal indicated in the bottom right corner (if any). The red cloud lists the set of players for which this vertex is in the co-B\"uchi condition.}
    \label{fig:qsat2SSEGame}
\end{figure}

    \paragraph{$\mathrm{\textsc{Parity}}$-hardness (with two players)}

    We show here that when $n$ is fixed to $2$ (and when $\phi = \top$), the fixed-payoff SSE problem is at least as hard as solving a two-player zero-sum parity game.
    This will show that our problem cannot be fixed-parameter tractable with only $n$ as a parameter, unless parity games can be solved in polynomial time.
    Let us recall that this problem is known to belong to $\NP \cap \coNP$.

    Let $\Game$ be a parity game $\Game_{\|v_0}$ with two players, \emph{Adélaïde} (written $\adel$) and \emph{Barthélémy} (written $\bart$), and such that $\kappa_\bart = \kappa_\adel + 1$ (i.e., Barthélémy's winning condition is the complement of Adélaïde's one).
    Let us now consider the game $\Game'_{\|v_0}$, with the same arena, and with color mappings $\kappa'_\adel = \kappa_\adel$ and $\kappa_\bart$ constantly equal to $2$.

    Let us assume Adélaïde has a winning strategy $\sigma_\adel$ in the game $\Game_{\|v_0}$: then, for any strategy $\sigma_\bart$, the strategy profile $\bsigma$ is, in the game $\Game'_{\|v_0}$, an SSE where both Adélaïde and Barthélémy get the payoff $1$.
    Indeed, Barthélémy gets the payoff $1$ as he always does, Adélaïde gets the payoff $1$ because $\sigma_\adel$ wins against $\sigma_\bart$, and that strategy profile is an SSE because no player can deviate to harm the other one: Barthélémy wins every play, and Adélaïde is playing a winning strategy.

    Conversely, if the strategy profile $\bsigma$ is an SSE in $\Game'_{\|v_0}$ in which both Adélaïde and Barthélémy get the payoff $1$, then consider some alternative strategy $\sigma'_\bart$ for Barthélémy: like every play, the play $\< \sigma_\adel, \sigma'_\bart \>$ is won by Barthélémy in $\Game'$.
    Consequently, it is also won by Adélaïde, otherwise it would constitute a harmful deviation for Barthélémy.
    It is therefore also won by Adélaïde in $\Game$.
    This proves that $\sigma_\adel$ is a winning strategy in $\Game_{\|v_0}$.

    As a consequence, the problem of solving a two-player zero-sum parity game reduces to the SSE fixed-payoff existence problem in parity games with only two players; and therefore, the SSE fixed-payoff existence problem in parity games cannot be solved in polynomial time if only $n$ is fixed (unless parity games are solvable in $\P$).
\end{proof}

\section{In \texorpdfstring{$\omega$}{ω}-regular games}

When the payoff functions are defined by parity automata, those ones must be incorporated in the arena of the game, entailing an exponential blowup.
However, there is no exponential blowup in the formula that defines the winning conditions of Prover and Challenger, hence the game can be solved by using algorithms that exists in the literature, in exponential time.

\begin{thm}
    In $\omega$-regular games, the fixed-payoff SSE problem is $\ExpTime$-complete.
    Hardness holds even if $\phi = \top$ and if all automata are Büchi or co-Büchi automata.
    When both the numbers of players and colors are fixed, the problem becomes $\P$-easy.
\end{thm}

\begin{proof}
    \paragraph{$\ExpTime$-easiness, and $\P$-easiness with fixed numbers of colors and players.}
        
Let $\Game_{\|v_0}$ be an $\omega$-regular game, and let $\bx \in \QQ^\Pi$.
    Let $W = \{i\in\Pi ~|~ x_i=1\}$ be the set of players winning according to payoff vector $\bx$.
Let $\left(\auto_{i}\right)_{i\in\Pi\cup\{\phi\}}$ be the parity automata defining the payoff of player $i$ and correctness constraint over $\Game_{\|v_0}$.
Let $Q_{i}$ (resp. $\delta_{i}$, $\kappa_{i}$, $q_{0i}$) be the set of states (resp. transition function, coloring mapping, initial state) of $\auto_{i}$.
We assume all these automata use at most $k$ colors, i.e. the co-domain of every color function $\kappa_{i}$ is included in $\{0,\dots,k\}$.
Let: \[m=\max\left\{2,|V|,\max_{i\in\Pi\cup\{\phi\}}\left|Q_{i}\right|\right\}\] be the maximal size of the graph and automata provided as input (assumed to be at least 2 for complexity purposes).

We build the \emph{extended deviator game} as the product of the deviator game with all automata $\auto_{i}$.
Formally, the extended deviator game is the initialized game:
    \[\EDev_{\bx} (\Game)_{\|v_0^\ecd} = \left( \{\Prover, \Challenger\}, V^\ecd, E^\ecd, \left(V^\ecd_\Prover, V^\ecd_\Challenger\right), \mu^\ecd\right)_{\|v_0^\ecd},\]
    where:

     \begin{itemize}
         \item the players are Prover and Challenger.

         \item Prover controls the set $V^\ecd_\Prover$ and Challenger the set $V^\ecd_\Challenger$, where:
         $$V^\ecd_\Prover = V \times 2^\Pi \times \prod_{i\in\Pi\cup\{\phi\}} Q_{i}
             \and\text{ and }\and
             V^\ecd_\Challenger= E \times 2^\Pi \times \prod_{i\in\Pi\cup\{\phi\}} Q_{i}$$
         In the sequel, we write $\bq = (q_i)_{i\in \Pi \cup \{\phi\}}$ for elements of $\prod_{i\in\Pi\cup\{\phi\}} Q_{i}$.

         \item The initial vertex is $v_0^\ecd = (v_0, \emptyset, (q_{0i})_{i\in\Pi\cup\{\phi\}})=(v_0,\emptyset,\bq_0)$.

         \item The edge set $E^\ecd$ is defined as follows: from the vertex $(u, D, \bq)$, Prover can go to every vertex of the form $(uv, D, \bq)$ with $uv \in E$ (she \emph{proposes} the edge $uv$).
         From the vertex $(uv, D, \bq)$, Challenger can go to the vertex $(v, D, \bq')$ where $q_{i}'=\delta_{i}(q_{i},v)$ for each $i$ (he \emph{accepts} the edge $uv$), or to every vertex $(w, D \cup \{j\}, \bq')$, with $w \neq v$, $uw \in E$, and $q_{i}'=\delta_{i}(q_{i},w)$ for each $i$, and where $j$ is the player controlling $u$ (he \emph{deviates} from Prover's proposal).
         
         \item We define the coloring functions $\kappa'_{i}$ for $i\in \Pi$ as:
         $$\kappa'_{i}(u, D, \bq)=\kappa_{i}(q_{i})$$
         and:
         $$\kappa'_{i}(uv, D, \bq) = k+1$$
         (thus effectively ignoring proposal vertices).

         \item Given a play $\chi=(u_0, D_0, \bq_0)(u_0v_0, D_0,\bq_0)(u_1, D_1,\bq_1) (u_1v_1, D_1,\bq_1)\dots$ in this game, we write $\D(\chi) = \bigcup_k D_k$, the set of players who deviated along the play $\chi$.

         \item Then, the Boolean payoff function $\mu^\ecd$ is defined as follows.
         A play $\chi$ is won by Challenger if and only if either:
         \begin{itemize}
    \item we have $\D(\chi) \neq \emptyset$, and:
    $$
        \chi \in \left( \bigcup_{i \in W} \overline{\Parity}(\kappa'_{i}) \right) \cap \left( \bigcap_{i \in D(\chi)\cap W} \Parity(\kappa'_{i}) \right)
    $$
    (i.e., the play $\chi$ gives to at least one player $i$ a payoff lower than $x_i$, and to every player who deviated at least the same payoff as in $\bx$);
    
    \item or we have $\D(\chi) = \emptyset$, and:
    $$
        \chi \in \left(\bigcup_{i \in W} \overline{\Parity}(\kappa'_{i})\right) \cup \left(\bigcup_{i \notin W} \Parity(\kappa'_{i})\right) \cup \overline{\Parity}(\kappa_\phi')
    $$
    (i.e. the play $\chi$ gives to at least one player $i$ a payoff different from $x_i$ or the correction constraint is not satisfied).
    \end{itemize}
    \end{itemize}
    Note that we can convert all parity conditions into a chain of Rabin conditions: for $i\in \Pi\cup\{\phi\}$ and $c \in \{0, \dots, k\}$, let:
    \[\Xi^{i}_c=\left\{(u, D, \bq)\mid \kappa_{i}(q_{i})\leq c\right\}.\]
    Then $\Parity(\kappa'_{i})=\bigcup_{\substack{0\leq c \leq k\\c \text{ is even}}} \B(\Xi^{i}_c) \cap \overline{\B}(\Xi^{i}_{c-1})$ is the set of runs where $\Xi^{i}_c$ appears infinitely often and $\Xi^{i}_{c-1}$ only finitely often for an even color $c$.
    Remark that here we can (and do) ignore the color $k+1$ added to the game.
    We add extra sets:
    $$
        \noDev=\left\{(u, \emptyset, \bq)\mid u\in V, \bq \in \prod_{i\in\Pi\cup{\phi}} Q_{i}\right\}
        \text{~and~}
        \Delta_i=\left\{(u, D, \bq)\mid u\in V, i \in D, \bq \in \prod_{i\in\Pi\cup\phi} Q_{i}\right\}
    $$ that tracks whether no one (resp. player $i$) deviated.
    The winning condition for Challenger therefore becomes:
    $$
       \left(\overline{\B}(\noDev) \cap \left( \bigcup_{\substack{i \in W\\0\leq c \leq k\\c \text{ is odd}}} \B(\Xi^{i}_c) \cap \overline{\B}(\Xi^{i}_{c-1}) \right) \cap \left( \bigcap_{i\cap W} \overline{\B}(\Delta_i) \cup \bigcup_{\substack{0\leq c \leq k\\c \text{ is even}}} \B(\Xi^{i}_c) \cap \overline{\B}(\Xi^{i}_{c-1}) \right) \right)
        \and\cup\and$$
        $$\left(\B(\noDev) \cap \left(
            \left(\bigcup_{\substack{i \in W\\0\leq c \leq k\\c \text{ is odd}}} \B(\Xi^{i}_c) \cap \overline{\B}(\Xi^{i}_{c-1})\right)
            \cup \left( \bigcup_{\substack{i \notin W\\0\leq c \leq k\\c \text{ is even}}} \B(\Xi^{i}_c) \cap \overline{\B}(\Xi^{i}_{c-1})\right)
            \right.\right.\left.\left.
            \cup \left(\bigcup_{\substack{0 \leq c\leq k\\c \text{ is odd}}} \B(\Xi^{\phi}_c) \cap \overline{\B}(\Xi^{\phi}_{c-1})\right)
        \right)\right)
    $$

The extended deviator game therefore has $2 m 2^{n+1} m^{n}=O(m^n)$ vertices and a winning condition that is provided by an Emerson-Lei condition with $(n+1) k$ colors, and a formula of size polynomial in $n$ and $k$.
Solving this game with the $\PSpace$ algorithm used in the proof of Theorem~\ref{thm:sseExistencePspaceEasy} would provide an $\ExpSpace$ algorithm.
However, using the algorithm from~\cite[Corollary 24]{HausmannLehautPiterman23}, this game can be solved in $O((n  k)!  (m^n)^{n  k+2})=O((n  k)!  m^{n^2  k+2n})$.
As a result the fixed-payoff SSE problem for $\omega$-regular games is in $\ExpTime$.
When both $n$ and $k$ are fixed, this complexity boils down to $\P$.

    \paragraph{$\ExpTime$-hardness.}

We first prove the following lemma.

\begin{lem}\label{lem:conjunctionParityGameExptimeHard}
    Deciding a two-player zero-sum game where one player's winning condition is defined as the intersection of the languages of Büchi automata is $\ExpTime$-hard, and similarly with co-Büchi automata.
\end{lem}

\begin{proof}
We proceed by reduction from the emptiness problem for the intersection of deterministic top-down tree automata.

\subp{Tree automata} We call here \emph{tree automata} what is called more precisely \emph{deterministic top-down tree automata} in the literature.

\begin{defi}[Tree automata]
    A \emph{tree automaton} is a tuple $\TA = (Q, (F_k)_{k \in \{0, \dots, k_{\max}\}}, q_0, \Delta)$, with:
    \begin{itemize}
        \item a finite set $Q$ of states;
        \item a finite family $(F_k)_{k \in \{0, \dots, k_{\max}\}}$, where each set $F_k$ is a finite set of \emph{symbols of arity $k$} (and the sets $F_k$ are pairwise disjoint);
        \item an initial state $q_0$;
        \item and a set $\Delta \subseteq \bigcup_k Q \times F_k \times Q^k$ of transitions, such that for each arity $k$ and each pair $(q, f) \in Q \times F_k$, there exists at most one transition $(q, f, q_1, \dots, q_k) \in \Delta$.
    \end{itemize}
    
    A \emph{$(F_k)_k$-tree}, or simply \emph{tree} when the context is clear, is a tuple $T = (V, E, \preceq, r, \lambda)$ with:
    \begin{itemize}
        \item a graph $(V, E)$ where for every two vertices $u \neq v \in V$, there is at most one path from $u$ to $v$;
        \item a \emph{root} $r \in V$;
        \item a total order $\preceq$ on the set $V$;
        \item a mapping $\lambda: V \to \bigcup_k F_k$ that labels each vertex $v$ with a symbol $\lambda(v)$ that has arity $\card E(v)$.
    \end{itemize}

    A tree $T$ is \emph{recognized} by the tree automaton $\TA$ if and only if there exists a mapping $v \mapsto q_v$ such that:
    \begin{itemize}
        \item we have $q_r = q_0$,
        \item and for every $v$, we have $(q_v, \lambda(v), q_{w_1}, \dots, q_{w_k}) \in \Delta$, where $w_1 \preceq \dots \preceq w_k$ are the elements of the set $E(v)$.
    \end{itemize}
\end{defi}

A natural decision problem about tree automata is the following one.

\begin{prob}[Emptiness problem for the intersection of tree automata]
    Given $n$ tree automata $\TA_1, \dots, \TA_n$ with the same family $(F_k)_k$, is there a tree $T$ that is recognized by all of them?
\end{prob}

That problem is known to be $\ExpTime$-complete.

\begin{slem}[\cite{TATA07}]
    The emptiness problem for the intersection of tree automata is $\ExpTime$-complete.
\end{slem}

\subp{Reduction}
Let $\TA_1, \dots, \TA_n$ be $n$ tree automata with the same family $(F_k)_k$.
We construct in polynomial time a Boolean zero-sum game $\Game_{\|v_0}$ with two players, Adélaïde and Barthélémy, where Adélaïde's winning condition is defined as the intersection of the languages of $n$ Büchi automata (or $n$ co-Büchi automata), and where Adélaïde has a winning strategy if and only if there is a tree recognized by all tree automata.

Let $F = \cup_k F_k$.
We define the set of Barthélémy's vertices as:
$$V_\bart = F \cup \{\bot\},$$
and Adélaïde's vertices as:
$$V_\adel = \left\{(f, \l) ~\left|~ \begin{matrix}
    k \in \NN,\\
    f \in F_k,\\
    \l \in \{1, \dots, k\}
\end{matrix}\right.\right\} \cup \{\top\}.$$
From each vertex $f$ where the symbol $f$ has positive arity, there is an edge to each vertex $(f, \l)$, and from each vertex $(f, \l)$, there is an edge to each vertex $f' \in F$.
From each vertex $f$ where the symbol $f$ has arity $0$, there is an edge to the vertex $\bot$.
From the vertex $\top$, there is an edge to every vertex $f \in F$, and the vertex $\top$ is the initial vertex.
Intuitively, Adélaïde and Barthélémy define together a branch of a tree, from the root to a leaf: for each vertex, Adélaïde chooses a label, and Barthélémy chooses the index of the child that will be considered.

Let us now construct the automata that will define Adélaïde's winning condition.
For each tree automaton $\TA_i$, we define an automaton $\auto_i$.
The states of the automaton $\auto_i$ are the same as the tree automaton $\TA_i$, plus two sink states $\triangledown$ and $\triangle$.
The initial state is the same.
As for transitions, when reading a vertex of the form $f$ or $\top$, the automaton $\auto_i$ remains in the same state.
When it reads a vertex of the form $(f, \l)$ from state $q$, it switches to state $q_\l$ such that there is a transition $(q, f, q_1, \dots, q_k) \in \Delta_i$, if such a transition exists---and to state $\triangledown$ otherwise.
When it reads the vertex $\bot$, it switches to state $\triangle$.
From the states $\triangle$ and $\triangledown$, reading any vertex, the automaton remains in the same state.

Colors are defined so that the automaton $\auto_i$ accepts a play if and only if the state $\triangle$ is reached.
Note that this can be defined with the colors $0$ and $1$ as well as with the colors $1$ and $2$, i.e., the automaton can be defined as a Büchi automaton or as a co-Büchi automaton.

A play is therefore accepted by the automaton $\auto_i$ if and only if along the branch that Adélaïde and Barthélémy have constructed together, the labeling $v \mapsto q_v$ compatible with the tree automaton $\TA_i$ can be properly defined.

\subp{Equivalence}
Each strategy $\sigma_\adel$ for Adélaïde defines a tree $T$, up to renaming vertices; and conversely, each tree $T$ defines a strategy for Adélaïde, up to ignoring what Adélaïde does in subgames where she has herself deviated.

Thus, if there exists a tree $T$ that is accepted by each tree automaton $\TA_i$, then consider the corresponding strategy for Adélaïde: against that strategy, Barthélémy chooses a branch of $T$.
For each automaton $\auto_i$, the mapping $v \mapsto q_v$ testifying of the acceptance of the tree $T$ by the tree automaton $\TA_i$ is well defined, and in particular is well defined along the branch chosen by Barthélémy.
Hence the play that is generated is accepted by the automaton $\auto_i$, and is therefore won by Adélaïde.

If now there is no such tree, then for every strategy $\sigma_\adel$ for Adélaïde, let us consider the corresponding tree $T$.
There exists, then, a tree automaton $\TA_i$ that does not recognize the tree $T$, and a vertex $v$ of $T$ on which the corresponding mapping $v \mapsto q_v$ cannot be defined.
If Barthélémy chooses a branch that traverses that vertex, the automaton $\auto_i$ will then switch to the state $\triangledown$, making Adélaïde lose.
\end{proof}

    Now that we know that this problem is $\ExpTime$-hard, we can prove our result by reducing it to the SSE fixed-payoff problem with $\phi = \top$.
    Let $\Game_{\|v_0}$ be a two-player zero-sum Boolean game with players Barthélémy and Adélaïde, where Adélaïde's winning condition is defined as the intersection of the languages of $n$ parity automata $\auto_1,\dots,\auto_n$.

\subp{Reduction}
    We build the $(n+2)$-player game $\Game^\dagger_{\|v_1}$ that has an SSE with payoff $(1)_{i \in \Pi}$ if and only if Barthélémy has a winning strategy in $\Game$.
    The $n+2$ players are $\Pi = \{1,\dots,n,\bart,\adel\}$.
    The game arena consists of the original game $\Game_{\|v_0}$ with a prepended path where each player from $1$ to $n$ can either choose to continue or reach a sink state $\triangledown$, as depicted by \Cref{fig:conjParity2SSEGame}.
    Player $i \in \{1, \dots, n\}$ wins if either $\auto_i$ accepts in $\Game_{\|v_0}$ or the play ends up in $\triangledown$; otherwise he loses.
    This is ensured in the automata $\auto^\dagger_{i}$ by adding two states to $\auto_i$ to take into account this prefix (see \Cref{fig:conjParity2sseAutomata}---the numbers indicated in each state is the color of that state).
    Barthélémy gets payoff $1$ if the play ends up in $\triangledown$ and gets $0$ if it reaches $\Game_{\|v_0}$.
     Adélaïde always wins in this game.

     \begin{figure}
    \centering
    \begin{tikzpicture}[node distance=2cm]
        \node[literalState=$1$,initial] (q1) at (0,0) {$v_1$};
        \node[literalState=$2$,right of=q1] (q2) {$v_2$};
        \node[right of=q2] (qdots) {$\cdots$};
        \node[literalState=$n$,right of=qdots] (qn) {$v_n$};
        \node[state,rectangle,rounded corners=7pt,dotted,minimum size=1.5cm,right of=qn,node distance=3cm] (orig) {$\Game_{\|v_0}$};
        \node[nobodyState,below of=qdots] (sink) {$\triangledown$};
\begin{pgfonlayer}{arrows}
        \path[->] (q1) edge (q2);
        \path[->] (q2) edge (qdots);
        \path[->] (qdots) edge (qn);
        \path[->] (qn) edge (orig);
        \path[->] (q1) edge[bend right] (sink);
        \path[->] (q2) edge[bend right,in=150] (sink);
        \path[->] (qn) edge[bend left] (sink);
        \path[->] (sink) edge[loop below] (sink);
        \end{pgfonlayer}
    \end{tikzpicture}
    \caption{The game $\Game^\dagger_{\|v_1}$}
    \label{fig:conjParity2SSEGame}
\end{figure}
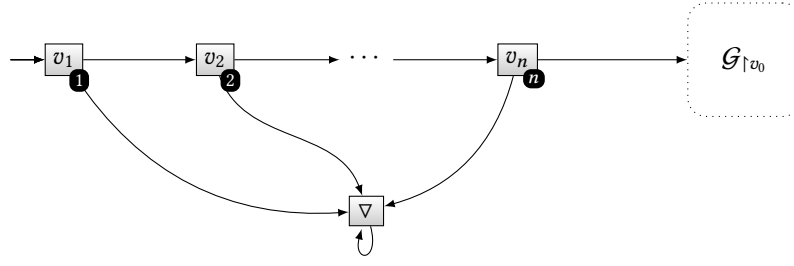

     \begin{figure}
    \centering
    \begin{subcaptionblock}{0.9\textwidth}
    \centering
    \begin{tikzpicture}[auto,node distance=3cm]
        \node[automatonState,initial] (init) at (0,0) {$0$};
        \node[automatonState,below right of=init] (sink) {$1$};
        \node[automatonState,rectangle,rounded corners=7pt,dotted,minimum size=1.5cm,above right of=sink] (autom) {$\auto_{i}$};
        \path[->] (init) edge[loop above] node{$q_1,\dots,q_n,\triangledown$} (autom);
        \path[->] (init) edge node{$v_0$} (autom);
        \path[->] (init) edge node[swap]{$v\in V\setminus\{v_0\}$} (sink);
        \path[->] (autom) edge node{$q_1,\dots,q_n,\triangledown$} (sink);
        \path[->] (sink) edge[loop above] node{$*$} (sink);
    \end{tikzpicture}
    \caption{Parity automata $\auto^\dagger_{i}$ for player $i$ to win}
    \label{fig:conjParity2sseAutomataRegular}
    \end{subcaptionblock}
    
    \begin{subcaptionblock}{0.4\textwidth}
    \centering
    \begin{tikzpicture}[auto,node distance=3cm]
        \node[automatonState,initial] (init) at (0,0) {$0$};
        \node[automatonState,right of=init] (lose) {$1$};
        \path[->] (init) edge[loop above] node{$q_1,\dots,q_n,\triangledown,v\in V\setminus\{v_0\}$} (init);
        \path[->] (init) edge node{$v_0$} (lose);
        \path[->] (lose) edge[loop above] node{$*$} (lose);
    \end{tikzpicture}
    \caption{Parity automaton $\auto^\dagger_{\bart}$.}
    \label{fig:conjParity2sseAutomataBarthélémy}
    \end{subcaptionblock}
    \hfill
    \begin{subcaptionblock}{0.4\textwidth}
    \centering
    \begin{tikzpicture}[auto,node distance=3cm]
        \node[automatonState,initial] (init) at (0,0) {$1$};
        \node[automatonState,right of=init] (lose) {$0$};
        \path[->] (init) edge[loop above] node{$q_1,\dots,q_n,\triangledown,v\in V\setminus\{v_0\}$} (init);
        \path[->] (init) edge node{$v_0$} (lose);
        \path[->] (lose) edge[loop above] node{$*$} (lose);
    \end{tikzpicture}
    \caption{Parity automaton $\auto^\dagger_{\adel}$.}
    \label{fig:conjParity2sseAutomataAdélaïde}
    \end{subcaptionblock}

    \caption{Automata associated to the game $\Game^\dagger$}
    \label{fig:conjParity2sseAutomata}
\end{figure}

\subp{If Barthélémy has a winning strategy in $\Game_{\|v_0}$, then there is an SSE in $\Game_{q_1}$ where all players get payoff $1$}
    Assume that  Barthélémy has a winning strategy in $\Game_{\|v_0}$.
    Let $\bsigma$ be the following profile in $\Game^\dagger$: for $i\in \{1,\dots,n\}$, from $q_i$ go to $\triangledown$;  Barthélémy plays his winning strategy in $\Game_{\|v_0}$, and  Adélaïde plays an arbitrary strategy.
    Since $\triangledown$ is reached in the play induced by $\bsigma$, the payoff vector is $(1)_{i \in \Pi}$.
    We will show that $\bsigma$ is an SSE.

    Let $C$ be a coalition of players and assume $\bsigma'$ is a harmful deviation from players in $C$.
    First, since  Adélaïde cannot be harmed by a deviation (she always wins), we can assume that $\adel\in C$.
    In addition, if one of the players $i\in\{1,\dots,n\}$ does not effectively deviate, then the play ends up in $\triangledown$ and the payoff vector does not change; therefore all of these players must be in $C$.
    On the other hand, if the play changes the payoff by not ending in $\triangledown$, then  Barthélémy strictly decreases his payoff, and then cannot be part of the coalition.
    Therefore, any harmful deviation must be from the coalition $C=\{1,\dots,n,\adel\}$.
    As the deviation reaches $\Game_{\|v_0}$,  Barthélémy is indeed harmed; for the deviation to be deemed harmful it remains to show that no player in the coalition is themself harmed.
    However, since  Barthélémy has a winning strategy and plays it in $\Game_{\|v_0}$, the play is not accepted by (at least) one automaton $\auto_i$, so the corresponding player $i$ loses and thus decreases their payoff.
    As a result, player $i$ cannot be part of the coalition, which is a contradiction, and $\bsigma$ is an SSE.

\subp{If Adélaïde has a winning strategy, then there is no such SSE}
    Now assume that  Adélaïde has a winning strategy in $\Game_{\|v_0}$.
    Let $\bsigma$ be a profile with payoff vector $(1)_{i \in \Pi}$.
    We will show this profile is not an SSE.
    For this payoff to occur, that means the play ends up in $\triangledown$.
    Consider the coalition $C=\{1,\dots,n,\adel\}$ and the deviation (which may not be actually effective for some players) $\bsigma'$ defined as follows: for $i \in \{1,\dots,n\}$, from $q_i$ go to $q_{i+1}$ (or $v_0$ if $i=n$); then, Adélaïde plays her winning strategy.
    The play thus built will reach $\Game_{\|v_0}$ where the created play is accepted by all $\auto_i$, so the payoff for player $i$ remains $1$ while the payoff for  Barthélémy strictly decreases from $1$ to $0$.
    Therefore $\bsigma'$ is a harmful deviation and $\bsigma$ is not an SSE.    

\subp{Conclusion}
    As a result there is an SSE with payoff $(1)_{i \in \Pi}$ if and only if  Barthélémy has a winning strategy in $\Game_{\|v_0}$ for the winning condition $\neg \bigcap_{i=1}^n \Lang(\auto_i)$.
    Since solving a two-player game with such a winning condition is $\ExpTime$-hard, by \cref{lem:conjunctionParityGameExptimeHard}, even if all automata are Büchi automata or co-Büchi automata, our hardness result follows.
\end{proof}

%% file: 3bRSE.tex
\section{Introduction}

\subsection{About randomness in games}

It is now time to highlight the fact that the formalism we have considered up to this point implicitly rules out any form of randomness.
However, it is common in game theory to consider games where chance plays a role.
This can be done in two ways: first, by introducing randomness into the game itself, for example, with \emph{stochastic vertices}, i.e. vertices not controlled by any player, where the outgoing edge is chosen randomly according to a fixed probability distribution; and second, by allowing players to randomize their strategies, for instance, by tossing a coin when undecided between two edges.

In the real world, especially in the context of computer-related systems (quantum computers being set aside), generating true randomness is exceedingly difficult, and the question of whether it can even exist is highly debated. 
However, randomness in games can be understood more broadly as an abstraction for the absence of information, in line with Andrey Kolmogorov's conception of probabilities: if player $i$ makes a randomized choice between actions $a$ and $b$, each with probability $\frac{1}{2}$, this simply means that the other players have no way of knowing which action player $i$ will actually choose, or even which one is more likely.

It is a well-known fact (a proof can be derived from Martin's proof of determinacy in Blackwell games \cite{DBLP:journals/jsyml/Martin98}) that such randomization is unnecessary in (Boolean) two-player zero-sum games within our formalism: if players are allowed to randomize their strategies, pure (i.e., non-randomized) strategies are just as effective as randomized ones.

However, this result no longer holds when the formalism is extended to concurrent games (where players make simultaneous choices) \cite{Everett1957}. In this case, a player (say Adélaïde) may have an incentive to randomize her strategy so that her opponent (say Barthélémy) cannot adapt to the action she plans. A canonical example is the rock-paper-scissors game: if Adélaïde plays a given action deterministically (e.g., rock), Barthélémy can easily counter it (in this case, by playing paper). To prevent this, an optimal strategy for Adélaïde is to play each action with probability $\frac{1}{3}$---which might be interpreted as playing a given action, but ensuring Barthélémy has no information about it.
Another interesting example arises when analyzing penalty kicks in soccer: the kicker must choose between sending the ball to the left or right side of the goal, while the goalkeeper simultaneously (and ideally quickly) chooses to defend one of those sides. The probability of scoring is much higher when the goalkeeper chooses the wrong side. Data analysis from the German Bundesliga appears to show that players' choices closely align with those suggested by theory \cite{Schumann2007}.
Similar phenomena arise in turn-based (i.e., non-concurrent) games with \emph{imperfect information} \cite{DBLP:journals/lmcs/RaskinCDH07}.

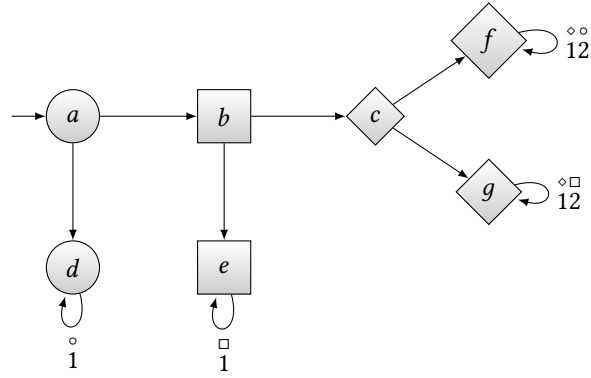
\begin{figure}
			\centering
			\begin{tikzpicture}
				\node[vert, initial left] (a) at (0, 0) {$a$};
				\node[vert, rectangle] (b) at (2, 0) {$b$};
                \node[vert, diamond] (c) at (4, 0) {$c$};
                \node[vert] (d) at (0, -2) {$d$};
                \node[vert, rectangle] (e) at (2, -2) {$e$};
                \node[vert, diamond] (f) at (5.5, 1) {$f$};
                \node[vert, diamond] (g) at (5.5, -1) {$g$};
				\path (a) edge (b);
				\path (b) edge (c);
				\path (c) edge (f);
				\path (c) edge (g);
                \path (a) edge (d);
                \path (b) edge (e);
				\path (d) edge[loop below] node[below] {$\stackrel{\circ}{1}$} (d);
				\path (e) edge[loop below] node[below] {$\stackrel{\Box}{1}$} (e);
				\path (f) edge[loop right] node[right] {$\stackrel{\diamond}{1}\stackrel{\circ}{2}$} (f);
				\path (g) edge[loop right] node[right] {$\stackrel{\diamond}{1}\stackrel{\Box}{2}$} (g);
			\end{tikzpicture}
			\caption{A game where randomization might be useful}
			\label{fig:ne_randomized}
		\end{figure}

To return to our framework, randomization might also be useful for constructing an equilibrium in a multiplayer game.
In this case, its purpose is no longer to prevent a player, viewed as an opponent, from using some information, but rather to use randomness to distribute payoffs among several players in such a way that they are satisfied with the \emph{expected payoff} they receive.

\begin{exa}
Consider the mean-payoff game depicted in \Cref{fig:ne_randomized} (with all non-specified rewards being zero), and let us examine whether there exists a Nash equilibrium in which player $\Diamond$ receives a payoff of 1.
If randomization is not allowed, the answer is clearly \emph{no}: to give player $\Diamond$ a payoff of 1, the play must necessarily reach either vertex $f$ or vertex $g$. In both cases, player $\Circle$ (in the first case) or player $\Box$ (in the second case) can deviate profitably by moving to vertex $d$ or $e$, respectively.
However, if randomization is allowed, player $\Diamond$, starting from vertex $c$, can take both edges $cf$ and $cg$ with probability $\frac{1}{2}$.
In this case, the expected payoffs for both player $\Circle$ and player $\Box$ are equal to 1, and they have no incentive to deviate---assuming that the measure used to define a profitable deviation is the expected payoff.
\end{exa}

\begin{figure}
        \centering
		\begin{tikzpicture}
		\node[vert] (a) at (0, 0) {$a$};
		\node[vert] (c) at (-2, 0) {$c$};
		\node[vert, rectangle] (b) at (2, 0) {$b$};
		\node[vert, rectangle] (d) at (4, 0) {$d$};
		\path[->] (a) edge (c);
		\path (a) edge[bend left] node[above] {$\stackrel{\circ}{0} \stackrel{\Box}{3}$} (b);
		\path (b) edge[bend left] node[below] {$\stackrel{\circ}{0} \stackrel{\Box}{3}$} (a);
		\path[->] (b) edge (d);
		\path (d) edge [loop right] node {$\stackrel{\circ}{2} \stackrel{\Box}{2}$} (d);
		\path (c) edge [loop left] node {$\stackrel{\circ}{1} \stackrel{\Box}{1}$} (c);
		\end{tikzpicture}
		\caption{A game without SPE} 
		\label{fig:sans_spe2}
\end{figure}
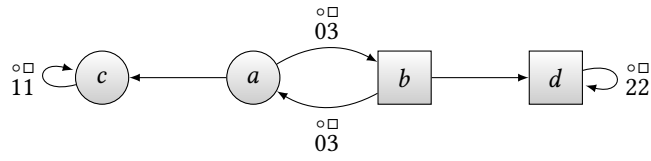

Finally, randomization may also be used to punish infinite deviations, up to relaxing our equilibrium notions.

\begin{exa}
Consider the mean-payoff game depicted by \cref{fig:sans_spe2}: we have shown in the proof of \cref{thm:no_spe} that, when the players are not allowed to play randomized strategies, this game contains no SPE, and even no $\epsilon$-SPE with $\epsilon < 1$.
On the other hand, consider the following randomized stationary strategy profile: from the vertex $a$, player $\Circle$ goes to the vertex $c$ with probability $\epsilon > 0$, and to the vertex $b$ with probability $1-\epsilon$.
From the vertex $b$, player $\Square$ always goes to the vertex $d$.
This strategy profile is an $\epsilon$-SPE (in every subgame, no player can improve their expected payoff by more than $\epsilon$), and is defined for every $\epsilon > 0$.
\end{exa}

In line with this last example, it is already known~\cite{Kuipers2021} that randomized $\epsilon$-SPEs always exist in deterministic mean-payoff games where all rewards are zero, except in self-loops on sink vertices (which corresponds to what we call later in this chapter \emph{simple stochastic games}, but with no stochastic vertex).
We conjecture that this remains true in all mean-payoff games.

\begin{conj}\label{conj:epsilon_spes}
    Let $\Game_{\|v_0}$ be a mean-payoff game (with no stochastic vertices).
    Then, for every $\epsilon > 0$, there exists a randomized $\epsilon$-SPE in $\Game_{\|v_0}$.
\end{conj}

\subsection{Constrained existence of randomized NEs (and SPEs)}

The constrained existence problem in stochastic games with randomized strategies has been studied by Michael Ummels and Dominik Wojtczak, who showed that it is undecidable in all the settings we have considered, as soon as randomness is introduced.

More precisely, in~\cite{Ummels2011}, they prove the undecidability of the constrained existence problem for randomized Nash equilibria in \emph{simple quantitative games}, i.e., games with \emph{terminal vertices}, from which no other vertex is accessible, and where each player's payoff depends solely on the terminal vertex eventually reached (each player receives a payoff of $0$ if no terminal vertex is ever reached).
Clearly, simple games can be seen as a subclass of mean-payoff games.
Notably, this result does not require the presence of stochastic vertices.
On the other hand, in~\cite{UW11}, they show that undecidability still holds even when payoffs are restricted to $0$ and $1$, i.e., in Boolean simple games, provided stochastic vertices are introduced.
Moreover, in that setting, undecidability persists even when players are not allowed to randomize their strategies.
Boolean simple games can be viewed as a subclass of both energy and parity games (and even Büchi and co-Büchi games).
Thus, the work of Michael Ummels and Dominik Wojtczak already covers all the game classes studied in this document, except for discounted-sum games, where decidability remains an open problem even in the absence of randomness (see \cref{thm:ds_ne_hardness}).

Furthermore, their proofs rely on variations of a single reduction from the halting problem of two-counter machines.
With careful refinements, the same reduction can also be used to establish the undecidability of the constrained existence problem for SPEs in these game classes.

This existing body of work thus provides a computational argument for exploring alternative equilibrium notions when randomness is involved, in the hope of identifying problems that are decidable with reasonable complexity.
Another motivation comes from the limitations of Nash equilibria, defined using the classical expected payoff, in capturing intuitively rational behavior.

\subsection{Randomness and risk}

\subp{Motivating example}
Let us consider a $1$-player game where a protagonist is proposed two options: (a) earning \texteuro1; (b) playing a lottery in which, with probability $\frac{1}{40}$, she gets \texteuro40, and with probability $\frac{39}{40}$, she does not earn anything.
Classically, rational strategies would be maximizing the expected payoff. From this perspective, both options yield an expected payoff of \texteuro1, making them equivalent.
This approach is particularly justified when the game represents a scenario that can be repeated many times: the law of large numbers ensures that, in the long run, the average payoff will converge to the expected payoff. 

However, when the game is played only once, the protagonist may prioritize immediate needs. If she urgently requires \texteuro1, the guaranteed option (a) becomes preferable.
Conversely, if she is a risk-taker, or finds herself in a situation where only the \texteuro40 can make a significant difference, she may prefer the high-risk option (b).
Although this choice might appear irrational, it mirrors the behavior of millions of people who participate everyday in games that even have a negative expected payoff, driven by the allure of a potentially life-changing win, and generating an annual turnover of USD 536 billions~\cite{GamblingNewspaper23} for the gambling industry.
That industry, on the other hand, operates on a large scale where expected payoff becomes the key metric. 
This contrast underscores the importance of alternative measures to expected payoff that account for each agent's risk tolerance.

\subp{Risk measures}
A \emph{risk measure} captures the perception that a player has of what their payoff will be. In that sense, they generalize the notion of expected payoff.
Various risk measures exist in the literature, and have been used extensively in the field of economics and finance. 
Some of these risk measures include expected shortfall (ES), value at risk (VaR)~\cite{Aue18}, variance~\cite{Bra99}, entropic risk measure (ER)~\cite{FS02}.

A lot of work has been done in considering these risk measures over Markov decision processes (MDPs) which use variance (along with mean) as a risk-measure~\cite{FK89, PSB22,MT11}, ES~\cite{RRS15,KM18,Meg22} (also referred to as conditional value at risk (CVaR), average value at risk (AVaR), expected tail loss (ETL), and superquantile in literature) and ER~\cite{How72,BR14,BCMP24}.
Studying the entropic risk measure in MDPs appears more practical compared to expected shortfall  or using variance-penalized risk-measures. This impracticability of ES and variance-penalized measure in particular is due to the intractable exponential memory~\cite{HK15} and time required to compute optimal strategies~\cite{PSB22}, even for the one agent system of Markov decision processes (MDPs). On the other hand, when the risk measure used is ER, players have optimal positional strategies in MDPs~\cite{How72}, which makes it a prime candidate for consideration in multi-agent settings.

\subp{Entropic risk measure}
The entropic risk measure is computed by assigning to each agent a \emph{risk parameter}, i.e., a value $\rho \in \RR$.
The entropic risk measure of a random variable $X$ is then defined as
$\M_\rho[X] = -\frac{1}{\rho} \log_e \left( \EE \left[ e^{-\rho X}\right] \right)$.
Assume the random variable $X$ is a player's payoff.
If the risk parameter $\rho$ is positive, then more weight will be given to the bad payoffs: the corresponding player can then be considered as risk-averse.
Conversely, players with a negative $\rho$ are more risk-loving.
When $\rho$ tends to $0$, the entropic risk measure converges to the classical expectation $\EE[X]$.

\begin{figure}
            \centering
			\begin{tikzpicture}
				\node[initial left, stoch] (a) at (0, 0) {$a$};
				\node[vert] (b) at (1.5, 0) {$b$};
                \node[stoch] (c) at (2.5, 1) {$c$};
                \node (t1) at (4, 2) {$t_1:~\stackrel{\circ}{40}$};
                \node (t2) at (4, 0) {$t_2:~\stackrel{\circ}{0}$};
                \node (t3) at (3, -1) {$t_3:~\stackrel{\circ}{1}$};
                \path (a) edge[loop above] node[above] {$\frac{1}{2}$} (a);
				\path (a) edge node[above] {$\frac{1}{2}$} (b);
                \path (b) edge[blue, thick] (c);
                \path (b) edge[red, thick] (t3);
				\path (c) edge node[above] {$\frac{1}{40}$} (t1);
				\path (c) edge node[below] {$\frac{39}{40}$} (t2);
			\end{tikzpicture}
			\caption{A stochastic MDP}
			\label{fig:example_gamma}
            \end{figure}
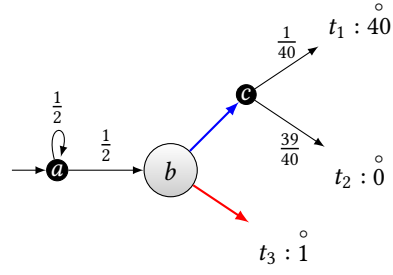

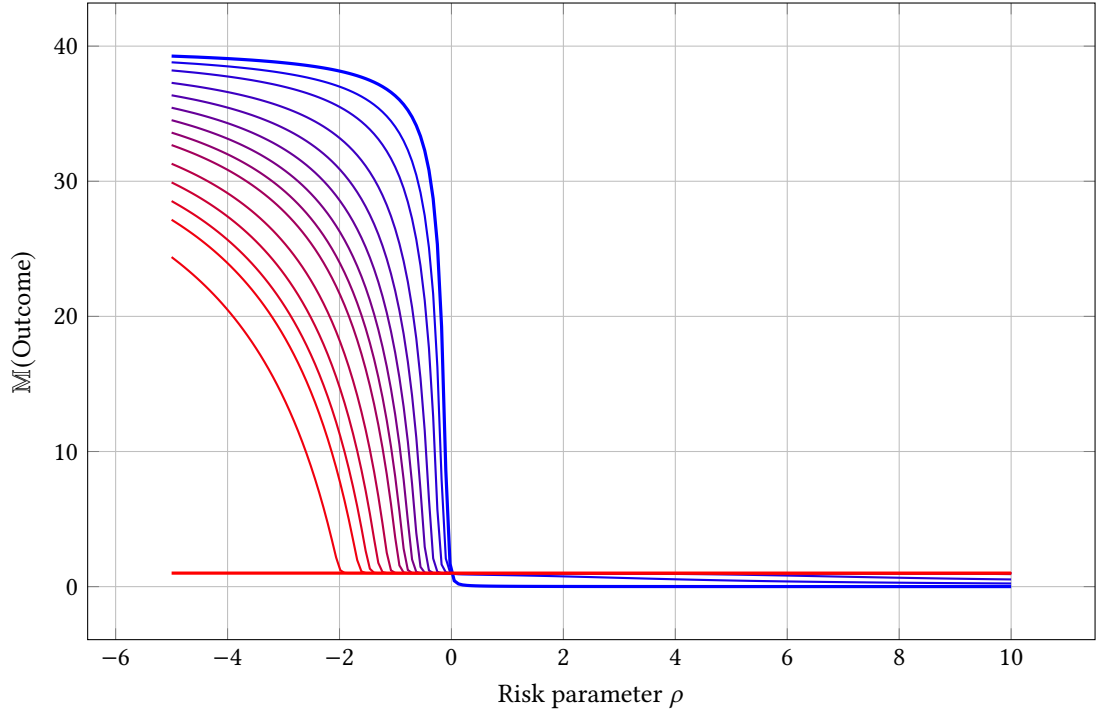
\begin{figure}
            \centering
			\begin{tikzpicture}
              \begin{axis}[
                xlabel={Risk parameter $\rho$},
                ylabel={$\M(\text{Outcome})$},
                domain=-5:10,
                samples=200,
                    width=\textwidth,
                   height=10cm,
                grid=major,
                ]
                \addplot [
                  red!8!blue,
                  thick
                ]
                {-1/x * log2((9/10)*e^(-1*x) + (1/400) * e^(-40*x) + (39/400))/log2(e)};
                \addplot [
                  red!16!blue,
                  thick
                ]
                {-1/x * log2((199/200)*e^(-1*x) + (1/8000) * e^(-40*x) + (39/8000))/log2(e)};
                \addplot [
                  red!24!blue,
                  thick
                ]
                {-1/x * log2((19999/20000)*e^(-1*x) + (1/800000) * e^(-40*x) + (39/800000))/log2(e)};
                \addplot [
                  red!32!blue,
                  thick
                ]
                {-1/x * log2((1999999/2000000)*e^(-1*x) + (1/80000000) * e^(-40*x) + (39/80000000))/log2(e)};
                \addplot [
                  red!40!blue,
                  thick
                ]
                {-1/x * log2((199999999/200000000)*e^(-1*x) + (1/8000000000) * e^(-40*x) + (39/8000000000))/log2(e)};
                \addplot [
                  red!48!blue,
                  thick
                ]
                {-1/x * log2((19999999999/20000000000)*e^(-1*x) + (1/800000000000) * e^(-40*x) + (39/800000000000))/log2(e)};
                \addplot [
                  red!56!blue,
                  thick
                ]
                {-1/x * log2((1999999999999/2000000000000)*e^(-1*x) + (1/80000000000000) * e^(-40*x) + (39/80000000000000))/log2(e)};
                \addplot [
                  red!64!blue,
                  thick
                ]
                {-1/x * log2((199999999999999/200000000000000)*e^(-1*x) + (1/8000000000000000) * e^(-40*x) + (39/8000000000000000))/log2(e)};
                \addplot [
                  red!72!blue,
                  thick
                ]
                {-1/x * log2((199999999999999999/200000000000000000)*e^(-1*x) + (1/8000000000000000000) * e^(-40*x) + (39/8000000000000000000))/log2(e)};
                \addplot [
                  red!80!blue,
                  thick
                ]
                {-1/x * log2((199999999999999999999/200000000000000000000)*e^(-1*x) + (1/8000000000000000000000) * e^(-40*x) + (39/8000000000000000000000))/log2(e)};
                \addplot [
                  red!88!blue,
                  thick
                ]
                {-1/x * log2((199999999999999999999999/200000000000000000000000)*e^(-1*x) + (1/8000000000000000000000000) * e^(-40*x) + (39/8000000000000000000000000))/log2(e)};
                \addplot [
                  red!94!blue,
                  thick
                ]
                {-1/x * log2((199999999999999999999999999/200000000000000000000000000)*e^(-1*x) + (1/8000000000000000000000000000) * e^(-40*x) + (39/8000000000000000000000000000))/log2(e)};
                \addplot [
                  red!94!blue,
                  thick
                ]
                {-1/x * log2((199999999999999999999999999999999/200000000000000000000000000000000)*e^(-1*x) + (1/8000000000000000000000000000000000) * e^(-40*x) + (39/8000000000000000000000000000000000))/log2(e)};
                \addplot [
                  blue,
                  very thick
                ]
                {-1/x * log2((1/40) * e^(-40*x) + (39/40))/log2(e)};
                \addplot [
                  red,
                  very thick
                ]
                {+1};
              \end{axis}
            \end{tikzpicture}
			\caption{Each curve represents the perceived reward of a player choosing only blue strategy, only red, or randomizing between both strategies.}
			\label{fig:example_plot}
            \end{figure}

The game depicted by Figure~\ref{fig:example_gamma} extends the lottery example we discussed earlier. 
Black vertices are stochastic, and the circle vertex is controlled by player $\Circle$.
A play can be seen as a sequence of moves of a token along the edges of the graph, starting from $a$: from a stochastic vertex, it takes one of the outgoing edges with the probabilities indicated on those, and from a vertex controlled by the player, she chooses which edge it takes.
The payoff $40$, $0$, or $1$ is obtained when the terminal vertex $t_1$, $t_2$, or $t_3$ is reached, respectively.
If no terminal vertex is reached, then the payoff is $0$.
Taking the red edge corresponds to option (a): then, her risk entropy is always $1$, for every risk parameter $\rho$.
But if she chooses option (b), that is, if she takes the blue edge, her risk entropy is $\M_\rho[\mu_{\circ}] = -\frac{1}{\rho} \log \left( \EE \left[ e^{-\rho \mu_{\circ}}\right] \right) = -\frac{1}{\rho} \log \left(  \frac{1}{40} e^{-40\rho } + \frac{39}{40} \right)$.
Both cases are illustrated with red and blue curves in \cref{fig:example_plot}.
The curves cross at abscissa $\rho = 0$, where the entropic risk measure corresponds to the expectation. Note that other strategies are possible if \emph{randomization} is allowed---the player could, for example, toss a coin and participate in the lottery if the outcome is heads. The perceived reward of randomizing between outermost red and blue edges are illustrated in the intermediate cases with mixtures of red and blue in \cref{fig:example_plot}.

Unfortunately, even for two player zero-sum stochastic games with total-reward objectives (payoff is the sum of the rewards seen along the way), computing optimal strategies can only be done in $\PSpace$, when the base $e$ is replaced by a rational number; and if $e$ is the base of the exponent, then it is only in $\ExpTime$~\cite{gallegohernandez_et_al:LIPIcs.STACS.2025.37}.
Solving the two-player zero-sum case is a specific case of finding equilibria in two-agent systems where the payoffs of the two agents are exactly the negation of each others and so are the risk parameters of each of the agents.
Therefore, reasoning about multi-agent systems with ER also has potential to be computationally intractable.

\section{Definitions}

We provide here additional definitions that will be used in this chapter and the following.

\subsection{Probabilities}\label{ssec:proba}
Given a (finite or infinite) set of outcomes $\Omega$ and a probability measure $\PP$ over $\Omega$, let $X$ be a random variable over $\Omega$, that is, a mapping $X: \Omega \to \RR$. We then write $\EE^\PP[X]$, or simply $\EE[X]$, for the expectation of $X$, when it is defined.
Given a finite set $S$, a \emph{probability distribution} over $S$ is a mapping $d: S \to [0,1]$ that satisfies the equality $\sum_{x \in S} d(x) = 1$.
We write $\Supp (d)$ for the \emph{support} of the distribution $d$, that is, the set of elements $x \in S$ such that $d(x) > 0$.

\subsection{Risk measures}
Given a set $\Omega$ of outcomes, a \emph{risk measure} over $\Omega$ is a mapping $M$ which maps a probability measure $\PP$ over $\Omega$ and a random variable $X$ to a real value $M^\PP[X]$.

Sometimes, in the literature, risk measures are expected to have the following three properties: (1) they are \emph{normalized}, i.e., we have $M^\PP[0] = 0$; (2) they are  \emph{monotone}, i.e., the pointwise inequality $X \leq Y$ implies $M^\PP[X] \leq M^\PP[Y]$; and (3) they are \emph{translative}, i.e., we have $M^\PP[X + c] = M^\PP[X] + c$ for every constant $c$.
In particular, the expectation of a random variable $\EE$ satisfies those properties.
Sometimes also, the word \emph{translative} refers to the opposite of this definition, i.e., to the condition $M^\PP[X + c] = M^\PP[X] - c$ for every constant $c$.
We will not need any of those properties in the sequel---we simply note that they are satisfied (with the first definition of translativity) by the risk measures we consider.

\subsection{Simple stochastic games}

Since we now want to consider games in which players must deal with stochastic phenomena, we need a new definition that allows for stochastic vertices.
At the same time, since the equilibrium notions we study in this chapter and the next are new, we do not want this study to be influenced by the properties of sophisticated payoff functions, such as those examined in the previous chapters.
Therefore, we restrict our work to \emph{simple stochastic games}, i.e., quantitative games in which each player's payoff is entirely determined by the \emph{terminal vertex} that is eventually reached (or not).

\begin{defi}[Simple stochastic game]
    A \emph{simple stochastic game} is a tuple $\Game = (V, E, \Pi, (V_i)_{i \in \Pi}, \p, \mu)$, where we have:
    \begin{itemize}
        \item a directed graph $(V, E)$, called the \emph{underlying} graph of $\Game$;

        \item a finite set $\Pi$ of \emph{players};

        \item a partition $(V_i)_{i \in \Pi \cup \{?\}}$ of the set $V$, where $V_i$ denotes the set of vertices \emph{controlled} by player $i$, and the vertices in $V_?$ are called \emph{stochastic vertices};

        \item a \emph{probability function} $\p: E(V_?) \to [0, 1]$, such that for each stochastic vertex $u$, the restriction of $\p$ to $E(u)$ is a probability distribution;

        \item a mapping $\mu: T \to \RR^\Pi$ called \emph{payoff function}, where $T$ is the set of \emph{terminal vertices}, i.e. vertices of the graph $(V, E)$ that have no outgoing edges.
        We also write $\mu_i$, for each player $i$, for the function that maps a terminal vertex $t$ to the $i^\text{th}$ coordinate of the tuple $\mu(t)$.
        The mapping $\mu$ extends to the set $(V \setminus T)^\omega \cup (V \setminus T)^* T$ by defining $\mu(v_1 \dots v_k t) = \mu(t)$, and $\mu(v_1 v_2 \dots) = (0)_{i \in \Pi}$ (if no terminal vertex is reached, everyone gets the payoff $0$).
    \end{itemize}
\end{defi}

Note that we dropped the assumption according to which every vertex has an outgoing edge: this was necessary to have \emph{terminal vertices}, defining the players' payoff.
On the other hand, for computational reasons, we will often assume that every vertex, except possibly the initial one when defined, has an ingoing edge: thus, we always have $\card E \geq \card V - 1$.

In this chapter and the following, we often use the word \emph{game} for simple stochastic games.
Moreover, we use the vocabulary and the notations that we have introduced for games as defined in \cref{defi_game}, for simple stochastic games, with analogous definitions.
For instance, we will sometimes consider \emph{initialized} simple stochastic games (without always mentioning that they are initialized).
We often assume that we are given a simple stochastic game $\Game_{\|v_0}$ and implicitly use the same notations as in the definition above.
In such games, \emph{histories} are still defined as finite paths in the graph $(V, E)$.
\emph{Plays} are now paths that are either infinite, or end in a terminal vertex.

Finally, note that simple stochastic games can also be seen as a particular case of mean-payoff games (with stochastic vertices), up to adding a self-loop on terminal vertices---those loops are then the only edges with possibly nonzero rewards.

\emph{Markov decision processes} and \emph{Markov chains} can be defined as particular cases of stochastic games.

\begin{defi}[Markov decision process, Markov chain]
    A (initialized or not) \emph{Markov decision process} is a simple stochastic game with one player.
    A \emph{Markov chain} is a simple stochastic game with zero player.
\end{defi}

\subsection{Strategies, and strategy profiles}

We now give a new definition for strategies, allowing for randomization.
This definition will always be the one used in the sequel.

\begin{defi}
    In a game $\Game_{\|v_0}$, a \emph{strategy} for player $i$ is a mapping $\sigma_i$ that maps each history $hu \in \Hist_i\Game_{\|v_0}$ to a probability distribution over $E(u)$.
\end{defi}

Again, the vocabulary and the notations that have been defined around strategies in the sense of \cref{def:strategies} are naturally adapted.
For instance, a \emph{strategy profile} is still a tuple of strategies indexed by players.

A path $\alpha_0 \alpha_1 \dots$ (be it a history or a play) is now said to be \emph{compatible} with the strategy $\sigma_i$ if for each $k$ such that $\alpha_k \in V_i$, the probability that the strategy $\sigma_i$ outputs the vertex $\alpha_{k+1}$ after the history $\alpha_{\leq k}$ is positive, that is, if we have $\sigma_i(\alpha_{\leq k})(\alpha_{k+1}) > 0$.
That definition naturally extends to strategy profiles.

A strategy profile $\bsigma_{-i}$ in the game $\Game_{\|v_0}$ defines an initialized Markov decision process $\Game_{\|v_0}[\bsigma_{-i}]$, where the vertices of the (infinite) underlying graph are the histories of $\Game_{\|v_0}$ and the edges are added from $hu$ to each the history $huv$ if and only if we have $uv \in E$.
Similarly, a strategy profile $\bsigma$ for $\Pi$ defines an initialized Markov chain $\Game_{\|v_0}[\bsigma]$.
Thus, it also defines a probability measure $\PP_\bsigma$ over plays---which turns the payoff functions $\mu_i$ into random variables.

\begin{rem}
    A history $h$ is compatible with $\bsigma$ if and only if it has positive probability of being generated.
    This equivalence fails for plays: a play $\pi$ may be compatible with $\bsigma$ but generated with probability $0$, if it is infinite (and therefore never reaches any terminal vertex) and if along $\pi$, infinitely many randomized choices occur, be it by the players or by stochastic vertices.
\end{rem}

We say that a strategy $\sigma_i$ is \emph{pure} when for each history $hu$, there is a vertex $v$ such that $\sigma_i(hu)(v) = 1$. Then, we often just write $\sigma_i(hu) = v$.
The strategy $\sigma_i$ is \emph{positional} when it is pure and stationary.
Those concepts are naturally generalized to strategy profiles.

\begin{rem}
    Pure strategies can be seen as strategies in the sense of \cref{def:strategies}.
\end{rem}

\subsection{Risk-sensitive equilibria}

Let us first give a new definition of Nash equilibria, taking into account the existence of stochastic vertices and of randomized strategies.
For the sake of readability, we write $\EE(\bsigma)$ for $\EE^{\PP_\bsigma}$.

\begin{defi}[Nash equilibrium]
    Let $\Game_{\|v_0}$ be a game.
    Let $\bsigma$ be a strategy profile in $\Game_{\|v_0}$, let $i$ be a player, and let $\sigma'_i$ be a deviation for player $i$.
    The deviation $\sigma'_i$ is \emph{profitable} if we have $\EE(\bsigma_{-i}, \sigma'_i)[\mu_i] > \EE(\bsigma)[\mu_i]$
    The strategy profile $\bsigma$ is a Nash equilibrium, or NE, if no player has a profitable deviation from $\bsigma$.
\end{defi}

As we saw in the introduction to this chapter, Nash equilibria in stochastic games fail to capture notions of risk aversion (or tolerance). This is why we aim to study generalizations in which the expectation is replaced by other risk measures.
Such generalizations are known as \emph{risk-sensitive equilibria}~\cite{Now05}. We define them here for games played on graphs.
Similarly to what we did with the expectation, when $M$ is a risk measure and $\bsigma$ is a strategy profile, we write $M(\bsigma)$ to denote $M^{\PP_\bsigma}$.

\begin{defi}[Risk-sensitive equilibrium]
    Let $\Game_{\|v_0}$ be a game, and let $\bM = (M_i)_{i \in \Pi}$ be a profile of risk measures.
    Let $\bsigma$ be a strategy profile in $\Game_{\|v_0}$, let $i$ be a player, and let $\sigma'_i$ be a strategy for player $i$, called \emph{deviation} of player $i$ from $\bsigma$.
    The deviation $\sigma'_i$ is \emph{profitable} with regards to the risk measure $M_i$ if we have $M_i(\bsigma_{-i}, \sigma'_i)[\mu_i] > M_i(\bsigma)[\mu_i]$.
    The strategy profile $\bsigma$ is a $\bM$-\emph{risk-sensitive equilibrium}, or $\bM$-RSE, if no player $i$ has a profitable deviation from $\bsigma$ with regards to $M_i$.
\end{defi}

We can now define the following variant of the constrained existence problem, where each player's payoff is replaced by a given risk measure of that payoff.

\begin{prob}[Constrained existence of risk-sensitive equilibria]
    Given a game $\Game_{\|v_0}$, a profile of risk measures $\bM$, and two payoff vectors $\bx, \by \in \QQ^\Pi$, does there exist a $\bM$-RSE $\bsigma$ in $\Game_{\|v_0}$ such that for each $i \in \Pi$, we have $x_i \leq M_i(\bsigma)[\mu_i] \leq y_i$?
\end{prob}

To turn this problem into an algorithmic decision problem, we still need to restrict it to some specific sets of risk measures that can be finitely encoded. 
That is what we do in the sequel of this chapter, with the \emph{entropic risk measure}, and in the next one, with the \emph{extreme risk measure}.

    \section{Entropic risk measure}

The entropic risk measure is a measure of the perceived payoff, which depends on the aversion or inclination of the player toward risk through the exponential utility function. 
It is defined using a \emph{risk parameter}, i.e. a real value $\rho\in\RR \setminus \{0\}$: large positive values indicate risk-averseness, large negative values risk-inclination.
To see a visual representation of the entropic risk measure, see \cref{fig:example_plot} in the introduction of this chapter.

\begin{defi}[Entropic risk measure]\label{def:er}
Given a risk parameter $\rho$, the \emph{entropic risk measure} is defined, for every probability measure $\PP$ and random variable $X$, by:
$$\M_{\rho}^\PP[X] = -\frac{1}{\rho} \log_e \left( \EE^\PP \left[ e^{-\rho X}\right] \right).$$
We generalize this definition by allowing every base $\beta > 1$ instead of Euler's number. The \emph{entropic risk measure with base $\beta$} is then defined by: 
$$\M^\PP_{\beta\rho}[X] = -\frac{1}{\rho} \log_\beta \left( \EE^\PP \left[ \beta^{-\rho X}\right] \right).$$
\end{defi}

The probability measure $\PP$, the parameter $\rho$ and the base $\beta$ may be omitted when they are clear from the context.

\begin{rem}
\begin{itemize}
    \item For every $\beta$ and $\rho$, the entropic risk measure $\M_{\beta\rho}$ is a monotone, normalized and translative risk measure.

    \item By enabling any base $\beta$, we obtain a definition that is more general only on a computational level, since handling Euler's number may not be equivalent to handling rational values.
    Baring computational concerns, these definitions with different bases are equivalent, since for every $\beta$, we have $\M_{\beta\rho} = \M_{e\rho'}$, where $\rho' = \rho \log_e(\beta)$.

    \item The above definition implies that for $\rho = 0$, the function is not defined.
    However, it is known that for all $\PP$, $\beta$ and $X$, the quantity $\M_{\rho}[X]$ converges to $\EE[X]$ when $\rho$ tends to $0$ (see e.g.~\cite{PDM20}).
    Therefore, we henceforth assume that $\M_{0}[X] = \mathbb{E}[X]$ to make risk entropy defined for all finite risk parameters $\rho$.
\end{itemize}
\end{rem}

When we are given a profile $\brho = (\rho_i)_{i \in \Pi}$ of risk parameters, we will sometimes write $\M_{\beta\brho}[\mu]$ for the tuple $\left(\M_{\beta\rho_i}[\mu_i]\right)_{i \in \Pi}$.
Risk entropy defines a family of RSEs, namely the $(\M_{\beta\rho_i})_i$-RSEs, that we also call \emph{$(\beta, \brho)$-entropic risk-sensitive equilibria}, or $(\beta, \brho)$-ERSEs.

As we will see, entropic risk-sensitive equilibria do not resolve the undecidability issue that affects Nash equilibria, for a simple reason: they are merely a generalization of the latter.
However, they represent a first step toward our next notion of equilibrium, \emph{extreme risk-sensitive equilibria}, which we will study in the next chapter.
This chapter must therefore be understood as a warm-up for the next one, with a few first results.
In \cref{sec:ERSE_existence}, we give a sufficient condition for ERSEs to exist, and in \cref{sec:ERSE_complexity}, we give some complexity results.

\section{The existence of ERSEs}\label{sec:ERSE_existence}

The following theorem states the existence of such an RSE that uses no randomness in its strategy profile, in cases where all payoffs are non-negative. 

\begin{thm}[Existence of ERSE]\label{thm:existanceRSE}
    Let $\Game_{\|v_0}$ be a simple stochastic game with only non-negative payoffs.
    Then, there exists a (pure) $(\beta,\rho)$-ERSE in $\Game_{\|v_0}$.
\end{thm}

\begin{proof}
 Pure Nash equilibria always exist in a stochastic multi-player game with prefix-closed Boolean objectives~\cite[Theorem 3.10]{Umm10} (a correction of an existing proof~\cite{CMJ04}). It is known that simple stochastic games where rewards are all positive (or all negative) can be converted into a game with reachability objectives such that if there is an NE in the latter, then there is an NE in the converted game with the reachability objective. Indeed, if all the rewards are positive, we can always scale the rewards for each player of a stochastic game to ensure they are in the unit interval $[0,1]$.
 Then, we can replace terminal vertices by the gadget given by \cref{fig:reachability} (we assume $x_1 \leq \dots \leq x_n$, and in the reachability game, each vertex $r_j$ belongs to the set player $i_j$ wishes to reach).
Therefore, with the same result, Nash equilibria always exist in simple stochastic games with non-negative rewards on the terminals.

\begin{figure}
            \centering
            \begin{subfigure}[b]{0.3\textwidth}
            \label{fig:reachability_terminal}
            \centering
			\begin{tikzpicture}
				\node[initial left] (t) at (0, 0) {$t:~\stackrel{i_1}{x_1}\stackrel{i_2}{x_2} \dots \stackrel{i_n}{x_n}$};
            \end{tikzpicture}
            \caption{A terminal vertex}
            \end{subfigure}
            \begin{subfigure}[b]{0.6\textwidth}
            \centering
            \label{fig:reachability_stochastic}
            \begin{tikzpicture}
                \node[stoch] (a) at (4, 0) {$t$};
                \node[vert] (r1) at (6, 3) {$r_1$};
                \node[vert] (r2) at (6, 1) {$r_2$};
                \node (r3) at (6, -1) {\dots};
                \node[vert] (rn) at (6, -3) {$r_n$};
                \path (a) edge node[above left] {$x_1$} (r1);
				\path (a) edge node[below right] {$x_2-x_1$} (r2);
                \path (a) edge node[below left] {$x_n - x_{n-1}$} (rn);
			\end{tikzpicture}
            \caption{An equivalent gadget}
            \end{subfigure}
			\caption{Converting simple quantitative games into reachability games}
			\label{fig:reachability}
            \end{figure}
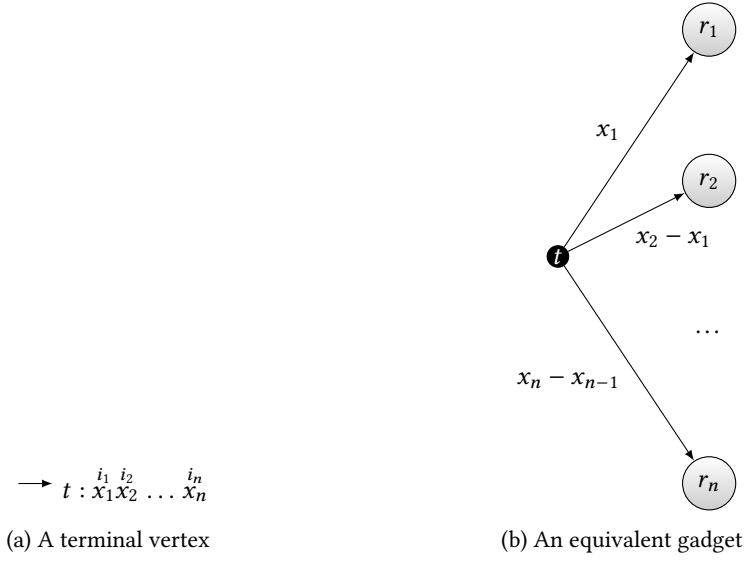

Then, we can conclude our theorem using the following lemma.

\begin{lem}\label{lemma:RSEtoQSSG}
Given a game $\Game_{\|v_0}$ and a tuple $\brho \in \RR^\Pi$, there exists a game $\Game'_{\|v_0}$ with the same underlying graph, player set, and probability function (but possibly different payoff function), such that the $(\beta,\rho)$-ERSEs in $\Game_{\|v_0}$ are exactly the Nash equilibria in $\Game'_{\|v_0}$.
\end{lem}

\begin{proof}
    Consider the simple stochastic game $\Game_{\|v_0} = \tpl{V,E,\Pi,(V_i)_{i\in \Pi},\p ,\mu}$. We will define a payoff  function $\mu'$ over the same set of terminals for game $\Game'_{\|v_0}$ such that the game $\Game'_{\|v_0} = \tpl{V,E,\Pi,(V_i)_{i\in \Pi},\p ,\mu'}$ has a Nash equilibrium if and only if the game $\Game$ has a $(\beta,\brho)$-ERSE.
    For a terminal vertex $t$, we simply define $\mu_i'(t) = 1-\beta^{-\rho_i\mu_i(t)}$ if $\rho>0$, and $\mu_i'(t) = \beta^{-\rho_i\mu_i(t)}-1$ if $\rho<0$.
    
    Consider the function:
    $$\modifiedreward{\beta}{\rho}\colon x\mapsto \begin{cases}
        1-\beta^{-\rho x} &\text{if } \rho>0 \\
        \beta^{-\rho x}-1 &\text{if } \rho<0
    \end{cases}$$
    as the modified reward function.  This function is similar to the negative utility function defined in the work of Baier et al.~\cite{BCMP24}, where they replace terminal rewards with the negative value of $\beta^{-\rho\mu_i(t)}$ (as they assume $\rho>0$), in order to compute the winner in a two-player zero-sum game with risk-averse players. We additionally add or subtract $1$ from their value  to ensure that besides monotonicity, this function also maps the play that does not reach a terminal in the original game to the payoff $0$, and therefore in the modified game to preserve that such plays are still mapped to $0$. 

Since we have $\M_{\beta\rho}\left[X\right] = \frac{-1}{\rho}\log_\beta\tpl{\EE[\beta^{-\rho X}]}$, we immediately obtain the following result.

    \begin{slem}\label{inequality:REvsExp}  
        For any random variable $X$ and constant $r$, for a value $\rho>0$, we have: 
        $$\M_{\beta\rho}\left[X\right] \geq r\text{ if and only if }\EE\left[\modifiedreward{\beta}{\rho}(X)\right] \geq 1-\beta^{-\rho r}$$
    \end{slem}

    Therefore, any Nash equilibrium in the game $\Game'_{\|v_0}$ implies that there is a strategy profile $\bsigma$ such that, for all players $i\in\Pi$, in the MDP induced by $\bsigma_{-i}$, the strategy $\sigma_i$ of player $i$ is an optimal strategy. 

    We first consider the case of a player $i$ where $\rho_i>0$. The case $\rho_i<0$ is analogous, so we omit it. 
    For every strategy $\tau_i$ of player $i$, where $\rho_i>0$, if we write $\btau = (\bsigma_{-i}, \tau_i)$, we have 
    $\EE(\bsigma)[\mu_i']\geq \EE(\btau)[\mu_i']$, since $\bsigma$ is a Nash equilibrium. 
    Since the payoffs of $\Game'$ at a terminal $t$ is just $\modifiedreward{\beta}{\rho}(\mu_i())$, we therefore have $\EE(\bsigma)[\modifiedreward{\beta}{\rho_i}(\mu_i)]\geq \EE(\btau)[\modifiedreward{\beta}{\rho_i}(\mu_i)]$. From \Cref{inequality:REvsExp}, we have:
    $$\EE(\bsigma)\left[\modifiedreward{\beta}{\rho_i}(\mu_i)\right]\geq \EE(\btau)\left[\modifiedreward{\beta}{\rho_i}(\mu_i)\right]$$
    if and only if we have: 
    $$\EE(\bsigma)\left[1-\beta^{-\rho_i \mu_i}\right]\geq \EE(\btau)\left[1-\beta^{\rho_i \mu_i}\right],$$
    that is if and only if we have:
    $$\EE(\bsigma)\left[-\beta^{-\rho_i \mu_i}\right]\geq \EE(\btau)\left[-\beta^{-\rho_i \mu_i}\right].$$
           
Taking $\frac{1}{\rho}\log_\beta$ on both sides,   we get the above is true if and only if:
$$\M_{\beta\rho_i}(\bsigma)[\mu_i] \geq -\frac{1}{\rho_i}\log_\beta\tpl{-\EE\left[\modifiedreward{\beta}{\rho_i}(\btau)[\mu_i]\right] },$$
i.e., if and only if $\M_{\beta\rho_i}[\bsigma](\mu_i) \geq \M_{\beta\rho_i}[\btau](\mu_i)$.
      Therefore, the strategy profile $\bsigma$ is an ERSE. \qedhere \qedhere
\end{proof}
\end{proof}

We conjecture that this result remains true when we remove the guarantee that rewards are non-negative.

\begin{conj}\label{conj:existence_erse}
    Let $\Game_{\|v_0}$ be a simple stochastic game.
    Then, there exists a (pure) $(\beta,\rho)$-ERSE in $\Game_{\|v_0}$.
\end{conj}

\section{Constrained existence problem}\label{sec:ERSE_complexity}

We now turn to the constrained existence problem of $\tpl{\beta,\brho}$-ERSEs.
Unfortunately, it is undecidable in the general case, but we will see that it becomes decidable if we consider restricted classes of strategies.

\subsection{Undecidability in the general case}

\begin{thm}\label{thm:ERSE_undec}
    The constrained existence problem of $\tpl{\beta,\brho}$-ERSEs with $\brho \in \QQ^{\Pi}$ is undecidable, even for any fixed value of $\beta$, for $\brho = \bzero$, and with only nonnegative payoffs.
\end{thm}

\begin{proof}
         The undecidability of the constrained existence problem follows from the work of Ummels and Wojtczak~\cite[Theorem 4.9]{UW11} where they show the undecidability of the constrained existence problem for Nash equilibria in the setting with 10 or more players. Since Nash equilibria constitute a specific instance of the setting of ERSEs where $\brho = \bzero$, the undecidability of our setting follows. 
\end{proof}

\subsection{Restrictions on strategies}

We therefore turn our attention to the constrained existence problem when the class of strategies considered is restricted.
Before we state our theorem, we need to define the existential theory of the reals (with exponentiation in the second case), and the associated complexity class.

\begin{defi}[Existential theory of the reals (with exponentiation)]
    A \emph{formula of the existential theory of the reals}, or \emph{ETR} for short, is a formula of the form $\exists x_1 \dots \exists x_k~\phi(x_1, \dots, x_k)$, where $\phi$ is a formula written with the symbols $0$, $1$, $=$, $\leq$, $<$, $+$, $-$, $\times$, $\wedge$, $\Leftrightarrow$, $\neg$, and parentheses, with the expected syntactic rules and semantics.
    The \emph{existential theory of the reals with exponentiation}, or \emph{ETRE} for short, is defined similarly with one additional symbol: Euler's number $e$.
\end{defi}

We write $\exists\RR$ for the complexity class of problems that can be reduced in polynomial time to the problem of deciding the validity of an ETR formula.
That class is known to be included in $\PSpace$, by the following lemma.

\begin{lem}[\cite{Can88}]\label{lem:etr}
    Deciding the validity of a formula in ETR is $\PSpace$-easy.
\end{lem}

Similarly, we write $\exists\RR(e)$ for the class of problems that can be reduced in polynomial time to the validity problem of an ETRE formula.
That class is known to be included in $3\ExpTime$.

\begin{lem}[\cite{gallegohernandez_et_al:LIPIcs.STACS.2025.37}]\label{lem:etre}
    Deciding the validity of a formula in ETRE is $3\ExpTime$-easy.
\end{lem}

We can now state our theorem.

\begin{thm}\label{thm:ERRSErestricted}
The constrained existence problem of $(\beta,\brho)$-ERSEs, in quantitative simple stochastic games:
\begin{enumerate}
    \item remains undecidable when players are restricted to pure strategies;\label{itm:ERRSEitmundec}
    \item is decidable when players are restricted to stationary strategies, or to positional strategies:\label{itm:ERRSEdecidable}
\begin{enumerate}
        \item in $3\ExpTime$ if $\beta = e$ and the risk-parameters $\rho_i$ are rational;\label{itm:ERRSEitm3Exp}
        \item and is $\exists\RR$-easy if the risk parameters and the base $\beta$ are rational.
        In the stationary case, the problem is $\exists\RR$-complete, and in the positional case, it is $\NP$-hard.\label{itm:ERRSE:ETR}
    \end{enumerate}
\end{enumerate}
\end{thm}

\begin{proof}
\paragraph{Undecidability result}

The undecidability of the pure case is inherited from Nash equilibria~\cite[Theorem~4.9]{UW11}, since the reduction for undecidability uses only pure strategies.

\paragraph{Decidable subcases}

\subparagraph*{Decidability lowerbounds.}
$\NP$-hardness has been shown in the case of positional NEs, i.e. when each player's risk parameter is $0$, by Ummels and Wojtczak~\cite[Theorem 4.4]{UW11}.
Similarly, $\exists\RR$-hardness has been proved for the constrained existence problem of stationary NEs by Kristoffer Arnsfelt Hansen and Steffan Christ Sølvsten~\cite{DBLP:conf/mfcs/HansenS20}.

\subparagraph*{Decidability upperbounds.}

Our proof is similar to the one by Ummels and Wojtczak~\cite[Theorem 4.5]{UW11}.
However, we need to do slightly more work to encode the payoff expressed by the entropic risk measure.

First, observe that it is enough to verify if there is a stationary Nash equilibrium in the modified game obtained where all the terminal rewards $\mu_i(v)$ are replaced instead with $1-\beta^{\rho_i\mu_i(v)}$. This follows from \cref{lemma:RSEtoQSSG}. 

Since the players are restricted to strategies that are stationary, we give a non-deterministic algorithm that uses the solution to sentences in $\exists\RR$ if the values of $\beta$ and $\rho_i$s are rational.  Since we have $\NPSpace = \PSpace$ by Savitch's theorem, this does not change the complexity.
     
     For a game $\Game_{\|v_0} = \tpl{V,E,\Pi,(v_i)_{i\in \Pi}, \p,\mu}$ and two payoff vectors $\bx$ and $\by$, our algorithm guesses, first, the support $S \subseteq E$ of the strategies that will be considered; that is, the set of edges that will be used with positive probability.

\begin{lem}\label{lem:existR}
            For any $z$, which requires $\ell$ bits to encode, there is a formula in ETR that uses only polynomially many variables in $\ell$ to encode $\modifiedreward{\beta}{\rho_i}(z) = 1-\beta^{-\rho_i z}$, where $\beta$ and $\rho_i$ can also be represented in ETR using a polynomial formula.
            If $\beta = e$ and $\rho_i$ is rational, then $\modifiedreward{e}{\rho}(z) = 1-\beta^{-\rho z}$ can be expressed in ETRE using a formula of most polynomial length.
\end{lem}

\begin{proof}
            In the first case, we assume without loss of generality that $\beta$ is a natural number. If $\beta$ is rational instead, and is represented by a value $\frac{a}{b}$, then we can individually find $a_1  = a^{\rho_i z}$ and $b_1 = b^{\rho_i z}$, and just find $\frac{b_1}{a_1}$, which is written in ETR with the formula $\exists  t_r \: \exists a_1' \: (a_1'\times a_1  = 1) \land (t_r = a_1'\times b_1)$.
        Now, under this assumption, we deal with fixed finite exponentiation with rational values. Similarly, we can assume without loss of generality that $\rho_i z$ is a natural number. Otherwise, we can write $r^\frac{a}{b} = r^a\times z^\frac{1}{b}$ and $z^\frac{1}{b}$ can be defined by the formula $\exists y \colon  y^b = z$.

        It now suffices to show that for two values $b,a$, both natural numbers, the quantity $b^a$ can be expressed in ETR succinctly, using only a formula that has length that is not more a poly-log of $b$ or $a$. 
        Let $a = \sum_{i=0}^{\log_2{a}}a_i 2^i$, where $a_i\in\{0,1\}$. 
        This follows from the following observations. 
        \begin{itemize}
            \item First, we can write $b^a = \prod_{i=1}^{\log_2{a}}\tpl{b^{a_i}b^{2^i}}$.
            \item Second, the quantity $2^i$ can be expressed in a formula with at most $i+1$ many variables.
            \item Third, the quantity $b^{2^i}$ requires at most $i$ many variables to express, because if $b_i$ represents $b^{2^{i}}$, then we have $b^{2^{i+1}} = b^{2^i}\times b^{2^i}$.
            \item Finally, using a similar trick, the quantity $b$ itself can be represented using at most $\log_2{b}$-many variables. 
        \end{itemize}
        
        Finally, if $\beta = e$ and $\rho_i$ is rational, the quantity $e^{-\rho z}$ can be expressed following the same reasoning by a succinct formula that uses the symbol $e$.
\end{proof}

    \cref{lem:existR} ensures that we can efficiently represent the variables used for the payoffs of the modified game.
    We now can write an equation assuming that all terminal rewards are available to us as constants. 
    This will write the equation in three parts. Since we have guessed the support, we first ensure that, in fact, there are variables corresponding to the probabilities of the strategy that only take positive values on the edges corresponding to the support set that we guessed.  
    Then, we write equations using variables that compute the values of the induced Markov chain from this strategies. Finally, we also have a formula whose solution corresponds to the values of the MDP obtained for each player when playing against the strategies of all other players. Then we compare if the value of the MDP is at least as large as the underlying Markov chain for each player, to ensure that it is indeed an equilibrium. To write all of this in ETR, we introduce the following variables:
     \begin{itemize}
         \item one variable  $p_{vw}$ for each pair of vertices $vw$, which corresponds to the probabilities associated with the strategy profile;
         \item a variable $r^i_v$ which corresponds to the entropic risk measure of player $i$ from vertex $v$ if they follow the strategy defined by the probabilities above; 
         \item a variable $m^i_v$ which corresponds to the value obtained by player $i$ if the game is treated as an MDP against other players.
     \end{itemize}

We can now define our formulae.
 
     \begin{lem}\label{prop:ETRformula}
        Given a game $\Game_{\|v_0}$, two payoff vectors $\bx$ and $\by$, and a subset $S \subseteq E$:
         \begin{itemize}
             \item \emph{if $\beta$ and all $\rho_i$ are rational values:} there exists a formula in ETR, computable in polynomial time, that is satisfied if and only if there exists a stationary ERSE $\bsigma$ with $x_i \leq \M_{\rho_i}(\bsigma)[\mu_i] \leq y_i$ for each $i \in \Pi$ that uses exactly (with positive probability) the edges in $S$;
             
             \item \emph{if $\beta = e$, and if all $\rho_i$ are rational values:} there exists a formula in ETRE, computable in polynomial time, that is satisfied if and only if there exists a stationary ERSE $\bsigma$ with $x_i \leq \M_{\rho_i}(\bsigma)[\mu_i] \leq y_i$ for each $i \in \Pi$ that uses exactly (with positive probability) the edges in $S$.
             
             \item Those results remain true if the restriction to stationary XRSEs is replaced by a restriction to positional XRSEs.
         \end{itemize}
    \end{lem}
    
    \begin{proof}
    This proof is similar to the one found in Ummels and Wojtczak~\cite[Theorem 4.5]{UW11}, but we provide it to suit our setting, for the sake of completeness. 
    First, we have a formula that states that the values $p_{vw}$ indeed describe a strategy. We further ensure that for stochastic vertices, the value $p_{vw}$ encodes exactly the value dictated by the probability function $\p$ by the stochastic vertex:
    \begin{align*}
    \Phi_S(\Bar{p}) =& \bigwedge_{vw\in S} \left(\left( p_{vw}> 0\right) \wedge \left(p_{vw} \leq 1\right)\right)
    \land \bigwedge_{vw\in E \setminus S} \left( p_{vw}= 0\right) \\
    &\land \bigwedge_{i\in \Pi}\bigwedge_{v\in V_i} \tpl{\sum_{w\in E(v)} p_{vw}=1}\land 
    \bigwedge_{v\in V_?} \left( p_{vw} =  \p(vw)\right).
    \end{align*}
    For a fixed support $S$ of a strategy $\bsigma$, it is possible to compute the set $T_S$ of terminal vertices that have non-zero probability of being reached in the underlying Markov chain that is formed, and the vertices $V_S$ from which such terminals can be reached with non-zero probability.
    For a given terminal vertex $t$ and player $i$, we write $\mu'_i(t)$ for the quantity $1-\beta^{\rho_i \mu(t)}$.
    We can now write a statement that define the values of the variables $r^i_v$.
    \[\Omega_S^i(\Bar{p},\Bar{r}^i)\:= \bigwedge_{t\in T_S} \left(r^i_t =\mu_i'(t)\right) \land
                 \bigwedge_{v\notin V_S} \left( r^i_v = 1\right) \land
                 \bigwedge_{v\in V_S\setminus T_S} \tpl{r^i_v = \sum_{w\in E(v)}p_{vw} r^i_v}.
                 \]     
    To compute the values of the variables $m^i_v$, we construct a similar first-order statement:
    \[\Psi_S^i(\Bar{p},\Bar{m}^i) \:= \bigwedge_{t\in T} \left( m^i =\mu_i'(t) \right) \land
                 \bigwedge_{v\in V_i, w\in E(v)} \left(m_v^i\geq m_w^i\right) \land
                 \bigwedge_{v\notin V\setminus V_i} \tpl{m^i_w = \sum_{w\in E(v)} p_{vw} m^i_v}.\]
Then, our statement is:
$$\exists \Bar{p}\:\exists\Bar{r}\:\exists\Bar{m}~ \Phi(\Bar{p})\land \bigwedge_{i\in\Pi}\left(\tpl{x^i_{v_0}\leq r^i_{v_0}}\land \tpl{r^i_{v_0}\leq y^i_{v_0}}\land\Omega_S^i(\Bar{p},\Bar{r}^i)\land \Psi_S^i(\Bar{p},\Bar{m}^i)\land \left( m_{v_0}^i\leq r_{v_0}^i \right) \right).$$
Using \cref{lem:existR}, this formula belongs to ETR if $\beta$ is rational, and to ETRE if $\beta = e$.

Finally, this result can be extended to the positional case by slightly changing the definition of the formula $\Phi_S(\bar{p})$:
\begin{align*}
    \Phi_S(\Bar{p}) = &\bigwedge_{vw\in S} \left( p_{vw}=1\right)
    \land \bigwedge_{vw\in E \setminus S} \left( p_{vw}= 0\right) \\
    &\land \bigwedge_{i\in \Pi}\bigwedge_{v\in V_i} \tpl{\sum_{w\in E(v)}\p_{vw}=1}\land 
    \bigwedge_{v\in V_?} \left( p_{vw} =  \p(vw)\right)
    \end{align*}
    \end{proof}

The conclusion follows from \cref{lem:etr,lem:etre}.
\end{proof}

%% file: 3cXRSE.tex
In \cref{chap:RSE}, we introduced the formalism of simple stochastic games, randomized strategies, and the notion of \emph{risk-sensitive equilibria} for an abstract profile of risk measures.
We studied a first risk measure: the \emph{entropic risk measure}, which is widely used in economics and finance.
We showed that, unless strategies are restricted to stationary or positional ones, the constrained existence problem remains undecidable in this setting.
A natural question then arises: can alternative risk measures be found that make this problem decidable?
For this reason, in this chapter, we introduce a risk measure derived from the limit cases of the entropic risk measure: the \emph{extreme risk measure}.

Let us first examine what such a measure means in a two-player zero-sum game.
Consider a pirate attempting to hack a computer system, such as a bank server.
The interactions between these two agents may involve randomized actions---for instance, when the bank server generates a security key or when the hacker tries to guess a user's password. They may also include external stochastic processes, such as the number of users attempting to connect to the bank server at any given time.
To safeguard users' funds, the bank server must be programmed in a \emph{pessimistic} manner, assuming that any security breach with a nonzero probability is a security breach that will inevitably occur. 
Conversely, the pirate can be seen as an \emph{optimistic} agent, content with even a small probability of success---he may be able to retry his attempts indefinitely, or to execute them in parallel using a large network of computers.
Such interactions are often modeled using non-stochastic games, by merging the pirate and random events into a single \emph{environment} player.
However, this approach is not entirely equivalent to a probabilistic one, as the latter allows discarding events that have probability zero---such as consistently winning heads-or-tails-like interactions over an infinite horizon.

More importantly, the adversarial approach cannot be generalized to multiplayer settings, in which it may be necessary to account for the existence of stochastic phenomena that are perceived differently by each agent.
Consider, for example, two secured computer systems interacting with each other---say, a bank server and an online selling platform.
Their behavior cannot be properly modeled without accounting for the fact that both are pessimistic: each has a specification to uphold and cannot tolerate even a minimal probability of being compromised.
If a pirate discovers a method that grants a nonzero probability of breaching either the bank server or the selling platform (but never both simultaneously), then each system will anticipate its own failure.
The bank server will assume that it will be hacked, while the selling platform will assume the same for itself---even though these two scenarios are mutually exclusive.

This example highlights the importance of studying equilibria where all players are either extreme optimists or extreme pessimists---that is, players who, when faced with a random experiment, always expect either the best or the worst possible outcome. In other words, we consider risk-sensitive equilibria defined by risk measures corresponding to the extreme cases of risk entropy---what we refer to as the extreme risk measure.

\section{The extreme risk measure}

\subsection{Definition}\label{ssec:def_xr}

Let us consider a random variable $X$ that ranges over $\RR$.
The \emph{pessimistic risk measure} of $X$ is the highest value $x$ such that $X$ almost-surely takes a value above $x$.
When $X$ takes finitely many values, that corresponds to the least value that it takes with positive probability.
In probability theory, that measure is sometimes referred to as \emph{essential infimum}, written $\essinf$.
The definition of \emph{optimistic risk measure} is symmetric.

\begin{defi}[Optimistic, pessimistic risk measure]\label{def:opt_pess}
The \emph{pessimistic risk measure} of a random variable $X$ is defined by
$\PM^\PP[X] = \essinf(X) = \sup \{x \in X ~|~ \PP(X \geq x) = 1\}$.
Analogously,  the \emph{optimistic risk measure} of $X$ is $\OM^\PP[X] = \esssup(X) =  \inf \{x \in X ~|~ \PP(X \leq x) = 1\}$.    
\end{defi}

When we are given a game $\Game_{\|v_0}$, we can assign a risk measure to each player by defining a partition $(P, O)$ of $\Pi$, where the set $P$ represents the set of players that are \emph{pessimists}, whose perceived payoffs are defined by the pessimistic risk measure, while $O$ represents the \emph{optimists}, who intend to maximize their optimistic risk measure.
For convenience, we group both measures under the umbrella term \emph{extreme risk measure (XR)}, and often assume that the pair $(P, O)$ is given;
then, we write $\X_i$ for $\PM$ when $i \in P$, and for $\OM$ when $i \in O$.
Since each player $i$ is usually interested only in the risk measure of their own payoff, we will also write $\X_i(\bsigma)$ for the quantity $\X_i(\bsigma)[\mu_i]$.
We define \emph{extreme risk-sensitive equilibria}, or XRSEs for short, as $(\X_i)_i$-RSEs.

\subsection{Link with the entropic risk measure}

We show that our definition of extreme risk measure corresponds to the limit cases of entropic risk measure.
Observe that in \cref{fig:example_gamma}, following the  blue strategy, the only payoffs that are obtained with positive probability were $40$ and $0$, which are also the limits of the risk entropy when $\rho$ tends to infinite values.
On the other hand, in the red strategy, the only payoff obtained with positive probability is the payoff $1$. Although the payoff $0$ is possible since the play $a^\omega$ is compatible with every strategy, this outcome must be ignored since it is realized with probability $0$.

\begin{thm}\label{thm:RE=PEorOE}
    Let $X$ be a random variable that ranges over $\RR$, and let $\beta > 1$.

    \begin{itemize}
        \item The limit risk entropy of $X$ when $\rho$ tends to $+\infty$ exists and is equal to the pessimistic risk measure, that is, we have $\lim_{\rho \to +\infty} \M_{\beta\rho} [X] = \PM[X]$.    
       \item Similarly, the limit risk entropy of $X$ when $\rho$ tends to $-\infty$ exists and is equal to the optimistic risk measure, that is, we have $\lim_{\rho \to -\infty} \M_{\beta\rho} [X] = \OM[X]$.
    \end{itemize}
\end{thm}

\begin{proof}
    \paragraph{Limit in $+\infty$}
    
    First, let us note that for every $\rho$, we always have $\M_{\beta\rho}[X] \geq \PM[X]$.
        Let now $\epsilon > 0$.
        We want to prove that there exists $\rho_0 \in \RR$ such that for every $\rho \geq \rho_0$, we have $\M_{\beta\rho}[X] \leq \PM[X] + \epsilon$.

        Let us first notice that we have:
       \begin{align*}
           \M_{\beta\rho}[X] &= -\frac{1}{\rho} \log_\beta \left( \int_{x \in \RR} \beta^{-\rho x} \d \PP(X = x) \right)\\
            &= -\frac{1}{\rho} \log_\beta \left( \int_{x \in \RR} \beta^{-\rho \PM[X]} \beta^{-\rho (x-\PM[X])} \d \PP(X = x) \right)\\
            &= \PM[X] -\frac{1}{\rho} \log_\beta \left( \int_{x \in \RR} \beta^{-\rho (x-\PM[X])} \d \PP(X = x) \right)\\
            &= \PM[X] -\frac{1}{\rho} \log_\beta \Bigg( \int_{x \leq \PM[X] + \frac{\epsilon}{2}} \beta^{-\rho (x-\PM[X])} \d \PP(X = x) \\
            &\qquad\qquad\qquad+ \int_{x \geq \PM[X] + \frac{\epsilon}{2}} \beta^{-\rho (x-\PM[X])} \d \PP(X = x) \Bigg)\\
             &\leq \PM[X] - \frac{1}{\rho} \log_\beta \left( \int_{x \leq \PM[X] + \frac{\epsilon}{2}} \beta^{-\rho \frac{\epsilon}{2}} \d \PP(X = x) + 0 \right)\\
            &= \PM[X] - \frac{1}{\rho} \log_\beta \left( \PP\left(X \leq \PM[X] + \frac{\epsilon}{2}\right) \beta^{-\rho \frac{\epsilon}{2}} \right)\\
            &= \PM[X] - \frac{1}{\rho} \log_\beta \left( \PP\left(X \leq \PM[X] + \frac{\epsilon}{2}\right)\right) + \frac{\epsilon}{2}.
        \end{align*}

        For $\rho$ large enough, this quantity is indeed smaller than $\PM[X] + \epsilon$.

        \paragraph{Limit in $-\infty$}
        
        Let us first notice that for every $\beta, \rho, X$, we have the equality $\M_{\beta\rho}[X] = -\M_{\beta(-\rho)}[-X]$.
        Thus, we can apply the previous result, and find:
        \begin{align*}
        \lim_{\rho \to -\infty} \M_{\beta\rho} [X] 
        &= \lim_{\rho \to -\infty} -\M_{\beta(-\rho)} [-X]
        \\&= -\lim_{\rho \to +\infty} \M_{\beta\rho} [-X]
        \\& =  - \PM[-X]
        \\ &= - \inf \{x \in \RR ~|~ \PP(-X \leq x) > 0\}
        \\ &= - \inf \{x \in \RR ~|~ \PP(X \geq -x) > 0\}
        \\ &= \sup \{x \in \RR ~|~ \PP(X \geq x) > 0\}
        \\ &= \OM[X].
        \end{align*}
\end{proof}

\subsection{A technical lemma}

Before moving to our result, we give the following lemma, that simply rephrases a well-known result adapted to our context, but that will be extensively used in the sequel.

\begin{lem}\label{lm:secretlemma}
    Let $\Game_{\|v_0}$ be a game with two players, called $i$ and $j$.
    We assume given a partition $(P, O)$ of $\{i, j\}$.
    Then, the quantity:
    $$\inf_{\sigma_j \in \Strat_j\Game_{\|v_0}} \sup_{\sigma_i \in \Strat_i\Game_{\|v_0}} \X_i(\bsigma)$$
    can be computed in time $O(m)$, where $m$ is the number of edges in $\Game$.
    Moreover, the infimum is reached with a positional strategy of player $j$; and there is a positional strategy of player $i$ that realizes the supremum for every strategy of player $j$.

    Consequently, the optimality of positional strategies and the $O(m)$ upper bound also hold in Markov decision processes, and in Markov chains; and, on the other hand, it holds when $j$ is a fictional player that represents a coalition of players who all have as unique objective to minimize player $i$'s risk measure.
\end{lem}

\begin{proof}
    The $O(m)$ upper bound holds by a slight adaptation of the classical attractor algorithm~\cite[Chapter~5.3]{AG11}.
    Note that those algorithms run in time $O(m+n)$, where $n$ is the number of vertices; but here, we assumed that each vertex (except possibly $v_0$) has at least one ingoing edge, hence $n \leq m+1$ and $m+n = O(m)$.
    That algorithm immediately induces positional optimal strategies.
    Another way to obtain that second result, however, is the following: once the quantity $x = \inf_{\sigma_j} \sup_{\sigma_i} \X_i(\bsigma)$ is known, strategies that realize the infimum and the supremum can be seen as optimal strategies in the Boolean zero-sum game in which player $i$ wants with positive probability (if they are optimist) or with probability $1$ (if they are pessimist) to reach the set of terminals yielding them at least payoff $x$ (if $x > 0$) or to avoid the set of terminals yielding them less than payoff $x$ (if $x \leq 0$).
    This is then a reachability game (seen either from player $i$'s of from player $j$'s perspective), and it is well-known~\cite{AG11} that in such a game, for both players, positional strategies suffice to maximize the probability of winning.
    In particular, if one has a strategy to win that game with positive probability, or with probability $1$, there is also such a strategy that is positional.
\end{proof}

\section{The existence of XRSEs}

We now answer a fundamental question about every notion of equilibrium, which is the condition under which it exists. 
We show that (stationary) XRSEs are guaranteed to exist in games with only non-negative rewards, similarly as ERSEs.
But our proof does not rely on the same arguments, and we instead give a proof that goes along with an algorithm.

\begin{thm}\label{thm:XRSEexists}
    Let $\Game_{\|v_0}$ be a game with only non-negative rewards, and let $(P,O)$ be a partition of $\Pi$.
    Then, there exists a stationary XRSE in $\Game_{\|v_0}$.
    Moreover, there exists an algorithm that, given such a game, outputs the representation of such an XRSE in time $O(m^2 p)$, where $m$ is the number of edges, and $p$ the number of \emph{pessimistic} players.
\end{thm}

\begin{proof}
    \paragraph{Example}

    Our algorithm generates an XRSE by constructing a decreasing sequence $E = E_0, E_1, \dots$ of sets of edges, and considering, for each $k$, the stationary strategy profile that randomizes between all the outgoing edges in $E_k$ from all vertices.
    
    Let us illustrate it with the game depicted by Figure~\ref{fig:ex_extreme1}, which involves two pessimists, player $\Circle$ and player $\Square$.
    In that game, both players want to leave the cycle, but each of them would prefer the other player to leave.
    If we first consider the strategy profile that always randomizes between all the available edges, then both terminal vertices are reached with positive probability, and it is almost sure that one of them is reached: each player, as a pessimist, will therefore expect that they will leave the cycle first, and get payoff $1$.
    Thus, their risk measure is $1$.
    Then, player $\Square$ (and symmetrically player $\Circle$) has a profitable deviation by refusing to leave the cycle, and by always going back to the vertex $a$: then, it is almost sure that player $\Circle$ will eventually leave the cycle, and player $\Square$'s risk measure is now $2$.
    Note that player $\Circle$ cannot detect such a deviation of strategy, since she does not have access to the internal coins tossed by player $\Square$.
    
    Then, we remove the edge $bt_2$ (or $at_1$).
    This results in a set of edges such that, if we consider again the strategy profile where both players always randomize, from each vertex, between the remaining outgoing edge, then player $\Square$ gets risk measure $2$, and player $\Circle$ cannot get more than $1$ by deviating.
    That new strategy profile is a (stationary) XRSE.

    \paragraph{Algorithm}

    Throughout this proof, for a given set of edges $F \subseteq E$, we write $\Game^F$ for the game obtained from $\Game$ by removing all the edges that do not belong to $F$.
    In that game, we define $\bsigma^F$ as the stationary strategy profile that maps each vertex $v$ to some probability distribution whose support is $F(v)$.
    Note that the probabilities do not matter here: we are only interested in the support of the distribution of the strategy profile.

    \begin{figure}[h] 
			\centering
            \begin{subfigure}[t]{0.3\textwidth}
            \centering
			\begin{tikzpicture}
				\node[initial above, vert] (a) at (0, 2) {$a$};
				\node[vert, rectangle] (b) at (2, 2) {$b$};
                \node (t1) at (0, 0.75) {$t_1:~\stackrel{\circ}{1}~\stackrel{\square}{2}$};
                \node (t2) at (2, 0.75) {$t_2:~\stackrel{\circ}{2}~\stackrel{\square}{1}$};
                \path (a) edge[bend left] (b);
				\path (b) edge[bend left] (a);
                \path (a) edge (t1);
                \path (b) edge (t2);
			\end{tikzpicture}
			\caption{Two pessimists}
			\label{fig:ex_extreme1}
            \end{subfigure}
            \hfill
            \begin{subfigure}[t]{0.3\textwidth}
            \centering
			\begin{tikzpicture}
                \node[initial above, stoch] (c) at (1, 3) {$c$};
				\node[vert] (a) at (0, 2) {$a$};
				\node[vert, rectangle] (b) at (2, 2) {$b$};
                \node (t1) at (0, 0.75) {$t_1:~\stackrel{\circ}{1}~\stackrel{\square}{2}$};
                \node (t2) at (2, 0.75) {$t_2:~\stackrel{\circ}{2}~\stackrel{\square}{1}$};
                \path (c) edge (a);
                \path (c) edge (b);
                \path (a) edge[bend left] (b);
				\path (b) edge[bend left] (a);
                \path (a) edge (t1);
                \path (b) edge (t2);
			\end{tikzpicture}
			\caption{Two pessimists with a common coin}
			\label{fig:ex_extreme2}
            \end{subfigure}
            \hfill
            \begin{subfigure}[t]{0.3\textwidth}
            \centering
			\begin{tikzpicture}
                \node[initial above, stoch] (c) at (1, 4.25) {$c$};
                \node[vert] (d) at (-0.5, 3.25) {$d$};
                \node[vert, rectangle] (e) at (2.5, 3.25) {$e$};
				\node[vert] (a) at (0, 2) {$a$};
				\node[vert, rectangle] (b) at (2, 2) {$b$};
                \node (t1) at (0, 0.75) {$t_1:~\stackrel{\circ}{1}~\stackrel{\square}{2}$};
                \node (t2) at (2, 0.75) {$t_2:~\stackrel{\circ}{2}~\stackrel{\square}{1}$};
                \node (t3) at (1, 3.25) {$t_3:~\stackrel{\circ}{2}~\stackrel{\square}{2}$};
                \path (c) edge (d);
                \path (c) edge (e);
                \path (d) edge (t3);
                \path (e) edge (t3);
                \path (d) edge (a);
                \path (e) edge (b);
                \path (a) edge[bend left] (b);
				\path (b) edge[bend left] (a);
                \path (a) edge (t1);
                \path (b) edge (t2);
			\end{tikzpicture}
			\caption{Two pessimists with a common coin and some temptation}
			\label{fig:ex_extreme3}
            \end{subfigure}
        \caption{Some games involving two pessimistic players}\label{fig:ex_extreme}
\end{figure}
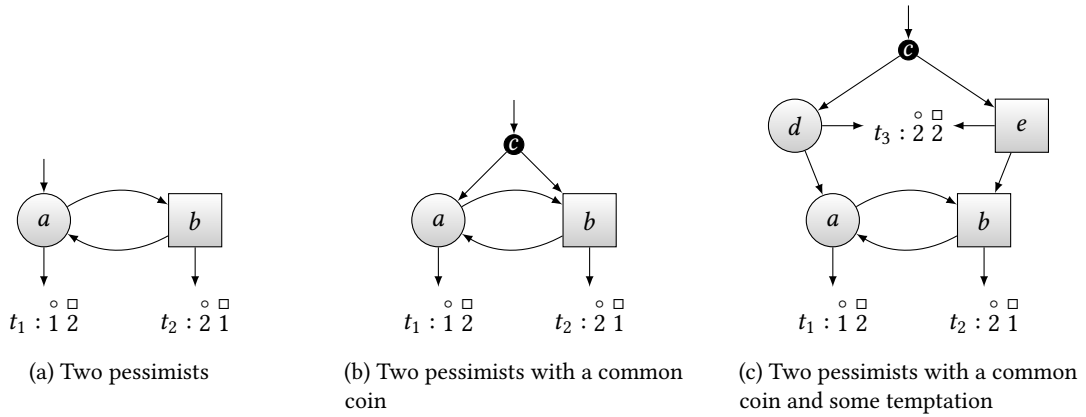
    
    We proceed by presenting the algorithm, \cref{algo:existence}, that takes as an input the game $\Game_{\|v_0}$ and the partition $(P, O)$, and returns a subset $F \subseteq E$ such that, as we will show, the strategy profile $\bsigma^F$ is always an XRSE.
    That algorithm defines a decreasing sequence $E_0, E_1, \dots$ of subsets of $E$, where $E_0 = E$.
    At each step $k$, for each pessimist $i$, it computes the risk measure $z_i^k$ of player $i$ in $\bsigma^{E_k}$, and then the set $W_i^k$ of vertices $v$ such that, from $v$, whatever player $i$ does, that player almost surely gets a payoff smaller than or equal to $z_i^k$.
    If we have $v_0 \in W_i^k$ for each $i$, then the algorithm stops there and returns the set $E_k$ (and we will show below that it means that $\bsigma^{E_k}$ is an XRSE).
    Otherwise, we pick player $i$ such that $v_0 \not\in W_i^k$ (a player who provably has a profitable deviation), and define $E_{k+1}$ by removing all the edges accessible from $v_0$ leading from $V \setminus W_i^k$ to $W_i^k$.

            \begin{algorithm}
            \begin{algorithmic}\caption{Exhibition of one stationary XRSE}\label{algo:existence}
                \Procedure{Existence}{$\Game_{\|v_0}, P, O$}
                    \State $k \gets 0$
                    \State $E_k \gets E$
                    \While{$\top$}
                        \State Compute $A^k = \{v \in V \mid v \text{ is accessible from } v_0 \text{ in } (V, E_k)\}$
                        \ForAll{$i \in P$}
                            \State Compute $z^k_i = \X_i(\bsigma^{E_k})$
                            \State Compute $W^k_i = \left\{v \in V ~\left|~ \forall \tau_i \in \Strat_i \Game^{E_k}_{\|v_0}, \text{we have } \PP_{\bsigma^{E_k}_{-i}, \tau_i}(\mu_i \leq z^k_i) > 0\right.\right\}$
                        \EndFor
                        \If{$\exists i$ such that $v_0 \not\in W_i^k$}
                            \State Pick one such $i$
                            \State $E_{k+1} \gets E_k \setminus ((A^k \setminus W_i^k) \times W_i^k)$
                            \State $k \gets k+1$
                        \Else
                            \State \Return $E_k$
                        \EndIf
                    \EndWhile
                \EndProcedure
            \end{algorithmic}
        \end{algorithm}

    \paragraph{Correctness}
    
    A first quick invariant that we need to prove is the following one, which will guarantee that the games $\Game^{E_k}$ and the strategies $\bsigma^{E_k}$ are well-defined.

    \begin{inv}\label{inv:outgoingedges}
        For each $k$, each stochastic vertex $v$, we have $E(v) \subseteq E_k$, and for each non-stochastic vertex $v$, we have $E(v) \cap E_k \neq \emptyset$.
    \end{inv}
    
\begin{proof}
    The set $E_0 = E$ trivially satisfies the invariant.

    Now, let us assume that $E_k$ satisfies the invariant.
    At step $k$, an edge is removed if and only if it goes from a vertex $u \in A^k \setminus W^k_i$ to a vertex $v \in W^k_i$.
    Consider a stochastic vertex $u \in A^k$: if it has an edge that leads to vertex $v \in W^k_i$, then whatever player $i$ plays from $u$, with positive probability, the vertex $v$ is reached; and then, if the other players play the strategy profile $\bsigma^{E_k}$, then with positive probability, player $i$ gets the payoff $z^k_i$ or less.
    Hence $u \in W_i^k$, and the edge $uv$ is not removed, and remains in the set $E_{k+1}$.
    Similarly, if $u$ is not a stochastic vertex, but all its outgoing edges lead to a vertex that belong to $W_i^k$, then the vertex $u$ itself belongs to $W_i^k$, hence the outgoing edges of $u$ will not all be removed.
    The invariant is therefore still true at step $k+1$, and by induction, is true for all $k$.
\end{proof}
    
Each step of the algorithm is then also properly defined.
Moreover, we have termination.

\begin{prop}
    \cref{algo:existence} terminates.
\end{prop}

\begin{proof}
    With \cref{inv:outgoingedges}, we now know that \cref{algo:existence} successfully constructs a sequence $E_0, E_1, \dots$ of sets of edges until it stops and returns the last of those sets.
    Termination is an immediate consequence of the fact that this sequence is decreasing.

    Indeed, for each step $k$ at which nothing is returned, there exists a player $i$ with $v_0 \not\in W_i^k$.
    On the other hand, the set $W_i^k$ is necessarily accessible from $v_0$:

    \begin{lem}
        The set $W_i^k$ is nonempty, and accessible from $v_0$ in the graph $(V, E_k)$.
    \end{lem}

    \begin{proof}
        If $z^k_i$ is obtained by reaching a terminal vertex $t$, then we have $t \in W_i^k$, and $t$ is accessible from $v_0$.
        If now $z^k_i = 0$ is obtained by reaching no terminal vertex, then when following $\bsigma^k$, with positive probability, no terminal is reached.
        Then, there is in particular a vertex $u$ that has positive probability of being visited infinitely often.
        And when playing $\bsigma^k$ from $u$, the probability that some terminal is ever reached is actually $0$, since if it was some constant $q > 0$, then the probability of visiting $u$ infinitely often would be $\lim_\l (1-q)^\l = 0$.
        In other words, no terminal vertex is accessible from $u$ in $(V, E_k)$, and then, we have $u \in W_i^k$.
    \end{proof}

    Now, along a play that starts from $v_0 \not\in W_i^k$ and visits $W_i^k$, there exists at least one edge that goes from a vertex that does not belong to $W_i^k$, to a vertex that does.
    Such an edge is then removed in the set $E_{k+1}$, which is therefore strictly included in the set $E_k$.
    This holds for every $k$, ensuring termination.
\end{proof}

We now know that the algorithm terminates, i.e., constructs a finite decreasing sequence $E = E_0, E_1, \dots, E_n$, and then returns the set $E_n$, as a succinct representation of the stationary strategy profile $\bsigma^{E_n}$.
What remains to be proven is that this strategy profile is an XRSE.
Before proving that it is an XRSE in the game $\Game_{\|v_0}$, we first prove that it is one in the game $\Game^{E_n}_{\|v_0}$, i.e., when the edges that have been removed cannot be used to deviate.

\begin{prop}\label{prop:xrseGEn}
    The strategy profile $\bsigma^{E_n}$ is an XRSE in the game $\Game^{E_n}_{\|v_0}$.
\end{prop}

\begin{proof}
    Consider a player $i$, and a deviation $\sigma'_i$ of player $i$ from the strategy profile $\bsigma^{E_n}$ in the game $\Game_{\|v_0}^{E_n}$.
    Let $x = \X_i(\bsigma^{E_n}_{-i}, \sigma'_i)$.

    \subparagraph*{If player $i$ is an optimist.}
    If $x = 0$, then since all rewards are non-negative, we have $x \leq \X_i(\bsigma^{E_n})$.
    If $x > 0$, then the payoff $x$ is obtained by reaching a terminal vertex $t$.
    But then, that terminal vertex is accessible from $v_0$ in the graph $(V, E_n)$, and is therefore also reached with positive probability when all players follow the strategy profile $\bsigma^{E_n}$.
    Hence, again, the inequality $x \leq \X_i(\bsigma^{E_n})$.
    
    \subparagraph*{If player $i$ is a pessimist.}
    Then, since the algorithm terminated at step $n$, player $i$ is such that $v_0 \in W_i^k$.
    The strategy $\sigma'_i$, like every strategy $\tau_i$ for player $i$, satisfies therefore the inequality $\PP_{\bsigma^{E_n}_{-i}, \sigma'_i}(\mu_i \leq z_i^n) > 0$.
    Consequently, we have $x \leq z_i^k = \X_i(\bsigma^{E_n})$.

    In both cases, the deviation $\sigma'_i$ is not profitable, hence the conclusion.
\end{proof}

Let us now prove that putting back the removed edges does not change that result, and therefore conclude the correctness proof.

\begin{prop}
    The strategy profile $\bsigma^{E_n}$ is an XRSE in the game $\Game_{\|v_0}$.
\end{prop}

\begin{proof}
    Let $i$ be a player, and let $\sigma'_i$ be a deviation from $\bsigma^{E_n}$ for player $i$ in $\Game_{\|v_0}$.
    Since $\bsigma^n$ is stationary, we can assume that $\sigma'_i$ is positional by \cref{lm:secretlemma}.
    If the strategy $\sigma'_i$ uses only edges of $E_n$, then it can be considered as a deviation from $\bsigma^{E_n}$ in the game $\Game^{E_n}$, hence by \cref{prop:xrseGEn}, it is not a profitable deviation.
    
    Let us now assume that the strategy $\sigma'_i$ uses an edge that does not belong to $E_n$, i.e. there exists a vertex $v$ that is visited with positive probability in the strategy profile $(\bsigma^{E_n}_{-i}, \sigma'_i)$ and an edge $vw \in E \setminus E_n$ such that $w = \sigma'_i(v)$.
    Since only edges controlled by pessimists have been removed, we can immediately deduce that player $i$ is a pessimist.
    
    Now, among such edges, let us choose one whose removal occurred the earliest, that is, let us choose it in order to minimize the index $k$ such that $vw \in E_k \setminus E_{k+1}$.
    Thus, in the strategy profile $(\bsigma^{E_n}_{-i}, \sigma'_i)$, it is almost sure that only edges of $E_k$ are used.

    The fact that the edge $uv$ has been removed at step $k$ means that we had $u \not\in W_i^k$ and $v \in W_i^k$.
    Thus, from the vertex $v$, if player $i$ uses only edges of $E_k$ (which is the case when they follow $\sigma'_i$), and if the other players follow the strategy profile $\bsigma^{E_k}_{-i}$, player $i$ gets the payoff $z_i^k$ or less with positive probability.
    Let us show that it is also the case when the other players follow the strategy profile $\bsigma^{E_n}$ instead of $\bsigma^{E_k}$.

\begin{lem}
    From the vertex $v$, we have $\PP_{\bsigma^{E_n}_{-i}, \sigma'_i}(\mu_i \leq z_i^k) > 0$.
\end{lem}

\begin{proof}
    We proceed by proving that when the players follow, from the vertex $v$, the strategy profile $(\bsigma^{E_k}, \sigma'_i)$, there is a positive probability that player $i$ gets the payoff $z_i^k$ or less \emph{and} that the set $W_i^k$ is never left.

    Indeed, if in that strategy profile there is a positive probability that player $i$ gets a payoff smaller than or equal to $z_i^k$ by reaching a terminal vertex $t$ that yields such a payoff, then we have $t \in W_i^k$, and with positive probability the terminal vertex $t$ is reached without leaving $W_i^k$.
    Similarly, if such a payoff is obtained by reaching no terminal vertex, and therefore getting payoff $0$, then, using a reasoning that has already been used above, with positive probability a vertex $w$ is reached without leaving $W_i^k$, such that from $w$, no terminal vertex is accessible anymore in $(V, E_k)$; and, therefore, all the vertices accessible from $w$ in that graph belong to $W_i^k$, hence once $w$ is reached it is almost sure that $W_i^k$ is never left.

    Then, the set $E_{k+1}$ was defined so that $W_i^k$ is no longer accessible from $v_0$ in the graph $(V, E_{k+1})$.
    Therefore, those vertices are not accessible at any step $\l > k$, and therefore no outgoing edge of a vertex of $W_i^k$ is ever removed in the sequel, i.e. $E_n \cap (W_i^k \times V) = E_k \cap (W_i^k \times V)$.
    Consequently, since $\sigma'_i$ uses only edges of $E_k$, when the strategy profile $(\bsigma^n, \sigma'_i)$ is played from $v$, it is also true that with positive probability player $i$ gets the payoff $z_i^k$, or less.
\end{proof}
    
    This lemma proves that we have $\X_i(\bsigma^{E_n}_{-i}, \sigma'_i) \leq z_i^k$.
    To conclude that the deviation $\bsigma'_i$ is not profitable, we still need to prove that this quantity $z_i^k$ is smaller than or equal to (actually strictly smaller) the risk measure $\X_i(\bsigma^n)$.
    That inequality is an immediate consequence of the following lemma.

\begin{lem}
    For every pessimistic player $j$, the sequence $(z_j^\l)$ of player $j$'s risk measures is non-decreasing.
\end{lem}

\begin{proof}
    Let $\l$ be a step index, and let us prove that we have $z_j^\l < z_j^{\l+1}$.
    The quantity $z_j^{\l+1}$ is the pessimistic risk measure of player $j$ in the strategy profile $\bsigma^{E_{\l+1}}$: there is therefore a positive probability that player $j$ gets the payoff $z_j^{\l+1}$ when that strategy profile is followed.
    Player $j$ obtains that payoff either by reaching a terminal vertex to which that payoff is assigned, or by reaching no terminal vertex at all.

    Let us first show that the second case is actually impossible.
    If player $j$ gets the payoff $z_j^{\l+1} = 0$ by reaching no terminal vertex, with the same reasoning as above, there is, in particular, a vertex $v$ that has positive probability of being visited infinitely often when $\bsigma^{\l+1}$ is played from $v_0$, and therefore such that no terminal vertex is accessible from $v$ in $(V, E_{\l+1})$.
    But then, let $j'$ be the player that was controlling the edges that were removed at step $\l$.
    Let us consider a strategy $\tau_{j'}$ of player $j'$ that uses only edges of $E_\l$.
    Then, when the strategy profile $(\bsigma^\l_{-j'}, \tau_{j'})$ is played from the vertex $v$, it will almost surely be true that either no terminal vertex is reached, leading to the payoff $0$, or an edge of $E_\l \setminus E_{\l+1}$ is taken, leading therefore to a vertex of $W_i^\l$, and to a risk measure of $z_{j'}^\l$ or less.
    Thus, since all rewards are non-negative and therefore $z_{j'}^\l \geq 0$, the vertex $v$ belongs to the set $W_{j'}^\l$, which is impossible since it should then have been made unaccessible in the graph $(V, E_{\l+1})$.

    Therefore, player $j$ gets payoff $z_j^{\l+1}$ by reaching a terminal giving them that payoff, which means that such a terminal is accessible from $v_0$ in the graph $(V, E_{\l+1})$.
    Then, it is also accessible from $v_0$ in the graph $(V, E_\l)$, and therefore it is reached with positive probability when following the strategy profile $\bsigma^{E_\l}$.
    Consequently, we have $z_j^\l \leq z_j^{\l+1}$.
\end{proof}

    Consequently, with $j = i$, we have $z_i^k \leq z_i^n$, and therefore $\X_i(\bsigma^{E_n}_{-i}, \sigma'_i) \leq z_i^k \leq z_i^n = \X_i(\bsigma^{E_n})$.
    The strategy $\sigma'_i$ is not a profitable deviation from the strategy profile $\bsigma^{E_n}$, which is therefore a (stationary) XRSE.
\end{proof}

    \paragraph{Complexity}

    We finally show that \cref{algo:existence} runs with time $O(m^2p)$.
    At each iteration of the while loop, at least one edge is removed; we therefore have at most $m$ iterations of said loop.

    Now, during the $k^\text{th}$ iteration, the algorithm computes the set $A^k$, and for each pessimistic player $i$, the algorithm also computes the quantity $z_i^k = \X_i(\bsigma^{E_k})$, and then the set $W^k_i$.
    All of those computations can be done in time $O(m)$ using \cref{lm:secretlemma}. Since there are $p$ players, this step therefore takes $O(mp)$ time.
    Finally, checking whether $v_0 \in W^k_i$ for each player $i$ takes time $O(p)$, and removing all the edges leading to $W_i^k$ to define the set $E_{k+1}$ takes time $O(m)$.
    Hence, the complexity of the algorithm is $O(m^2 p)$.
\end{proof}

Like in the case of ERSEs, we conjecture that existence, and even existence of a stationary strategy profile, remain true in the general case.

\begin{conj}\label{conj:existence_xrse}
    Let $\Game_{\|v_0}$ be a simple stochastic game, and a partition $(P, O)$ of the player set $\Pi$.
    Then, there exists a (stationary) XRSE in $\Game_{\|v_0}$.
\end{conj}

\section{Constrained existence problem}

We now study the computational complexity of the constrained existence problem of XRSEs.
The main result of this section is the following theorem, which proves that, contrary to the same problem with ERSEs, it is a decidable fragment of the constrained existence of RSEs.

\begin{thm}\label{thm:XRSE_NPc}
    The constrained existence problem for XRSEs is $\NP$-complete and is $\NP$-hard even when all players are pessimistic and all rewards are non-negative.
\end{thm}

To prove \cref{thm:XRSE_NPc}, we first prove \cref{thm:XRSEmemorysmall}, which shows that if there is an XRSE then there is one that uses finite memory. We show that this finite-memory strategy profile can be described only using polynomial size, which in turn proves $\NP$ membership (\cref{lm:XRSE_NPeasy}).

Later, we consider the problem of XRSEs when the players are restricted to pure, stationary or positional strategies. We show that in all the above cases, the problem remains $\NP$-complete. The upperbound is similar to the general case, but the lower bound is shown in \cref{lm:XRSE_NPhard} by showing a reduction from the problem $\THREESAT$ to the constrained existence problem. 

\begin{thm}\label{thm:infinite_rho_restricted_strategy_np_easy}
    The constrained existence problem of XRSEs is also $\NP$-complete when the players are restricted to positional, stationary,  or pure strategies. 
\end{thm}

Finally, we show in the following theorem that when all players are optimistic, the problem can even be solved in polynomial time.

\begin{thm}\label{thm:XRSE_Pcomplete}
    The constrained existence problem of XRSE is $\PTime$-complete when all players are optimists, that is, when we have $P=\emptyset$.
\end{thm}

We dedicate the rest of the section to proving these three results.

\subsection{Membership in $\NP$}

$\NP$ membership is a consequence of the fact that when an XRSE exists, there also exists one with the same extreme risk measures that uses finite memory, with a number of states that is polynomial in the size of the game.
Let us therefore illustrate, with examples, how and why memory is required in such XRSEs.
We consider the following constrained existence question and analyse the same question on three example graphs. 

\begin{quote}$(*)$
   Is there an XRSE in the game in which both players have exactly the risk measure $1$?
\end{quote}

\subparagraph*{Game in \cref{fig:ex_extreme1}.}Let us consider the game in \cref{fig:ex_extreme1} again.
The answer to Question~$(*)$ here is \emph{no}.
Intuitively, such an XRSE would require at least two plays of positive probability: one that ends in $t_1$, and one that ends in $t_2$.
For those two plays to occur with positive probability, the strategy profile must proceed to a randomized action at vertex $a$ or $b$: i.e., one of the players, at some point of time, must toss a coin to give the payoff $1$ to player $\Circle$ in one case and to $\Square$ in the other.
But then, since that player is the only one that can see that coin, they have a profitable deviation by lying about the outcome, and always choose the option that gives them the best payoff.
More randomization will not help: as long as one of the players randomizes, be it once, several times, or infinitely often, they have an incentive to deviate and stay in the cycle and wait until the other player leaves.

\subparagraph*{Game in \cref{fig:ex_extreme2}.} Consider now a slight modification, as shown in~\cref{fig:ex_extreme2}.
There, the first player that plays is determined at random by the edge that is taken from an initial stochastic vertex.
The answer to Question $(*)$ for this game is \emph{yes}. 
The random choice on which player gets payoff $1$ is decided by the stochastic vertex. Since both players can see which edge is taken from there, this serves as a source of unbiased randomness based on which they act.
For example, it can be decided that if the play visits the vertex $a$ immediately after $c$, then player $\Circle$ must visit the terminal $t_1$, and similarly, if it visits the vertex $b$, then player $\Square$ must visit the terminal $t_2$. If the edge $ab$ is taken, player $\Square$ punishes player $\Circle$ by always going back to $a$, and vice versa.
In other words, the stochastic vertex provides the players with a common coin.

\subparagraph*{Game in \cref{fig:ex_extreme3}.} Finally, consider the game depicted in~\cref{fig:ex_extreme3}.
Here, both players $\Circle$ and $\Square$ have the possibility of deviating to a terminal with payoff $2$ in one play.
The stationary strategy profile in which from vertex $d$, player $\Circle$ goes from $d$ to $a$ and then to $t_1$, and in which player $\Square$ goes from $e$ to $b$ and then to $t_2$, is therefore not an XRSE: both players have a profitable deviation that goes to terminal $t_3$.
But the answer to Question~$(*)$ still remains \emph{yes}!
If, from vertex $d$, player $\Circle$ goes from vertex $d$ to $a$ and then to $b$, from which player $\Square$ leaves to $t_2$, and symmetrically, from vertex $e$, player $\Square$ goes to $a$ through $b$ from which player $\Circle$ goes to $t_1$, then that strategy profile is an XRSE, in which everyone gets the risk measure $1$. 
This is because a player has a profitable deviation only if they can play in a way that guarantees them a risk measure better than $1$, i.e., that guarantees them \emph{almost surely} a payoff greater than~$1$.
If there remains a play that occurs with nonzero probability and offers a lower reward, then the player does not increase their risk measure.  
Therefore, an XRSE where player $\Circle$ gets the extreme risk measure $1$ only needs to have one play with positive probability in which she gets the payoff $1$, \emph{and} in which she cannot increase her payoff by deviating. We say that such a play \emph{anchors} that player. In our example, the play $cebat_1$ anchors player $\Circle$. 

We see in this last example that memory is required to remember either the subset of players that are being anchored, or if a player has deviated from the strategy and must be punished. 
Given one or more players that are being anchored, the memory state of any of the players does not change unless either a player deviates or, more importantly, randomization occurs. When randomization occurs, the set of players that are anchored in each of the plays is a subset of the set of players anchored before this play \emph{split}.
In our examples, the set of players that are anchored at $c$ is both $\Circle$ and $\Square$, and it immediately splits.
After the splits, when we have only one player to anchor, the players can follow a positional strategy profile; and similarly when one player deviates and must be punished, the players can follow a positional strategy profile.

Following this intuition, we prove a theorem that bounds the amount of memory required by a strategy to a polynomial in the number of players and vertices in the game.

\begin{thm}\label{thm:XRSEmemorysmall}
    Let $\bsigma$ be an XRSE in the game $\Game_{\|v_0}$ with $n$ vertices and $p$ players,  and a partition $(P, O)$ of the set $\Pi$.
    Then, there exists a finite-memory XRSE $\bsigma^\star$ with at most $3np-2n+p+1$ many memory states, and such that $\X(\bsigma^\star) = \X(\bsigma)$. Furthermore, if $\bsigma$ is pure, then there is such a strategy profile $\bsigma^\star$ that is pure.
\end{thm}

\begin{proof}
The XRSE $\bsigma$ being given, we first define the labeling $\Lambda$ which, intuitively, defines which players are being anchored after every given history.

\paragraph{Definition of $\Lambda$.}

\begin{lem}[The labeling $\Lambda$]\label{lm:Lambda}
        There exists a labeling $\Lambda$ that maps each history $h \in \Hist\Game_{\|v_0}$ compatible with $\bsigma$ to a set $\Lambda(h) \subseteq \Pi$, such that for each such $h$, if we write $\{v_1, \dots, v_k\} = \Supp(\bsigma(h))$, we have the following properties.
        \begin{enumerate}
            \item\label{itm:splitsetsanchorwithouti} If the vertex $\last(h)$ is stochastic, or belongs to some player $i \not\in \Lambda(h)$, then the sets $\Lambda(hv_1), \dots, \Lambda(hv_k)$ form a partition of the set $\Lambda(h)$.

            \item\label{itm:splitsetsanchorwithi} If the vertex $\last(h)$ belongs to some player $i \in \Lambda(h)$, then the sets $\Lambda(hv_1) \setminus \{i\}, \dots, \Lambda(hv_k) \setminus \{i\}$ form a partition of the set $\Lambda(h) \setminus \{i\}$, and player $i$ belongs to all sets $\Lambda(hv_1), \dots, \Lambda(hv_k)$.

            \item\label{itm:optimistanchor} For each optimistic player $i \in \Lambda(h)$, we have $\X_i(\bsigma_{\|h}) = z_i$.
            
            \item\label{itm:pessimistanchor} For each pessimistic $i \in \Lambda(h)$, for all strategies $\tau_i$ of player $i$, we have $\X_i(\bsigma_{-i\|h}, \tau_i) \leq z_i$.

            \item\label{itm:nosplit} If there is a successor $v_\l$ such that $\Lambda(hv_\l) = \Lambda(h)$, then all other successors $v_{\l'}$ are such that $\X_i(\bsigma_{\|hv_{\l'}}) < z_i$ for each optimist $i \in \Lambda(h)$, and there exists $\tau_i$ with $\X_i(\bsigma_{-i\|hv_{\l'}}, \tau_i) > z_i$ for each pessimist $i \in \Lambda(h)$.
        \end{enumerate}
    \end{lem}

\begin{proof}
    We define the labeling $\Lambda$ inductively. 

    \subparagraph*{Base case.}
    First, on the one-vertex history $v_0$, we define $\Lambda(v_0) = \Pi$: at the start, all players must be anchored.
    Let us notice that the history $v_0$ satisfies Property~\ref{itm:optimistanchor}, which states that the optimists get the optimistic expectation they are supposed to get, and Property~\ref{itm:pessimistanchor}, which states that the pessimists have no profitable deviations.
    The other properties will be checked in the inductive case.

    \subparagraph*{Inductive case.}
    Suppose $\Lambda(hv)$ has already been defined, where $hv$ is a history compatible with the strategy profile $\bsigma$, and that the five properties are satisfied by $\Lambda$ on all histories on which it is already defined.
    Let $w_1, \dots, w_k$ be the successors of $v$ that are chosen by the strategy $\bsigma(hv)$ with non-zero probability, that is, the support of $\bsigma(hv)$.
    If $k=1$, then we define $\Lambda(hvw_1) = \Lambda(hv)$.
    Note that Properties~\ref{itm:splitsetsanchorwithouti},~\ref{itm:splitsetsanchorwithi}, and~\ref{itm:nosplit} are immediately satisfied, and that Properties~\ref{itm:optimistanchor} and~\ref{itm:pessimistanchor} are satisfied by induction hypothesis.
    
    If $k>1$, we need to partition the set $\Lambda(h)$ between the $k$ successors.
    To do so, we will use the following result.
    
    \begin{slem}\label{claim:successorAnchor}
    For each player $i \in \Lambda(hv)$ that does not control the vertex $v$, the following holds.
    \begin{itemize}
        \item If player $i$ is an optimist, and $\X_i(\bsigma_{\|hv}) = z_i$, then there is at least one successor $w_\l$ such that we have $\X_i(\bsigma_{\|hvw_\l}) = z_i$.
    
        \item If player $i$ is a pessimist, then there is a successor $w_\ell \in \Supp(\bsigma(hv))$, such that for every  strategy $\tau_i^\ell$ from $w_\l$, we have $\X_i(\bsigma_{-i\|hvw_\ell}, \tau_i) \leq z_i$.
    \end{itemize}
    \end{slem}
    
    \begin{proof}
        The first case follows from Property~\ref{itm:optimistanchor} in the induction hypothesis.
        As for the second case, we proceed by contradiction.
        Let us assume that for each $w_\l$, there exists a strategy $\tau_i^\l$ such that $\X_i(\bsigma_{-i\|hvw_\l}, \tau^\l_i) > z_i$.
        Then, the strategy $\tau_i$ defined by $\tau_{i\|vw_\l} = \tau^\l$ for each $\l$ is such that $\X_i(\bsigma_{-i\|hv}, \tau_j) > z_i$, which is impossible since $\Lambda(hv)$ is assumed to satisfy Property~\ref{itm:pessimistanchor}.
    \end{proof}

    We define each set $\Lambda(hvw_\l)$ by iterating through each element of $\Lambda(hv)$ as follows:
    \begin{itemize}
        \item \emph{Initialization.} For all $w_\l$, declare $\Lambda(hvw_\l) = \emptyset$.
        
        \item \emph{Iteration over players.} Consider each player $i\in \Lambda(hv)$ sequentially and proceed as follows:
        \begin{itemize}
            \item if player $i$ controls the vertex $v$, then add $i$ to every set $\Lambda(hvw)$.

            \item If player $i$ does not control $v$, then add $i$ to the set $\Lambda(hvw_\l)$ where $w_\l$ is defined by \cref{claim:successorAnchor}.
        \end{itemize}
    \end{itemize}
    
    Not that the first four properties are thus guaranteed to be satisfied.
    Moreover, when there are several successors $w_\l$ possible, we always favour those such that, at the moment where the decision is taken, the sets $\Lambda(hvw_\l)$ are the smallest.
    This suffices to guarantee Property~\ref{itm:nosplit}.
    \end{proof}

\paragraph{Construction of the strategy profile $\bsigma^\star$}

Based on $\Lambda$ as in \cref{lm:Lambda}, we construct a finite-memory strategy profile. 
Formally, the definition of the strategy profile $\bsigma^\star$ will be done by defining its memory structure.

The memory states are the following:
    \begin{itemize}    
        \item for each player $i$, the state $\punish_i$;

        \item for each vertex $v$ and each subset $A \subseteq \Pi$ of players such that there exists $h$ with $A = \Lambda(h)$, the state $\anchor_{Av}$;

        \item the state $\anchor_{\Pi\bot}$.
    \end{itemize}
    
We now define the transitions from each of those states and from each vertex. Observe that as long as the memory state does not change, the memory structure follows a positional strategy profile. So, we describe such stationary strategy profiles and also describe when the memory state changes. 
For each set $A \subseteq \Pi$, we write $W_A$ for the set of vertices that $\bsigma$ may visit while anchoring the set $A$, i.e., the set of vertices $v$ such that there exists a history $hv$ with $\Lambda(hv) = A$.

    \subparagraph*{Punishing memory states $\punish_i$.}
    First, let us define what $\bsigma^\star$ does when in state $\punish_i$, for some player $i$. Those memory states will correspond to the \emph{punishing strategies}, followed when player $i$ deviates from the assigned strategy with the other memory states. 
    By \cref{lm:secretlemma}, there is a positional strategy profile $\btau^{\dag i}_{-i}$ that minimizes, from every vertex of the game, the payoff that player $i$ can enforce.
    In addition, we pick an arbitrary positional strategy $\sigma^{\dag i}_i$.
    Then, when the strategy profile $\bsigma^\star$ is in the memory state $\punish_i$ and reads a vertex $v$, the memory structure outputs the vertex $\btau^{\dag i}(v)$ and remains in the same memory state $\punish_i$.

    \subparagraph*{Anchoring states with no player to anchor.}
    Let us now define what happens in memory state $\anchor_{\emptyset v}$.
    Consider the objective of achieving a payoff vector that has positive probability of being achieved in $\bsigma$, while visiting only vertices of $W_\emptyset$.
    Using classical attractor-based proofs (or \cref{lm:secretlemma}), there exists a positional strategy profile that achieves that objective with probability $1$ from every vertex from which that is possible: let us call it $\btau^{\anch \emptyset}$.
    Then, when in memory state $\anchor_{\emptyset v}$ and reading the vertex $w$, we distinguish two cases.
    \begin{itemize}
        \item If $vw$ is an edge that is compatible with the strategy profile $\btau^{\anch \emptyset}$, then the strategy profile $\bsigma^\star$ outputs the vertex $\btau^{\dag i}(w)$, where $i$ is the player controlling $v$, and shifts to the memory state $\punish_i$.

        \item Otherwise, it outputs the vertex $\btau^{\anch \emptyset}(w)$ and moves to the memory state $\anchor_{\emptyset w}$.
    \end{itemize}

    \subparagraph*{Anchoring state with one player to anchor.}
    We can now move to singletons, and define what happens in the states of the form $\anchor_{\{i\} v}$.
    In such a state, we define a strategy profile that gives player $i$ exactly the extreme risk measure $z_i$.
    More precisely, we want player $i$ to receive payoff $z_i$ with positive probability, and never leave the set of vertices $W_{\{i\}}$ with probability~$1$.
    Using \cref{lm:secretlemma}, there exists a positional strategy profile $\btau^{\anch i}$ that satisfies that property from every vertex from which it is possible.
    Note that this objective is in particular satisfiable, and therefore satisfied by $\btau^{\anch i}$, from every vertex $v \in W_{\{i\}}$.
    Similar to the previous step, we define the strategy profile $\bsigma^\star$ in the states of the form $\anchor_{\{i\} v}$ so that it follows the strategy profile $\btau^{\anch i}$, remembers the last vertex that was visited and uses that memory to switch to the corresponding punishing state when some player $j$ deviates.

        \subparagraph*{Anchoring states with two or more players to anchor.}
    Now, let us consider the states of the form $\anchor_{A v}$, where $A$ has cardinality at least $2$.
    The existence of $\anchor_{A v}$ implies that there is a history $h$ such that $\Lambda(h) = A$.
    Moreover, since each randomization splits the label of histories in sets that have at most one element in common (Properties~\ref{itm:splitsetsanchorwithouti} and~\ref{itm:splitsetsanchorwithi}), there is only one side of each split that can contain $A$, which implies that among such histories $h$, we can choose one that is a prefix of all others.
    After history $h$, the histories labelled by $A$ form a sequence $h, hv_1, hv_1v_2, \dots$ which may be infinite, end in a terminal vertex, or end with a new split.

    \begin{itemize}

    \item \emph{If that sequence ends with a split,} then there is a longest history $hv_1 \dots v_q$ with $\Lambda(hv_1 \dots v_q) = A$ and $k \geq 2$ vertices $w_1, \dots, w_k \in \Supp(\bsigma(hv_1 \dots v_q))$ such that we have $\Lambda(hv_1 \dots v_q w_\l) \neq \emptyset, A$ for each $\l$.
    We can then define a simple history $h'v_q$ that also goes from the vertex $\last(h)$ to the vertex $v_q$, with $\Occ(h'v_q) \subseteq \Occ(\last(h) v_1 \dots v_q)$.
    We then define the strategy profile $\bsigma^\star$ in each state $\anchor_{Av}$ so that it follows the history $h' v_q$ and remembers the last vertex visited, and switches to the state $\punish_i$ and follows the strategy profile $\btau^{\dag i}$ when a given player $i$ deviates and takes an edge that they are not supposed to take.
    Moreover, when an edge is taken that does belong to the history $h'v_q$, but not because a player deviated (it is then necessarily because of a stochastic vertex), the memory switches to the state $\anchor_{\emptyset v}$ (where $v$ is the last vertex seen) and immediately follows the corresponding strategy.
    Finally, when the vertex $v_q$ is reached and the memory is in state $\anchor_{A \last(h')}$, the strategy profile $\bsigma^\star$ chooses randomly between the edges $v_q w_1$, \dots, and $v_q w_k$, all with positive probability.
    Such action will often be referred to as a \emph{split}.

        \item \emph{If that sequence is infinite or ends in a terminal vertex,} then $\pi^A = v_1 v_2 \dots$ is a play, and satisfies $\Lambda(h\pi^A_{< k}) = A$ for each $k$.
        We can then consider a play $\pi^{A\star}$ with $\Occ(\pi^{A\star}) \subseteq \Occ(\pi^A)$ and $\Inf(\pi^{A\star}) \subseteq \Inf(\pi^A)$ that is either a simple path from $\pi^A_0$ to the terminal reached by $\pi^A$, or, if $\pi^A$ is infinite, a simple lasso (let us recall that those are defined as plays of the form $h'c^\omega$, where the history $h'c$ is simple).
        We can moreover choose $\pi^{A\star}$ so that the set of vertices visited infinitely often (if there are any) in $\pi^{A\star}$ is included in the set of vertices visited infinitely often in $\pi^A$.
Then, we can define $\bsigma^\star$ in the states of the form $\anchor_{A v}$ as following the play $\pi^{A\star}$, and remembering the last vertex seen.
    When a player $i$ deviates and takes an edge that should not be taken, the memory switches to the state $\punish_i$ and follows the strategy profile $\btau^{\dag i}$.
    Finally, when an edge is taken that does not belong to $\pi^{A\star}$ but does not correspond to a deviation either, we switch to the state $\anchor_{\emptyset w}$ where $w$ is the last vertex seen, and to the corresponding strategy profile.
\end{itemize}

\subparagraph*{Initialization.}
    The strategy profile $\bsigma^\star$  has the state $\anchor_{\Pi\bot}$ as the initial memory state.
    In this state, it behaves exactly as in any state of the form $\anchor_{\Pi v}$, but without having memorized a last visited vertex $v$, since there is no such vertex.
    From that memory state therefore, it necessarily reads the vertex $v_0$, and starts acting as described in the previous case.

\subparagraph*{The pure case.}
In this construction, the vertices on which the strategy profile $\bsigma^\star$ proceeds to an actual randomization (i.e., the vertices $v$ such that there exists a history $hv$ such that the support of the distribution $\bsigma^\star(hv)$ contains more than one element) are vertices on which $\bsigma$ also proceeds to such a randomization.
Therefore, if $\bsigma$ is pure (i.e., if randomizations occur only on stochastic vertices), so is $\bsigma^\star$.

\paragraph{A combinatorial break: counting states}

    Now that the strategy $\bsigma^\star$ is defined, let us bound the memory it uses.
    There are, obviously, exactly $p$ states of the form $\punish_i$, and one state $\anchor_{\Pi\bot}$.
    To prove that there are at most $3np-2n$ states of the form $\anchor_{Av}$, we need to prove that there are at most $3p-2$ sets $A$ such that there is a history $h$ with $\Lambda(h) = A$.

    Let us call \emph{$\Lambda$-anchored} all such sets $A$.
    By analogy with strategies, we write $\Lambda_{\|hv}$ for the labeling that maps each history $h' \in \Hist\Game_{\|v}$ compatible with $\bsigma_{\|hv}$ to the set $\Lambda(hh')$, and we will also use the notion of anchoredness for each of those labelings $\Lambda_{\|hv}$.
    We proceed by proving the following stronger result.
    
    \begin{prop}\label{prop:combinatorial}
        For every history $h$ compatible with $\bsigma$, if $\Lambda(h)$ contains at least two elements, then there are at most $3\card\Lambda(h)-2$ sets that are $\Lambda_{\|h}$-anchored.
    \end{prop}

\begin{proof}
    For each history $h$, we write $f(h)$ for the number of $\Lambda_{\|h}$-anchored sets that have cardinal at least $2$.
    There are $\card\Lambda(h)+1$ subsets of $\Lambda(h)$ that have cardinality $0$ or $1$: the result will therefore be proved if we prove $f(h) \leq 2\card\Lambda(h) - 3$.
    The proof goes by induction on $m = \Lambda(h) \geq 2$.

    \subparagraph*{Base case.}
    If $m = 2$, the set $\Lambda(h)$ is a pair $\Lambda(h) = \{i, j\}$.
    Then, since the $\Lambda_{\|h}$-anchored sets are all subsets of $\Lambda(h)$, the only set of cardinality at most $2$ that is $\Lambda_{\|h}$-anchored is the pair $\{i, j\}$ itself, hence we have $f(h) = 2 \times 2 - 3 = 1$, as desired.

    \subparagraph*{Inductive case.}
    If $m > 2$, and if we assume that the result is true for every history $h'$ with $2 \leq \card\Lambda(h') \leq m-1$, then let $\{v_1, \dots, v_k\} \subseteq \Supp(\bsigma(h))$ be the set of possible next vertices $v$ such that $\card\Lambda(hv) \geq 2$.

    If $k = 1$, i.e., if $\Lambda(hv_1) = \Lambda(h)$, then we have $f(h) = f(hv_\l)$ and the result for $h$ will be proved if we prove it for $hv_1$.
    Following that reasoning, we can extend the history $h$ until we are not in that case: if we always are, then the only $\Lambda_{\|h}$-anchored sets are $\Lambda(h)$ itself, and possibly the empty set and some singletons, hence $f(h) = 1$ and the result is immediate.
    We can therefore assume that $k > 1$.

    Let $i$ be the player controlling the vertex $\last(h)$; we set $i = \bot$ if $\last(h)$ is a stochastic vertex.
    Then, by Properties~\ref{itm:splitsetsanchorwithouti} and~\ref{itm:splitsetsanchorwithi} of \cref{lm:Lambda}  guaranteed during the construction of $\Lambda$, the sets $\Lambda(hv_1) \setminus \{i\}, \dots, \Lambda(hv_k) \setminus \{i\}$ form a partition of $\Lambda(h) \setminus \{i\}$.
    Therefore, no set of cardinality $2$ or more can be simultaneously $\Lambda(hv_\l)$-anchored and $\Lambda(hv_{\l'})$-anchored for $\l \neq \l'$, hence the equality $f(h) = 1+\sum_\l f(hv_\l)$.
    Now, since each set $\Lambda(hv_\l)$ has at least $2$ and less than $m$ elements, we can apply the induction hypothesis to deduce:
    $$f(h) \leq 1 + \sum_{\l=1}^k (2\card\Lambda(hv_\l) - 3).$$
    Moreover, we have $\sum_\l \card\Lambda(hv_\l) \leq m + k - 1$ (each element of $\Lambda(h)$ occurs in one of the sets $\Lambda(hv_\l)$, except possibly one that would occur in all of them), hence the inequality above becomes:
    $$f(h) \leq 1 + 2(m+k-1) -3k$$
    $$= 2m - 1 - k$$
    and since we have assumed $k \geq 2$, we obtain $f(h) \leq 2m-3$.
\end{proof}    

Let us recall that $\Lambda(v_0) = \Pi$.
As a particular case of this result, we obtain, if $p \geq 2$, that there are at most $3p-2$ sets that are $\Lambda$-anchored, as desired.
In the case $p = 1$, the game $\Game_{\|v_0}$ is an MDP, and using \cref{lm:secretlemma}, we can immediately construct $\bsigma^\star$ as a positional strategy (which has therefore $1 \leq 3n \times 1 - 2n + 1 + 1$ memory states) with the same risk measure as $\bsigma$.

    \paragraph{The strategy profile $\bsigma^\star$ has the desired extreme risk measures.}

We now show that $\X(\bsigma^\star) = \X(\bsigma)$.
Let us recall that we defined $\bz = \X(\bsigma)$.
    
\begin{prop}\label{prop:ActualPayoff}
    The strategy profile $\bsigma^\star$ satisfies the equality $\X(\bsigma^\star) = \bz$.
\end{prop}

\begin{proof}
    Let $i$ be a player: we want to prove that $\X(\bsigma^\star) = z_i$.
    Let us first see how $z_i$ has positive probability of being obtained in the strategy profile $\bsigma$, and we will then show that no larger (respectively smaller) payoff, if $i$ is optimistic (respectively pessimistic), has a positive probability of being obtained with the same strategy profile.

    \subparagraph*{Player $i$ gets payoff $z_i$ with positive probability.}
    Let $A \subseteq \Pi$ be one of the smallest sets (for the inclusion relation) containing $i$ such that there exists a history $hu$ with $\Lambda(hu) = A$.
    Then, by construction of $\Lambda$, there exists a finite sequence of sets $\Pi = A_0, A_1, \dots, A_m = A$ and of histories $h_1 v_1 w_1, \dots, h_m v_m w_m$ where for each $k$, the history $h_{k+1}$ starts from $w_k$, the history $h_1 v_1 \dots h_k v_k w_k$ is compatible with $\bsigma$, and we have $\Lambda(h_1 v_1 \dots h_k v_k) = A_{k-1}$ and $\Lambda(h_1 v_1 \dots h_k v_k w_k) = A_k$.
    We can then write $hu = h_1 v_1 \dots h_m v_m w_m$.
    
    Consider the strategy profile $\bsigma^\star$, which initially follows  the positional strategy profile $\btau^{\anch \Pi}$. This strategy profile generates, with nonzero probability, a history $h'_1 v_1$ starting from vertex $v_0$ to vertex $v_1$, based on our construction. 
    From that vertex $v_1$, it proceeds to a randomized action and, with positive probability, moves to the vertex $w_1$ and switches to the positional strategy profile $\btau^{\anch A_1}$, and so on: there is, therefore, a history $h'_1 v_1 h'_2 v_2 \dots h'_m v_m w_m$ that is compatible with the strategy profile $\bsigma^\star$ and after which the collective memory is in the state $\anchor_{A v_m}$, and plays accordingly.

    Since $A_m= A$ is the  subset of $\Pi$ where $i\in A = \Lambda(hu)$, we are in the case where the set $A$ is no longer split further by our labeling. That is, there is a play $\pi$ from $w_m$ such that $\Lambda(h \pi_{\leq k}) = A$ for every $k$ such that this is defined.
    Then, in the construction of the strategy profile $\bsigma^\star$ we have distinguished two cases: the one where $A$ was a singleton, and the one where it had at least two elements (the empty case is excluded, since $A$ contains player $i$).

\begin{itemize}
    \item \emph{If $A$ is a singleton,} then after the history $hu$, without any player deviating, all players are following the strategy profile $\btau^{\anch i}$.
    By its definition, that strategy profile achieves the payoff $z_i$ for player $i$ with positive probability.

   \item \emph{If $A$ has at least two elements,} then after that same history, all players are following the play $\pi^{A\star}$, which yields the same payoffs as $\pi^A$.
    However, we must still prove that player $i$ actually gets the payoff $z_i$ in $\pi^A$, and that the play $\pi^{A\star}$ is generated with positive probability (i.e. that it does not cross infinitely many stochastic vertices---which we must first show for $\pi^A$).
We do so in the following result, which we will use again later.
\end{itemize}

\begin{slem}\label{claim:piA}
    The play $\pi^A$ is (eventually) generated with positive probability when the players follow the strategy profile $\bsigma$.
    Similarly, the play $\pi^{A\star}$ is generated with positive probability when they follow $\bsigma^\star$.
    Both plays yield to each player $j \in A$ the payoff $z_j$.
\end{slem}

\begin{proof}
    Let $j \in A$.
    Let us proceed by case disjunction according to the risk measure used by player $j$.

\begin{itemize}
\item \emph{If player $j$ is an optimist,} then, by Property~\ref{itm:optimistanchor} of \cref{lm:Lambda}, we have $\X_j(\bsigma_{\|h}) = z_j$, and therefore $\PP_{\bsigma_{\|h}}(\mu_j = z_j) > 0$, i.e., by the law of total probability:
        $$\PP_{\bsigma_{\|hu}}(\pi^A) \PP_{\bsigma_{\|h}}(\mu_j = z_j \mid \pi^A) + \sum_k \sum_{w \in E(\pi_k) \setminus \{\pi_{k+1}\}} \PP_{\bsigma_{\|hu}}(\pi^A_{\leq k} w) \PP_{\bsigma_{\|hu}}(\mu_j = z_j \mid \pi^A_{\leq k}w) > 0.$$
        But using Property~\ref{itm:nosplit}, all the terms of the summation on the right are zero, hence the product:
        $$\PP_{\bsigma_{\|h}}(\pi^A) \PP_{\bsigma_{\|h}}(\mu_j = z_j \mid \pi^A)$$
        is positive, i.e. the play $\pi^A$ has  a positive probability of being generated and $\mu_j(\pi^A) = z_j$.

   \item \emph{If player $j$ is a pessimist,} then because of Property~\ref{itm:nosplit} again, for every $k \geq 0$ and each $w \in \Supp\left(\bsigma\left(h\pi^A_{\leq k}\right)\right)$, there exists a strategy $\tau_j^{kw}$ such that $\X_j(\bsigma_{-j\|h\pi^A_{\leq k}w}, \tau_j^{kw}) > z_j$.
        By composing all those strategies, we obtain a deviation $\tau_j$ of the strategy $\sigma_{j\|h}$; which, by Property~\ref{itm:pessimistanchor}, satisfies the inequality $\X_j(\bsigma_{-j\|h}, \tau_j) \leq z_j$.
        Therefore, either:
        \begin{itemize}
            \item we have:
            $$\min_k \min_{w \in \Supp\left(\bsigma\left(h\pi^A_{\leq k}\right)\right) \setminus \{\pi^A_{k+1}\}} \X_j(\bsigma_{-j\|h\pi_{\leq k}w}, \tau^{kw}_j) \leq z_j,$$
            which is impossible by definition of the strategies $\tau_j^{kw}$;

            \item or we have $\PP_{\bsigma_{-j\|h}, \tau_j}(\pi^A) = \PP_{\bsigma_{\|h}}(\pi^A) \neq 0$ and $\mu_j(\pi^A) \leq z_j$, and then actually $\mu_j(\pi^A) = z_j$.
        \end{itemize}
        \end{itemize}

We have thus proven that player $j$ gets the payoff $z_j$ in $\pi^A$, and that the play $\pi^A$ is generated with positive probability in $\bsigma$.
The analogous results about $\pi^{A\star}$ follow using the equalities $\Occ(\pi^{A\star}) = \Occ(\pi^A)$ and $\Inf(\pi^{A\star}) = \Inf(\pi^A)$.
\end{proof}

    In those two cases (if $A$ is a singleton or has several elements), we obtain that the strategy profile $\bsigma^\star$ is such that, with some positive probability, player $i$ gets the payoff $z_i$.

    \subparagraph*{Player $i$ gets risk measure $z_i$.}
    We still have to prove that player $i$ has zero probability of getting a lower payoff (if they are a pessimist) or a higher payoff (if they are an optimist).
    To show both cases, we prove the following result:

    \begin{slem}
        Every payoff vector that has a positive probability of being achieved in the strategy profile $\bsigma^\star$ also has a positive probability of being achieved in the strategy profile $\bsigma$.
    \end{slem}

\begin{proof}
    Let $\bz'$ be such a payoff vector.
    Then, there is a history $hw$ compatible with $\bsigma^\star$ and a set $A \subseteq \Pi$ such that, after the history $hw$, the strategy profile $\bsigma^\star$ is in state $\anchor_{A \last(h)}$, and from that point it has a nonzero probability of achieving the payoff vector $\bz'$ while staying in states of the form $\anchor_{A v}$.

    \begin{itemize}
        \item \emph{If $A$ is empty}, then the strategy profile $\btau^{\anch \emptyset}$ has been defined as a strategy profile that almost surely generates a payoff vector that is generated with positive probability by $\bsigma$, from every vertex from which that is possible.
    That requirement is satisfiable, and therefore satisfied by $\btau^{\anch \emptyset}$, from the vertex $w$, since that vertex is itself reached with positive probability in the strategy profile $\bsigma$.
    Therefore, the payoff vector $\bz'$ is also achieved with positive probability in $\bsigma$.
    
    \item\emph{If $A$ is a singleton,} say $A = \{j\}$, then the strategy profile $\btau^{\anch j}$ has been defined so that from every vertex from which that is possible, on the one hand, it generates the payoff $z_j$ with positive probability, and on the other hand, it is almost sure that the payoff vector that will be generated has also positive probability of being generated in $\bsigma$.
    Similarly as above, that requirement is satisfiable from the vertex $w$, since $\bsigma_{\|hw}$ satisfies it.
    Therefore, again, the payoff vector $\bz'$ is also achieved with positive probability in $\bsigma$.
    
    \item\emph{If $A$ has at least two elements}, then the strategy profile $\bsigma^\star$ stays in states of the form $\anchor_{A v}$ only along one play, namely $\pi^{A\star}$, and that play generates a payoff vector that was also associated with the play $\pi^A$ that by \cref{claim:piA}, has positive probability of being generated in $\bsigma$, hence the same conclusion. \qedhere
        \end{itemize}
\end{proof}

This proves the equality $\X_i(\bsigma^\star) = z_i$.
\end{proof}

\paragraph{The strategy profile $\bsigma^\star$ is an XRSE.}

We have now constructed the finite-memory strategy profile $\bsigma^\star$, showed that it had the expected number of memory states, and that it generates the expected risk measures.
We must now give the final argument for our construction: that strategy profile is also an extreme risk-sensitive equilibrium.
We will prove that result by showing separately that optimists have no profitable deviations, and then that neither do pessimists.

\begin{prop}\label{prop:NodeviationOpt}
    No optimist has a profitable deviation in $\bsigma^\star$.
\end{prop}

\begin{proof}
    Let $i$ be an optimist, and let us consider a deviation $\sigma'_i$ of that player from $\bsigma^\star$. Let us write $z'$ for the risk measure $z' = \X_i(\bsigma^\star_{-i}, \sigma'_i)$.

    Let us notice that along every play compatible with $\bsigma^\star_{-i}$, the transitions that are possible in the memory structure of the strategy profile $\bsigma^\star$ can be classified as follows:
    \begin{itemize}
        \item transitions among states of the form $\anchor_{A v}$ for a fixed $A$;
        
        \item transitions from a state of the form $\anchor_{A v}$ to a state of the form $\anchor_{B w}$ with $B \subset A$;

        \item transitions from a state of the form $\anchor_{A v}$ to the state $\punish_i$;

        \item and transitions from $\punish_i$ to itself.
    \end{itemize}
    Therefore, any such play stabilizes either in the state $\punish_i$, or among the states of the form $\anchor_{A v}$ for a fixed set $A$.
    Consequently, if in the strategy profile $(\bsigma^\star_{-i}, \sigma'_i)$ player $i$ gets the payoff $z'$ with positive probability, then we can also say that either:
    \begin{itemize}
        \item with positive probability, player $i$ gets the payoff $z'$ \emph{and} the state $\punish_i$ is reached;

        \item or there exists a set $A \subseteq \Pi$ such that with positive probability, player $i$ gets the payoff $z'$, and the collective memory remains in states of the form $\anchor_{A v}$.
    \end{itemize}

    \emph{In the first case,} let us consider a history $hv$ compatible with $\bsigma^\star_{-i}$ such that the collective memory is in an anchoring state after $h$ and in state $\punish_i$ after $hv$.
    If player $i$ can obtain the risk measure $z'$ by going to $v$ from that vertex against $\bsigma^\star_{-i\|h}$, and therefore, against the punishing strategy profile $\btau^{\dag i}_{-i}$, it means that they can enforce that risk measure against every possible strategy profile from $\last(h)$.
    On the other hand, if the collective memory is in an anchoring state after $h$, it means that the vertex $\last(h)$ is also visited with positive probability in the strategy profile $\bsigma$ (otherwise we would have switched to a punishing state earlier).
    There is therefore a history $h'$ compatible with $\bsigma$ such that $\last(h) = \last(h')$; and after that history, against the strategy profile $\bsigma_{\|h'}$, player $i$ also has the possibility of getting with positive probability the payoff $z'$.
    Since $\bsigma$ is an XRSE, that implies $z' \leq z_i$.

    \emph{In the second case,} let us notice that the strategy profiles of the form $\btau^{\anch A}$ are pure, and therefore that any deviation of player $i$ is immediately detected and leads to a switch to state $\punish_i$.
    Therefore, if the collective memory remains in states of the form $\anchor_{A v}$, it means that player $i$ is actually following the strategy $\sigma^\star_i$.
    Thus, we also have $z' \leq z_i$.
    
    The strategy $\sigma'_i$ is not a profitable deviation from $\bsigma^\star$.
\end{proof}

We can now end the proof with the dual proposition.

\begin{prop}
    No pessimist has a profitable deviation in $\bsigma^\star$.
\end{prop}

\begin{proof}
    Let $i$ be a pessimist, and consider a deviation $\sigma'_i$ of that player from $\bsigma^\star$.
    We intend to prove that the deviation $\sigma'_i$ is not profitable, that is, when following the strategy profile $(\bsigma^\star_{-i}, \sigma'_i)$, there is still a positive probability that player $i$ receives a payoff smaller than or equal to $z_i$.
    Using \cref{lm:secretlemma}, we can assume without loss of generality that $\sigma'_i$ is pure.

    First, we observe that for each history $hv$ compatible with $\bsigma^\star$ such that, after $hv$, the collective memory is in state $\anchor_{A \last(h)}$ with $i \in A$, the vertex $v$ is such that there also exists a history $h'v$ compatible with $\bsigma$ with $\Lambda(h'v) = A$.
    By Property~\ref{itm:pessimistanchor} of \cref{lm:Lambda}, we have $\X_i(\bsigma_{-i\|h'v}, \tau_i) \leq z_i$ for every $\tau_i$.
    Therefore, if player $i$ accepts to follow the history $hv$ and, then, deviates and takes an edge that makes the collective memory switch to the state $\punish_i$, then with positive probability player $i$ gets a payoff lesser than or equal to $z_i$.
    If such an action is ever performed, then the deviation $\sigma'_i$ is not profitable.

    Let us now assume that $\sigma'_i$ performs no such action: after every history $hv \in \Hist_i \Game_{\|v_0}$, if the collective memory is in a state of the form $\anchor_{A \last(h)}$ with $i \in A$, the vertex $\sigma'_i(hv)$ belongs to the set $\Supp(\sigma^\star_i(hv))$.
    Then, by Property~\ref{itm:splitsetsanchorwithi} of \cref{lm:Lambda}, we also have $i \in \Lambda(hv\sigma'_i(hv))$.
    Thus, there still exists a set $A$ with $i \in A$ such that, with positive probability, when following the strategy profile $(\bsigma^\star_{-i}, \sigma'_i)$, the strategy profile $\bsigma^\star_{-i}$ stabilizes among memory states of the form $\anchor_{A v}$; and then, the strategy profile $\bsigma^\star$ only proceeds to pure actions, hence the strategy $\sigma'_i$ is actually following $\sigma^\star_i$.

    Using the same arguments as in the proof of \cref{prop:ActualPayoff} (definition of $\btau^{\anch i}$ in case $A = \{i\}$, and \cref{claim:piA} in case $A$ has more elements), we can then conclude that player $i$ gets the payoff $z_i$ with positive probability and therefore that the deviation $\sigma'_i$ is not profitable.
    \end{proof}

    The strategy profile $\bsigma^\star$ is an XRSE, satisfies the equality $\X(\bsigma^\star) = \X(\bsigma)$, and uses the desired number of memory states.
    Furthermore, if $\bsigma$ is pure, so is $\bsigma^\star$.
\end{proof}

Finally, using \cref{thm:XRSEmemorysmall}, we can show the following lemma.

\begin{lem}\label{lm:XRSE_NPeasy}
    The constrained existence problem of XRSEs is in $\NP$. The same problem when players are restricted to pure strategies is still in $\NP$.
\end{lem}

\begin{proof}
    Let $\Game_{\|v_0}$ be a simple quantitative stochastic game.
    Let $(P,O)$ be a partition of $\Pi$, and let $\bx$ and $\by$ be threshold vectors.
    By \cref{thm:XRSEmemorysmall}, if there exists a (pure) XRSE with $\bx \leq \X(\bsigma) \leq \by$, then there exists one with at most $3np-2n+p+1$ memory states, where $p$ is the number of players and $n$ is the number of vertices.
    Such a strategy profile can be guessed in polynomial time.
    
    We now show that, once such a finite-memory strategy profile $\bsigma$ is guessed, one can check in polynomial time whether it is an XRSE, and satisfies the constraint $\bx \leq \X(\bsigma) \leq \by$.
    
    \begin{itemize}
        \item First, given $\bsigma$, for each player $i$, the quantity $\X_i(\bsigma)$ can be computed in polynomial time, since it reduces to computing player $i$'s risk measure in the Markov chain induced by $\bsigma$ (which has polynomial size) (by \cref{lm:secretlemma}).

        \item Second, checking that $\bx \leq \X(\bsigma) \leq \by$ can be done in polynomial time.

        \item Third, for each player $i$, one must check that player $i$ has no profitable deviation.
        This can also be done in polynomial time (by \cref{lm:secretlemma}) by computing the best risk measure player $i$ can get in the MDP induced by $\bsigma_{-i}$ (which has polynomial size).\qedhere
    \end{itemize}
\end{proof}

\subsection{Restrictions on strategies}

We now consider subcases where the space of a strategies is restricted. 
We show in \cref{thm:infinite_rho_restricted_strategy_np_easy} that restricting the memory or amount of randomness of the strategy still renders the problem only in $\NP$.
Later in this section, we prove that all these problems, including the general problem, are $\NP$-hard. This subsection therefore completes the proof of \cref{thm:XRSE_NPc,thm:infinite_rho_restricted_strategy_np_easy}.

We restrict the set of strategies of each player to stationary, positional or pure. 
We show that the problem is in $\NP$ for each of these cases.

\begin{lem}\label{lm:restrictionsNPeasy}
    The constrained existence problem, when all the players are restricted to positional, stationary, or pure strategies, is in $\NP$. 
\end{lem}

\begin{proof}
    We show that we can still guess a strategy profile, and verify in polynomial time if it is indeed an XRSE.
    For the cases of positional and stationary strategies, guessing a strategy profile is straightforward, since such a strategy profile $\bsigma$ can be represented using polynomially many bits.
    We can  then verify that a given strategy profile $\bsigma$ gives risk measures within the constraints, and also is an XRSE in polynomial time (\cref{lm:secretlemma}). 

    However, for pure strategies, memory might be required. But we showed with \cref{thm:XRSEmemorysmall} that if there is a pure strategy profile, then there is one that requires polynomial memory, and therefore our result follows.  
\end{proof}

We now prove $\NP$-hardness of the constrained existence problem for the general setting as well as the cases where strategies are restricted. 

\begin{lem}\label{lm:XRSE_NPhard}
    The constrained existence problem of XRSEs
    is $\NP$-hard, even when all players are pessimists and all rewards are non-negative.
    It remains $\NP$-hard when the strategies are reduced to stationary, pure, or positional ones.
\end{lem}

\begin{proof}
  We prove $\NP$-hardness by reducing from the problem $\THREESAT$. Consider a $\THREESAT$ formula $\phi$, over the variables  $x_1,\dots,x_n$, where $\phi = C_1\land C_2\land \dots\land C_m$, where for each $i$ we have $C_i = (\ell_{i1}\lor \ell_{i2}\lor \ell_{i3})$ and for $j=1,2,3$, we have $\ell_{ij} = x_k$ or $\ell_{ij} = \neg x_k$ for some $k\in \{1, \dots, n\}$. 
  We construct  a game $\Game_\phi$ with two players for each literal $\ell$, denoted by $\Circle \ell$ and $\Square \ell$. The game is depicted in \cref{fig:NPhard}.
  For convenience, some terminal vertices have been represented several times.
  Each player $\Circle \ell$ controls one vertex, the vertex $\Circle \l$, of circled shape, and symmetrically, each player $\Square \ell$ controls the square-shaped vertex $\Square \l$.
  Further, we add a player $C_i$, who controls the vertex $C_i$, for each clause $C_i$. Finally, there is a player $\Diamond$ who does not control any vertex. There are also stochastic vertices, that are represented by the black circles.
  In each terminal vertex, the symbol $\forall$ should be understood as "every (other) player".

 We assume all players are pessimistic, and ask if there is an XRSE where player $\Diamond$'s risk measure is exactly $2$.
We give the formal definition of the game $\Game_\phi$ below.

    \begin{figure}
        \centering
        
        \begin{tikzpicture}[node distance=1.7cm and 2.2cm, on grid, auto]

            \node[vert] (qnc1) at (0, 0) {$\neg x_{1}$};
            \node[vert, initial,initial text=] (qc1) at (0, 1) {$x_{1}$};

            \node[scale=0.6] (tpunish1) at (0,-1) {$t_\dag:~\stackrel{\forall}{0}$};
            
            \node[stoch, scale=0.75] (stoc2) at (1, 0) {$s_{\neg x_1}$};
            \node[stoch, scale=0.75] (stoc1) at (1, 1) {$s_{x_1}$};

            \node[scale=0.6] (reward1) at (1,2) {$f_{x_1}:~\stackrel{\circ x_1}{1}$$\stackrel{\forall}{2}$};
            \node[scale=0.6] (reward2) at (1,-1) {$f_{\neg x_1}:~\stackrel{\circ \neg x_1}{1}$$\stackrel{\forall}{2}$};
            
            \node[vert, rectangle] (qns1) at (2, 1) {$\neg x_{1}$};
            \node[vert, rectangle] (qs1) at (2, 0) {$x_{1}$};

            \node[scale=0.6] (punishW1) at (2,2) {$t_\diamond:~\stackrel{\diamond}{0}$$\stackrel{\forall}{2}$};
            \node[scale=0.6] (punishW2) at (2,-1) {$t_\diamond:~\stackrel{\diamond}{0}$$\stackrel{\forall}{2}$};
            
            \node[vert] (qnc2) at (3.5, 0) {$\neg x_{2}$};
            \node[vert] (qc2) at (3.5, 1) {$x_{2}$};
            \node[stoch, scale=0.75] (stoc3) at (4.5, 1) {$s_{x_2}$};
            \node[stoch, scale=0.75] (stoc4) at (4.5, 0) {$s_{\neg x_2}$};

            \node[scale=0.6] (reward3) at (4.5,2) {$f_{x_2}:~\stackrel{\circ x_2}{1}$$\stackrel{\forall}{2}$};
            \node[scale=0.6] (reward4) at (4.5,-1) {$f_{\neg x_2}:~\stackrel{\circ \neg x_2}{1}$$\stackrel{\forall}{2}$};
            
            \node[vert, rectangle] (qns2) at (5.5, 1) {$\neg x_{2}$};
            \node[vert, rectangle] (qs2) at (5.5, 0) {$x_{2}$};

            \node[scale=0.6] (punishW3) at (5.5,2) {$t_\diamond:~\stackrel{\diamond}{0}$$\stackrel{\forall}{2}$};
            \node[scale=0.6] (punishW4) at (5.5,-1) {$t_\diamond:~\stackrel{\diamond}{0}$$\stackrel{\forall}{2}$};

            \node[scale=0.6] (tpunish2) at (3.5,-1) {$t_\dag:~\stackrel{\forall}{0}$};
            \node (qc3) at (6.5, 1) {};

            \node (dots) at (6.8,0.5) {$\dots$};

            \node[vert, rectangle, initial,initial text=] (qnsn) at (8, 1) {$\neg x_{n}$};
            \node[vert, rectangle, initial,initial text=] (qsn) at (8, 0) {$x_{n}$};

            \node[scale=0.6] (punishW5) at (8,2) {$t_\diamond:~\stackrel{\diamond}{0}$$\stackrel{\forall}{2}$};
            \node[scale=0.6] (punishW6) at (8,-1) {$t_\diamond:~\stackrel{\diamond}{0}$$\stackrel{\forall}{2}$};
            
            \node[stoch, scale=0.75] (stochfin) at (9,0.5) {$s_\mathsf{r}$};
            \node (fakenode) at (9,0.5) {};
            
            \node (c1) at (10,1.8) {$C_1$};
            \node (c2) at (10,1) {$C_2$};
            \node (cdots) at (10,0.4) {$\vdots$};
            \node (c3) at (10,-0.6) {$C_m$};

            \node[scale=0.6] (ter1) at (11.2, 1.7) {$t_{x_2}:~\stackrel{\Box x_2}{1}$ $\stackrel{\forall}{2}$};
            \node[scale=0.6] (ter2) at (11.2, 1) {$t_{x_4}:~\stackrel{\Box x_4}{1}$ $\stackrel{\forall}{2}$};
            \node[scale=0.6] (ter3) at (11.2, -0.3) {$t_{\neg x_{11}}:~\stackrel{\Box\neg x_{11}}{1}$ $\stackrel{\forall}{2}$};

          \path[->]
              (qc1) edge (stoc1)
              (qc1) edge (qnc1)
              (qnc1) edge (stoc2)
              (stoc1) edge (qns1)
              (stoc2) edge (qs1)
              (qns1) edge (qc2)
              (qs1) edge (qc2)
              (qc2) edge (stoc3)
              (qc2) edge (qnc2)
              (qnc2) edge (stoc4)
              (stoc3) edge (qns2)
              (stoc4) edge (qs2)
              (qns2) edge (qc3)
              (qs2) edge (qc3)
              (qnsn) edge (stochfin)
              (qsn) edge (stochfin)
              (fakenode) edge (c1)
              (fakenode) edge (c2)
              (fakenode) edge (c3)
              (c2) edge (ter1)
              (c2) edge (ter2)
              (c2) edge (ter3)
              (qnc1) edge (tpunish1)
              (qnc2) edge (tpunish2);

        \path[->]
            (stoc1) edge (reward1)
            (stoc2) edge (reward2)
            (stoc3) edge (reward3)
            (stoc4) edge (reward4)
            (qs1) edge (punishW2)
            (qns1) edge (punishW1)
            (qs2) edge (punishW4)
            (qns2) edge (punishW3)
            (qsn) edge (punishW6)
            (qnsn) edge (punishW5);
        
        \end{tikzpicture}
        \caption{Construction of a game $\Game_\phi$ from a $\THREESAT$ formula $\phi$}
        \label{fig:NPhard}
    \end{figure}

\paragraph{Construction of the game $\Game_\phi$: vertices, edges and payoffs}

For each literal $\ell$, we define two players $\Square\ell$ and $\Circle\ell$. We add one other player $C_i$ for each clause $C_i$, and an additional constraining player $\Diamond$.
All players are pessimists.

Each player owns at most one vertex in the game, and therefore, we will refer to the player and vertex interchangeably. There is one vertex for each of the players mentioned above other than $\Diamond$, who owns no vertices. Further, there are $2n + 1$ many stochastic vertices: one for each literal $s_{x_1},s_{x_2},\dots,s_{x_n}$,  $s_{\neg x_1},s_{\neg x_2},\dots,s_{\neg x_n}$, and finally one clause-randomizer $s_\mathsf{r}$. 
There are also $2n + 2$ terminal vertices, written $f_{\ell}$ and $t_{\ell}$ for each literal $\ell$, and further the terminal vertices $t_\Diamond$ and $t_\dag$.

We now define the edges between the vertices of the graph for all $i\in \{1, \dots, n\}$:  there are edges from $\Circle x_i$ to $\Circle\neg x_i$, and edges from $\Circle\neg x_i$ to $t_\dag$.
    Further, for every literal $\ell = x_i$ or $\neg x_i$, there are edges:
    \begin{itemize}
        \item from $\Circle\ell$ to $s_{\ell}$;
        \item from $s_{\ell}$ to $f_\ell$ and to $\Square\Bar{\ell}$, where $\Bar{\ell} = \neg x_i$ if $\ell = x_i$ and $\Bar{\ell} = x_i$ if $\ell = \neg x_i$;
        \item from $\Square\ell$ to $t_\Diamond$;
        \item from $\Square\ell$ to $\Circle x_{i+1}$ if  $i<n$,  and  to $s_\mathsf{r}$ if  $i=n$.
    \end{itemize}
    Finally, for all clauses $C_j$, there are edges from $s_\mathsf{r}$ to $C_j$ and from $C_j$ to $t_{\ell}$ such that $\ell$ occurs positively in the clause $C_j$.

    The terminal vertices yield the following payoffs.
\begin{itemize}
    \item In terminal $t_\ell$, all players get payoff $2$, except the player $\Square \ell$ who gets payoff $1$. 
    \item In terminal $f_\ell$, all players get payoff $2$, except player $\Circle \ell$ who gets payoff $1$.
    \item In terminal $t_\dag$, all players get payoff $0$.
    \item In terminal $t_\Diamond$, all players get payoff $2$, except player $\Diamond$ who gets payoff $0$.
\end{itemize}    

Finally, we let the constraints be that player $\Diamond$ gets a risk measure of exactly $2$.
Equivalently, we define $\bx$ and $\by$ by $\by = (2)_{i \in \Pi}$, $x_i = 0$ for each $i \in \Pi \setminus \{\Diamond\}$, and $x_\Diamond = 2$.

    \paragraph{If $\phi$ is satisfiable, then there is an XRSE satisfying the constraints.}
    
    Consider a satisfying valuation $\nu$ of the $\THREESAT$ formula $\phi$.
    
    For each $i$, let $\ell_i$ denote the literal, among $x_i$ and $\neg x_i$, which is set to true by 
    the satisfying valuation $\nu$.
    Let us define the (positional) strategy profile $\bsigma^\nu$.
    
    \begin{itemize}
        \item Player $\Circle \ell_i$ goes to $s_{\ell_i}$.
        \item Player $\Circle x_i$ goes to $s_{x_i}$ if $\nu(x_i) = 1$, and to $\Circle \neg x_i$ otherwise.
        \item Player $\Circle\neg x_i$ goes to $s_{\neg x_i}$ if $\nu(x_i) = 1$ and to $t_\dag$ otherwise.
        \item For each player $\Square \ell$, the strategy is to chose the edge that does \emph{not} lead to $t_\Diamond$. That is, the edge to $\Circle x_{i+1}$ if $\ell = x_i$ or $\neg x_{i}$ and  $i<n$,  and  the edge to $s_\mathsf{r}$ if  $i=n$.
        \item Each clause player $C_i$ takes the edge to the vertex $\ell_j$ such that the litteral $\ell_j$ was set to true by the satisfying valuation $\nu$. 
    \end{itemize}
     We now show that this is an XRSE that satisfies the constraint. First, we verify if the constraints are satisfied. Observe that following the strategy profile $\bsigma^\nu$, it is almost sure that none of the terminals where player $\Diamond$ has payoff less than $2$ will be reached. Therefore this satisfies the constraints. 

     We now argue that $\bsigma^\nu$ is an XRSE, i.e. that no player can get a better risk measure by deviating.
     The result is immediate for player $\Diamond$ and for the clause players, who all get risk measure $2$, the best they could hope for.
     
     For each literal $\l$, player $\Circle\ell$ gets risk measure $1$ if $\l$ is set to true, and risk measure $2$ if $\ell$ is set to false.
     The same argument as above holds therefore in the second case.
    In the first case, she gets risk measure $0$, but she has no profitable deviation, since the only deviation available leads to $t_\dag$ and to the payoff $0$.
     
     Player $\Square \ell$  has also risk measure  $2$ when $\ell$ is set to false.
     Otherwise, he gets payoff $1$. In that second case, the vertex owned by the player is not visited in any history of the game, hence he has no possibility of deviating.

     The (positional) strategy profile $\bsigma^\nu$ is therefore an XRSE.
     
    \paragraph{If there is an XRSE satisfying the constraints, then $\phi$ is satisfiable.} 
    
    Let us assume that there exists an XRSE $\bsigma$ in the game $\Game_\phi$, such that player $\Diamond$ gets the risk measure $2$.
    We prove, first, that we can assume that $\bsigma$ is pure (and therefore positional, since there is then only one history leading to each vertex).

\begin{slem}
    There exists a positional XRSE $\bsigma^\star$ in $\Game_{\phi\|x_1}$ where player $\Diamond$ gets risk measure $2$.
\end{slem}

\begin{proof}
    Let us first focus on what happens in vertices that have positive probability of being reached.

    If the vertex $\Circle\neg x_i$ has a positive probability of being reached in $\bsigma$, then any strategy of the player $\Circle \neg x_i$ that goes to $t_\dag$ with positive probability gives the player $\Diamond$ the risk measure $0$.
    Therefore, necessarily, the strategy $\sigma_{\circ \neg x_i}$ consists of deterministically going to $s_{\neg x_i}$.
    The same argument holds for the vertices of the form $\Square \l$.
    
    If now the vertex $\Circle x_i$ has a positive probability of being reached and if the player $\Circle x_i$ randomizes between the two edges available, then she gets the risk measure $1$, since the terminal vertex $f_{x_i}$ is reached with positive probability and $t_\dag$ with probability zero.
    But then, if she deviates and goes to the vertex $\Circle \neg x_i$ with probability $1$, she avoids the terminal vertex $f_{x_i}$, and the other players will not react since they do not detect the deviation.
    She therefore gets the risk measure $2$, and the deviation is profitable.
    Consequently, the strategy $\sigma_{\circ x_i}$ can only  deterministically select one of those two edges.

    At the end of the game, for each $j$, the player $C_j$ could play a randomized strategy. In such a case, her strategy can be replaced by a pure strategy that takes, deterministically, one of the edges that she was previously taking.
    Such a modification in her strategy can only increase the risk measure of some players (namely, those of the form $\Square \l$) without impacting player $\Diamond$'s risk measure or giving any player the possibility of profitably deviating.

    Finally, if one of those vertices is reached after a history that is not compatible with $\bsigma$, i.e. if one of those players deviates: it is necessarily due to a deviation of a player of the form $\Circle x_i$, since any other deviation would immediately lead to a terminal vertex.
    If she went to $s_{x_i}$ instead of $\Circle \neg x_i$, what the other players do afterwards does not matter, since such a deviation cannot be profitable: with positive probability, the terminal vertex $f_{x_i}$ is reached, and she gets payoff $1$.
    If she went to $\Circle x_i$ instead of $s_{x_i}$, then we can assume that player $\Circle \neg x_i$'s strategy consists of going to the terminal vertex $t_\dag$, giving her the payoff $0$.
    Those modifications do not impact the fact that $\bsigma$ is an XRSE.
\end{proof}

We therefore assume that $\bsigma$ is positional.
Let us now define the valuation $\nu$ as follows: for each variable $x_i$ we have $\nu(x_i) = 1$ if $\sigma_{\circ x_i}(\Circle x_i) = s_{x_i}$, and $\nu(x_i) = 0$ if $\sigma_{\circ x_i}(\Circle x_i) = \Circle \neg x_i$.
Let then $C_j$ be a clause, and let us prove that it is satisfied by $\nu$.
Let $t_\l = \sigma_{C_j}(C_j)$.
Then, the player $\Square \l$ gets risk measure $1$ in the XRSE $\bsigma$.
Consequently, the vertex $\Square \l$ is never reached: otherwise, the only play compatible with $\bsigma$ in which player $\Square \l$ gets payoff $1$ would traverse the vertex $\Square \l$, and player $\Square \l$ would have a profitable deviation by going to the terminal vertex $t_\Diamond$.
If that is the case, then the definition of $\nu$ given above implies that the literal $\l$ is true.

The valuation $\nu$ satisfies therefore the formula $\phi$.

\paragraph{Conclusion.}
We have defined an instance of the constrained existence problem of XRSEs from an instance of $\THREESAT$ and proved that one is a positive instance if and only if the other is.
This proves the $\NP$-hardness of the constrained existence problem of XRSEs, since the game $\Game_\phi$ can clearly be constructed in polynomial time.
Moreover, the game $\Game_\phi$ is such that if an XRSE where player $\Diamond$ gets risk measure $2$ exists, then there also exists such an equilibrium that it positional, which proves also $\NP$-hardness when the players are restricted to pure, stationary or positional strategies.
\end{proof}

This lemma, along with \cref{lm:XRSE_NPeasy}, proves \cref{thm:XRSE_NPc}; and along with \cref{lm:restrictionsNPeasy}, it proves \cref{thm:infinite_rho_restricted_strategy_np_easy}.

\subsection{Things get easier when everyone is optimistic}

Since our $\NP$-hardness results involved only pessimistic players, we now show that
the constrained existence problem of XRSEs becomes $\PTime$-complete when the perceived reward of each player is computed based on the risk measure $\OM$, thus proving \cref{thm:XRSE_Pcomplete}.
We first show an upperbound by giving a polynomial-time algorithm.  

\begin{lem}\label{lm:ptimeupperbound}
    If all players are optimists, then the constrained existence problem of XRSEs is in $\PTime$, and there is an algorithm that decides it in time $O(pm^2)$, where $m$ is the number of edges in $\Game$ and $p$ the number of players.
    Moreover, the algorithm can be modified to output an XRSE that satisfies the constraints, if one exists, in time $O(pm^2 + m^3)$---or in time $O(pm^2)$ if all upper thresholds are non-negative.
\end{lem}

\begin{proof} \paragraph{Preliminary remarks}

We are given the game $\Game_{\|v_0}$ and two threshold vectors $\bx, \by \in \QQ^{\Pi}$; we wish to find an XRSE $\bsigma$ such that $\bx \leq \X(\bsigma) \leq \by$. 

    Throughout the proof, when $W \subseteq V$ is a set of vertices and $F \subseteq E$ is a set of edges, we write $\Attr(W, F)$ for the \emph{positive probabilistic attractor} of $W$ in the graph $(V, F)$, i.e. the set of vertices $v$ such that for every strategy profile $\bsigma$ in $\Game_{\|v}$ that uses only edges of $F$, there is a positive probability of reaching $W$.
    As a consequence of \cref{lm:secretlemma} (replacing the vertices of $W$ with terminal vertices), we have the following.
    
    \begin{slem}\label{claim:positiveattractorLinear}
      Given $W$, the set $\Attr(W, F)$ can be computed in time $O(m)$.
    \end{slem}

Similarly, \cref{lm:secretlemma} enables us to compute the \emph{adversarial values} of each vertex, i.e., the best risk measure that the player controlling that vertex can ensure from that vertex when the other players are fully hostile.

    \begin{slem}\label{claim:adverserialXRLinear}
        For each $i$ and $v \in V_i$, the quantity:
        $$\val(v) = \inf_{\btau_{-i} \in \Strat_{-i}\Game_{\|v}} \sup_{\tau_i \in \Strat_i\Game_{\|v}} \X_i(\btau)$$
        can be computed in time $O(m)$. 
    \end{slem}

Then, computing all those values can be done in time $O(m^2)$.
We can therefore assume that those quantities $\val(v)$ are given with the input.

    \paragraph{Cycle-friendly and cycle-averse cases}

We differentiate two types of instances. 
    If there exists a player $i$ such that we have $y_i < 0$, then the requirement $\bx \leq \X(\bsigma) \leq \by$ implies that $\bsigma$ must almost surely reach a terminal vertex: we call that case the \emph{cycle-averse} case.
    If there is no such player, we are in the \emph{cycle-friendly} case.
    Our algorithm will work slightly differently in those two cases.
    However, the fundamental idea is still the same in both cases: we prune iteratively the set of edges, and each of the subsets $F \subseteq E$ which we obtain will induce a strategy profile $\bsigma^F$, in which the profitable deviations will be detected and used to prune new edges.
    However, the definition of $\bsigma^F$ differs in the cycle-averse and the cycle-friendly case.

    \paragraph{Algorithm in the cycle-friendly case}
    In the cycle-friendly case, for a given set of edges $F$, the strategy profile $\bsigma^F$ in the game $\Game_{\|v_0}$, is defined as follows: from each non-stochastic vertex $v$, when $v$ is seen for the first time, the strategy profile randomizes uniformly between all the edges $vw \in F$.
    Later, when $v$ is visited again, it always repeats the same choice.
    Equivalently, each player initially chooses, at random, a positional strategy, and then follows it.
    If some player $i$ deviates and takes an edge that they are not supposed to take (be it an edge that does not belong to $F$ or an outgoing edge of a vertex from which a different edge has already been taken), then all the players switch to the positional strategy profile $\btau^{\dag i}$, where $\btau^{\dag i}_{-i}$ minimizes the best risk measure that player $i$ can get (a positional such strategy profile exists by \cref{lm:secretlemma}), and $\tau^{\dag i}_i$ is some positional strategy.

    Our algorithm in the cycle-friendly case is presented in~\cref{algo:cyclefriendly}.
    Each step $k$ consists of identifying a new set of vertices $V_\bad^k$ that must be avoided.
    At step $k=0$, it is the set of terminal vertices that give some player $i$ a payoff that is larger than $y_i$, which would then make them have an off-constraints risk measure.
    At step $k \geq 1$, it is the set of vertices $v$ whose adversarial value $\val(v)$ is greater than the risk $z_i^k = \X_i(\bsigma^{E_k})$, where $i$ is the player controlling $v$.
    In other words, the vertices from which that player can have a profitable deviation.
    Note that it that second case, the computation of $V_\bad^k$ requires the computation of $z_i^k$, which can be done in time $O(m)$ by computing the set of terminals that are accessible from $v_0$ in $(V, E_k)$, and by deciding whether the probability of reaching no terminal is positive: that will be the case if and only if there exists a positional strategy profile that uses only edges of $E_k$ (and therefore that $\bsigma^{E_k}$ is following with positive probability) such that with positive probability no terminal vertex is reached, which can be decided in time $O(m)$ using \cref{lm:secretlemma}.

    Then, the positive probabilistic attractor $A_k = \Attr(v_\bad^k, E_k)$ is computed.
    If $k \geq 1$ and $v_0 \in A_k$, i.e., if it is not possible to avoid reaching the set $V_\bad^k$, the answer $\No$ is returned.
    Otherwise, the set $E_{k+1}$ is defined from $E_k$ by removing all the edges that lead from a vertex that does not belong to $A_k$ to a vertex that does, thus making sure that $V_\bad^k$ will never be reached.
    The algorithm stops when there is no more edge to remove.
    Then, if we have $z_i^k \geq x_i$ for each $i$, the algorithm answers $\Yes$ and outputs the set $E_k$ as a succinct representation of the strategy profile $\bsigma^{E_{k+1}}$.
    Otherwise, it answers $\No$.

            \begin{algorithm}
            \begin{algorithmic}\caption{Constrained existence problem with optimists in the cycle-friendly case}\label{algo:cyclefriendly}
                \Procedure{CycleFriendly}{$\Game, \Bar{x},\Bar{y}$}
                    \State $k \gets 0$
                    \State $E_k\gets E$
                    \State $V_\bad^k = \{t\in T \mid \mu_i(t)> y_i\}$ 
                    \State $A_k \gets \Attr(V^k_\bad, E_k)$
                    \If{$v_0 \in A_k$}
                        \Return{$\No$}
                    \Else
                        \State $E_{k+1} \gets E_k \setminus \{uv\in E_k\mid u\not\in A_k\text{ and }v\in A_k\}$
                    \While{$k=0$\text{ or }$E_{k+1}\neq E_k$}
                        \State $k\gets k+1$
                        \State Compute $z^k_i = \X_i(\bsigma^{E_k})$ for each $i \in \Pi$
                        \State $V^k_\bad \gets \{v \mid \val(v) > z_i^k \text{ for } i \in \Pi \text{ such that } v \in V_i\}$
                        \State $A_k \gets \Attr(V^k_\bad, E_k)$
                        \If{$v_0 \in A_k$}
                            \Return{$\No$}
                        \Else
                            \State $E_{k+1} \gets E_k \setminus \{uv\in E_k\mid u\notin A_k\text{ and }v\in A_k\}$
                        \EndIf
                    \EndWhile
                    \EndIf
                    \If{$z_i^k\geq x_i$ for all players $i$}
                        \Return $(\Yes, E_{k+1})$
                    \Else{\text{ }}\Return{$\No$}
                    \EndIf
                \EndProcedure
            \end{algorithmic}
        \end{algorithm}

     \subparagraph*{Correctness in the cycle-friendly case.}
To prove the correctness of \cref{algo:cyclefriendly}, we first need to prove that the edge removals are such that all vertices always keep at least one outgoing edge, and that the stochastic ones always keep all of them, so that the strategy $\bsigma^{E_k}$ is always properly defined.

\begin{inv}
    At each step $k$, every vertex $v \not\in V_?$ is such that $E_k(v) \neq \emptyset$, and every vertex $v \in V_?$ is such that $E_k(v) = E(v)$.
\end{inv}

The proof is an immediate induction.
We also need termination.

\begin{prop}
    \cref{algo:cyclefriendly} terminates.
\end{prop}

\begin{proof}
    At each step $k \geq 1$, we either have that the algorithm terminates, or that an edge is removed.
    The sequence $E_1, E_2, \dots$ is therefore strictly decreasing (note that we might have $E_0 = E_1$), hence it cannot be infinite.
\end{proof}

Now, to prove correctness, we will first prove the following result.

        \begin{slem} \label{claim:zik}
            For each player $i$ and each index $k$, every strategy profile $\bsigma'$ that uses only edges of $E_k$ is such that $\X_i(\bsigma') \leq z_i^k$.
        \end{slem}

        \begin{proof}
            This result is a consequence of the fact that every payoff vector that can be obtained with positive probability in the strategy profile $\bsigma'$ is obtained with positive probability in the strategy profile $\bsigma^{E_k}$.
            
            Indeed, consider some payoff vector $\bz$ that has a positive probability of being generated in $\bsigma'$.
            If $\bz$ is obtained by reaching a terminal vertex, then that terminal vertex is accessible from $v_0$ in the graph $(V, E_k)$, and it therefore has a positive probability of being reached in $\bsigma^{E_k}$.
            
            If $\bz = (0)_i$ is obtained by reaching no terminal vertex, then by \cref{lm:secretlemma}, there exists a positional strategy profile $\btau$ that uses only edges of $E_k$ such that with positive probability, no terminal vertex is reached.
            Then, when following the strategy profile $\bsigma^{E_k}$, there is a positive probability that the players actually follow $\btau$.
            And therefore, there is also a positive probability to get the payoff vector $\bz = (0)_i$ in the strategy profile $\bsigma^{E_k}$.

            Since all players are optimists, the result follows.
        \end{proof}

Note that this result implies that the sequence $(z_i^k)_k$, for each $i$, is nondecreasing.

We can now prove correctness.
To do so, we need to prove two propositions: the algorithm recognizes only positive instances, and recognizes all of them.

\begin{prop}
    The algorithm recognizes only positive instances.
\end{prop}

\begin{proof}
        Let us assume that the algorithm answers $\Yes$ at step $k$: let us show that the strategy profile $\bsigma^{E_k}$ is an XRSE that satisfies the desired constraints.
        Note that the algorithm does never switch to the final refinements at step $0$, hence we necessarily have $k \geq 1$.

        \subparagraph*{The strategy profile $\bsigma^{E_k}$ satisfies $\bx \leq \X({\bsigma^{E_k} }) \leq \by$.}
        
        The lower bound is immediate since the algorithm answers $\Yes$ at step $k$ only if the strategy profile $\bsigma^{E_k}$ satisfies that constraint.

            Regarding the upper bound, observe that the set $E_1$ has been defined so that the set $\Attr(V_{\frownie}^0, E)$, and therefore the set $V_{\frownie}^0$, is not accessible from $v_0$ in the graph $(V, E_1)$, and therefore not in the graph $(V, E_{k+1})$.
            Thus, it is almost sure in $\bsigma^{E_{k+1}}$ that no vertex of $V_{\frownie}^0$ will ever be reached.
            In other words, all terminals that have a positive probability of being reached give each player $i$ a lower payoff than $y_i$.
            Now, if there is a positive probability that the play never reaches a terminal, that also does not give any player $i$ such a payoff, since we are in the cycle-friendly case.

        \subparagraph*{The strategy profile $\sigma^{E_k}$ is an XRSE.}
        
        Let $i$ be a player, and let $\sigma'_i$ be a deviation of player $i$ from $\bsigma^{E_k}$.
        We can assume without loss of generality that $\sigma'_i$ is pure.        
        Let $z' = \X_i({\bsigma^{E_k}_{-i}, \sigma'_i})$ be the extreme risk measure obtained by player $i$.
        We want to prove that the deviation $\sigma'_i$ is not profitable, that is, we have $z' \leq z_i^k$.

        If the deviation $\sigma'_i$ uses only the edges of $E_k$, then it cannot be profitable by \cref{claim:zik}.
        But if it does use more edges, let us show that it cannot be a profitable deviation either.
 
\begin{slem}
    If there is a history $hv$ compatible with $\bsigma^{E_k}$ such that $v\sigma'_i(hv) \not\in E_k$, then we have $\X_i(\bsigma^{E_k}_{\|hv}, \sigma'_{i\|hv}) \leq z_i^k$.
\end{slem}

\begin{proof}
    After such a history, the strategy profile $\bsigma^{E_k}_{-i}$ follows the positional strategy profile $\btau^{\dag i}_{-i}$.
    By the definition of that strategy profile, we have $\X_i(\bsigma^{E_k}_{\|hv}, \sigma'_{i\|hv}) \leq \val(v)$.
    On the other hand, the vertex $v$ is accessible from $v_0$ in $(V, E_k)$, since it is visited with a positive probability in $\bsigma^{E_k}$.
    Therefore, it does not belong to the set $A_k$, and in particular not to the set $V_\bad^k$, which means that we have $\val(v) \leq z_i^k$.
    Hence, the conclusion follows. 
\end{proof}

In the general case, the payoffs that player $i$ obtains with positive probability in the strategy profile $(\bsigma^{E_k}_{-i}, \sigma'_i)$ are obtained either by using only edges that belong to $E_k$, or by using an edge that does not. In both cases, we have shown that player $i$ cannot get a payoff greater than $z_i^k$, which proves that the strategy profile $\bsigma^{E_k}$ is an XRSE.
\cref{algo:cyclefriendly} answers $\Yes$ only on positive instances, and outputs in that case a succinct representation of an XRSE matching the constraints.
\end{proof}

It now remains to prove the converse.
 
            \begin{prop}
                \cref{algo:cyclefriendly} recognizes all positive instances. 
            \end{prop}
    
            \begin{proof}
           Let us assume that we have a positive instance, i.e., that there exists an XRSE $\bsigma$ with $\bx \leq \X(\bsigma) \leq \by$.
        Let us show that the algorithm will answer $\Yes$.
        To do so, we first prove the following invariant: if an edge is removed at some step, then it is never taken by the XRSE $\bsigma$.

        \begin{inv} \label{inv:edgesnotused}
            For each $k \geq 0$, every edge that has positive probability of being eventually taken in $\bsigma$ belongs to $E_k$.
        \end{inv}

        \begin{proof}
        We prove the invariant by induction. 
        
        \subparagraph*{Base cases.} The case $k=0$ is immediate, since we have $E_0 = E$.
        
        Further, at step $k=1$, if the strategy profile $\bsigma$ uses eventually, with positive probability, an edge that does not belong to $E_1$, then it goes with positive probability to a vertex $v \in \Attr(V^0_\bad, E_0)$.
        Then, with positive probability, a terminal vertex will be reached that gives to some player $i$ a payoff greater than $y_i$, which is impossible.
        Therefore, such an edge cannot be taken in $\bsigma$.

        \subparagraph*{Induction step.} Let us assume that the invariant is true until step $k \geq 1$, and let us show that it holds at step $k+1$.    
        Let $uv$ be an edge that is used with positive probability when following $\bsigma$, and let us assume toward contradiction that it does not belong to $E_{k+1}$.
        Since the invariant is true at each step until $k$, we can assume that $uv$ has been removed at step $k$, i.e., that we have $uv \in E_k \setminus E_{k+1}$.
        Then, we have $u \not\in A_k$ and $v \in A_k$.
        The strategy profile $\bsigma$ has therefore positive probability of visiting the set $A_k$, and therefore the set $V_\bad^k$.
        Then, from a vertex of $V_\bad^k$, i.e., a vertex $v$ with $\val(v) > z_i^k$, player $i$ can deviate and get a risk measure strictly better than $z_i^k$.
        But since the invariant is true at step $k$, the strategy profile $\bsigma$ uses only vertices of $E_k$, and therefore, by \cref{claim:zik}, we have $\X_i(\bsigma) \leq z_i^k$: player $i$ has a profitable deviation in $\bsigma$, which is impossible.
\end{proof}

        We are now able to conclude.
        The answer $\No$ can be given in the two following cases:
            \begin{itemize}
                \item \emph{If at step $k$, we have $v_0 \in \Attr(V^k_\bad, E_k)$.}
                Then, the strategy profile $\bsigma$ visits the set $V^k_\bad$ with positive probability.
                With the same arguments that were used in the proof of \cref{inv:edgesnotused}, that is not possible.
                
                \item \emph{If during step $k$, no edge is removed, but we have $z^k_i < x_i$ for some player $i$.}
                Since $\bsigma$ uses only edges of $E_k$ by \cref{inv:edgesnotused}, we can apply \cref{claim:zik}, and obtain $\X_i(\bsigma) \leq z^k_i$, and therefore $\X_i(\bsigma) < x_i$: that case is therefore also impossible by definition of $\bsigma$.
            \end{itemize}
        None of those cases is possible, hence our algorithm will eventually answer $\Yes$.
    \end{proof}

    \paragraph{Algorithm in the cycle-averse case}

    The algorithm and the structure of the proof will be similar.
    However, we need some significant modifications, especially in the definition of the strategy profiles $\bsigma^F$.

    In the cycle-averse case, for a given set of edges $F$, the strategy profile $\bsigma^F$, in the game $\Game_{\|v_0}$, is defined as follows: from each vertex $v \not\in V_?$, it randomizes uniformly between all the edges $vw \in F$. Contrary to the cycle-friendly case, the outcome of such a randomization has no influence on what will happen if $v$ is seen again.
    If some player $i$ deviates and takes an edge that they are not supposed to take (an edge that does not belong to $E_k$, then), then all the players switch to the positional strategy profile $\btau^{\dag i}$, where $\btau^{\dag i}_{-i}$ minimizes the best risk measure that player $i$ can get (a positional such strategy profile exists by \cref{lm:secretlemma}), and $\tau^{\dag i}_i$ is some positional strategy.
 
    Our algorithm in the cycle-averse case is presented in~\cref{algo:cycleaverse}.
    Again, each step $k$ identifies a new set of vertices that must be avoided.
    Their definition depends now on the parity of $k$.
    When $k$ is even, it is the same as in the cycle-friendly case: the set $V^k_\bad$ is the set of vertices $v$ such that $\val(v) > z_i^k$, where $i$ is the player controlling $v$, and $A_k$ is the positive probabilistic attractor of $V_\bad^k$.
    When $k$ is odd, we define directly $A_k$ as the set of vertices from which whatever the players play, there is a positive probability of reaching no terminal vertex.
    
    Again, the computation of $V_\bad^k$ for an even step $k \geq 2$ requires the computation of $z_i^k$, which can be done in time $O(m)$ by computing the set of terminals that are accessible from $v_0$ in $(V, E_k)$, and by deciding whether the probability of reaching no terminal is positive: that will be the case, now, if and only if there exists a vertex from which no terminal vertex is accessible, which can also be decided in time $O(m)$.
    As for odd steps, the computation of $A_k$ can also be done in $O(m)$ using \cref{lm:secretlemma}.
    
    If $k \geq 1$ and $v_0 \in A_k$, i.e., if it is not possible to avoid reaching the set $V_\bad^k$, the answer $\No$ is returned.
    Otherwise, the set $E_{k+1}$ is defined from $E_k$ by removing all the edges that lead from a vertex that does not belong to $A_k$ to a vertex that does, thus making sure that $V_\bad^k$ will never be reached.

    The loop stops when there is no more edge to remove, i.e., when we get $E_{k+2} = E_k$.
    Then, the algorithm answers $\No$ if we have $z_i^k < x_i$ for some $i$.
    Otherwise, it  performs \emph{final refinements}, defined as follows: first, it defines $F_0 = E_k$.
    Then, once $F_\l$ is defined for some $\l$, it checks whether there exists an edge $uv$ that matches the following conditions in the graph $(V, E_\l)$:
        \begin{enumerate}
            \item\label{itm:cuttableedge} the vertex $u$ is not stochastic and has several outgoing edges;
            \item\label{itm:nolessterminals} all the terminal vertices accessible from $v$ are also accessible from $v_0$ without using $uv$;
            \item\label{itm:nocycle} at least one terminal vertex is accessible from $u$ without using $uv$.
        \end{enumerate}
    In the following, we will refer to those conditions as Conditions~\ref{itm:cuttableedge}, \ref{itm:nolessterminals}, and \ref{itm:nocycle}.
    If there exists such an edge, then we define $F_{\l+1} = F_\l \setminus \{uv\}$.
    If there is no such edge, the algorithm stops there, answers $\Yes$, and returns $F_\l$ as a succinct representation of $\bsigma^{F_\l}$.

            \begin{algorithm}[t]
        \begin{algorithmic}\caption{Constrained existence problem with optimists in the cycle-averse case}\label{algo:cycleaverse}
                \Procedure{CycleAverse}{$\Game, \Bar{x},\Bar{y}$}
                    \State $k \gets 0$
                    \State $E_k\gets E$
                    \State $V_\bad^k = \{t\in T \mid \mu_i(t)> y_i\}$ 
                    \State $A_k \gets \Attr(V^k_\bad, E_k)$
                    \If{$v_0 \in A_k$}
                        \Return{$\No$}
                    \Else
                        \State $E_{k+1} \gets E_k \setminus \{uv\in E_k\mid u\not\in A_k\text{ and }v\in A_k\}$
                    \While{$E_{k+2}\neq E_k$\text{ or }$k \leq 1$}
                        \State $k\gets k+1$
                        \If{$k$ is even}
                            \State Compute $z^k_i = \X_i(\bsigma^{E_k})$ for each $i \in \Pi$
                            \State $V^k_\bad \gets \{v \mid \val(v) > z_i^k \text{ for } i \in \Pi \text{ such that } v \in V_i\}$
                            \State $A_k \gets \Attr(V^k_\bad, E_k)$
                        \Else
                            \State $A^k_\bad \gets \{v \mid \forall \btau \in \Strat_\Pi\Game_{\|v}, \PP_\btau(\Occ \cap T = \emptyset) > 0\}$
                        \EndIf
                        \If{$v_0 \in A_k$}
                            \Return{$\No$}
                        \Else
                            \State $E_{k+1} \gets E_k \setminus \{uv\in E_k\mid u\notin A_k\text{ and }v\in A_k\}$
                        \EndIf
                    \EndWhile
                    \EndIf
                    \If{$z_i^k < x_i$ for some player $i$}
                        \Return{$\No$}
                    \Else
                        \State $\l \gets 0$
                        \State $F_\l \gets E_k$
                        \Comment{Final refinement steps}
                        \While{there exists $uv$ satisfying Conditions~\ref{itm:cuttableedge}, \ref{itm:nolessterminals}, and \ref{itm:nocycle}}
                            \State $\l \gets \l+1$
                            \State $F_{\l+1} \gets F_\l \setminus \{uv\}$
                        \EndWhile
                        \State \Return{$(\Yes, F_\l)$}
                    \EndIf
                \EndProcedure
            \end{algorithmic}
        \end{algorithm}

     \paragraph{Correctness in the cycle-averse case}
The fact that $\bsigma^{E_k}$ and $\bsigma^{F_\l}$ are always correctly defined can be proved with arguments similar as those that were used for the cycle-friendly case.
We now focus on correctness properly said.
We first need the following properties.

\begin{inv}\label{inv:terminalsaccessible}
    For every even $k > 0$, and for every $\l$, the graph $(V, E_k)$, or $(V, F_\l)$, contains no vertex that is accessible from $v_0$ and from which no terminal vertex is accessible.
\end{inv}

\begin{proof}
    In the graph $(V, E_k)$ (for $k>0$ even), no induction is required: the set $E_k$ has been obtained after an odd step, in which the set $A_{k-1}$ has been made inaccessible.
    Thus, if we have a vertex $v$ from which no terminal vertex is accessible, it means that in the graph $(V, E_{k-1})$, all paths from $v$ to a terminal vertex were traversing a vertex of $A_{k-1}$, which implies that $v$ itself belonged to $A_{k-1}$, and is therefore not accessible from $v_0$ in $(V, E_k)$.

    This also proves that the invariant is true during the final refinements at step $\l = 0$.
    If now we assume that it is true at some step $\l$, then Condition~\ref{itm:nocycle} guarantees that it remains true at step $\l+1$.
\end{proof}

\begin{inv} \label{inv:finishingtouchesconstantpayoff}
    If the algorithm switches to final refinements after step $k$, then for each step $\l$ of final refinements and for each player $i$, we have $\X_i(\bsigma^{F_\l}) = z_i^k$.
\end{inv}

\begin{proof}
    The invariant is immediate for $\l=0$, since we have $F_\l = E_k$.
    Then, if it is true at step $\l$, it remains true at step $\l+1$.
    Indeed, Condition~\ref{itm:nolessterminals} guarantees that the set of terminal vertices accessible from $v_0$ in $(V, F_\l)$ is the same as in $(V, F_{\l+1})$.
    In other words, the terminal vertices that are reached with positive probability in $\bsigma^{F_\l}$ and $\bsigma^{F_{\l+1}}$ are the same.
    Moreover, \cref{inv:terminalsaccessible} guarantees that it is almost sure that some terminal vertex will be reached, in $\bsigma^{F_\l}$ as well as in $\bsigma^{F_{\l+1}}$.
    Therefore, the set of payoff vectors that have positive probability of being obtained is the same in both strategy profiles, hence the risk measures are the same.
\end{proof}

We can now prove correctness.
To do so, we need to prove two propositions: the algorithm recognizes only positive instances, and recognizes all of them.

\begin{prop}
    The algorithm recognizes only positive instances.
\end{prop}

\begin{proof}
        Let us assume that the algorithm answers $\Yes$ at step $\l$ of the final refinements, after having switched to the final refinements loop at step $k$: let us show that the strategy profile $\bsigma^{F_\l}$ is an XRSE that satisfies the desired constraints.
        Note that the algorithm does never answer $\Yes$ at step $0$, hence we necessarily have $k \geq 1$.

 \subparagraph*{The strategy profile $\bsigma^{F_\l}$ satisfies $\bx \leq \X({\bsigma^{F_\l} }) \leq \by$.}
        
        The algorithm switches to the final refinements at step $k$ only if the strategy profile $\bsigma^{F_\l}$ satisfies $\X(\bsigma^{E_k}) \geq \bx$.
        Then, by \cref{inv:finishingtouchesconstantpayoff}, we also have $\X(\bsigma^{F_\l}) \geq \bx$.

        As for the upper bound, observe that the set $E_1$ has been defined so that the set $\Attr(V_{\frownie}^0, E)$, and therefore the set $V_{\frownie}^0$, is not accessible from $v_0$ in the graph $(V, E_1)$, and therefore not in the graph $(V, F_\l)$ either.
        Thus, it is almost sure in $\bsigma^{F_\l}$ that no vertex of $V_{\frownie}^0$ will ever be reached.
        In other words, all terminals that have positive probability of being reached give to each player $i$ a payoff smaller than $y_i$.
        That is sufficient to prove the lower bound, because it is almost sure, when following $\bsigma^{F_\l}$, that some terminal vertex will eventually be reached, by \cref{inv:terminalsaccessible}.

        \subparagraph*{The strategy profile $\bsigma^{F_\l}$ is an XRSE.}
        
        Let $i$ be a player, and let $\sigma'_i$ be a deviation of player $i$ from $\bsigma^{F_\l}$.
        Let $z' = \X_i({\bsigma^{F_\l}_{-i}, \sigma'_i})$ be the extreme risk measure obtained by player $i$.
        We want to prove that the deviation $\sigma'_i$ is not profitable, i.e., that we have $z' \leq z_i^k$ (since we have $\X_i(\bsigma^{E_\l}) = z_i^k$ by \cref{inv:finishingtouchesconstantpayoff}).
        To do so, we first show that player $i$ cannot obtain a payoff better than $z_i^k$ after using an edge that does not belong to $F_\l$.

        We first show that if the deviation $\sigma'_i$ uses only edges of $F_\l$, then it cannot be profitable.

\begin{slem}\label{claim:finishingtouchescycleimpossible}
    If it is almost sure, when following $(\bsigma^{F_\l}_{-i}, \sigma'_i)$, that only edges of $F_\l$ will be used, then the deviation $\sigma'_i$ is not profitable.
\end{slem}

\begin{proof}
    First, let us note that as long as player $i$ uses only edges that belong to $F_\l$, the strategy profile $\bsigma^{F_\l}$ behaves in a stationary way, and we can therefore assume without loss of generality that $\sigma'_i$ is positional.

    The payoff $z'$ may be obtained by reaching a terminal vertex: in that case, that terminal vertex is accessible from $v_0$ in $(V, F_\l)$, and therefore also reached with positive probability when following the strategy profile $\bsigma^{F_\l}$, hence $z' \leq z_i^k$.

    Let us show that it cannot be obtained by reaching no terminal.
    We proceed by contradiction: if, in the strategy profile $(\bsigma^{F_\l}, \sigma'_i)$, there is a positive probability of reaching no terminal when following that strategy profile, then there is a vertex that has positive probability of being visited infinitely often.
    We can then define the set $W$ of such vertices, i.e., the set $W = \{v \in V \mid \PP_{\bsigma^{F_\l}, \sigma'_i}(v \in \Inf) > 0\}$.
    Thus, when the strategy profile $(\bsigma^{F_\l}, \sigma'_i)$ is followed from a vertex of $W$, it is almost sure that no terminal vertex is reached, and that the set $W$ will never be left.
    We can then choose $w \in W$ such that it has positive probability of being reached without visiting any other vertex of $W$ before, i.e., such that there exists a history $hw$ from $v_0$ with $\Occ(h) \cap W = \emptyset$ (note that $h$ can be empty).
    
    On the other hand, in the graph $(V, F_\l)$, there is at least one terminal vertex accessible from $w$: all vertices from which no terminal is accessible are made themselves inaccessible at odd steps, the switch to final refinements loop happens only if there is no more edge to remove in that perspective, and Condition~\ref{itm:nocycle} guarantees that the final refinements loop leave at least one terminal vertex accessible from every vertex accessible from $v_0$.

    From each terminal $t$ accessible from $w$, we pick a simple path $h_0^t \dots h_{q_t}^t$ from $h_0^t = w$ to $h_{q_t}^t = t$ in the graph $(V, F_\l)$.
    Those paths define a directed acyclic graph (DAG) $D = (V_D, E_D)$ rooted at $w$, where all non-terminal vertices have at least one outgoing edge, with $V_D \subseteq V$ and $E_D \subseteq F_\l$.
    Now, since the strategy $\sigma'_i$ guarantees that no terminal vertex will be reached, each branch $h^t$ of that DAG is such that there exists a (smallest) index $j$ with $h^t_j \in W \cap V_i$, and $\sigma'_i(h^t_j) \neq h^t_{j+1}$.
    It may be the case that $h^t_j \sigma'_i(h^t_j) \in E_D$, i.e., that from $h^t_j$, player $i$ proceeds to an undetectable deviation and takes another branch of the DAG.
    But that cannot be the case for all $t$: otherwise, there would be a branch $h^t$ that would be followed with positive probability when following $(\bsigma^{F_\l}_{-i}, \sigma'_i)$ from $w$, and therefore a terminal vertex $t$ that would be reached with positive probability, which contradicts the definition of $w$.

    There must therefore exist an edge $uv \in F_\l \setminus E_D$, with $u \in V_D \cap V_i \cap W$.
    We will show that such an edge should have been removed during the final refinements loop.
    First, it immediately satisfies Condition~\ref{itm:cuttableedge}.
    Moreover, the vertex $u$ is necessarily on a branch $h^t$ of $D$ that leads to a terminal vertex $t$, hence it satisfies Condition~\ref{itm:nocycle}.
    Finally, since $v$ is accessible from $w$ in $(V, F_\l)$, the terminal vertices that are accessible from $v$ in $(V, F_\l)$ are all accessible from $w$ in that same graph, and therefore are accessible from $w$ in the DAG $D$.
    Since $w$ is accessible from $v_0$ without visiting any vertex of $W$, and it particular without visiting $u$, it means that the terminal vertices accessible from $v$ are also accessible from $v_0$ without using the edge $uv$.
    In other words, the edge $uv$ satisfies Condition~\ref{itm:nolessterminals}, and should have been removed during the final refinements.

    This case is therefore impossible: when the players follow the strategy profile $(\bsigma^{F_\l}, \sigma'_i)$, it is almost sure that some terminal vertex will be reached, and that concludes the proof.
\end{proof}
        
But now, if the strategy $\sigma'_i$ does use edges that do not belong to $F_\l$, let us show that it cannot be a profitable deviation either.

\begin{slem}
    If there is a history $hv$ compatible with $\bsigma^{F_\l}$ such that we have $v\sigma'_i(hv) \not\in F_\l$, then we have $\X_i(\bsigma^{F_\l}_{\|hv}, \sigma'_{i\|hv}) \leq z_i^k$.
\end{slem}

\begin{proof}
    After such a history, the strategy profile $\bsigma^{F_\l}_{-i}$ follows the positional strategy profile $\btau^{\dag i}_{-i}$.
    By definition of that strategy profile, we have $\X_i(\bsigma^{F_\l}_{\|hv}, \sigma'_{i\|hv}) \leq \val(v)$.
    On the other hand, the vertex $v$ is accessible from $v_0$ in $(V, F_\l)$, since it is visited with positive probability in $\bsigma^{F_\l}$.
    Therefore, it does not belong to the set $A_k$ (if $k$ is even) or $A_{k-1}$ (if $k$ is odd), and in particular not to the set $V_\bad^k$ or $V_\bad^{k-1}$, which means that we have $\val(v) \leq z_i^k$.
    Hence the conclusion.
\end{proof}

In the general case, the payoffs that player $i$ obtains with positive probability in the strategy profile $(\bsigma^{F_\l}_{-i}, \sigma'_i)$ are either obtained using only edges that belong to $F_\l$, or by using an edge that does not: in both cases, we have shown that player $i$ cannot get a payoff greater than $z_i^k$, which proves that the strategy profile $\bsigma^{F_\l}$ is an XRSE.
\cref{algo:cyclefriendly} answers $\Yes$ only on positive instances, and outputs in that case a succinct representation of an XRSE matching the constraints.
\end{proof}

We will now prove the converse.

            \begin{prop}
                \cref{algo:cycleaverse} recognizes all positive instances. 
            \end{prop}
    
            \begin{proof}
           Let us assume that we have a positive instance, i.e., that there exists an XRSE $\bsigma$ with $\bx \leq \X(\bsigma) \leq \by$.
        Let us show that the algorithm will answer $\Yes$.
        To do so, we first prove the following invariant: if an edge is removed at some step \emph{before the final refinements}, then it is never taken by the XRSE $\bsigma$.

        \begin{inv} \label{inv:edgesnotused_bis}
            For each $k \geq 0$, every edge that has positive probability of being eventually taken in $\bsigma$ belongs to $E_k$.
        \end{inv}

        \begin{proof}
        We prove the invariant by induction. 
        
        \subparagraph*{Base case.} The case $k=0$ is immediate, since we have $E_0 = E$.
        
        Further, at step $k=1$, if the strategy profile $\bsigma$ uses eventually, with positive probability, an edge that does not belong to $E_1$, then it goes with positive probability to a vertex $v \in \Attr(V^0_\bad, E_0)$.
        Then, with positive probability, a terminal vertex will be reached that gives to some player $i$ a payoff greater than $y_i$, which is impossible.
        Therefore, such an edge cannot be taken in $\bsigma$.

        \subparagraph*{Induction step.} Let us assume that the invariant is true until step $k \geq 1$, and let us show that it holds at step $k+1$.    
        Let $uv$ be an edge that is used with positive probability when following $\bsigma$, and let us assume toward contradiction that it does not belong to $E_{k+1}$.
        Since the invariant is true at each step until $k$, we can assume that $uv$ has been removed at step $k$, i.e., that we have $uv \in E_k \setminus E_{k+1}$.
        Then, we have $u \not\in A_k$ and $v \in A_k$.
        The strategy profile $\bsigma$ has therefore positive probability of visiting the set $A_k$.
        
        We must now distinguish the cases where $k$ is even or odd.
        If $k$ is even, then the strategy profile $\bsigma$ has therefore positive probability of visiting a vertex $v \in V_\bad^k$.
        If player $i$ is the player controlling $v$, then that player has a deviation in which they get risk measure at least $\val(v) > z_i^k$.
        Let us now note that when the players follow the strategy profile $\bsigma$, it is almost sure that a terminal vertex will eventually be reached, and that all the terminal vertices that have positive probability of being reached also have positive probability of being reached in $\bsigma^{E_k}$, since the invariant is true at step $k$: therefore, we have $z_i^k \geq \X_i(\bsigma)$, and player $i$ has a profitable deviation after $v$, which contradicts the fact that $\bsigma$ is an XRSE.

        If $k$ is odd, then, by definition of $A_k$, when the strategy profile $\bsigma$ is followed, there is a positive probability of reaching no terminal.
        But that is impossible in the cycle-averse case.

        The invariant is therefore necessarily still true at step $k+1$.
\end{proof}

        We are now able to conclude.
        The answer $\No$ can be given in the two following cases:
            \begin{itemize}
                \item \emph{If at step $k$, we have $v_0 \in \Attr(V^k_\bad, E_k)$.}
                Then, the strategy profile $\bsigma$ visits the set $V^k_\bad$ with positive probability.
                With the same arguments that were used in the proof of \cref{inv:edgesnotused_bis}, that is not possible.
                
                \item \emph{If during steps $k-1$ and $k$, no edge is removed, but we have $z^k_i < x_i$ for some player $i$.}
                Since $\bsigma$ uses only edges of $E_k$ by \cref{inv:edgesnotused_bis} and reaches almost surely a terminal (since we are in the cycle-averse case), we have $\X_i(\bsigma) \leq z^k_i$, and therefore $\X_i(\bsigma) < x_i$: that case is therefore also impossible by definition of $\bsigma$.
            \end{itemize}
        None of those cases is possible, hence our algorithm will eventually answer $\Yes$.
    \end{proof}

\paragraph{Complexities}
We consider here the complexity of the two algorithms.
Since at least one edge is removed every two steps, there are $O(m)$ steps.
In each of them, we need $O(p)$ calls to simple algorithms: computation of $z_i^k$, of $V_\bad^k$, of $A_k$.
Hence the complexity $O(pm^2)$.

In the cycle-averse case, when an output is asked, we need to add the final refinements loop, which consist of $O(m)$ additional steps in which we check, for each of the $O(m)$ remaining edges, whether they satisfy Conditions~\ref{itm:cuttableedge},~\ref{itm:nolessterminals}, and~\ref{itm:nocycle}: that can be done in time $O(m)$.
Hence the complexity $O(pm^2 + m^3)$.
\end{proof}

Finally, we show that the problem is $\PTime$-hard, even when there are only two players.

\begin{lem}\label{lm:ptimelowerbound}
    The constrained existence problem of XRSEs with optimistic players is $\PTime$-hard even with only two players.
\end{lem}

\begin{proof}
        Given a deterministic two-player (between players $\Circle$ and $\Square$) zero-sum reachability game $\Game_{\|v_0}$ with target set of vertices $T$, we construct a simple stochastic game (with no stochastic vertices) where there is an XRSE $\bsigma$ satisfying $\X_\circ(\bsigma) = 1$ and $\X_\Box(\bsigma) = -1$ if and only if player $\Circle$ wins the game.  

        The game is simply obtained by assigning rewards on the zero-sum two player game as follows: we make all nodes in the target set $T$ of the reachability game as a terminal node where player $\Circle$ gets reward $1$ and player $\Square$ the reward $-1$. Recall that if no terminal is reached, both players get reward $0$.  

        If $\Circle$ has a strategy to win the reachability game, then the same strategy for $\Circle$, along with any strategy for $\Square$, will be an XRSE in that new game, and then satisfies the constraint.
        Similarly, if on the other hand, player $\Square$ has a strategy to avoid the states $T$, then no strategy of $\Circle$ that gives her payoff $+1$ and gives player $\Square$ the payoff $-1$ will be an equilibrium, since $\Square$ can always deviate to the winning strategy in the reachability game that offers him the better payoff of $0$.
\end{proof}

%% file: 4Discussion.tex
Throughout this document, we have introduced new techniques and complexity results concerning Nash equilibria and SPEs in parity, mean-payoff, discounted-sum, and energy games.
We have also shown how these results extend to rational verification.
We have studied strong secure equilibria in parity and $\omega$-regular games, demonstrating that this concept effectively meets key criteria for secure protocols.
Regarding stochastic games and the use of randomized strategies, we considered the entropic risk measure, proving that the constrained existence problem for equilibria defined with this measure is undecidable.
However, we established decidability results under certain restrictions on players' strategies.
We then introduced the extreme risk measure and demonstrated that, when equilibria are defined using this measure, the constrained existence problem becomes decidable.

We believe these results provide novel insights into the study of multi-player games and their applications in computer science.

Several questions remain open, such as the complexity of the SPE constrained existence problem in Büchi games (\cref{op:spe_buchi}), its fixed-parameter tractability in mean-payoff games (\cref{op:spe_mp_fpt}), its recursive enumerability in energy games (\cref{op:spe_energy_re}), the rational synthesis problem in mean-payoff games (\cref{op:rat_syn_mp}), the existence of randomized $\epsilon$-SPEs in mean-payoff games (\cref{conj:epsilon_spes}), or the existence of entropic or extreme risk-sensitive equilibria in simple stochastic games with negative rewards (\cref{conj:existence_erse,conj:existence_xrse}).
More broadly, interesting research directions emerge from studying alternative or more sophisticated classes of games.
For instance, multiplayer games where players pursue a mix of parity and mean-payoff objectives, as in \cite{DBLP:conf/lics/ChatterjeeHJ05}, or games where some players have parity objectives while others have mean-payoff ones.
Additionally, the unconventional equilibrium notions we introduce, such as extreme risk-sensitive equilibria, could be explored within the game classes where we have studied NEs and SPEs: parity games, mean-payoff games, discounted-sum games, and energy games.
Such equilibria could also be examined in concurrent or imperfect information games, or combined with other equilibrium properties, leading to concepts such as \emph{extreme risk-sensitive subgame-perfect equilibria} or \emph{strong secure extreme risk-sensitive equilibria}.

The application of these results must, however, be approached with caution.
Like all mathematical results, the proofs presented in this work offer the significant advantage of being non-ambiguous and fully verifiable, but they deal with abstract objects.
Modeling real-life situations through such abstractions is inherently a simplification, carrying the risk of omitting crucial elements.

Applications of game theory, particularly in economics, have faced criticism~\cite{Rubinstein2013,Guerrien2000} for their ambition to describe---and even predict---human behavior, despite empirical observations often contradicting theoretical predictions.
Indeed, when multiple individuals must make decisions based on their beliefs about others' choices, neither empirical studies nor theoretical proofs guarantee that they will select strategies forming a Nash equilibrium or any other equilibrium notion.
In such situations, the conceptual tools of game theory primarily serve to clarify the dilemmas that agents face rather than to predict their behavior---though it is worth noting that designing game-theoretic notions that better align with human behavior is an active area of research (see, for instance, Herbert A. Simon's work on the \emph{satisficing} concept~\cite{simon1956rational}).

In contrast, computer-science-related applications may represent some of the most relevant uses of game theory, as they often have a more \emph{normative} than \emph{descriptive} objective, in Bernard Guerrien's terminology~\cite{Guerrien2000}.
In scenarios involving multiple agents, each with a set of possible strategies and well-defined objectives, the existence of a strong secure equilibrium provides a concrete guarantee: if the agents are given such an equilibrium as a protocol to follow, no coalition can disrupt the protocol to harm another agent without also harming one of its own members.
Such a protocol could, therefore, be reasonably proposed as a \emph{safe} protocol.

However, the (legitimate) desire to explore all the potential applications of game theory in our field may also lead to interpretations that attribute to it more predictive power than it actually possesses.
For instance, the underlying approaches of rational verification or rational synthesis may be more debatable.
Is it reasonable to consider a system safe if it guarantees a safety property only against responses that form a Nash equilibrium or an SPE?
Such claims should be carefully examined within specific application contexts, supported by both solid theoretical foundations and empirical validation.
However, we believe that developing a deeper understanding of these concepts through algorithmic results can only contribute positively in that regard.